  \documentclass[3p]{elsarticle}   

\usepackage[T1]{fontenc}
\usepackage{url}

\usepackage[pagewise,displaymath]{lineno}

\usepackage{multirow}            
\usepackage{hyperref}
 \usepackage[utf8]{inputenc}
 \usepackage{slashed}
  \usepackage{amsmath}
  \usepackage{amssymb}
\usepackage[dvipsnames]{xcolor}
\usepackage{xcolor}
\usepackage{graphicx}
\usepackage{hyperref}
  \usepackage{amsfonts}
\usepackage{multicol}
 \usepackage{tikz}
\usepackage{cancel}
 \usepackage{mathtools}

\biboptions{sort&compress}

\definecolor{nicered}{rgb}{0.7,0.1,0.1}
\definecolor{nicegreen}{rgb}{0.1,0.5,0.1}
\definecolor{red}{rgb}{1.0, 0, 0}
\definecolor{darkblue}{rgb}{.0,.0,.8}
\definecolor{niceblue}{rgb}{0,0,1}
\definecolor{niceviolet}{rgb}{0.5,0,1.0}
\definecolor{blue}{rgb}{0,0,1}
\hypersetup{colorlinks=true,citecolor=nicegreen,linkcolor=nicered,urlcolor=nicered}

\usepackage[normalem]{ulem}    
\modulolinenumbers[1]                

\journal{Physics Reports}

\newcommand{\beq}{\begin{equation}}
\newcommand{\eeq}{\end{equation}}
\newcommand{\bea}{\begin{eqnarray}}
\newcommand{\eea}{\end{eqnarray}}
\newcommand{\fa}{f_a}

\newcommand{\va}{v_a}
\def\Tr{\,\mbox{Tr}\,}

\def\diag{\mbox{diag}\,}
\def\vev#1{\left\langle #1\right\rangle}
\def\abs#1{\left| #1\right|}
\def\Im{\mbox{Im}\,}
\def\Re{\mbox{Re}\,}
\def\fig#1{{Fig.~\ref{#1}}}

\def\Table#1{{Table~\ref{#1}}}

\def\sect#1{{Section~\ref{#1}}}
\def\sects#1#2{{Sections~\ref{#1}--\ref{#2}}}

\def\o{{\rm osc}}
\def\e{{\rm eq}}
\def\r{{\rm rad}}

\newcommand{\mP}{m_{\rm Pl}}

\newcommand{\Q}{\mathcal{Q}}
\newcommand{\mX}{\mathcal{X}}
\newcommand{\mN}{\mathcal{N}}
\newcommand{\bX}{\textbf{\textit{X}}}

\newcommand{\eqn}[1]{Eq.~(\ref{#1})}
\newcommand{\eqns}[2]{Eqs.~(\ref{#1})--(\ref{#2})}
\newcommand{\Eqn}[1]{Eq.~(\ref{#1})}
\newcommand{\Eqns}[2]{Eqs.~(\ref{#1})--(\ref{#2})}

\newcommand{\gag}{g_{a\gamma}}
\newcommand{\gae}{g_{ae}}
\newcommand{\gap}{g_{ap}}
\newcommand{\gan}{g_{an}}

\newcommand{\gaN}{g_{aN}}

\newcommand{\PkO}{\Delta^2_{\mathcal{R}}(k_0)}
\newcommand{\Pk}{\Delta^2_{\mathcal{R}}(k)}
\newcommand{\NDW}{N_\mathrm{DW}}

\renewcommand{\[}{\left[}
\renewcommand{\]}{\right]}
\renewcommand{\(}{\left(}
\renewcommand{\)}{\right)}

\def \lsim{\mathrel{\vcenter
     {\hbox{$<$}\nointerlineskip\hbox{$\sim$}}}}
\def \gsim{\mathrel{\vcenter
     {\hbox{$>$}\nointerlineskip\hbox{$\sim$}}}}

\begin{document}

\begin{frontmatter}

\title{The landscape of QCD axion models}

\begin{flushright}
DESY 20-036
\end{flushright}



\author{Luca Di Luzio\corref{corr1}}
\ead{luca.diluzio@desy.de}
\address{Deutsches Elektronen-Synchrotron DESY, Notkestra\ss e 85, D-22607 Hamburg, Germany}
\author{Maurizio Giannotti\corref{corr2}}
\ead{MGiannotti@barry.edu}
\address{Physical Sciences, Barry University, 11300 NE 2nd Ave., Miami Shores, FL 33161, USA}
\author{Enrico Nardi\corref{corrauth}}
\cortext[corrauth]{Corresponding author}
\ead{enrico.nardi@lnf.infn.it}
\address{INFN, Laboratori Nazionali di Frascati, C.P. 13, I-00044 Frascati, Italy}
\author{Luca Visinelli\corref{corr3}\,}
\ead{l.visinelli@uva.nl} 
\address{Gravitation Astroparticle Physics Amsterdam (GRAPPA),\\
		Institute for Theoretical Physics Amsterdam and Delta Institute for Theoretical Physics,\\ 
		University of Amsterdam, Science Park 904,
		1098 XH Amsterdam,
		The Netherlands 
		%
}

\begin{abstract}
We review the landscape of QCD axion models.  Theoretical constructions  that 
extend the window for the axion mass and couplings  beyond  conventional 
regions are highlighted and classified.  Bounds from cosmology,  astrophysics and 
experimental searches are reexamined and updated. 
\end{abstract}



\end{frontmatter}

\newpage
\tableofcontents
\newpage


\section{Introduction}
\label{sec:intro}

At the dawn of the third decade of the third millennium, 
particle physics seems stranded in an awkward intrigue.
The celebrated theoretical construction known as the 
Standard Model (SM) has been established as the correct 
description of fundamental phenomena down to scales 
of the order of  $10^{-16}\,$cm. However, a certain number of observations including  
dark matter (DM), neutrino masses $m_\nu $, 
and the cosmological matter-antimatter asymmetry $\eta_b$, 
remain unaccounted   within the SM,  and constitute  indisputable evidences that 
the present theory must be extended. 
Besides this, the SM discomfits particle physicists because of a certain number 
of theoretical distresses generically  related with exceedingly small numbers,
that are usually referred to as problems of naturalness, as for example the value of the 
cosmological constant (dark energy)  in units of the  
Planck mass $\Lambda\sim 10^{-31} \mP$, 
 the electroweak breaking vacuum expectation value (VEV)  $v\sim 10^{-17}\mP$, 
 the  CP violating QCD angle $\theta \lsim 10^{-10}$.\footnote{These  small number problems 
do not stand  on the same footing. 
For example a tiny value of $\theta$ is {\it technically natural}, 
in the sense that it does not  get lifted by quantum corrections. On the other hand, 
while anthropic or environmental selection arguments can provide  explanations for 
the values of  $\Lambda$~\cite{Weinberg:1987dv} and $v$~\cite{Agrawal:1997gf,Agrawal:1998xa},  
a value  of $\theta_{\rm QCD}$  many orders of magnitude larger than the 
experimental limit would still leave our Universe basically unaffected~\cite{Ubaldi:2008nf,Dine:2018glh}.}

Theoretical constructions that extend the SM are clearly more appealing 
 when  they are able to solve more than one of the previous issues with the same amount of theoretical input. 
 A well known case is supersymmetry equipped with an $R$-parity symmetry to forbid fast 
 proton decay,  which protects the value of  $v/ \mP$ from quantum corrections,  and at the same time predicts that a new 
stable particle, which shares all the properties of a good DM candidate, must exist.
Another example is the type-I seesaw model for neutrino masses which, besides accounting for 
the suppression of the neutrino 
mass scale~\cite{Minkowski:1977sc,Yanagida:1979as,GellMann:1980vs,Mohapatra:1980yp},  can also 
yield   quite naturally  a cosmological baryon asymmetry of the correct  size~\cite{Fukugita:1986hr}.
The serious drawbacks  of these two theories are that supersymmetry has not been found at the LHC, 
while the experimental  verification of type-I seesaw leptogenesis remains well outside the reach of all 
current experiments~\cite{Davidson:2002qv,Buchmuller:2004nz,Davidson:2008bu,Fong:2013wr,Chun:2017spz}.

A third example of a `two birds with one stone' theory is provided  by the 
axion~\cite{Weinberg:1977ma,Wilczek:1977pj}. 
In an effective field theory (EFT) description, the SM is extended by introducing a 
single new massless pseudo-scalar particle $a$, the axion, for which only  one  coupling 
is mandatory,  namely an effective coupling  to the CP violating topological gluon density  
$(a/f_a+\theta) G\tilde G$,  where $f_a$ is the scale suppressing the effective operator,  
$G=G_{\mu\nu}$ is the gluon field strength tensor,  $\tilde G_{\mu\nu}$ its dual, and we have added  
to the axion-gluon operator the infamous CP violating $\theta$ term. 
Such a simple extension has astonishingly far reaching consequences: 
the strong CP problem is solved because the minimum of the 
vacuum energy occurs when the coefficient of $G\tilde G$ vanishes~\cite{Vafa:1984xg}. 
Thus, by acquiring a suitable VEV, the axion disposes of the perilous CP violating  operator. 
In performing this task,  the axion acquires a tiny mass, and  in this process a  
cosmological population of zero momentum excitations, which nowadays still sums 
 to the energy density of the Universe, is unavoidably produced. Hence  axions  definitely  
 contribute to  the DM. Whether they can wholly account for it remains, for the time being,
an open question.  
Solving the strong CP problem  and providing a natural DM candidate  by no means 
exhausts the role of axions in fundamental physics. Axion phenomenology  crosses 
boundaries between particle physics, astrophysics and cosmology,  it is replete 
with  interdisciplinary connections which have already provided fruitful insights into different 
domains of physics. Axions have unusually intriguing  features, although deeply interwoven 
with QCD, they  interact more feebly than all SM particles, and although their typical mass is much 
smaller than the mass of at least two types of neutrinos, they might dominate the matter content of our Universe.
Moreover, differently from the case of supersymmetry, current experimental searches have so far only been 
able to cut out relatively small regions of the parameter space in which the QCD axion can naturally live, 
but differently from leptogenesis, the axion hypothesis is within the reach of experimental verification, and 
 it is  conceivable, for example, that the canonical axion DM mass window could be thoroughly explored within the 
next one or two decades. If ever discovered, there is little doubt that the existence of axions would 
reshape more than one branch of fundamental physics. 

The axion has been so far introduced by postulating a non-renormalizable axion-gluon operator.
In quantum field theory (QFT) there is a simple prescription  for constructing a renormalizable 
completion whose low-energy limit matches the required form of the effective action density. 
What is needed is a Lagrangian equipped with a global $U(1)_{\rm PQ}$  
symmetry, exact at the classical level but broken   at the quantum level by a colour anomaly, 
that undergoes spontaneous breaking at some high energy scale. 
Such a symmetry is known as Peccei-Quinn (PQ) 
symmetry after its  proposers~\cite{Peccei:1977hh,Peccei:1977ur}.\footnote{Historically, axion 
theory developed in the reverse order: the PQ symmetry was invented first, and only subsequently 
it was  realised that the PQ mechanism implied the existence 
of a very light pseudo-scalar boson~\cite{Weinberg:1977ma,Wilczek:1977pj}.}
 As we will see in the next sections, the pseudo-Nambu Goldstone  Boson (pNGB)  resulting 
from such a broken symmetry exhibits precisely the properties required for an  axion.

The result of dressing the QFT  prescription with a complete model is, however,  far from being 
unique.  For example, in many cases different realisations of the PQ symmetry give rise to axions 
that do not interact only with the gluons,  but that couple  also to  other SM particles.
This indeed enriches in many aspects the subject of axion phenomenology.
In particular,  
it provides additional important channels for  experimental axion searches.
However, the continuous proliferation of new theoretical constructions, that
has received  a major boost especially in recent years,  
has brought to a state of affairs in which it is rather arduous for experimental 
and theoretical researchers to  attain a  sufficiently complete and reliable acquaintance  with the 
vast  literature on axion models. Hence, we believe that an updated account of 
new (and old) theoretical ideas in axion model building,  which we feel is lacking in the literature,  could 
be timely and useful.

The aim of this work is to review the landscape of QFT  realisations of 
the PQ symmetry.\footnote{We will only target  genuine QCD axion models, that is, 
models that  generate an effective axion-gluon operator and that  solve the strong CP problem.  
Other types of very light pseudo-scalar particles that share some of the properties of the QCD axion, 
but do not solve the strong CP problem, and that are commonly denoted as axion-like 
particles (ALPs), are not touched on in this Review.} 
A reasoned classification of the babel  of axion models is accomplished   
by pinpointing theoretical constructions  that predict unusual  properties of the axion, 
especially in relation to the different experimental approaches that could be pursued for 
their  detection,  as for example enhanced or suppressed couplings to specific SM states, 
or unconventional  mass regions where the axion could saturate the DM density.  
Although presently axion searches rely almost exclusively on  axion couplings to photons, 
a number of  novel detection concepts which exploit cutting-edge techniques has been recently put forth 
with the aim of searching for axions through their couplings to nucleons or electrons. Even if in most cases 
only pathfinder or demonstrative small-scale setups have been commissioned,  which generally  have  
projected sensitivities that hardly  reach  into the  parameter space regions hinted by 
most popular axion models, other less known theoretical constructions could be probed, 
constrained or ruled out by these experiments, which  can then effectively contribute 
to  circumscribe the realm of viable axion models.
We thus expect that the experimental community of axion hunters could benefit from the 
classification scheme  that we have adopted. 

This Review is self contained, it  includes pedagogical, but at the same time sufficiently  detailed introductions 
to axion theory, cosmology, astrophysics and experimental searches, that are  intended to provide guidance to the 
neophyte, whether she is a young student planning  to orient her researches towards  axion physics, 
or an experienced colleague active in a different domain of physics, but willing to get insights in a 
field that is currently experiencing  a blooming phase. Experts in the field can instead browse 
quickly through the most pedagogical parts, and focus directly on the sections of their interest.

We start in \sect{sec:strongCPprob} with a  description of the origin of the strong 
CP problem and of the PQ mechanism that solves it. 
We then review model-independent properties of the axion (mass and couplings) 
illustrating how they can be derived  from a chiral Lagrangian formulation. 
Model-dependent features are addressed here only for two 
popular benchmark constructions,  universally known as KSVZ and DFSZ axion models.  
An introduction to flavour violating  axion couplings and a brief discussion of CP violating  
couplings are also included. 
We conclude this section with some remarks about possible  sources of explicit  breaking
of the PQ symmetry  and the way they could endanger the effectiveness of the PQ mechanism.
Excellent reviews exist in the literature that address in more depth some of these topics.
Coleman's Erice lectures~\cite{Coleman:1985rnk} contain an enlightening treatment 
of the QCD vacuum and of the strong CP problem.  Early reviews on axion theory can be 
found in Refs.~\cite{Kim:1986ax,Cheng:1987gp,Peccei:1988ci}. More recent accounts  
are given in Ref.~\cite{Peccei:2006as} and in the review of Kim and Carosi~\cite{Kim:2008hd}.

\sect{sec:section2}  addresses  axion cosmology.  
After a basic introduction to the physics of the early Universe,  
we recall the properties of the axion potential and of the axion mass 
at temperatures around the QCD phase transition. Then we describe the 
misalignment mechanism  as a source of a relic density of axions. 
The issue of cosmic topological defects which arise during axion-related 
phase transitions is also briefly addressed, as well as the constraints from 
axion isocurvature fluctuations that apply when the PQ symmetry is broken 
before inflation.  Next we discuss the canonical mass window within  which axions can 
saturate the DM relic density.  We conclude this section with a brief overview 
of axion miniclusters and axion stars. 
An excellent review on the theory and cosmology of axions can be found 
in Ref.~\cite{Sikivie:2006ni}.  Another review of  the cosmological role of axions, 
which also addresses  the cosmology of  axions superpartners in supersymmetric models, is 
Ref.~\cite{Kawasaki:2013ae}. 
A more recent and rather complete account of axion cosmology  
is given  in Ref.~\cite{Marsh:2015xka}.

\sect{sec:Astro_bounds}  is devoted to a thorough description of 
the role of axions in astrophysics. The layout of the discussion 
analyzes  axion couplings to SM states one at the time, and 
describes in which particular stellar environment and for which reasons 
each coupling becomes particularly relevant. This section also contains 
an updated summary of  astrophysical bounds  on the different types 
of couplings. 
The astrophysics of axions has been the subject of  thorough investigations since the 
time the axion was invented. 
An early compilation of  astrophysical bounds from stars can be found in Ref.~\cite{Cheng:1987ff}. 
The early reviews of Turner~\cite{Turner:1989vc} and Raffelt~\cite{Raffelt:1990yz}  are still actual 
as concerns many  qualitative aspects of axion astrophysics, and remain important references.  
Two accounts of astrophysical axion bounds dating around 
year 2006 can be found  in Refs.~\cite{Raffelt:2006cw} and~\cite{Asztalos:2006kz}. A recent 
review which also includes an assessment of  the astrophysical hints for the existence 
of axions which can be inferred from some anomalies observed in specific phases of 
stellar evolution  can be found in Ref.~\cite{Giannotti:2017hny}.

\sect{sec:Experiments}  contains an  account of the status of  axion experimental searches 
(helioscopes, haloscopes, light shining through  wall) and a summary of existing experimental 
constraints and projected limits that, at the time of writing, is up to date.  However,  in view of the 
continuous and rapid evolution of the experimental landscape, this part  will likely  
 become outdated in the not too distant future.
Early accounts of experimental searches for invisible axions have been presented   
in Refs.~\cite{Rosenberg:2000wb,Battesti:2007um} and later in Ref.~\cite{Graham:2015ouw}.  
A  recent review, which is also remarkably complete, can be found in  Ref.~\cite{Irastorza:2018dyq}.
Simultaneously with the present work,   Ref.~\cite {Sikivie:2020zpn}  appeared which  
contains a  review of  proposed methods to search  for the axion.

 \sect{sec:axion_landscape_beyond_benchmarks}  is the central part of this Review. 
We begin with a systematic  classification of  models that  predict sizeable 
enhancements in the axion coupling to photons, electrons and nucleons.  We describe the 
mechanisms at the basis of these enhancements, and we confront the resulting 
enlarged parameter space with current bounds.  
We then focus on models that predict flavour violating axion couplings 
to quarks and leptons, and we review the role played by  existing limits 
on Flavour Changing Neutral Currents (FCNC) in constraining  constructions of this type.   
Mechanisms  that allow to  extend the mass region in which axions 
can account for the whole of DM  deserve particular attention, in view of the fact that 
the best experimental sensitivities to the axion-photon coupling are attained by  
haloscope experiments, which however can only probe rather narrow and pre-defined 
axion-DM mass intervals.
We review models that implement the possibility of saturating the DM energy density 
for values of the axion mass both larger and smaller than the conventional values, and 
we explain through which mechanisms this  result can be obtained. 
For completeness,  we include at the end of this section a review of models
in which the axions are  `super-heavy',   namely  with masses in excess of 100 keV.   

In \sect{sec:beyond QCD}, we extend the discussion 
to a  different set of axion-related topics. We first review constructions that attempt  
to connect axion physics to other unsolved SM issues, like neutrino masses, 
the cosmological baryon asymmetry,  inflation, and the possibility of detecting gravitational waves 
originating from the PQ phase transition. 
Next we  discuss  available  solutions to a couple of well known problems that 
can generally affect model realisations of the PQ symmetry,  namely how
to maintain under control dangerous sources of explicit PQ breaking, 
and how to ensure that axion-related domain walls will not represent 
cosmological threats.   The possibility that axions are composite states arising from a 
new strong dynamics is an old idea that has been recently revived, 
hence we present a survey of the related literature. 
We include a brief account of attempts to embed axions in 
Grand Unified Theories (GUTs), 
and we quickly touch on a last topic, that by itself would deserve 
a dedicated review, that is, axion arising from string theory. 

In the Appendix 
the reader can find a table with the  symbols and notations that have 
been used in the mathematical expressions including an  explanation of their meaning, a table  
containing  the definition of the acronyms used in the text, and a table 
containing a list of   current and planned axion experiments,  with  the relevant 
reference where the experimental setup  is described.

\section{From the strong CP problem to the QCD axion}
\label{sec:strongCPprob}

This Section is devoted to the physical foundations of the QCD axion 
as a solution of the strong CP problem. We start by reviewing the  non-trivial vacuum structure 
of Yang-Mills theories in \sect{sec:thetavacinYM} and the $\theta$ dependence of the QCD vacuum 
energy in \sect{sec:thetadepphys}.
Next, we discuss the $\theta$ contribution to the 
neutron electric dipole moment (EDM) in \sect{sec:nEDM} and 
give a critical assessment of the strong CP problem and its possible solutions 
in \sect{sec:strongCP}, among which the axion solution via the PQ 
mechanism. The rest of the Section is devoted to the study of standard 
axion properties, starting from the axion effective Lagrangian in \sect{sec:AxionEFT}, 
including a general description of model-dependent axion couplings in \sect{sec:modeldepax}, 
and continuing with a pedagogical derivation of the so-called 
benchmark axion models in \sect{sec:UVcomp}. 
In \sect{sec:summaryaxioncoupl} we provide a concise summary of standard axion properties, 
while \sects{sec:IntroFlavourViolating}{sec:CPvaxioncoupl} are devoted to 
a basic introduction to flavour and CP violating axions. 
We conclude in \sect{sec:PQquality} 
with the so-called PQ quality problem.  

\subsection{QCD vacuum structure}
\label{sec:thetavacinYM}
Until the mid of the 70's,  when the formulation of Quantum Chromodynamics (QCD) was 
being developed,  the so-called $U(1)$ problem \cite{Weinberg:1975ui} 
was thought to be one of its major difficulties, 
while the absence of strong CP violation was believed to be one of its main successes 
\cite{Nanopoulos:1973wz,Weinberg:1973un}. 
Few years later, with the discovery of Yang-Mills instantons \cite{Belavin:1975fg}
and the non-trivial QCD vacuum structure \cite{Callan:1976je,Jackiw:1976pf}, 
this point of view was unexpectedly turned around.  
The solution of the $U(1)$ problem brought as a gift the so-called strong CP problem. 
In order to present this story, which also provides the physical foundations of axion physics, 
let us start from the QCD Lagrangian\footnote{We adopt the following conventions: 
$D_\mu = \partial_\mu - i g_s T^a A^a_\mu$, 
$G^a_{\mu\nu} = \partial_\mu A^a_\nu - \partial_\nu A^a_\mu + g_s f^{abc} A^b_\mu A^c_\nu$ 
and $\tilde{G}^{a}_{\mu\nu} = \frac{1}{2} \epsilon_{\mu\nu\rho\sigma} G^{a\,\rho\sigma}$,  
with $\epsilon^{0123} = -1$. 
The latter convention is used in \cite{diCortona:2015ldu}, 
while for instance Ref.~\cite{Irastorza:2018dyq} employs $\epsilon^{0123} = +1$. 
\label{foot:epsilon}
} 
%
\beq
\label{eq:QCDLag}
\mathcal{L}_{\rm QCD} = 
\sum_q \bar q \(i \slashed{D} - m_q e^{i\theta_q\gamma_5}\) q 
-\frac{1}{4} G^{a\,\mu\nu} G^{a}_{\mu\nu} 
+\theta \frac{g_s^2}{32\pi^2} G^{a\,\mu\nu} \tilde{G}^{a}_{\mu\nu} \, ,  
\eeq
where $\gamma_5$ acts on the chiral components $q=q_L+q_R$ as $\gamma_5\, q_{L,R} =\mp \, q_{L,R}$.
This Lagrangian  contains two potential sources of CP violation:  
the phases of the quark masses $\theta_q$,  
and the so-called topological term, proportional to $\theta$ (in short $G\tilde G$). 
In fact, both $\theta_q$ and $\theta$ violate P and T (and hence CP).
On the other hand, the $G \tilde G$ operator can be written as a   
total derivative
%
\beq 
\label{eq:GGtotalder}
G^{a\,\mu\nu} \tilde{G}^{a}_{\mu\nu} = 
\partial_\mu K^\mu = 
\partial_\mu \epsilon^{\mu\alpha\beta\gamma} \( A^{a}_{\alpha} G^{a}_{\beta\gamma} 
- \frac{g_s}{3} f^{abc} A^{a}_{\alpha} A^{b}_{\beta} A^{c}_{\gamma}\) \, , 
\eeq
in terms of the Chern-Simons current, $K^\mu$,  
and hence it bears no effects in perturbation theory. 
However, classical configurations do exist for which the effects of this term 
cannot be ignored. These configurations are topological in nature, and 
can be identified  by going to Euclidean space and writing the volume integral 
of the $G \tilde G$ term as 
\begin{equation}
\label{eq:GtGeuclid}
 \int d^4x \, G^{a}_{\mu\nu} \tilde{G}^{a}_{\mu\nu} =
  \int d^4x \, \partial_\mu K_\mu =
   \int_{S_3} d\sigma_\mu K_\mu \,,
\end{equation}
where $S_3$ is the three-sphere at infinity and $d\sigma_\mu$ an element of its hypersurface.
In order for these configurations to contribute to the path integral, we
require  that the gauge potentials are such that the field strength tensor $G_{\mu\nu}^a$ vanishes as $|x|\to \infty$ 
so that  the action is finite. 
Besides $A^a_\mu \big |_{S_3} = 0$, other configurations that can be  obtained from this by a gauge 
transformation also satisfy $G^a_{\mu\nu}=0$ at the boundary. 
In terms of  the Lie algebra valued potential $A_\mu = A_\mu^a T^a$ where  $T^a$ are the group generators, they  read  
 $A'_\mu =  U^{-1} A_\mu U +ig_s^{-1} U^{-1}\partial_\mu U$, so that at the boundary  $A'_\mu =ig_s^{-1}  U^{-1}\partial_\mu U $.  
Configurations of this  type are  called {\it pure gauges}. 
We are interested in pure gauges for which $U$ cannot be continuously deformed into the identity 
in group space. To argue that such configurations exist, let us consider 
an $SU(2)$ subgroup of $SU(3)$ and let us restrict the gauge potentials 
defining $K_\mu$ in the surface integral in~\eqn{eq:GtGeuclid}  to this subgroup.
Since $SU(2)$  has  $S_3$ as  group manifold, these  potentials provide 
a mapping $S_3 \to S_3$. 
It can be shown that for mappings of non-trivial topology the integral in~\eqn{eq:GtGeuclid}  
counts the number   of times the hypersphere at infinity is wrapped around 
the $S_3$ group manifold. More precisely 
$ \int d^4x \, G^{a}_{\mu\nu} \tilde{G}^{a}_{\mu\nu} = \frac{32 \pi^2}{g_s^2} \nu$,
where $\nu\in \mathbb{Z}$ is called winding number or Pontryagin index.
Thus,  in Euclidean space   $SU(2)$ field configurations of finite action  
fall in homotopy classes of different winding number. An important point is that   it is not possible 
to deform a  field configuration into another of different winding number while maintaining the action finite.
 As regards general $SU(3)$ gauge field configurations,
they can be classified  in the same $SU(2)$ homotopy classes, the reason being that any mapping from  $S_3$ into any
simple Lie group $G$ can be  deformed  into a mapping to a $SU(2)$ subgroup of $G$ 
in a continuous way \cite{bott:1956}, 
hence with no change of  homotopy class.
Configurations of unit winding number were explicitly  constructed by  Belavin, Polyakov,  Schwartz 
and  Tyupkin~\cite{Belavin:1975fg} who also showed that 
their finite action $S_{1} = \frac{8 \pi^2}{g^2_s}$ corresponds to a minimum, 
which implies  that they are solutions of the classical equation of motion in Euclidean space. 
Being of finite action, these gauge configuration are  localised in all the four dimensions,  which  
justifies the name {\it instantons}.  

To assess the  relevance of instantons let us return to physical Minkowski space
and let us  consider a gauge field configuration of winding number $\nu$.   
Choosing the temporal gauge $A^a_0=0$ so that  $K_i=0$ allows to rewrite    
\eqn{eq:GtGeuclid} as 
\begin{equation}
\label{eq:GtGmink}
\nu =
\frac{g_s^2}{32\pi^2}  \int d^4x \, G^{a}_{{\!\scriptscriptstyle{(\nu)}}\mu\rho} \tilde{G}_{\!\scriptscriptstyle{(\nu)}}^{a\mu\rho} \  \to \ 
 \frac{g_s^2}{32\pi^2}   \int d^4x \,  \partial_0 K^0_{\!\scriptscriptstyle{(\nu)}} 
= \frac{ g_s^2}{32\pi^2} \int d^3x \, K_{\!\scriptscriptstyle{(\nu)}}^0(\text{\textit{\textbf{x}}},t) \big|^{t=+\infty}_{t=-\infty} =  \nu \,. 
\end{equation}
This shows  that  $\int d^4x\, G_{\!\scriptscriptstyle{(\nu)}} \tilde G_{\!\scriptscriptstyle{(\nu)}}$  corresponds to an 
interpolation from a pure gauge  with  winding 
number  $n$ at $t=-\infty$,  
to a different pure gauge configuration at $t=+\infty$  with winding  
number $m=n+\nu$.  
More precisely, one can interpret  (multi)instanton solutions as  tunnelling from one $G^a_{\mu\nu}=0$  vacuum state 
$ |n\rangle $ to a gauge-rotated one with different winding 
number $ |m\rangle$~\cite{tHooft:1976rip,tHooft:1976snw}. 
In the semiclassical approximation the tunnelling  probability is  given by 
the exponential of the (multi)instanton action $e^{-S_\nu} $ where  
$S_\nu =   \frac{8 \pi^2}{g^2_s}|\nu|$~ \cite{Callan:1976je,Jackiw:1976pf}, 
so that the effects of solutions with higher  winding number $|\nu|>1$~\cite{Witten:1976ck,Jackiw:1976fs} 
are strongly suppressed with respect to instanton effects with action $S_1$, and hence  of little interest.   

An important remark is now in order.  While 
the integral  in \eqn{eq:GtGmink} is  gauge invariant, and hence 
the difference $m-n=\nu$  is a physically meaningful number,  
the Chern-Simons current $K^\mu$  by itself is not gauge invariant, 
which means that $n$ and $m$ labelling the vacuum states have no real physical meaning. This is also 
evidenced by the the fact that the action of a gauge transformation of non-trivial winding number 
amounts to a relabelling  $U_{\!\scriptscriptstyle{(1)}}|n\rangle=|n+1\rangle$.
Clearly a more consistent definition of the  physical vacuum is called for. 
Let us consider the  linear combination
\beq 
\label{eq:thetavacdef}
|\theta\rangle = \sum_{n=-\infty}^{+\infty} e^{in\theta} |n\rangle \, ,
\eeq
where $\theta \in [0,2\pi)$ is an angular parameter, which is known as 
the \emph{$\theta$ vacuum}.\footnote{A less conventional and more intuitive way of 
introducing the $\theta$ vacuum that relies on general quantum-mechanical
 principles applied to a Yang-Mills theory can be found in  Ref.~\cite{Jackiw:1979ur}.} 
%

This vacuum state has the important property of being  an eigenstate of the unitary 
operator of the gauge transformation 
\beq 
\label{eq:actionOm1}
U_{\!\scriptscriptstyle{(1)}}|\theta\rangle = \sum_{n=-\infty}^{+\infty} e^{in\theta} |n+1\rangle = e^{-i\theta} |\theta\rangle \, , 
\eeq
so that is  physically well-defined.  The introduction of the $\theta$ vacuum is 
also necessary  to preserve locality and cluster decomposition \cite{Callan:1976je,Weinberg:1996kr}. 
To see this, let us consider the expectation value of a local operator $\mathcal{O}$ 
within a large Euclidean volume $\Omega$
\beq 
\label{eq:clusterdec1}
\vev{O}_\Omega = 
\frac{\sum_\nu f(\nu) \int_\nu  \mathcal{D}\phi \, e^{-S_{\Omega}[\phi]} \mathcal{O}[\phi]}
{\sum_\nu f(\nu) \int_\nu \mathcal{D}\phi \, e^{-S_{\Omega}[\phi]}} \, ,
\eeq
where $\phi$ denotes all the fields of the theory, 
$S_{\Omega}$ is the integral of the Lagrangian restricted to the 
volume $\Omega$ and 
we have included the sum over all topological sectors $\nu$, 
with a general weight factor $f(\nu)$.  
Suppose now the volume $\Omega$ is 
split into two large regions, $\Omega = \Omega_1 + \Omega_2$, with 
$\cal{O}$ localised in $\Omega_1$. 
The integral in \Eqn{eq:clusterdec1} can hence be split into   
\beq 
\label{eq:clusterdec2}
\vev{O}_\Omega = 
\frac{\sum_{\nu_1,\nu_2} f(\nu_1 + \nu_2) 
\int_{\nu_1} \mathcal{D}\phi \, e^{-S_{\Omega_1}[\phi]} \mathcal{O}[\phi] 
\int_{\nu_2} \mathcal{D}\phi \, e^{-S_{\Omega_2}[\phi]}}
{\sum_{\nu_1,\nu_2} f(\nu_1+\nu_2) \int_{\nu_1} \mathcal{D}\phi \, e^{-S_{\Omega_1}[\phi]}
\int_{\nu_2} \mathcal{D}\phi \, e^{-S_{\Omega_2}[\phi]}} \, ,
\eeq
with the constraint $\nu = \nu_1 + \nu_2$. 
The principle of cluster decomposition 
(which states that distant enough experiments 
must yield uncorrelated results)
requires that the physics 
in the volume $\Omega_2$ cannot affect the average of an observable localised in 
$\Omega_1$. In order for this to be true one needs 
\beq 
\label{eq:condclusterdec1}
f(\nu_1 + \nu_2) = f(\nu_1) f(\nu_2) \, , 
\eeq
so that the volume $\Omega_2$ cancels out in the ratio of \Eqn{eq:clusterdec2}. 
Remarkably, this fixes the form of $f(\nu)$ to be 
\beq 
\label{eq:condclusterdec2}
f(\nu) = e^{i\theta \nu} \, ,
\eeq
where $\theta$ is a free parameter. 

A notable property of $\theta$ is the fact that its value 
cannot be changed via the action of a gauge 
invariant operator. This can be seen by considering 
the time ordered product of a set of gauge invariant operators 
$\mathcal{O}_1\mathcal{O}_2\ldots$
between two different vacuum states 
\beq 
\label{eq:thetathetap1}
\langle \theta' | T (\mathcal{O}_1\mathcal{O}_2\ldots) | \theta \rangle 
= \sum_{m,n} e^{i (n\theta-m\theta')} \langle m | T (\mathcal{O}_1\mathcal{O}_2\ldots) | n \rangle
= \sum_{m,n} e^{i (n\theta-m\theta')} F(\nu) \, , 
\eeq 
where in the last step we have emphasised that the matrix element depends only on the 
difference $\nu = n - m$, because 
$\Omega_1 T(\mathcal{O}_1\mathcal{O}_2\ldots) \Omega_1^{-1} = T(\mathcal{O}_1\mathcal{O}_2\ldots)$
and hence both $n$ and $m$ are shifted by the same amount under a large gauge transformation.  
Then \Eqn{eq:thetathetap1} becomes: 
\beq 
\label{eq:thetathetap2}
\langle \theta' | T (\mathcal{O}_1\mathcal{O}_2\ldots) | \theta \rangle 
= \sum_{n} e^{i n(\theta-\theta')} \sum_{\nu} e^{i \frac{\nu}{2}(\theta+\theta')} 
F(\nu) = 2\pi \delta(\theta-\theta') \sum_\nu e^{i\nu\theta} F(\nu) \, ,
\eeq 
which is zero for $\theta \neq \theta'$.
This property is referred to as a super-selection rule: 
$\theta$ is a fundamental parameter that labels the Yang-Mills vacuum 
and each value of $\theta$ labels a different theory. 

To see explicitly how the $\theta$ term enters the QCD Lagrangian, 
let us consider  
the vacuum-to-vacuum transition in the presence of an external source $J$
\beq 
\label{eq:vactovacsumn}
\langle \theta_+ | \theta_- \rangle_J = 
\sum_{m,n} e^{in\theta} e^{-im\theta} \langle m_+ | n_- \rangle_J = 
\sum_{\nu} e^{i\nu\theta} \sum_{m} \langle m_+ | (\nu + m)_- \rangle_J \, .
\eeq
The vacuum amplitude is a sum over different vacuum transitions 
in which $\nu$ corresponds to the net change of winding number between 
$t=-\infty$ and $t=+\infty$, 
weighted by the factor $e^{i\nu\theta}$. 
The latter can be replaced, thanks to \Eqn{eq:GtGmink},   
by an effective contribution to the 
Yang-Mills Lagrangian
\beq 
\label{eq:vactovacpath}
\langle \theta_+ | \theta_- \rangle_J = 
\sum_\nu \int \mathcal{D} A\, 
e^{- \int d^4x\, \frac{1}{4} 
G G
+i \theta \frac{g^2_s}{32\pi^2} \int d^4x\, 
G \tilde G
+ \text{$J$-term}}
\delta\left( \nu - \frac{g_s^2}{32\pi^2} \int d^4x\, 
G \tilde G
\right) \, ,
\eeq
where the transition amplitude $\sum_m \langle m_+ | (\nu + m)_- \rangle_J$ 
has been expressed in terms of a path integral over all gauge field 
configurations $A$ \emph{with fixed $\nu$} (hence the delta function)
and the phase factor $e^{i\nu\theta}$ has been replaced by 
a $G\tilde G$ term in the Euclidean action. 

In summary, the non-trivial structure of the Yang-Mills vacuum  
requires that the path integral is extended to include gauge field 
configurations with non-trivial winding number, and in turn this  
requires that the  CP-violating  $G\tilde G$ term must be  
included  in the effective action. A strong argument in support of the correctness 
of this picture  comes from the fact  that in the hadronic spectrum  there are no 
signs of a light state that could correspond to the Goldstone
boson of  a $U(1)_A$ symmetry spontaneously broken by 
the quark condensates, a puzzle that was  dubbed  by Weinberg 
`the $U(1)_A$ problem'~\cite{Weinberg:1975ui}.
The  topologically non-trivial gauge configurations responsible  for the  non-vanishing 
of the surface integral in \eqn{eq:GtGeuclid}  provide the 
solution: the complex nature of the QCD vacuum makes $U(1)_A$  not a true
symmetry of QCD~\cite{tHooft:1976rip,tHooft:1976snw,tHooft:1986ooh},  and 
this  explains the heaviness of the $\eta'$ meson compared to the other pseudo 
Goldstone bosons of the spontaneously broken  chiral symmetry.


\subsection{Dependence of the QCD vacuum energy on $\theta$}
\label{sec:thetadepphys} 

We are interested in determining the $\theta$ dependence of 
the QCD vacuum energy density,  
$E(\theta)$. 
This is because the axion VEV
can be treated as an effective $\theta$ parameter, 
so that some exact results that can be established for  the QCD $\theta$ angle 
hold for the axion as well, 
and are especially important in the study of the axion potential. 
In the large 4-volume ($V_4$) limit, $E(\theta)$
this is related to the Euclidean functional generator, $Z(\theta)$, via (see e.g.~\cite{Coleman:1985rnk}) 
\beq 
\label{eq:ZAdef1}
Z(\theta) = \lim_{V_4 \to \infty} e^{-E(\theta) V_4} \, . 
\eeq 
The latter also admits a path integral representation given by\footnote{In passing to the 
Euclidean, $t = - i t_E$, the $G \tilde G$ operator picks up an imaginary part, 
which eventually leads to the periodic $\theta$-dependence of the QCD vacuum energy.}  
\beq 
\label{eq:ZAdef2}
Z(\theta) = \int \mathcal{D} A \, e^{-\frac{1}{4} \int d^4x \, GG 
+ i \theta \frac{g_s^2}{32\pi^2} \int d^4x \, G \tilde G} 
\sim
e^{- \frac{8 \pi^2}{g^2_s}} e^{i \theta} \, ,
\eeq
where in the last step we have taken the leading term 
in the semi-classical approximation $\hbar \to 0$, 
corresponding to the contribution of a $\nu = 1$ instanton. 
Since the instanton is translational invariant one still needs to integrate 
over its centre. This can be done within the dilute-instanton-gas approximation, 
which corresponds to summing-up the contribution of 
approximate solutions consisting of 
$n$ instantons and $\bar n$ anti-instantons 
with $n - \bar n = 1$ and   
with their centres widely separated. 
Using this approximation one gets (see e.g.~\cite{Coleman:1985rnk}) 
\beq 
\label{eq:EthetaQCD}
E(\theta) = -2K e^{- \frac{8 \pi^2}{g^2_s}} \cos\theta \, , 
\eeq
where $K$ is a positive constant encoding Jacobian factors due to the instanton zero modes 
(translations and dilatations) 
and a functional determinant originating from the gaussian integration over  
the quantum fluctuations on the 
instanton background. The latter are actually crucial for stabilising the zero mode associated 
to dilatations, since the integration over the instanton size $\rho$ formally diverges 
(at short distances) at the classical level, due to the classical scale invariance of QCD  
which is broken via radiative corrections.   
In practice, the breaking of scale invariance can be approximated by taking a running 
coupling $g_s (\mu = 1/\rho)$ in \Eqn{eq:EthetaQCD}, with 
\beq 
\label{eq:runninggs}
g_s^2(\mu) = \frac{8\pi^2}{\beta_{0} \log(\mu / \Lambda_{\rm QCD})} \, ,
\eeq
in terms of the one-loop QCD beta-function 
$\beta_0 = 11-2 n_f / 3$ with $n_f$ 
active flavours and the integration constant
$\Lambda_{\rm QCD} \approx 150$ MeV.  
Hence, the integration over the instanton sizes is dominated by 
values of $\rho$ corresponding to an unsuppressed exponential factor  
\beq 
\label{eq:expinstfact}
e^{- \frac{8 \pi^2}{g^2_s (1/\rho)}} = \( \rho \Lambda_{\rm QCD} \)^{\beta_0} \, ,
\eeq
namely for $\rho \sim 1 / \Lambda_{\rm QCD}$, which corresponds to the 
so-called \emph{large instantons}, 
in contrast to possible short-distance contributions which are exponentially suppressed due to the 
asymptotic freedom of the coupling $g_s$. 
It should be noted that the semi-classical approximation breaks down for 
$g_s (\mu = \Lambda_{\rm QCD}) \to \infty$, so that instanton calculus 
cannot be used for accurate predictions in QCD.\footnote{On the other hand, 
large instantons at finite temperature 
are suppressed by electric screening so that the 
semiclassical approximation is increasingly reliable at high $T \gg \Lambda_{\rm QCD}$, 
which serves as an infrared cut-off 
(see e.g.~\cite{Gross:1980br}).}

An alternative way to systematically deal with the $\theta$ dependence 
of the QCD vacuum is via chiral Lagrangian techniques, 
which will be reviewed in \sect{sec:AxionEFT} for the case of the axion. 
Before that, however, one needs to include massive quarks. 
In order to understand the role of quark fields in the problem, let us perform a global 
chiral transformation on a single quark field 
\beq 
\label{eq:chiraltransfq}
q \to e^{i \gamma_5 \alpha} q \, ,  
\eeq
(that is $q_R \to e^{i \alpha} q_R$ and $q_L \to e^{-i \alpha} q_L$). 
The associated axial current, $J^5_\mu = \bar \psi \gamma_\mu \gamma_5 \psi$, 
is not conserved because of the quark mass term and the chiral anomaly 
(note that the latter has a structure similar to the topological term) 
\beq 
\label{eq:J5anomaly}
\partial^\mu J^5_\mu = 2m_q \bar q i \gamma_5 q + \frac{g_s^2}{16 \pi^2} G \tilde G \, .
\eeq
Hence, we expect that both $\theta_q$ (see \eqn{eq:QCDLag}) and $\theta$ are shifted upon the 
transformation in \Eqn{eq:chiraltransfq}. 
One has $\theta_q \to \theta_q + 2 \alpha$, while it is less trivial to show that $\theta \to \theta - 2 \alpha$. 
This can be most easily understood in terms of the non-invariance of the path integral measure \cite{Fujikawa:1979ay}
under the transformation in \Eqn{eq:chiraltransfq}
\beq
\label{eq:measurePI}
\mathcal{D}q \mathcal{D}\bar q \to 
\( e^{-i\alpha \frac{g_s^2}{16 \pi^2} \int d^4x \, G \tilde G} \) \mathcal{D}q \mathcal{D}\bar q \, .
\eeq 
Hence, only the linear combination 
\beq 
\label{eq:defthetabar}
\bar \theta = \theta + \theta_q 
\eeq
is invariant under a quark chiral rotation, and hence  
physically observable. 
The generalisation of the $\bar \theta$ parameter in the 
electroweak theory (invariant under a generic chiral transformation involving 
an arbitrary set of quark fields) reads 
\beq 
\bar \theta = \theta + \text{Arg}\,\text{Det}\,Y_U Y_D \, ,
\eeq
in terms of the up and down Yukawa matrices. 

Some exact results regarding the $\theta$ dependence of the QCD vacuum 
energy density, $E(\theta)$, can be derived by using its path integral representation 
in the presence of fermions as well. 
Denoting collectively by $\mathcal{D} \phi \equiv \mathcal{D} A \, \mathcal{D} q \mathcal{D} \bar q$ the functional integration variables comprising gluons, 
quarks and anti-quarks fields,  
recalling the expression of the QCD vacuum energy 
density in terms of the functional $Z(\theta)$ defined in \eqn{eq:ZAdef1}, 
and being $\nu$ 
the winding number defined in \Eqn{eq:GtGmink},  
one can show the following properties: 
\begin{itemize}
\item $E(0) \leq E(\theta)$ 

This is a special case of the Vafa-Witten theorem \cite{Vafa:1984xg}, 
which states that parity cannot be spontaneously broken in QCD. 
To prove that, one exploits the following inequality 
\beq
\label{eq:VW1}
Z(\theta)
= \int  \mathcal{D} \phi \, e^{- S_{\theta=0} + i \theta \nu }
= \abs{\int  \mathcal{D} \phi \, e^{- S_{\theta=0} + i \theta \nu }} 
\leq \int \abs{ \mathcal{D} \phi \, e^{- S_{\theta=0} + i \theta \nu }} 
= \int  \mathcal{D} \phi \, e^{- S_{\theta=0} }
= 
Z(0) \, , 
\eeq
where we crucially exploited the fact that the path integral measure 
is positive definite, 
which is true for a vector-like theory like QCD \cite{Vafa:1983tf}. 
This, however, does not hold in chiral gauge theories like the SM. 
The consequences of this fact will be discussed in \sect{sec:CPvaxioncoupl}. 

\item $E(\theta) = E(\theta + 2\pi)$

This simply follows from the fact that $\theta$ is a global phase and $\nu$ 
an integer. 

\item $E(\theta) = E(-\theta)$ 

To show this let us perform a field redefinition $\phi \xrightarrow[]{\rm CP} \phi'$  
which leaves $\mathcal{D}\phi$ invariant. Hence, we have 
\beq
\label{eq:VW3}
Z(\theta)
= \int  \mathcal{D} \phi \, e^{- S(\phi)_{\theta=0} + i \theta \nu(\phi)}
= \int  \mathcal{D} \phi' \, e^{- S(\phi')_{\theta=0} + i \theta \nu(\phi')} 
= \int  \mathcal{D} \phi \, e^{- S(\phi)_{\theta=0} - i \theta \nu(\phi)} 
= Z(-\theta) \, , 
\eeq
where in the last but one step we used the fact that $S(\phi)_{\theta=0}$ 
is CP invariant, while the topological term is CP odd. 
Note, however, that $S(\phi')_{\theta=0} \neq S(\phi)_{\theta=0}$ in the SM, 
due to the CKM phase, so $E(\theta)$ picks up a small contribution odd in $\theta$. 
\end{itemize}




\subsection{Neutron EDM and the strong CP problem}
\label{sec:nEDM}


Among the CP violating observables induced by $\bar\theta$, the 
neutron EDM (nEDM) stands out as the most sensitive one.
The latter is defined in terms of the non-relativistic Hamiltonian 
\beq 
\label{eq:nEDMHLNR} 
H = - d_n \vec{E} \cdot \hat{S} \, ,
\eeq
and the current experimental limit
is $|d^{\rm exp}_n| < 3.0 \cdot 10^{-26} \text{\ $e$ cm} 
= 1.5 \cdot 10^{-12} \text{\ $e$ GeV}^{-1}$  ($90 \%$ CL) \cite{Afach:2015sja}.\footnote{In 
Feb 2020 the nEDM experiment at PSI has published  
a new improved limit $|d^{\rm exp}_n| < 1.8 \cdot 10^{-26} \text{\ $e$ cm}$ 
($90 \%$ CL) \cite{Abel:2020gbr}.} 
A new round of searches are actively underway with the goal of improving the sensitivity to CP violation 
by up to two orders of magnitude (see e.g.~\cite{Filippone:2018vxf}).
\Eqn{eq:nEDMHLNR} 
can be written in terms of a Lorentz invariant Lagrangian operator as follows
\beq 
\label{eq:nEDMHL} 
\mathcal{L} = - d_n \frac{i}{2} \bar n \sigma_{\mu\nu} \gamma_5 n F^{\mu\nu} \, . 
\eeq
The calculation of the nEDM has been performed using different kind of techniques, 
such as chiral perturbation theory \cite{Baluni:1978rf,Crewther:1979pi,Pich:1991fq}, 
QCD sum-rules \cite{Pospelov:1999mv}, holography~\cite{Bartolini:2016jxq} 
and lattice QCD \cite{Abramczyk:2017oxr,Dragos:2019oxn}   
(for a review of the technical challenges involved see e.g.~\cite{Pospelov:2005pr,Pospelov:1900zz}).  
These approaches show an overall agreement, albeit with uncertainties of $\mathcal{O} (50\%)$. 
Future inputs from the lattice could be crucial for reducing such error \cite{Dragos:2019oxn}. 
The natural size of the $\bar\theta$ contribution to the nEDM can be understood as follows.
The operator in \Eqn{eq:nEDMHL} is $d=5$ so one would naively expect its Wilson coefficient to be 
of $\mathcal{O}(1/m_n)$ size. However, in order to contribute to the nEDM one needs to pick-up 
an imaginary part which can only originate from the phase of a light quark mass 
(working in the basis where the $G\tilde G$ term is absent). 
Moreover, being a dipole, the operator must be generated via an EM loop. 
Hence, taking into account these two extra suppression factors,  
the effective contribution to $d_n$ can be estimated as 
\beq 
\label{eq:estimatenEDM1} 
\mathcal{L} \sim 
\frac{e}{16 \pi^2} \frac{m_q \, e^{i \bar\theta}}{m_n} \frac{1}{m_n} \bar n \sigma_{\mu\nu} \gamma_5 n F^{\mu\nu} 
\, ,
\eeq
Expanding linearly in $\bar \theta$ one gets 
\beq 
\label{eq:estimatenEDM2}
\abs{d_n} \sim \frac{1}{8 \pi^2} \frac{m_q}{m_n} \frac{\bar \theta \, e}{m_n} \approx 10^{-4}  \, \bar \theta \text{ $e$ GeV}^{-1} \, . 
\eeq
In fact, this naive estimate yields a somewhat smaller value compared 
to a real calculation.  
For instance, one of the most precise ones, based on QCD sum-rules, yields \cite{Pospelov:1999mv} 
\beq 
\label{eq:dntheory}
d_n = 2.4 \, (1.0) \cdot 10^{-16} \, \bar\theta \text{\ $e$ cm} = 1.2 \, (0.5) \cdot 10^{-2} \, \bar\theta \text{\ $e$ GeV}^{-1} \, ,
\eeq
thus implying the bound\footnote{Experiments searching for the EDM
of the electron in paramagnetic systems have recently achieved a remarkable sensitivity 
and they can be used to obtain novel independent
constraints on the QCD theta term at the level of 
$|\bar \theta| \lesssim 3 \cdot 10^{-8}$ \cite{Flambaum:2019ejc}.} 
\beq 
|\bar \theta| \lesssim 10^{-10} \, .   
\eeq  
Understanding the smallness of $\bar \theta$ 
consists in the so-called strong CP problem. 

\subsection{Musings on the strong CP problem}
\label{sec:strongCP}
The strong CP problem turns out to be qualitatively different from other ``small value'' 
problems in the SM. One first observation concerns the radiative stability 
of $\bar \theta$. Since CP is violated in the SM, one expect 
$\bar \theta$ to receive an infinite renormalisation due to the CKM phase \cite{Ellis:1978hq,Khriplovich:1993pf}. 
Based just on spurionic properties, the SM contribution to $\bar \theta$ must 
be proportional to the Jarlskog invariant \cite{Jarlskog:1985ht}, 
which is given in terms of the following CP-odd, 
flavour singlet (lowest order) combination of the quark Yukawas 
\begin{align}
\text{Im Det} \, [Y_UY_U^\dag, Y_DY_D^\dag] 
&= \( \frac{2^6}{v^{12}} \)
\displaystyle \prod_{i>j=u,c,t} (m_i^2 - m_j^2)
\prod_{k>l=d,s,b} (m_k^2 - m_l^2) 
\ J_{\rm CKM} \approx 10^{-20} \, , 
\end{align}
where $v = 246$ GeV and $J_{\rm CKM} = \Im V_{ud} V_{cd}^*V_{cs} V_{us}^* 
\approx \ 3 \times 10^{-5}$.\footnote{An alternative form of the Jarlskog 
invariant can be found in \cite{Kim:2014qia}.}
As shown in Ref.~\cite{Ellis:1978hq},  
this would correspond diagrammatically
to the insertion of 12 Yukawas 
(connected pairwise via 6 Higgs propagators)
in the propagator of a quark field and hence to a 6-loop diagram, 
whose imaginary part contributes to $\bar \theta$. 
On the other hand, 
the SM Yukawa Lagrangian features an accidental exchange symmetry: 
$H \leftrightarrow \tilde H$, $u_R \leftrightarrow d_R$, $Y_U \leftrightarrow Y_D$, 
under which $\text{Im Det} \, [Y_UY_U^\dag, Y_DY_D^\dag]$ is odd. 
Hence, the 6-loop contribution must vanish, and in order to get a non-zero contribution one has to 
insert e.g.~a $U(1)_Y$ gauge boson which breaks the $u_R \leftrightarrow d_R$ symmetry. 
Then a typical 7-loop contribution to the radiatively induced $\bar \theta$ will look like 
\beq 
\delta \bar \theta_{\rm div.} \sim \frac{g'^2 \(y^2_{u_R} - y^2_{d_R}\) \text{Im Det} \, [Y_UY_U^\dag, Y_DY_D^\dag] }{(4 \pi^2)^7} \log \Lambda_{\rm UV} \approx 10^{-33} \log \Lambda_{\rm UV}
\, ,
\eeq
where $y_{u_R,\, d_R}$ denotes the hypercharge of a given SM chiral quark and  
$\Lambda_{\rm UV}$ is an ultraviolet (UV) cut-off. 
Thus if we take the tree-level value of $\bar \theta$ to be small at some UV boundary 
(e.g.~the Planck scale), it will remain radiatively small when run down to the QCD scale. 
This has to be contrasted instead with the 
hierarchy problem of the electroweak scale, for which the Higgs mass parameter is quadratically 
sensitive to threshold effects from UV physics, $\delta \mu^2 \sim \text{(loop)} \times \Lambda^2_{\rm UV}$.  
Remarkably, the CP and flavour structure of the SM provides a non-trivial screening mechanism against 
radiative corrections to $\bar \theta$, which is not guaranteed in generic SM extensions.

Integrating out the heavy SM quarks can also lead to finite threshold 
corrections to $\bar\theta$, 
which arise at lower orders in perturbation theory. The size of the largest 
contribution was first estimated in \cite{Ellis:1978hq} to be\footnote{The different 
numerical value compared to Ref.~\cite{Ellis:1978hq} originates from 
employing up-to-date values for the SM parameters.} 
\beq 
\delta \bar\theta_{\rm fin.} \sim \(\frac{\alpha_s}{\pi} \)^4 \(\frac{\alpha_2}{\pi} \)^2 
\( \frac{m_s^2 m_c^2}{m_W^4} \) J_{\rm CKM}
\approx 10^{-18} \, ,  
\eeq
where long-distance QCD effects are such that 
the strong structure constant is $\alpha_s (\text{GeV}) / \pi \approx 1$ 
at the scale relevant for the calculation of the nEDM.  
A refined estimate \cite{Khriplovich:1985jr}, 
based on an actual 3-loop calculation, 
finds instead an $\mathcal{O}(\alpha_s G_F^2)$ contribution 
yielding $\delta \bar\theta_{\rm fin.} \approx 4 \times 10^{-19}$ 
(using a somewhat smaller value $\alpha_s \approx 0.2$). 
Although in both cases 
the contribution is only of 2-loop order in the electroweak 
structure constant, $\alpha_2$, it is still well below the nEDM experimental sensitivity. Generic SM extensions might however spoil this conclusion 
(see e.g.~\cite{deVries:2018mgf}).

 

Another fact that renders the strong CP problem different from other  small value problems of the SM 
such as that of the light Yukawas $(y_{u,\,d,\,e})$ or the cosmological constant, is the apparent 
lack of a possible  anthropic explanation. In fact, as long as $\bar \theta \lesssim 1\%$, nuclear physics 
and Big Bang Nucleosynthesis  are practically unaffected \cite{Ubaldi:2008nf},  so that a 
value $\bar \theta \lesssim 10^{-10}$ does  not seem to be connected to any  `catastrophic boundary'. 
A possible pathway to enforce explanations based   
on anthropic selection arguments might then be attempted  by trying to correlate the value 
of $\bar \theta$ with the value of some other small parameter for which an 
anthropic explanation does exist, as for example the   cosmological constant~\cite{Weiss:1987ns}. 
Attempts  in this direction have been carried out for example in Refs.~\cite{Takahashi:2009zzd,Kaloper:2017fsa},
However, the recent analysis of Ref.~\cite{Dine:2018glh}  found that  anthropic requirements on
the cosmological constant  rather favour values  of the CP violating angle   of $\mathcal{O}(1)$,  
reinforcing the idea that $\bar \theta\ll 1$ is  not related to anthropic selection.




\subsubsection{Solutions without axions}
\label{sec:solwoutaxions}

Before introducing the axion solution of the strong CP problem, 
we discuss for completeness three classes of solutions which do not rely on the axion.\footnote{Besides 
the three possibilities outlined below, 
more speculative solutions that do not invoke an axion also exist, 
see for example~\cite{Dvali:2005zk}.} 
 


\begin{itemize}
\item 
\emph{Massless quark solution}. 
If one of the quark fields (say the up quark) were massless, 
the QCD Lagrangian would 
feature a global $U(1)_u$ axial symmetry, which could be used to rotate the $\theta$ term to zero
(cf.~the discussion below \Eqn{eq:chiraltransfq}). 
For some time this was believed to be a possible solution of the strong CP problem, due to the difficulty 
in extracting the value of $m_u/m_d$ in chiral perturbation theory. 
Most notably, this was due to a second-order effect in the chiral Lagrangian,
known as Kaplan-Manohar ambiguity \cite{Kaplan:1986ru} 
(see also \cite{Georgi:1981be,Choi:1988sy,Banks:1994yg}). 
Nowadays this possibility 
has been ruled out by the fit to light quark masses on the lattice 
\cite{Tanabashi:2018oca}, 
which yields $m_u^{\overline{\rm MS}} (2 \ \text{GeV}) = 2.32 (10)$ MeV, 
that is more than 20$\,\sigma$ away from zero. 
An independent strategy in order to disprove the massless up 
quark solution 
without simulating light quarks on the lattice was put forth 
in Refs.~\cite{Cohen:1999kk,Dine:2014dga} and recently 
implemented on the lattice \cite{Alexandrou:2020bkd}, 
which confirmed the non-viability of the massless up-quark hypothesis. 

\item 
\emph{Soft P (CP) breaking}. 
It is conceivable that either P or CP are symmetries of the high-energy theory, 
thus setting $\bar \theta = 0$ in the UV. Models of this type were first proposed 
in \cite{Beg:1978mt,Georgi:1978xz,Mohapatra:1978fy} and later on in 
\cite{Nelson:1983zb,Barr:1984qx} in the context of grand-unified models. 
Eventually, P must be spontaneously broken 
in order to account for the SM chiral structure and 
similarly for CP in order to generate the CKM phase 
(and the extra CP violation that is needed for baryogenesis). 
In these setups the $\bar \theta$ term becomes calculable and 
the main challenge 
consists in generating the observed CP violation in the quark sector
 in such a way that threshold contributions to 
$\bar \theta$ are screened enough so that $\delta \bar \theta \lesssim 10^{-10}$. 
This can be done, however at the cost of some tuning or a somewhat exotic model building 
(for reviews, see e.g.~\cite{Dine:2015jga,Vecchi:2014hpa}). 

\item \emph{QCD solutions}. It is conceivable, at least in principle, 
that the solution of the strong CP problem might be hidden 
in the infrared (IR) dynamics of QCD. Attempts in this direction, e.g.~by `trivialising' 
the QCD vacuum by changing the topology of spacetime \cite{Khlebnikov:1987zg,Chaichian:2001nx,Khlebnikov:2004am,Bezrukov:2008da} 
or by invoking screening effects due to confinement \cite{Samuel:1991cm,Dowrick:1992gs,Gabadadze:2002ff}, 
often fail to provide a simultaneous solution to the $\eta'$ problem. 
\end{itemize}
It is fair to say that, while it is certainly worth looking for solutions of the strong CP problem 
within QCD, no convincing framework has emerged so far. 
As far as concerns the soft P (CP) breaking solutions instead, besides the 
model building complications involved, it is also unappealing 
the fact that the strong CP problem is solved by UV
dynamics, without a clear experimental way to test the mechanism.
From this point of view, the Peccei-Quinn solution, which is reviewed in the 
next section, is crucially different, 
since it delivers a low-energy experimental handle in the form of the axion.

\subsubsection{Peccei Quinn mechanism}
\label{sec:PQsolution}

From a modern perspective, 
the basic ingredient of the PQ 
solution \cite{Peccei:1977hh,Peccei:1977ur,Weinberg:1977ma,Wilczek:1977pj} 
of the strong CP problem
consists in the introduction of a new spin zero field $a(x)$,  
hereby denoted as the axion field, 
whose effective Lagrangian 
\beq 
\label{eq:Leffaxion}
\mathcal{L}_a = \frac{1}{2} (\partial_\mu a)^2 + \mathcal{L}(\partial_\mu a, \psi) 
+ \frac{g_s^2}{32 \pi^2} \frac{a}{f_a} 
G \tilde G
\eeq
is endowed with a quasi shift symmetry $a \to a + \kappa f_a$ 
(where 
$f_a$ is an 
energy scale called the axion decay constant)
that leaves the action 
invariant up to the term  
\beq
\label{eq:remnantqshift}
\delta S = \frac{g_s^2\kappa}{32\pi^2} \int d^4x\, G \tilde G
\, .
\eeq
The transformation parameter $\kappa$ is arbitrary and can be chosen to remove the $\bar \theta$ 
term, while the Vafa-Witten theorem \cite{Vafa:1984xg} (see \Eqn{eq:VW1} for the proof)
ensures that $\vev{a} = 0$ 
in a vector-like theory like QCD,\footnote{A crucial step of the proof 
relies on the positive definiteness of the fermionic determinant  
in the background 
of the gauge fields, 
i.e.~$\det (\slashed{D} + m) > 0$.
This is not ensured for a chiral theory where $m=0$, since it is not possible to write 
a bare mass term for fermions, and hence the Vafa-Witten theorem does not apply in such case. 
In fact, in the SM, which is chiral and 
contains an extra source of CP violation in the Yukawa sector, one expects  
an irreducible contribution to the axion VEV, as discussed in \sect{sec:CPvaxioncoupl}.
} 
thus solving dynamically the strong CP problem. Alternatively, one can explicitly compute the 
axion potential $V(a)$ with chiral Lagrangian techniques (which will be done in in \sect{sec:axionpotandmass})
and show that the absolute minimum 
is in $\vev{a} = 0$. 

Since the Lagrangian \Eqn{eq:Leffaxion} is non-renormalisable,  
it requires a UV completion at energies of the order of $f_a$. 
Historically, the first renormalisable 
model incorporating the axion solution 
of the strong CP problem was due to 
Peccei and Quinn \cite{Peccei:1977hh,Peccei:1977ur}, 
which postulated the existence of a $U(1)_{\rm PQ}$ global symmetry,  
spontaneously broken and anomalous under QCD. 
The presence of a pseudo-Goldstone boson $a(x)$ 
(dubbed axion, since it washes out the strong CP problem) was 
soon realised by Weinberg and Wilczek \cite{Weinberg:1977ma,Wilczek:1977pj}. 
However, before considering explicit models we will first discuss some general properties 
of the axion effective Lagrangian in \Eqn{eq:Leffaxion}, which is already sufficient 
for characterising some general aspects of axion physics. 

\subsection{Axion effective Lagrangian
}
\label{sec:AxionEFT}

The effective operator $aG\tilde G$
is the 
building block of the PQ solution of the strong CP problem, 
and provides some model-independent properties of the axion, which 
we discuss here with the help of the chiral Lagrangian \cite{Georgi:1986df,diCortona:2015ldu}. 
Let us consider for simplicity 2-flavor QCD, with 
$q^T = (u, d)$ and $M_q = \diag (m_u, m_d)$. The axion effective Lagrangian reads
\beq 
\label{eq:Leffaxion2}
\mathcal{L}_{a} = \frac{1}{2} (\partial_\mu a)^2 
+ \frac{a}{f_a} \frac{g^2_s}{32\pi^2} 
G \tilde G
+ \frac{1}{4} g^0_{a\gamma}  a 
F \tilde F
+ \frac{\partial_\mu a}{2f_a} \bar q c^0_q \gamma^\mu \gamma_5 q 
- \bar q_L M_q q_R + \text{h.c.} 
\, . 
\eeq
For later purposes, we have extended \Eqn{eq:Leffaxion} 
to include two model-dependent couplings:  
$g^0_{a\gamma}$ that couples the axion to $F\tilde F$    
and $c^0_q = \diag (c^0_u, c^0_d)$ that couples derivatively the axion to the quark axial 
current.\footnote{While the anomalous dimension of conserved currents vanishes, that  
is not the case for anomalous currents. In fact, as shown in 
Refs.~\cite{Kodaira:1979pa,Larin:1993tq} the iso-spin singlet axial current, 
$j^\mu_{\Sigma q} = \sum_q \bar q \gamma_\mu \gamma_5 q$,
renormalises multiplicatively. Taking this effect into account 
it is possible to connect the low-energy derivative axion couplings to quarks 
with their UV counterparts, which are understood to be the coefficients 
$c^0_q$ (for details see \cite{diCortona:2015ldu}).} 
Their origin will be clarified in \sect{sec:modeldepax}.
It is convenient to first eliminate the $aG\tilde G$ 
term via a field-dependent axial transformation of the quark fields:
\beq 
\label{eq:qtransfa}
q \to e^{i\gamma_5 \frac{a}{2f_a} Q_a} q \, , 
\eeq
where $Q_a$ is a generic matrix acting on the quark fields.  
This transformation has the effect of generating 
a term $-g^2_s\Tr Q_a /(32\pi^2) \frac{a}{f_a} G \tilde G$ which, by requiring   that 
 $\Tr Q_a = 1$, 
precisely cancels the axion-gluon term.
Since in general this transformation is anomalous under QED, 
it will also affect the $F\tilde F$ term. 
Moreover, extra axion-dependent terms are generated
by the quark mass 
operator and the quark kinetic term. 
Then \Eqn{eq:Leffaxion2} becomes
\beq 
\label{eq:Leffaxion3}
\mathcal{L}_{a} = \frac{1}{2} (\partial_\mu a)^2 
+ \frac{1}{4} g_{a\gamma} a
F \tilde F
+ \frac{\partial_\mu a}{2f_a} \bar q c_q \gamma^\mu \gamma_5 q 
- \bar q_L M_a q_R + \text{h.c.}
\, , 
\eeq
where we have defined axion-dressed parameters
\begin{align}
\label{eq:gagamma1}
g_{a\gamma} &= g^0_{a\gamma} - (2 N_c) \frac{\alpha}{2\pi f_a} \Tr (Q_a Q^2)  
\qquad 
\text{with} 
\qquad
Q=\diag(2/3,-1/3) \, , \\
\label{eq:cq1}
c_q &= c^0_q - Q_a \, , \\
\label{eq:Ma1}
M_a &= e^{i \frac{a}{2f_a} Q_a} M_q e^{i \frac{a}{2f_a} Q_a} \, , 
\end{align}
where $\alpha = e^2 / (4 \pi)$ and $N_c = 3$ is the number of colours. 
The axial quark current can be conveniently decomposed into an iso-singlet and 
an iso-triplet component
\beq 
\label{eq:Fierzdec}
\bar q c_q \gamma^\mu \gamma_5 q 
= \frac{1}{2} \Tr[c_q] \bar q \gamma^\mu \gamma_5 q 
+ \frac{1}{2} \Tr[c_q \sigma^a] \bar q \gamma^\mu \gamma_5 \sigma^a q \, ,
\eeq
where we have used the Fierz identity for Pauli matrices, 
$(\sigma^a)_{ij} (\sigma^a)_{kl} = 2 \delta_{il} \delta_{kj} - \delta_{ij} \delta_{kl}$. 
\Eqn{eq:Leffaxion3} should be compared with the chiral axion Lagrangian, 
including for simplicity only pions and axions\footnote{Nucleons 
can be included as well in chiral Lagrangian \cite{Georgi:1986df}. 
However, due to the lack of a mass gap between $\Lambda_{\chi} = 4 \pi f_\pi$ and $m_N$, 
the convergence of the EFT is not good. For this reason we are going to discuss 
axion-nucleon couplings separately, in the context of a non-relativistic EFT for nucleons 
\cite{diCortona:2015ldu} (cf.~\sect{sec:axionnucleon}). 
} 
\beq 
\label{eq:LeffaxionChiral}
\mathcal{L}^{\rm \chi PT}_{a} = 
\frac{f^2_\pi}{4} \[ \Tr ((D^\mu U)^\dag D^\mu U) 
+ 2 B_0 \Tr (U M_a^\dag + M_a U^\dag) \]
+ \frac{\partial^\mu a}{2 f_a} \frac{1}{2} \Tr[c_q \sigma^a] J_\mu^a \, ,
\eeq
where we neglected the iso-singlet current since it is associated to the heavy $\eta'$.\footnote{Interactions 
between the axion and the $\eta'$ meson can be taken into account in 
the large $N$ approximation by using the formalism of Refs.~\cite{Witten:1980sp,DiVecchia:1980yfw}.}
$B_0$ is related to the quark condensate and
\beq
\label{eq:defpioncurr}
J_\mu^a =  \frac{i}{2} f^2_\pi 
\Tr[\sigma^a (U D_\mu U^\dag - U^\dag D_\mu U)] \, , 
\eeq
is the pion iso-triplet axial-vector current 
(derived from the covariant derivative term in \eqn{eq:LeffaxionChiral})
that has the same transformation 
properties under 
$SU(2)_L \otimes SU(2)_R$ as the corresponding quark current in \Eqn{eq:Fierzdec}, 
and we have employed the standard parametrisation 
\beq 
\label{eq:Upionpar}
U = e^{i \pi^a \sigma^a / f_\pi} = \mathbb{I} \cos \frac{\pi}{f_\pi} 
+ i \frac{\sigma^a \pi^a}{\pi} \sin \frac{\pi}{f_\pi} \, ,
\eeq
with $\pi
= \sqrt{(\pi^0)^2 + 2 \pi^+ \pi^-}$, $f_\pi = 92.3$ MeV and 
$D_\mu U = \partial_\mu U + i e A_\mu [Q, U]$.  
In the following, we discuss the various terms  
arising from the axion chiral Lagrangian. 

\subsubsection{Axion potential and axion mass}
\label{sec:axionpotandmass}

Expanding the non-derivative part of the 
axion chiral Lagrangian, one obtains 
\begin{align} 
\label{eq:expB0}
2 B_0 \frac{f_\pi^2}{4} \Tr (U M_a^\dag + M_a U^\dag) &= 
B_0 f^2_\pi (m_u + m_d)
- \frac{1}{2} B_0 (m_u + m_d) \pi^2 \nonumber \\
&-\frac{i}{4} B_0 \frac{f_\pi^2}{f_a} a \Tr (U \{ Q_a,M_q \} ) + \text{h.c.}
+ \ldots \, .
\end{align}
It is customary to choose $Q_a = M^{-1}_q / \Tr M^{-1}_q$. 
Note that 
any linear coupling of the axion to an arbitrary number of pion fields 
is set to zero: for an odd number of pions because $\Tr \sigma^a = 0$, 
while for an even number there is a cancellation with the hermitian conjugate. 
In particular, this sets to zero a mass mixing term between 
$a$ and $\pi^0$ (ignoring possible kinetic mixing between $a$ and $\pi^0$, cf.~\sect{sec:axionpion}),  
while from the second term in \Eqn{eq:expB0} we obtain 
$m^2_\pi = B_0 (m_u + m_d)$ at the leading order (LO) in the chiral Lagrangian expansion. 
With the above choice of $Q_a$ the axion-pion potential turns out to be
\begin{align}
\label{eq:Vapi}
V(a,\pi^a) &= - 2 B_0 \frac{f_\pi^2}{4} \Tr (U M_a^\dag + M_a U^\dag) \nonumber \\
&= - \frac{m_\pi^2 f_\pi^2}{m_u + m_d} 
\left\{ \[
m_u \cos\(\frac{m_d}{m_u + m_d}\frac{a}{f_a}\) 
+ m_d \cos\(\frac{m_u}{m_u + m_d}\frac{a}{f_a}\) \] \cos\(\frac{\pi}{f_\pi}\) \right. \nonumber \\
& \left. \qquad \qquad \ \ + \frac{\pi^0}{\pi} \[ m_u \sin\(\frac{m_d}{m_u + m_d}\frac{a}{f_a}\)  
- m_d \sin\(\frac{m_u}{m_u + m_d}\frac{a}{f_a}\) \] \sin\(\frac{\pi}{f_\pi}\)
\right\} \, . 
\end{align}
Expanding for $a/f_a \ll 1$ we obtain
\beq
\label{eq:Vapiexp}
V(a,\pi^a) = 
- m_\pi^2 f_\pi^2 
\cos\(\frac{\pi}{f_\pi}\) 
+ \frac{1}{2}  \frac{m_u m_d}{(m_u + m_d)^2} \frac{m_\pi^2 f_\pi^2}{f_a^2} a^2  
\cos\(\frac{\pi}{f_\pi}\) + \mathcal{O}\( \frac{a^3}{f^3_a} \) \, ,  
\eeq
where linear terms in the axion field are absent by construction. 
The axion mass (squared) is then readily obtained 
by setting the pion on its ground state $\pi = 0$, which yields \cite{Weinberg:1977ma}
\beq 
m^2_a = \frac{m_u m_d}{(m_u + m_d)^2} \frac{m_\pi^2 f_\pi^2}{f_a^2}  \qquad \Longrightarrow \qquad 
m_a 
\simeq  5.7  \(\frac{10^{12} \ \text{GeV}}{f_a} \) \, \text{$\mu$eV}  \, .
\label{eq:axionmass}
\eeq
An alternative 
expression for the axion-pion potential, 
corresponding to the choice $Q_a = \frac{1}{2} \diag(1,1)$, is given by 
\cite{DiVecchia:1980yfw,diCortona:2015ldu} 
\begin{align}
\label{eq:Vapi0}
V(a,\pi^0) &= - m^2_\pi f_\pi^2 
\sqrt{1 - \frac{4m_u m_d}{(m_u + m_d)^2} \sin^2\( \frac{a}{2 f_a} \)}
\cos \( \frac{\pi^0}{f_\pi} - \phi_a\) \, ,
\end{align}
with 
\beq 
\label{eq:deftanphia}
\tan\phi_a = \frac{m_u - m_d}{m_u + m_d} \tan\( \frac{a}{2f_a} \) \, ,
\eeq
which clearly shows that the absolute minimum is in 
$(a,\pi^0) = (0,0)$.\footnote{QED corrections could in principle generate new minima \cite{DiLuzio:2019wsw}.  
However, 
this is prevented by the hierarchy $m_q/f_\pi \gg \alpha / \pi$,  
which makes the vacuum structure 
for the pion potential trivial.} 
In particular, on the pion ground state, $\pi^0 = \phi_a f_\pi$,
the chiral perturbation theory ($\chi$PT) axion potential takes the form 
\beq
\label{eq:VaChPT}
V(a) = - m^2_\pi f_\pi^2 
\sqrt{1 - \frac{4m_u m_d}{(m_u + m_d)^2} \sin^2\( \frac{a}{2 f_a} \)} \, .
\eeq
Note that while in an expansion  around   $a=0$ at the leading order the chiral potential 
and the one instanton cosine potential of \Eqn{eq:EthetaQCD} coincide,  for large field 
values, $a \sim f_a$ the two differ even qualitatively. 
This is because  in the regime of confinement,  where the estimate of the potential 
\eqn{eq:VaChPT}  based on chiral perturbation theory  is reliable, the instanton 
semiclassical approximation  breaks down, since  in that regime fluctuations of topologically 
non-trivial gauge configurations away  from the instanton  solution that extremise the classical action 
also become important. 
On the other hand, the chiral potential cannot be used above the chiral phase transition,  
and the one-instanton potential becomes more reliable at $T \sim 1$ GeV, 
which is the relevant regime for the calculation of the axion DM relic density 
(cf.~\sect{sec:section2}).

\subsubsection{Axion-pion coupling}
\label{sec:axionpion}

Next we inspect the derivative part of the axion-pion Lagrangian. 
This is obtained by expanding the iso-triplet current term in \Eqn{eq:LeffaxionChiral}:
\begin{align}
\label{eq:expJamu}
\frac{\partial_\mu a}{2 f_a} \frac{1}{2} \Tr[c_q \sigma^a] J^\mu_a &\simeq
- \frac{1}{2} 
\( \frac{m_d-m_u}{m_u + m_d} + c^0_d - c^0_u \)
\frac{f_\pi}{f_a} \partial_\mu a \partial^\mu \pi^0 \nonumber \\
& + \frac{1}{3} 
\( \frac{m_d-m_u}{m_u + m_d} + c^0_d - c^0_u \)
\frac{1}{f_a f_\pi} 
\partial_\mu a (2 \partial^\mu \pi^0 \pi^+ \pi^- - \pi^0 \partial^\mu \pi^+ \pi^- 
- \pi^0 \pi^+ \partial^\mu \pi^- ) \, .
\end{align}
Note that the first term in \Eqn{eq:expJamu} represents a kinetic mixing between the 
axion and the pion fields, which needs to be diagonalised in order to define the canonical 
axion and pion fields. 
The quadratic part of the axion-pion Lagrangian reads
\beq 
\label{eq:quadLagschem}
\mathcal{L}_a^{\rm quad.} = 
\frac{1}{2} 
\begin{pmatrix}
\partial_\mu a & \partial_\mu \pi^0
\end{pmatrix}
\begin{pmatrix}
1 & \epsilon \\
\epsilon & 1
\end{pmatrix}
\begin{pmatrix}
\partial^\mu a \\ 
\partial^\mu \pi^0
\end{pmatrix}
- 
\frac{1}{2} 
\begin{pmatrix}
a & \pi^0
\end{pmatrix}
\begin{pmatrix}
m_a^2 & 0 \\
0 & m_\pi^2
\end{pmatrix}
\begin{pmatrix}
a \\ 
\pi^0
\end{pmatrix} \, ,
\eeq
with $\epsilon = - \frac{1}{2} 
\( \frac{m_d-m_u}{m_u + m_d} + c^0_d - c^0_u \) \frac{f_\pi}{f_a}$ 
and $m_a / m_\pi = \mathcal{O}(\epsilon)$. 
In order to work with canonical propagators one can perform: 
$i)$ an orthogonal transformation to diagonalise the kinetic term, 
$ii)$ a rescaling to make the kinetic term canonical 
and $iii)$ an orthogonal transformation to re-diagonalise the mass term 
(which does not affect the canonical kinetic term). 
The net effect of these operations is 
to shift the current basis fields by 
$a \to a - \epsilon \pi^0$ and
$\pi^0 \to \pi^0 + (m_a^2/m_\pi^2) \epsilon a$. 
Since the axion component into the current pion field is 
suppressed at the level of $\epsilon^3$, this redefinition has no 
practical consequences for experimental sensitivities and astrophysical bounds,  
which are sensitive at most to $\mathcal{O}(\epsilon^2)$ effects.  
This justifies the fact that the correction due to kinetic mixing is generally ignored in the literature. 

The second addend in \Eqn{eq:expJamu} gives instead the axion-pion 
coupling (see also \cite{Chang:1993gm,Kim:2008hd}),  
defined via the Lagrangian term 
\beq 
\label{eq:apicoupdef}
\mathcal{L}_a^{\rm int} \supset
\frac{C_{a\pi}}{f_a f_\pi} 
\partial_\mu a (2 \partial^\mu \pi^0 \pi^+ \pi^- - \pi_0 \partial^\mu \pi^+ \pi^- 
- \pi_0 \pi^+ \partial^\mu \pi^- ) \, ,
\eeq
with
\beq 
\label{eq:Capidef}
C_{a\pi} = - \frac{1}{3} 
\( c^0_u - c^0_d - \frac{m_d-m_u}{m_u + m_d}  \) \, .
\eeq
Note that once the canonical axion and pion field are properly identified, 
the only linear coupling of the axion to the pions is the one in 
\Eqn{eq:apicoupdef}. The axion-pion coupling in \Eqn{eq:Capidef} 
generalises the expressions 
available in the literature in the case of KSVZ 
\cite{Chang:1993gm} 
and DFSZ \cite{Kim:2008hd} axions. 

\subsubsection{Axion-photon coupling}
\label{sec:axionphoton}

With the choice of $Q_a = M^{-1}_q / \Tr M^{-1}_q$ to  
ensure no axion-pion mass mixing, 
the LO axion-photon coupling in \Eqn{eq:gagamma1} becomes 
\beq 
\label{eq:gagamma2}
g_{a\gamma} = 
g^0_{a\gamma} 
- \frac{\alpha}{2\pi f_a} \( \frac{2}{3} \frac{4 m_d + m_u}{m_u + m_d} \) \, .
\eeq
The same result can be obtained via another choice of $Q_a$ (e.g.~the one leading to the $\chi$PT 
potential in \Eqn{eq:VaChPT}), but requires the inclusion of a non-zero axion-pion mixing. 

\subsubsection{Axion-nucleon coupling}
\label{sec:axionnucleon}

Following \cite{diCortona:2015ldu} we derive the 
axion coupling to nucleons (protons and neutrons), via an effective 
theory at energies $\ll \Lambda_{\rm QCD}$, relevant 
for momentum exchanges of the order of the axion mass, 
where the nucleons are non-relativistic.  
This approach turns out to yield a more reliable 
approximation than current algebra techniques \cite{Srednicki:1985xd}
or the chiral EFT for nucleons \cite{Kaplan:1985dv,Georgi:1986df}. 
Our goal is to match the quark current operator in \Eqn{eq:Leffaxion3} with 
a non-relativistic axion-nucleon Lagrangian. 
Using iso-spin as an active flavour symmetry and the 
axion as an external current, the LO effective axion-nucleon Lagrangian reads 
\begin{align}
\label{eq:LNaunder}
\mathcal{L}_{N} &= \bar N v^\mu \partial_\mu N + 
2 g_A \frac{c_u - c_d}{2} \frac{\partial_\mu a}{2 f_a} \bar N S^\mu \sigma^3 N 
+ 2 g^{ud}_0 \frac{c_u + c_d}{2} \frac{\partial_\mu a}{2 f_a} 
\bar N S^\mu N 
+ \ldots \nonumber \\
& = \bar N v^\mu \partial_\mu N + 2 g_A \frac{c_u - c_d}{2} \frac{\partial_\mu a}{2 f_a} \left( \bar p S^\mu p -  \bar n S^\mu n \right) 
+ 2 g_0^{ud} \frac{c_u + c_d}{2} \frac{\partial_\mu a}{2 f_a} \left( \bar p S^\mu p + \bar n S^\mu n \right) 
+ \ldots \, ,
\end{align}
where $N= (p,n)^T$ is the iso-spin doublet field, 
$v^\mu$ is the four-velocity of the non-relativistic nucleon and $S^\mu$ the spin operator. 
The couplings $g_A$ and $g_0^{ud}$ correspond respectively to the axial iso-vector and axial iso-scalar combinations, 
while the dots in \Eqn{eq:LNaunder} denote higher order terms, 
including non-derivative axion couplings which for $\vev{a}=0$ 
(no extra sources of CP violation) are at least quadratic in $a$.  
Matching the two effective Lagrangians over a single-nucleon matrix element, for example 
$\langle p | \mathcal{L}_a | p \rangle = \langle p | \mathcal{L}_N | p \rangle$, 
at the LO in the isospin breaking effects, we get
\begin{multline}
\frac{\partial_\mu a}{2 f_a} c_u \underbrace{\langle p | \bar u \gamma^\mu \gamma_5 u  | p \rangle}_{s^\mu \Delta u} +
\frac{\partial_\mu a}{2 f_a} c_d \underbrace{\langle p | \bar d \gamma^\mu \gamma_5 d  | p \rangle}_{s^\mu \Delta d} = 
\frac{\partial_\mu a}{2 f_a} g_A \frac{c_u - c_d}{2} \underbrace{2 \langle p | \bar p S^\mu p  | p \rangle}_{s^\mu} +
\frac{\partial_\mu a}{2 f_a} g_0^{ud} \frac{c_u + c_d}{2} \underbrace{2 \langle p | \bar p S^\mu p  | p \rangle}_{s^\mu} \, ,
\end{multline}
where we used the definition $2 \bar p S^\mu p = \bar p \gamma^\mu \gamma_5 p$, and $s^\mu$ is the spin of the 
nucleon at rest. Reshuffling the previous equation   
\begin{align}
g_A &= \Delta u - \Delta d \, , \\ 
g_0^{ud} &= \Delta u + \Delta d \, .
\end{align}
and substituting back into \Eqn{eq:LNaunder}, we get 
\beq 
\label{eq:LNaunder2}
\mathcal{L}_{N} \supset \frac{\partial_\mu a}{2 f_a} \left\{ 
\frac{c_u - c_d}{2} (\Delta u - \Delta d) 
\left( \bar p \gamma^\mu \gamma_5 p -  \bar n \gamma^\mu \gamma_5 n \right) +
\frac{c_u + c_d}{2} (\Delta u + \Delta d) 
\left( \bar p \gamma^\mu \gamma_5 p + \bar n \gamma^\mu \gamma_5 n \right) \right\} \, .
\eeq
The axion-nucleon coupling is defined in analogy to the axion-quark ones as 
\beq 
\frac{\partial_\mu a}{2 f_a} \bar N C_{aN} \gamma^\mu \gamma_5 N \, ,
\eeq
with $C_{aN} = \text{diag}(C_{ap},C_{an})$, 
for which we get (recall that $c_q = c^0_q - Q_a$)
\begin{align}
\label{eq:cpmatrixelem1}
C_{ap} &
= - \( \frac{m_d}{m_u + m_d} \Delta u + \frac{m_u}{m_u + m_d} \Delta d \) + 
c^0_u \Delta u + c^0_d \Delta d \, ,
\\
\label{eq:cnmatrixelem1}
C_{an} &
= - \( \frac{m_u}{m_u + m_d} \Delta u + \frac{m_d}{m_u + m_d} \Delta d \) + 
c^0_d \Delta u + c^0_u \Delta d \, , 
\end{align}
where 
$\Delta u = 0.897(27)$, 
$\Delta d = -0.376(27)$ and 
$m_u^{\rm \overline{MS}} (2 \ \text{GeV}) / m_d^{\rm \overline{MS}} (2 \ \text{GeV}) = 0.48 (3)$ 
\cite{diCortona:2015ldu}. 

\subsubsection{Axion-electron coupling}
\label{sec:axionelectron}

The axion-electron coupling is defined via the Lagrangian term 
\beq 
\label{eq:Caedef}
C_{ae} \frac{\partial_\mu a}{2 f_a} \bar e \gamma^\mu \gamma_5 e \, , 
\eeq
where $C_{ae} = c^0_e + \delta c_e$. 
In models where the tree-level contribution, $c^0_{e}$, is zero, 
the axion-electron coupling can still be generated radiatively. 
The relevant one-loop diagram is logarithmically divergent, and can be 
understood as an RGE effect on the $C_{ae}$ coefficient 
from the PQ scale down to the IR scale $\mu_{\rm IR}$
\cite{Georgi:1986df}. 
One finds \cite{Srednicki:1985xd,Chang:1993gm}\footnote{As pointed out 
by \cite{Chang:1993gm}, the original expression in Ref.~\cite{Srednicki:1985xd} contains a typo.} 
\beq 
\label{eq:Caeradiative}
\delta c_e = \frac{3\alpha^2}{4\pi^2} 
\[ 
\frac{E}{N}
\log\( \frac{f_a}{\mu_{\rm IR}} \)
- \frac{2}{3} \frac{4 m_d + m_u}{m_u + m_d} 
\log\( \frac{\Lambda_\chi}{\mu_{\rm IR}} \) \] \, . 
\eeq
where $E/N$ is related to $g^0_{a\gamma}$ 
via \Eqn{eq:g0agammadefPQ} as explained in the next Section. 
The part proportional to $E/N$ corresponds to the running 
between $f_a$ and $\mu_{\rm IR} < \Lambda_\chi$, while the second term 
arises from axion-pion mixing
and is cut-off at the chiral symmetry breaking scale, 
$\Lambda_\chi \simeq 1$ GeV, since for loop momenta 
larger than $\Lambda_\chi$ the effect of the colour anomaly is negligible.
The IR parameter $\mu_{\rm IR}$ should be taken 
of the order of the energy scale 
relevant to the physical process under consideration, 
typically $\mu_{\rm IR} = m_e$. 
\subsection{Origin of model-dependent axion couplings}
\label{sec:modeldepax}

Before discussing explicit axion models, it is useful to describe 
in a general way how the 
`model-dependent' axion couplings $g^0_{a\gamma}$ and $c^0_q$ 
introduced in the axion effective Lagrangian \Eqn{eq:Leffaxion2} 
arise from the point of view of a spontaneously broken $U(1)_{\rm PQ}$ symmetry. 
Let us denote by $J^{\rm PQ}_{\mu}$ the associated PQ current, which is 
conserved up to anomalies 
\beq 
\label{eq:dJPQAnomal}
\partial^\mu J^{\rm PQ}_{\mu} = 
\frac{g^2_s N}{16\pi^2} 
G \tilde G
+\frac{e^2 E}{16\pi^2} 
F \tilde F
\, , 
\eeq
where $N$ and $E$ are respectively the 
QCD and EM 
anomaly coefficients. 
From the Goldstone theorem $\langle 0 | J^{\rm PQ}_{\mu}  | a \rangle = i v_a \, p_\mu$,  
where the axion $a$ is the pseudo Goldstone boson of 
$U(1)_{\rm PQ}$ breaking and we introduced the order parameter $v_a$. 
The axion effective Lagrangian contains the terms
\beq 
\label{eq:Leffaxion3VVV}
\mathcal{L}_{a} \supset 
\frac{a}{v_a} \frac{g^2_s N}{16\pi^2} 
G \tilde G
+ \frac{a}{v_a} \frac{e^2 E}{16\pi^2} 
F \tilde F
+ \frac{\partial_\mu a}{v_a} J^{\rm PQ}_{\mu}
\, ,    
\eeq
where the first two terms are required by anomaly matching and the 
PQ current depends on the global charges of the fields transforming under $U(1)_{\rm PQ}$. 
E.g.~for a chiral SM fermion $f_L$ one has  
$J^{\rm PQ}_{\mu} |_{f_L} =- \bar f_L \mX_{f_L} \gamma^\mu f_L$, 
where $\mX_{f_L}$ denotes its PQ 
charge. Using the standard normalization of the $G\tilde G$ term in terms of $f_a$ 
as in \Eqn{eq:Leffaxion2} yields 
\beq 
\label{eq:favsvadef}
f_a = \frac{v_a}{2N} \, . 
\eeq
Hence \Eqn{eq:Leffaxion3VVV} can be rewritten as 
(taking for illustrative purposes just two chiral fermions $f_L$ and $f_R$)
\begin{align} 
\label{eq:Leffaxion4}
\mathcal{L}_{a} &\supset 
\frac{a}{f_a} \frac{g^2_s}{32\pi^2} 
G \tilde G
+ \frac{a}{f_a} \frac{e^2}{32\pi^2} \frac{E}{N}
F \tilde F
- 
\frac{\partial_\mu a}{2 f_a } \frac{1}{N} \[ \bar f_L \mX_{f_L} \gamma^\mu f_L  + \bar f_R \mX_{f_R} \gamma^\mu f_R \] \, , \nonumber \\
&= \frac{a}{f_a} \frac{g^2_s}{32\pi^2} 
G \tilde G 
+ \frac{1}{4} g^0_{a\gamma} a
F \tilde F
+ \frac{\partial_\mu a}{2f_a} \bar f c^0_f \gamma^\mu \gamma_5 f 
\, ,   
\end{align}
where in the second step we have dropped 
the coupling with the conserved vector current 
since the corresponding term vanishes upon integration by parts. 
The axion-photon couplings 
are thus defined as 
\beq 
\label{eq:g0agammadefPQ}
g^0_{a\gamma} = \frac{\alpha}{2 \pi f_a} \frac{E}{N} \, , 
\eeq
and the axion coupling to the fermion $f$ as 
\beq
\label{eq:c0fPQcharges}
c^0_f =
\frac{\mX_{f_L} - \mX_{f_R}}{2N}  
= \frac{\mX_{H_f}}{2N}
\, ,   
\eeq
where in the last step, assuming a Yukawa term $\bar f_L f_R H_f$, 
we have replaced the fermion PQ charges with the charge $\mX_{H_f}$ of the corresponding Higgs.
While the expressions above have a general validity in terms of 
the defining properties of the $U(1)_{\rm PQ}$ symmetry (i.e.~its anomalous content and 
the global charge assignments), in the following we will illustrate how to derive them in the context 
of specific UV models.



\subsection{Benchmark axion models}
\label{sec:UVcomp}

We now move to the discussion of explicit axion models, 
which provide a UV completion for the axion effective Lagrangian in \Eqn{eq:Leffaxion2}. 
The simplest realisation of the PQ mechanism is given by the 
Weinberg-Wilczek (WW) model \cite{Weinberg:1977ma,Wilczek:1977pj}, 
in which the QCD anomaly of the $U(1)_{\rm PQ}$ current 
is generated by SM quarks charged under the PQ symmetry, 
while the scalar sector is extended via an extra Higgs 
doublet in order to enforce the additional $U(1)_{\rm PQ}$ symmetry. 
In the WW model the axion decay constant, $f_a = (v/6)\sin2\beta$, 
with $\tan\beta=v_u/v_d$, is of  
the order of the 
electroweak scale $v \simeq 246$ GeV.   
Hence, being the axion coupling to SM fields not sufficiently suppressed, 
the WW model was soon ruled out by laboratory searches.\footnote{The original 
WW model was ruled out by a combination of 
beam dump experiments \cite{Donnelly:1978ty} and rare meson 
decays such as $K\to\pi a$ \cite{Hall:1981bc} and $\text{Quarkonia} \to \gamma a$ \cite{Wilczek:1977zn}. 
For a historical account see for instance Sect.~3 in Ref.~\cite{Davier:1986ps}.
However, this was under the assumption of universality of the PQ charges. 
Variant axion models of the WW type 
(i.e.~with non-universal PQ charges and the PQ breaking connected to the electroweak scale), 
took instead almost a decade to be ruled out from rare $\pi$ and $K$ meson 
decays \cite{Bardeen:1986yb}. 
\label{foot:WWruledout}}  
This led to the so-called ``invisible axion'' models, in which the 
PQ symmetry breaking is decoupled from the electroweak scale 
via the introduction of a SM singlet scalar field, 
acquiring a VEV $v_a \sim f_a \gg v$. 
Axion's interactions 
are then parametrically suppressed as $1/f_a \ll 1/v$. 

UV completions of the axion effective Lagrangian 
can be divided in two large classes,  
according to the way the 
QCD anomaly of the  
$U(1)_{\rm PQ}$ current is realised. 
In models of the 
Dine-Fischler-Srednicki-Zhitnitsky (DFSZ) type \cite{Zhitnitsky:1980tq,Dine:1981rt} 
the anomaly is carried by SM quarks (as in the WW model), 
while models of the Kim-Shifman-Vainshtein-Zakharov (KSVZ) 
type \cite{Kim:1979if,Shifman:1979if} require new coloured fermions. 
Since these two constructions provide the building blocks 
of most of the models considered in this report, 
we review them here in detail.

\subsubsection{KSVZ axion}
\label{sec:KSVZ}

The KSVZ model \cite{Kim:1979if,Shifman:1979if}
extends the SM field content with a vector-like fermion $\Q = \Q_L + \Q_R$
in the fundamental of colour, singlet under $SU(2)_L$, and neutral under hypercharge:
$\Q \sim (3,1,0)$,  
and a SM-singlet complex scalar $\Phi \sim (1,1,0)$.  
In the absence of a bare mass term for $\Q$,\footnote{In the spirit 
of having the PQ to arise as an accidental global symmetry (cf.~\sect{sec:PQquality}), 
this can be enforced e.g.~via 
the discrete gauge symmetry \cite{Kim:1979if}: 
$\Q_L \to -\Q_L$, $\Q_R \to \Q_R$, $\Phi \to - \Phi$.} the Lagrangian 
\beq 
\label{eq:LaKSVZ1}
\mathcal{L}_{\rm KSVZ} = \abs{\partial_\mu \Phi}^2 + \bar \Q i \slashed{D} \Q 
- \( y_\Q \bar \Q_L \Q_R \Phi + \text{h.c.} \) - V(\Phi) \, ,  
\eeq
features a $U(1)_{\rm PQ}$ symmetry
\beq 
\label{eq:KSVZPQtransf}
\Phi \to e^{i\alpha} \Phi \, , \qquad
\Q_L \to e^{i\alpha/2} \Q_L \, , \qquad 
\Q_R \to e^{-i\alpha/2} \Q_R \, .
\eeq 
The potential 
\beq 
\label{eq:VPhiKSVZ}
V(\Phi) = \lambda_\Phi \( \abs{\Phi}^2 - \frac{v_a^2}{2} \)^2 \, ,
\eeq
is such that the $U(1)_{\rm PQ}$ symmetry is spontaneously broken,  
with order parameter $v_a$. 
Decomposing the scalar field in polar coordinates
\beq 
\label{eq:PhidecKSVZ}
\Phi = \frac{1}{\sqrt{2}} (v_a + \varrho_a) e^{i a/\va} \, , 
\eeq
the axion field $a$ corresponds to the Goldstone mode (massless at tree level), 
while the radial mode $\varrho_a$ picks up a mass $m_{\varrho_a} = \sqrt{2\lambda_\Phi} v_a$. 
In the PQ broken phase also the fermion $\Q$ gets massive, with 
$m_\Q = y_\Q v_a / \sqrt{2}$. 

The Lagrangian term (where we neglected the heavy scalar radial mode)
\beq 
\label{eq:LaKSVZ2}
\mathcal{L}_{\rm KSVZ} 
\supset 
- m_\Q
\bar \Q_L \Q_R e^{ia/\va} 
+ \text{h.c.} \, ,
\eeq
is responsible for the generation of the $aG\tilde G$ operator in the effective theory below $m_\Q$. 
To see that, let us perform a field-dependent axial transformation:  
\beq 
\label{eq:transfQKSVZ}
\Q \to e^{-i\gamma_5\frac{a}{2 v_a}} \Q \, , 
\eeq
or, equivalently, 
$\Q_L \to e^{i\frac{a}{2 v_a}} \Q_L$ and 
$\Q_R \to e^{-i\frac{a}{2 v_a}} \Q_R$. 
In the transformed variables 
the field $\Q$ is now disentangled from the axion, so we can 
safely integrate it out. Moreover, being the transformation in \Eqn{eq:transfQKSVZ} 
anomalous under QCD, one gets (e.g.~from the non-invariance of the 
path integral measure \cite{Fujikawa:1979ay}): 
\beq 
\label{eq:deltaKSVZ1}
\delta \mathcal{L}_{\rm KSVZ} = \frac{g_s^2}{32\pi^2} \frac{a}{v_a} 
G \tilde G
\, , 
\eeq
where we have used the fact that $\Q$ is in the fundamental of colour. 
In such a case one can identify $v_a = f_a$ (cf.~\Eqn{eq:Leffaxion2}), and the only coupling of the axion 
with the SM fields is via the $aG\tilde G$ term (model-independent contribution)
discussed in \sect{sec:AxionEFT}.  

For later purposes (cf.~\sect{sec:KSVZ-like}) 
we discuss the KSVZ 
model in the more general setup in  which the heavy fermions $\Q$ reside in a generic reducible 
representation 
$\sum_\Q (\cal{C}_\Q,\cal{I}_\Q,\cal{Y}_\Q)$ 
of the $SU(3)_c\times SU(2)_L\times U(1)_Y$  gauge group. 
The only requirement for the PQ mechanism to work, 
is that at least one $\cal{C}_\Q$ is a non-trivial representation. 
The PQ current will have in general both a QCD and EM anomaly, 
represented by the anomaly coefficients $E$ and $N$ 
(see definition in \Eqn{eq:dJPQAnomal}) which read 
\beq 
\label{eq:EN}
N=\sum_\Q N_\Q \, , \qquad 
E=\sum_\Q E_\Q \, , 
\eeq
with $N_\Q$ and $E_\Q$ denoting the contributions to the anomalies of each irreducible representation 
(all taken to be left-handed, so in particular $\mX_{\Q^c_L} = - \mX_{\Q_R}$)
\begin{align}
\label{eq:NKSVZ}
N_\Q &= \mX_{\Q} d({\cal{I}}_{\Q}) T(\cal{C}_{\Q}) \, , \\
\label{eq:EKSVZ}
E_\Q &= \mX_{\Q} d({\cal{C}}_{\Q}) \Tr q^2_{\Q} = \mX_{\Q}
d({\cal{C}}_{\Q}) d({\cal{I}}_{\Q}) 
\( \frac{1}{12} \(d({\cal{I}}_{\Q})^2 -1\) + {\cal{Y}}_{\Q}^2 \) 
\, .
\end{align} 
Here $d(\cal{C}_{\Q})$ and $d(\cal{I}_{\Q})$ denote the dimension of the  
colour and weak isospin representations,  
$T(\cal{C}_\Q)$ is the colour Dynkin index 
(with standard normalisation $T(3)=1/2$, $T(6)=5/2$, $T(8)=3$, $T(15)=10$, etc.), 
$q_{\Q} = T_{\Q}^{(3)} + {\cal{Y}}_\Q$ denotes the $U(1)_{\rm EM}$ charge generator, 
and $\mX_{\Q_R}=-\mX_{\Q_L}  = \mp 1/2$ depending if 
in \Eqn{eq:LaKSVZ1}
the quark bilinear 
$\bar \Q_L \Q_R$ 
couples to $\Phi$ or $\Phi^\dagger$. 
Hence, after removing the axion field from the Yukawa Lagrangian 
via the transformation in \Eqn{eq:transfQKSVZ},
one gets the effective anomalous interactions (with $\alpha_s = g_s^2/(4\pi)$, etc.)
\beq 
\label{eq:deltaKSVZgen1}
\delta \mathcal{L}_{\rm KSVZ} = 
\frac{\alpha_s N}{4\pi} \frac{a}{v_a} 
G \tilde G
+\frac{\alpha E}{4\pi} \frac{a}{v_a} 
F \tilde F
\, . 
\eeq
It is customary to normalise the first term as in the axion effective Lagrangian 
\Eqn{eq:Leffaxion2}, so that 
\beq 
\label{eq:deltaKSVZgen2}
\delta \mathcal{L}_{\rm KSVZ} = 
\frac{\alpha_s}{8\pi} \frac{a}{f_a} 
G \tilde G
+\frac{\alpha}{8\pi} \frac{E}{N} \frac{a}{f_a} 
F \tilde F
\, ,  
\eeq
where the relation between the 
coefficient of the EM term and the effective coupling introduced in \Eqn{eq:Leffaxion2} 
is $g^0_{a\gamma} = \frac{\alpha}{2\pi f_a} \frac{E}{N}$ with $f_a$ defined in~\eqn{eq:favsvadef}. 
%
From \Eqn{eq:PhidecKSVZ} we see that the 
axion is defined as an angular variable over the domain $[0,2\pi v_a)$.\footnote{
For $\mX_\Phi \neq 1$ the axion domain would be instead $[0,2\pi v_a/\mX_\Phi)$.}
On the other hand, the QCD induced axion potential is periodic 
in $[0,2\pi f_a)$ (cf.~\Eqn{eq:Vapi0} or \Eqn{eq:Vapi}).  
We operatively define the domain wall (DW)   number,  will be important in cosmology, in terms  
of the QCD anomaly factor  
\beq 
 \label{eq:DWdef}
 N_{\rm DW} \equiv 2 N\,, 
\eeq
as the number of inequivalent  degenerate minima of the axion potential, that correspond to 
$\theta = 2\pi n/N_{\rm DW} $ with   $n\in \{0,1,\ldots,N_{\rm DW}-1\}$.
The consequences of DWs in cosmology will be discussed in \sect{sec:defects}, while the cosmological DW problem,  which arises in models with $N_{\rm DW} >1$, is reviewed in \sect{sec:AxionsDW}, together with possible solutions.
Here we just remark that the original KSVZ construction   
with  one vector-like pair of  heavy quarks, singlets under $SU(2)_L$ ($d({\cal{I}}_{\Q})=1$),  
  in the fundamental of $SU(3)$  ($T({\cal{C}}_{\Q})=1/2$),  
and with PQ charges  $\mX_{\Q_L}=-\mX_{Q_R} =1/2$ 
as follows from \eqn{eq:KSVZPQtransf},
belongs to the class of   $N_{\rm DW} =1$ models.


\subsubsection{DFSZ axion}
\label{sec:DFSZ}

The field content of the DFSZ model
includes two Higgs doublets 
$H_u \sim (1,2,-\tfrac{1}{2})$ and $H_d \sim (1,2,+\tfrac{1}{2})$
and a SM-singlet complex scalar field, $\Phi \sim (1,1,0)$.   
The latter extends the WW model, 
allowing to decouple the PQ breaking scale from the electroweak scale. 
We write the renormalisable scalar potential as\footnote{We couple the 
Higgs bilinear to $\Phi^\dag$ rather than to $\Phi$ to 
ensure positive values of the anomaly coefficients (see below) 
while maintaining the convenient charge normalisation $\mX_{\Phi} =1$,  
as  in \eqn{eq:fixedXDFSZ}. 
A different possibility for the Higgs coupling to the PQ breaking field 
is to replace the quartic coupling with  the super-renormalisable operator $H_u H_d \Phi^\dag$.  
This is a physically distinct choice 
as it implies $N_{\rm DW} = 3$, that is half of the one in standard DFSZ 
(cf.~discussion below \Eqn{eq:delLDFSZ}). 
Note, however, that none of the previous  two choices has effects on the 
axion couplings.\label{foot:HHPhi}} 
\beq
\label{eq:VDFSZ}
V(H_u, H_d, \Phi) = 
\tilde V_{\rm moduli} (\abs{H_u}, \abs{H_d}, \abs{\Phi}, \abs{H_u H_d}) + 
\lambda \, H_u H_d \Phi^{\dag 2} + \text{h.c.} \, . 
\eeq
%
\Eqn{eq:VDFSZ} contains all the 
moduli terms allowed by gauge invariance 
plus a non-hermitian 
operator 
which is responsible for the explicit breaking 
of the re-phasing symmetry of the three scalar fields 
into two linearly independent $U(1)$'s, to be identified with the 
hypercharge and the PQ symmetry\beq 
\label{eq:U(1)breakDFSZ}
U(1)_{H_u} \times U(1)_{H_d} \times U(1)_{\Phi} \to U(1)_Y \times U(1)_{\rm PQ} \, .      
\eeq
The action of the PQ symmetry on the fermion fields is taken to be the 
same for all the generations and, in the case of the DFSZ-I model, is determined by 
the following Yukawa Lagrangian 
\beq 
\label{eq:yukDFSZ}
\mathcal{L}_{\rm DFSZ-I}^Y = 
-Y_U \bar q_L u_R H_u  
-Y_D \bar q_L d_R H_d  
-Y_E \bar \ell_L e_R H_d + \text{h.c.} \, .  
\eeq
Alternatively, one can couple $\tilde H_u = i\sigma_2 H_u^*$ in the lepton sector,  
which goes under the name of DFSZ-II variant, 
\beq 
\label{eq:yukDFSZII}
\mathcal{L}_{\rm DFSZ-II}^Y = 
-Y_U \bar q_L u_R H_u  
-Y_D \bar q_L d_R H_d  
-Y_E \bar \ell_L e_R \tilde H_u  + \text{h.c.} \, .  
\eeq
By means of a proper scalar potential in \Eqn{eq:VDFSZ} one can ensure that 
all the three scalar fields pick up a VEV 
\beq
\label{eq:axiondirDFSZ}
H_u \supset 
  \frac{v_u}{\sqrt{2}} e^{i\frac{a_u}{v_u}} 
\, \binom{1}{0}
\, , \qquad
H_d \supset 
\frac{v_d}{\sqrt{2}} e^{i\frac{a_d}{v_d}} 
\,
\binom{0}{1}
\, , \qquad
\Phi \supset 
\frac{v_\Phi}{\sqrt{2}} e^{i\frac{a_\Phi}{v_\Phi}}
\, , 
\eeq
where $v_\Phi \gg v_{u,d}$ and we have neglected EM-charged and radial
modes that do not contain the axion. 
Note that  the parametrisation of the singlet field $\Phi$ in the last relation differs 
from the one used in \eqn{eq:PhidecKSVZ}  in that $a \to a_\Phi$ and $v_a \to v_\Phi$. 
This distinction is necessary whenever, as in DFSZ models,  besides the singlet angular mode  
$a_\Phi$,  Goldstone bosons of  other scalar multiplets concur to define the physical axion $a$, 
and additional VEVs contribute to its dimensional normalisation factor $v_a$.  
In order to identify the axion field $a$ in terms of   $a_{u,d,\Phi}$  let us  write down the 
PQ current 
\begin{align} 
\label{eq:JPQDFSZ}
J^{\rm PQ}_{\mu} &= 
-\mX_\Phi \Phi^\dag i \overset\leftrightarrow{\partial_\mu} \Phi  
-\mX_{H_u} H_u^\dag i \overset\leftrightarrow{\partial_\mu} H_u 
-\mX_{H_d} H_d^\dag i \overset\leftrightarrow{\partial_\mu} H_d 
+\ldots \ \ 
 \supset  \ \ J^{\rm PQ}_{\mu} |_a =
\sum_{i=\Phi,u,d} \mX_i v_i \partial_\mu a_i  \,, 
\end{align}
where  the dots stand for the fermion contribution to the current, 
while in $J^{\rm PQ}_{\mu} |_a$  we have retained  only  the $a_{u,d,\Phi}$ fields and 
defined  $\mX_{u,d} = \mX_{H_u,H_d}$  to compactify the result.
The axion field  is now defined as \cite{Srednicki:1985xd} 
\beq 
\label{eq:defaxionDFSZ}
a = \frac{1}{v_a} 
\sum_i \mX_i v_i a_i\,, 
\qquad 
v_a^2 = \sum_i \mX_i^2 v^2_i \,, 
%
\eeq
so that $J^{\rm PQ}_{\mu} |_a = v_a \partial_\mu a $ and, 
compatibly with Goldstone theorem, $\langle 0 | J^{\rm PQ}_{\mu} |_a | a \rangle = i v_a p_\mu$. 
Note, also, that 
under a PQ transformation $a_i \to a_i + \kappa \mX_i v_i$  
the axion field transforms as $a \to a + \kappa v_a$. 
The PQ charges in the scalar sector can be determined by requiring: 
$i)$ PQ invariance of  the operator $H_u H_d \Phi^{\dag 2}$, 
which implies $\mX_{H_u} + \mX_{H_d} - 2 \mX_\Phi = 0$, and $ii)$
orthogonality between  $J^{\rm PQ}_{\mu}|_a$ in \eqn{eq:JPQDFSZ} and the 
corresponding contribution to the hypercharge  current  $J^{\rm Y}_{\mu} |_a
= \sum_i Y_i v_i \partial_\mu a_i $,  
which implies $\sum_i  2 Y_i \mX_i v_i^2 = -\mX_{H_u} v_u^2 + \mX_{H_d} v_d^2 = 0$. 
The latter condition ensures that there is no kinetic mixing between the physical axion and 
the $Z$ 
boson.\footnote{This canonical form is required in order to formally integrate out the 
$Z$ boson, via a gaussian integration in the path-integral, when defining the axion EFT.} 
All charges are hence fixed up to an overall 
normalisation\footnote{Physical quantities 
such as axion couplings and the DW number 
do not depend on this normalisation, as it can be readily verified 
by repeating all the steps above for a generic $\mX_{\Phi}$.} 
that can be fixed by choosing 
a conventional  value  for 
$\mX_{\Phi}$: 
\beq 
\label{eq:fixedXDFSZ}
\mX_{\Phi} = 1 \, , \qquad 
\mX_{H_u} = 2 \cos^2\beta \, , \qquad 
\mX_{H_d} = 2 \sin^2\beta \, ,
\eeq
where we have defined $v_u/v = \sin\beta$, $v_d/v = \cos\beta$, 
with $v \simeq 246$ GeV. 
Substituting these expressions into \Eqn{eq:defaxionDFSZ} we obtain: 
\beq
\label{eq:acanonDFSZ}
v^2_a = v_\Phi^2 + v^2 (\sin2\beta)^2\, ,
\eeq
and given that $v_\Phi \gg v$  we have  $v_a\simeq v_\Phi $.
The axion coupling to SM fermions can be derived by 
inverting the first relation in \eqn{eq:defaxionDFSZ}
 to express  $a_{u,d}$ in terms of $a$ and select the $a$ dependent terms. This boils down 
 to replace  
$a_u / v_u \to \mX_{H_u} a / v_a$, 
$a_d / v_d \to \mX_{H_d} a / v_a$ and 
yields   
\begin{align}
\label{eq:axioncouplFERDFSZ}
\mathcal{L}_{\rm DFSZ-I} 
&\supset 
-m_U \bar u_L u_R e^{i  \mX_{H_u}  \frac{a}{v_a}}  
-m_D \bar d_L d_R e^{i \mX_{H_d} \frac{a}{v_a}}
-m_E \bar e_L e_R e^{i\mX_{H_d} \frac{a}{v_a}} + \text{h.c.} \, .
\end{align} 
The axion field can be now removed from the mass terms  
by  redefining the fermion fields according to  the field-dependent axial  
transformations:
\beq 
\label{eq:transfg5DFSZ}
u \to e^{-i\gamma_5 \mX_{H_u} \frac{a}{2 v_a}} u \, , \qquad 
d \to e^{-i\gamma_5 \mX_{H_d} \frac{a}{2v_a}} d \, , \qquad 
e \to e^{-i \gamma_5 \mX_{H_d} \frac{a}{2v_a}} e \, ,   
\eeq
which, because of the QCD and EM anomalies, 
induce an axion coupling to both $G\tilde G$ and $F\tilde F$.
Let us note in passing that since  the fermion charges satisfy the relations 
$\mX_{u_L}-\mX_{u_R}=\mX_{H_u}$, 
$\mX_{d_L}-\mX_{d_R}=\mX_{H_d}$, 
$\mX_{e_L}-\mX_{e_R}=\mX_{H_d}$ as dictated by PQ invariance of the Yukawa couplings, the transformations \eqn{eq:transfg5DFSZ} 
are equivalent to redefine the LR chiral fields with a phase transformation proportional to their PQ charges.
Specifying now \Eqns{eq:NKSVZ}{eq:EKSVZ} to the DFSZ-I case, 
one obtains
\begin{align}
\label{eq:NDFSZ1}
N &= n_g \(\frac{1}{2} \mX_{H_u} + \frac{1}{2} \mX_{H_d}\) = 3 \, , \\ 
\label{eq:EDFSZ1}
E &= n_g \( 3 \(\frac{2}{3}\)^2 \mX_{H_u} + 3 \(-\frac{1}{3}\)^2 \mX_{H_d} 
+ \(-1\)^2 \mX_{H_d}\) = 8 \, ,
\end{align}
where  $n_g = 3$  is the number of SM fermion generations 
while  $\mX_{H_{u,d}}$ are given in \eqn{eq:fixedXDFSZ}.
The anomalous part of the axion effective Lagrangian then reads
\beq
\label{eq:delLDFSZ}
\delta \mathcal{L}_{\rm DFSZ-I} 
= 
\frac{\alpha_s}{8\pi} \frac{a}{f_a} 
G \tilde G
+\frac{\alpha}{8\pi} \(\frac{E}{N}\) \frac{a}{f_a} 
F \tilde F
\, ,  
\eeq 
with $f_a = v_a/(2N)= v_a/6$ 
(hence the DW number is $N_{\rm DW} = 6$) and $E/N = 8/3$. 
The transformations in \Eqn{eq:transfg5DFSZ}, however, do not leave the  
fermion kinetic terms invariant, and their variation corresponds to  derivative  couplings of the axion to the SM 
fermion fields:
\begin{align} 
\label{eq:axuu}
\delta(\bar u i \slashed{\partial} u) &= \mX_{H_u}
 \frac{\partial_\mu a}{2 v_a} 
 \, \bar u \gamma^\mu\gamma_5 u = 
\( \frac{1}{3} \cos^2\beta \) 
\frac{\partial_\mu a}{2 f_a} 
\bar u \gamma^\mu\gamma_5 u \, , \\
\label{eq:axdd}
\delta(\bar d i \slashed{\partial} d) &=  
\mX_{H_d} \frac{\partial_\mu a}{2 v_a} 
 \, \bar d \gamma^\mu\gamma_5 d = 
\( \frac{1}{3} \sin^2\beta \) \frac{\partial_\mu a}{2 f_a}  \bar d \gamma^\mu\gamma_5 d \, , \\
\label{eq:axee}
\delta(\bar e i \slashed{\partial} e) & = 
\mX_{H_d}  \frac{\partial_\mu a}{2v_a} 
 \, \bar e \gamma^\mu\gamma_5 e = \( \frac{1}{3} \sin^2\beta \) 
\frac{\partial_\mu a}{2 f_a} 
\bar e \gamma^\mu\gamma_5 e \, , 
\end{align}
from which we can read out the effective axion-fermion couplings for DFSZ-I (cf.~the definition of the axion effective 
Lagrangian \Eqn{eq:Leffaxion2} with an  obvious extension to include the leptons):
\beq 
\label{eq:c0fDFSZdef}
c^0_{u_i} = \frac{1}{3} \cos^2\beta \, , \qquad 
c^0_{d_i} = \frac{1}{3} \sin^2\beta \, , \qquad 
c^0_{e_i} = \frac{1}{3} \sin^2\beta \, ,  
\eeq
where  $i = 1,2,3$ is a generation index. 
In the DFSZ-II model,   the leptons couple to the complex conjugate up-type Higgs $\tilde H_u$ instead than to $H_d$. This implies 
changing $\mX_{H_d} \to - \mX_{H_u} $  in 
the last (leptonic) term in equations (\ref{eq:axioncouplFERDFSZ}), (\ref{eq:transfg5DFSZ}) and     (\ref{eq:EDFSZ1})
which yields  $E=2$,   $E/N=2/3$ and $c^0_{e_i} = - \frac{1}{3} \cos^2\beta$. 
%
%

For DFSZ phenomenological studies  it is important to determine 
the allowed range for $\tan\beta=v_u/v_d$, which is set by the 
perturbative range of the top and bottom Yukawa couplings.   
A conservative limit is obtained by imposing a (tree-level) unitarity bound 
on Yukawa-mediated $2\to 2$ fermion scattering amplitudes 
at $\sqrt{s} \gg M_{H_{u,d}}$: $|\Re a_{J=0}| < 1/2$, where $a_{J=0}$ is the $J=0$ partial wave. 
Ignoring running effects, which would make the bound even stronger, 
and taking into account group theory factors \cite{DiLuzio:2016sur,DiLuzio:2017chi}, 
one gets \cite{Bjorkeroth:2019jtx}: $y_{t,b}^{\rm DFSZ} <  \sqrt{16\pi/3}$ 
(the strongest bound comes from the channel 
$Q_L u_R (d_R) \to Q_L u_R (d_R)$, 
with the initial and final states prepared into 
$SU(3)_c$ singlets). 
These translate into a lower and an upper bound on $\tan\beta$, 
that can be obtained via the relations 
$y_t^{\rm SM} = \sqrt{2} m_t / v = y_t^{\rm DFSZ} \sin\beta$, 
$y_b^{\rm SM} = \sqrt{2} m_b / v = y_b^{\rm DFSZ} \cos\beta$.
Using $m_t = 173.1$ GeV, $m_b = 4.18$ GeV and $v=246$ GeV 
yields the range 
\beq
\label{eq:pertrangetanbeta}
\tan\beta \in \[ 0.25, 170 \] \, .
\eeq
Note that the range in \Eqn{eq:pertrangetanbeta} holds both for DFSZ-I and DFSZ-II, 
since the $\tau$ Yukawa plays a sub-leading role for perturbativity.

\subsection{Summary of flavour and CP conserving axion couplings} 
\label{sec:summaryaxioncoupl}

Focussing on the most relevant axion couplings 
from the point of view of astrophysical constraints 
(see \sect{sec:Astro_bounds}) and 
experimental sensitivities (see \sect{sec:Experiments}), 
we collect here their numerical values  
including available higher-order corrections. 
Flavour and CP violating axion couplings will be discussed instead 
in the next two Sections. 

The relation between the axion mass and the axion decay constant  
(see \Eqn{eq:axionmass} for a LO expression), has been 
computed including QED and NNLO corrections in the chiral expansion \cite{Gorghetto:2018ocs} 
and reads
\beq 
\label{eq:axionmassfa}
m_a = 5.691(51) \( \frac{10^{12} \ \text{GeV}}{f_a} \) \, \text{$\mu$eV} \, . 
\eeq
A direct calculation of the topological susceptibility (see Eqs.~(\ref{eq:defK}) 
and (\ref{eq:chi0}) below) via QCD lattice techniques
finds a similar central value, with an error five times larger \cite{Borsanyi:2016ksw}. 

The axion interaction Lagrangian with photons,   
matter fields $f=p,n,e$, pions and the nEDM operator
can be written as
\begin{equation} 
\label{eq:Laint1}
\mathcal{L}^{\rm int}_a \supset \frac{\alpha}{8 \pi} \frac{C_{a\gamma}}{f_a} a F \tilde F
+ C_{af} \frac{\partial_\mu a}{2 f_a} \bar f \gamma^\mu \gamma_5 f 
+ \frac{C_{a\pi}}{f_a f_\pi} 
\partial_\mu a [\partial \pi\pi\pi]^\mu  
- \frac{i}{2} \frac{C_{an\gamma}}{m_n} \frac{a}{f_a}  \bar n \sigma_{\mu\nu} \gamma_5 n F^{\mu\nu} 
\, ,
\end{equation}
where we have schematically defined $[\partial \pi\pi\pi]^\mu 
= 2 \partial^\mu \pi^0 \pi^+ \pi^- - \pi_0 \partial^\mu \pi^+ \pi^- 
- \pi_0 \pi^+ \partial^\mu \pi^-$. 
The LO values of the $C_{ax}$ coefficients have been derived in \sect{sec:AxionEFT}. 
Taking into account 
Next-to-LO (NLO) chiral corrections for the axion-photon coupling 
and a LO non-relativistic effective Lagrangian approach for axion-nucleon 
couplings,\footnote{Axion-nucleon couplings in the framework of the NNLO 
chiral Lagrangian have been recently considered in \cite{Vonk:2020zfh}.} 
and including as well running effects,  
Ref.~\cite{diCortona:2015ldu} finds   
\begin{align}
\label{eq:Cagamma}
C_{a\gamma} &= \frac{E}{N} - 1.92(4) \, , \\
\label{eq:Cap}
C_{ap} &= -0.47(3) + 0.88(3) \, c^0_u - 0.39(2) \, c^0_d - C_{a,\, \text{sea}}
\, , \\
\label{eq:Can}
C_{an} &= -0.02(3) + 0.88(3) \, c^0_d - 0.39(2) \, c^0_u - C_{a,\, \text{sea}}
\, , \\ 
\label{eq:Casea}
C_{a,\, \text{sea}} &= 0.038(5) \, c^0_s 
+0.012(5) \, c^0_c + 0.009(2) \, c^0_b + 0.0035(4) \, c^0_t \, , \\
\label{eq:Cae}
C_{ae} &= c_e^0 + \frac{3\alpha^2}{4\pi^2} 
\[ \frac{E}{N} \log\( \frac{f_a}{m_e} \)
- 1.92(4)
\log\( \frac{\text{GeV}}{m_e} \) \] \, , \\
\label{eq:Capi}
C_{a\pi} &= 0.12(1) + \frac{1}{3} 
\( c^0_d - c^0_u \) \, , \\
\label{eq:Cangamma}
C_{an\gamma} &= 0.011(5) \, e \, , 
\end{align}
where we have added to the list of \cite{diCortona:2015ldu} also 
$C_{ae}$, $C_{a\pi}$ (at the LO in the chiral expansion) and 
$C_{an\gamma}$ (from the static nEDM result in \Eqn{eq:dntheory}). 

Sometimes the axion coupling to photons and matter field (first two terms in \Eqn{eq:Laint1}) 
is written as 
\beq 
\label{eq:Laint2}
\mathcal{L}^{\rm int}_a \supset 
 \frac{1}{4} g_{a\gamma} a F \tilde F 
+ g_{af} \frac{\partial_\mu a}{2 m_f} \bar f \gamma^\mu \gamma_5 f 
- \frac{i}{2} g_d \, a \, \bar n \sigma_{\mu\nu} \gamma_5 n F^{\mu\nu} \,, 
\eeq
where 
we defined 
\beq 
\label{eq:gagammagaf}
g_{a\gamma} = \frac{\alpha}{2 \pi} \frac{C_{a\gamma}}{f_a} \, , \qquad 
g_{af} = C_{af} \frac{m_f}{f_a} 
\, , \qquad 
g_d = \frac{C_{an\gamma}}{m_n f_a} 
\, . 
\eeq
The `model-independent' predictions for the axion couplings 
(namely those exclusively due to the $aG\tilde G$ operator) are obtained by setting 
$E/N \to 0$ and $c^0_i \to 0$ 
in \Eqns{eq:Cagamma}{eq:Cangamma}. 
The latter also correspond to the predictions of the simplest KSVZ model discussed 
in \sect{sec:KSVZ}, while the two DFSZ variants of \sect{sec:DFSZ} yield 
\begin{align}
\text{DFSZ-I}: \quad &E/N = 8/3 \, \quad 
c^0_{u_i} = \frac{1}{3} \cos^2\beta \, , \quad
c^0_{d_i} = \frac{1}{3} \sin^2\beta \, , \quad  
c^0_{e_i} = \frac{1}{3} \sin^2\beta \, , \\
\text{DFSZ-II}: \quad &E/N = 2/3 \, \quad 
c^0_{u_i} = \frac{1}{3} \cos^2\beta \, , \quad
c^0_{d_i} = \frac{1}{3} \sin^2\beta \, , \quad  
c^0_{e_i} = -\frac{1}{3} \cos^2\beta \, ,
\end{align}
with the index $i=1,2,3$ denoting generations and the perturbative unitarity domain 
$\tan\beta \in \[ 0.25, 170 \]$. 
In \sect{sec:axion_landscape_beyond_benchmarks} we will explore in depth how these `model-dependent' coefficients 
can be modified compared to the standard KSVZ/DFSZ benchmarks. 

For completeness, in the next two Sections we are going to discuss 
two other classes of model-dependent axion couplings
which can be of 
phenomenological interest, although they do not arise to 
a sizeable level in the standard KSVZ/DFSZ benchmarks. 
These are namely flavour violating axion couplings (\sect{sec:IntroFlavourViolating}) 
and CP-violating ones (\sect{sec:CPvaxioncoupl}).

\subsection{Flavour violating axion couplings}
\label{sec:IntroFlavourViolating}

Relaxing the hypothesis of the universality of the PQ current in 
DFSZ-like constructions leads to flavour violating axion couplings to quarks and leptons. 
This option will be explored in detail in \sect{sec:gaFCNC}.  
Here, we preliminary show how 
such couplings arise in a generalised DFSZ setup with 
non-universal PQ charges. 
Let us assume that quarks with the same EM charge but of different generations couple 
to different Higgs doublets, for definiteness $H_{1}$ or $H_{2}$, to which we assign  
the same hypercharge $Y_{H_1}=Y_{H_2}=-\frac{1}{2} $ but  different PQ charges $\mX_{1}\neq \mX_{2}$.
Let us start by considering the following Yukawa terms for the up-type quarks
\begin{equation}
\label{eq:H1H2}
\mathcal{L}^{Y_U}_{12}  = 
- (Y_U)_{11} \,\bar q_{1L} u_{1R} H_1
-(Y_U)_{22}\, \bar q_{2L} u_{2R} H_2
-(Y_U)_{12}\, \bar q_{1L} u_{2R} H_1 +\dots \, .
\end{equation}
The quark bilinear $ \bar q_{1L} u_{2R}$ in  the last term (or alternatively a similar term in the down-quark sector)  
is needed to generate the CKM mixing, and for the present discussion  it is irrelevant whether it  couples to $H_1$ or $H_2$. 
Note, also, that from PQ charge consistency  $\mX\(\bar q_{2L} u_{1R}\) = 
\mX\(\bar q_{2L} u_{2R}\) - \mX\(\bar q_{1L} u_{2R}\) + \mX\(\bar q_{1L} u_{1R}\) = -\mX_2 $ 
it follows  that  the  term  $\bar q_{2L} u_{1R} H_2$ is also allowed. 
However, being its structure determined by the first three terms we do not need to consider 
it explicitly.   
Projecting out from the Higgs doublets the neutral Goldstone bosons, as was done in \eqn{eq:axiondirDFSZ}, 
and identifying the axion field, we obtain the analogous of  \eqn{eq:axioncouplFERDFSZ} in the form
\begin{equation}
\label{eq:Lmu12}
\mathcal{L}_{12}^{m_U} = 
- (m_u)_{11} \,\bar u_{1L} u_{1R}\, e^{i\mX_1 \frac{a}{v_a}} 
- (m_u)_{22} \,\bar u_{2L} u_{2R} \,e^{i\mX_2 \frac{a}{v_a}}
- (m_u)_{12} \,\bar u_{1L} u_{2R} \,e^{i\mX_1 \frac{a}{v_a}} +\dots \, .
\end{equation}
Because of the presence of the mixing term, in this case it is not possible to remove 
the axion field  from the mass terms with a pure axial redefinition of the quark fields 
as in  \eqn{eq:transfg5DFSZ}, but it is necessary to introduce also a  
vectorial part in the field redefinition: 
\begin{equation}
\label{eq:u1u2Vector}
u_1 \to  e^{-i\(\gamma_5 \mX_1 +   \mX_2\) \frac{a}{2 v_a}} u_1 \, , \quad 
u_2 \to  e^{-i\(\gamma_5 \mX_2 +   \mX_1\) \frac{a}{2 v_a}} u_2 \, .
\end{equation}
%
By introducing a vector of the two quark flavours  $u = (u_1, u_2)^T$  and the two matrices of charges 
$\mX_{12} = {\rm diag}(\mX_1, \mX_2)$  and 
$\mX_{21} = {\rm diag}(\mX_2, \mX_1)$ 
the variation of the fermion kinetic terms 
due to the redefinitions in \eqn{eq:u1u2Vector} can be written as
%
%
\beq
\label{eq:Kinetic12}
\delta\(\bar u  i\slashed \partial u \) = \frac{\partial_\mu a}{2 v_a} \{
 \bar u  \mX_{12}    \gamma_\mu\gamma_5   u + \bar u \mX_{21}  \gamma_\mu u \}   \\
=   \frac{\partial_\mu a}{2 v_a}   \{ \bar u_L \mX_L \gamma_\mu u_L
+  \bar u_R  \mX_R \gamma_\mu u_R\}\,,   
\eeq
where 
$\mX_{R}=\mX_{21} + \mX_{12} = \( \mX_1 + \mX_2\) \begin{psmallmatrix}1& \\ & 1\end{psmallmatrix}$
and 
$\mX_{L}=\mX_{21} - \mX_{12} = \( \mX_2 - \mX_1\) \begin{psmallmatrix}1& \\ & -1\end{psmallmatrix}$.
The matrix of charges for the RH fields is proportional to the identity, 
and hence rotation to the mass eigenstates has no effects (this could have been 
argued already from \eqn{eq:H1H2} since the first and third Yukawa terms  imply  
$\mX(u_{1R})=  \mX(u_{2R}) = - \mX(\bar q_{1L} H_1)$).  
However,  since $\mX_L$ is not proportional to the identity, 
 once the LH  interaction states are 
rotated into the mass eigenstates  $u'_{L} = U^u_{L}\, u_{L}$ 
flavour violating couplings, controlled by  the matrix
$ \( \mX_2 - \mX_1\)  U^u_{L} \begin{psmallmatrix}1& \\ & -1\end{psmallmatrix} U^{u\dagger}_{L}$    
unavoidably appear.  Clearly, had we coupled the third term in~\eqn{eq:H1H2}
to $H_2$ rather than to $H_1$,  flavour violating couplings would have  instead 
appeared in the RH sector,  while  by extending the scheme to  three 
generations of up-type quarks, both the RH and LH sectors can be simultaneously affected.  
We will see in \sect{sec:gaFCNC} that interesting models which necessarily 
involve generation dependent PQ charges  are indeed characterised by 
flavour changing axion couplings of this type.

\subsection{CP-violating axion couplings}
\label{sec:CPvaxioncoupl}

Under some circumstances, discussed below, 
the axion field can develop CP-violating (scalar) couplings to matter fields, 
which can be parametrised via the following Lagrangian term 
\beq 
\label{eq:scalarCPvaxionPsi}
\mathcal{L}_a^{\rm CPV} 
= 
-g_{af}^S \, a \bar f f
\, .
\eeq 
In particular, 
CP-violating (scalar) axion couplings to nucleons $N={p,n}$, 
mediate new forces in the form 
of scalar--scalar (monopole--monopole) or scalar--pseudo-scalar (monopole--dipole) interactions \cite{Moody:1984ba}. 
This terminology comes from the fact that in the non-relativistic limit the scalar coupling is spin-independent, 
contrary to the case of the pseudo-scalar density. Let us consider, for instance, the monopole-monopole 
interaction. The non-relativistic potential between two nucleons $N_1$ and $N_2$
can be calculated in the inverse Born approximation (with $\vec q$ denoting the moment transferred)
\beq
\label{eq:mmpoltential}
V(r) = \int \frac{d^3 q}{(2\pi)^3} \frac{g_{N_1}^S g_{N_2}^S e^{i \vec q \cdot \vec r}}{\vec q^{\,2} + m^2_a} 
= -\frac{g_{N_1}^S g_{N_2}^S e^{-m_a r}}{4\pi r} \, ,
\eeq
which for $m_a \lesssim 1$ eV is subject to strong limits from e.g.~precision tests of Newton's 
inverse square law. Instead monopole-dipole interactions are at the base of new 
experimental setups sensitive to either $g_{aN}^S g_{an}$ or $g_{aN}^S g_{ae}$, 
that could eventually target the QCD axion,  
as discussed in \sect{sec:lab_searches}.   

CP-violating scalar axion couplings to nucleons 
are generated whenever the axion potential does not exactly 
relax the axion VEV to zero.\footnote{Since electrons always couple derivatively to the axion, 
the VEV of the latter does not generate a scalar axion coupling to electrons. Hence, 
only scalar axion couplings to nucleons are relevant for the QCD axion.} 
In the presence of extra sources of CP violation in the UV it is expected that 
$\theta_{\rm eff} = \vev{a} / f_a \neq 0$ and one finds 
\cite{Moody:1984ba}
\beq 
\label{eq:fromthetaefftogaN}
g_{aN}^S = \frac{\theta_{\rm eff}}{f_a} \frac{m_u m_d}{m_u + m_d}  \langle N | 
(\bar u u + \bar d d) | N \rangle \approx \theta_{\rm eff} \( \frac{17 \ \text{MeV}}{f_a} \) \, ,
\eeq 
where we used the lattice matrix element $(m_u + m_d) \langle N | 
(\bar u u + \bar d d) | N \rangle / 2 \approx 38$ MeV from \cite{Durr:2015dna}. 
Higher-order corrections to \eqn{eq:fromthetaefftogaN} were recently 
computed in Ref.~\cite{Bigazzi:2019hav} by using different approximation methods. 
Regarding instead the possible values of $\theta_{\rm eff}$, 
for instance, in the SM it is expected to be (based naive dimensional analysis \cite{Georgi:1986kr})
\beq 
\label{eq:thetaeffSM}
\theta_{\rm eff}^{\rm SM}  \sim G_F^2 f_\pi^4 J_{\rm CKM} \approx 10^{-18} \, ,
\eeq 
(note that $J_{\rm CKM} = \Im V_{ud} V_{cd}^*V_{cs} V_{us}^* $ has the same spurionic properties of $\theta$, being a CP-violating 
flavour singlet)
which leads, however, to axion scalar couplings that are far from being experimentally accessible. 
New physics above the electroweak scale 
might provide extra sources of CP violation and be responsible for a sizeable 
$\theta_{\rm eff}$ which could bear some phenomenological consequences. 
Consider for definiteness the CP-violating operator (colour EDM) 
\beq 
\label{eq:OCP}
\mathcal{O}_{\rm CPV} = \frac{i}{2} \tilde d_q g_s \bar q  T^a G^a_{\mu\nu} \sigma^{\mu\nu} \gamma_5 q \, , 
\eeq 
for any light quark flavour $q=u,d,s$, together with the 
QCD-dressed 1-point and 2-point axion functions, defined as 
\begin{align}
\label{eq:defK1}
\chi' &= i \int d^4 x 
\langle 0 | T \frac{\alpha_s}{8\pi} G\tilde G (x) \mathcal{O}_{\rm CPV} (0) | 0 \rangle \, , \\ 
\label{eq:defK}
\chi &= i \int d^4 x 
\langle 0 | T \frac{\alpha_s}{8\pi} G\tilde G (x) \frac{\alpha_s}{8\pi} G\tilde G (0) | 0 \rangle \, . 
\end{align}
The latter is known as topological susceptibility 
(sometimes also denoted as $K$ \cite{Shifman:1979if})
and can be also written as  the second derivative of the QCD generating 
functional \eqn{eq:ZAdef2}  (see also the QCD action density in Minkowski   \eqn{eq:QCDLag}) 
with respect to $\theta$. A comparison  with the effective  axion Lagrangian 
in \Eqn{eq:Leffaxion}  and the replacement $\frac{\partial^2}{\partial\theta^2}\big|_{\theta=0} \to f^2_a \; 
\frac{\delta^2}{\delta a^2}\big|_{a=0}$ highlights 
its relation with the axion mass:
\begin{equation}
\label{eq:chi0}
\chi = f^2_a m^2_a \,.
\end{equation}
A detailed computation that exploits  chiral Lagrangian techniques, chiral fits and 
lattice QCD results gives the value   $\chi = \(75.5(5)\,{\rm MeV}\)^2$~\cite{diCortona:2015ldu}. 
The two quantities $\chi'$ and $\chi$ enter the axion potential as 
\beq 
V(a) \approx \chi' \( \frac{a}{f_a} \) +  \frac{1}{2} \chi \( \frac{a}{f_a} \)^2 
\, ,
\eeq
where we neglected higher orders in $a/f_a \ll 1$.
Hence, the induced axion VEV is obtained by solving the tadpole equation 
\beq 
\label{eq:tadpoleaVEV}
\theta_{\rm eff} = - \frac{\chi'}{\abs{\chi}} \, .
\eeq
Using standard current algebra techniques one finds \cite{Pospelov:1997uv}
\beq 
\label{eq:thetaeffOCP}
\theta_{\rm eff} = - \frac{m_0^2}{2} \frac{\tilde d_q}{m_q} \, , 
\eeq
with $m_0^2 = \langle 0 | g_s \bar q (T^a G^a_{\mu\nu} \sigma^{\mu\nu}) q   | 0 \rangle / 
\langle 0 | \bar q q  | 0 \rangle \approx 0.8$ GeV$^2$. This shows that it is possible 
to generate values of $\theta_{\rm eff}$ which saturate the nEDM bound $\abs{\theta_{\rm eff}} \lesssim 10^{-10}$, 
e.g.~for $\tilde d_q \sim 1/(10^{12} \ \text{GeV})$. In terms of a SM gauge invariant effective 
operator above the electroweak scale, 
the scale of new physics is given by $v/\Lambda^2_{\rm NP} = \tilde d_q$, 
which yields $\Lambda_{\rm NP} \sim 10^{4}$ TeV.

\subsection{The dirty side of the axion}
\label{sec:PQquality}

One of the most delicate aspects of the PQ mechanism 
is the fact that it relies on a global $U(1)_{\rm PQ}$ symmetry, which has 
to be preserved to a great degree of accuracy in order for the axion VEV to 
be relaxed to zero,  a precision compatible with the non-observation of the neutron EDM. 
This issue is known as the \emph{PQ quality problem} 
\cite{Georgi:1981pu,Dine:1986bg,Barr:1992qq,Kamionkowski:1992mf,Holman:1992us,Ghigna:1992iv}, 
and it is reviewed below.  
 
Global symmetries are  generally considered not to be fundamental features in a QFT, 
and  this is particularly  well justified in the case of  anomalous symmetries 
which do not survive at the quantum level.  However, in some cases 
the field content of the theory together with the requirement of Lorentz 
and local gauge invariance restricts the  allowed renormalisable operators
to a set which remains invariant under some global redefinition of the fields. 
These symmetries are thus  {\it accidental} in the sense that they 
result from other first principle requirements. However, 
given that they are not imposed on the theory,  in general are 
not respected by higher-order non-renormalisable operators.
A  well known example of an accidental global symmetry   is baryon number ($B$) in the SM:  
operators carrying nonzero baryon number must have at least three quark fields to 
be colour singlets,  and then at least four fermion fields to form a Lorentz scalar. 
Thus their minimal dimension is $d=6$, whence the renormalisable Lagrangian conserves $B$.  
The high level of accuracy  of baryon number conservation required  
to comply with proton decay bounds implies that $B$-violating 
effective operators must be suppressed by a rather large  
scale  $\Lambda_{\slashed{B}} \gtrsim 10^{15}$ GeV which, however, can 
be considered natural  when  understood in terms  of some GUT dynamics.   
 
It would certainly be desirable to generate also the PQ symmetry as an accidental symmetry. 
The benchmark KSVZ and DFSZ constructions discussed in the previous  sections do not address 
this issue and, in particular, there is no first principle reason that  impedes writing in 
$\mathcal{L}_{\rm KSVZ}$  in  \eqn{eq:LaKSVZ1} a quark mass term   
$\mu_\mathcal{Q}  \bar{\mathcal{Q}}_L \mathcal{Q}_R$ or in  $\mathcal{L}_{\rm DFSZ}$   in 
\eqn{eq:VDFSZ} a direct coupling $\mu_H^2 H_u H_d$, both of which would destroy PQ 
invariance of the renormalisable Lagrangians. 
The problem becomes even more serious
once,  after  assuming  that an accidental PQ symmetry 
can be enforced in some way, one proceeds to estimate at which operator dimension the symmetry 
can  be first broken in order not to spoil the solution to the strong CP problem for a given 
value of the suppression scale.  Let us consider the following set of 
effective operators of dimension $d=2m+n$ that violate the PQ symmetry by $n$ units,
and let us assume for definiteness that they are suppressed  by 
the largest scale that can consistently appear in a QFT, the Planck scale $\mP$:
\begin{align} 
\label{eq:PQbreakMpl}
 -V^n_{\rm PQ-break} 
&= 
\frac{ \lambda_n  |\Phi|^{2m} \left(e^{-i\delta_n} \Phi^n+ e^{i\delta_n }{\Phi^\dagger}^n\right)}{\mP^{d-4}}  
\supset   \frac{ \lambda_n  f^4_a}{2} \!  \(\!\frac{f_a}{\sqrt{2} \mP}\!\)^{\!d-4}\cos\(\frac{na}{f_a} - \delta_n\) \nonumber \\
&\approx  m^2_{*}f^2_a\(\frac{\theta^2}{2}  -\frac{\theta}{n} \tan\delta_n\) \, ,
\end{align}
%
with $\lambda_n$ real and $\delta_n$ the phase of the coupling,\footnote{Precisely because these operators are not PQ invariant, 
the  PQ transformation required to remove  $\bar\theta$   from the QCD Lagrangian will in any case give a contribution    
$n\bar\theta$ to $\delta_n$. 
}
and  we have taken for simplicity a colour anomaly factor $2N=1$
so that  $\Phi = \frac{1}{\sqrt{2}} (f_a + \varrho_a) e^{i a/f_a}$. 
In the last relation we have expanded for  $\theta = \frac{a}{f_a}\ll1$ neglecting an irrelevant constant, and we have defined  
$m^2_{*} =  \frac{ \lambda_n f^2_a}{2}  \(f_a / (\sqrt{2} \mP)\)^{d-4}\!\!\cos\delta_n $.
The effect of the explicit breaking $ V^n_{\rm PQ-break}$ 
is to move the minimum of the  axion potential  away 
from the CP conserving minimum $\vev{\theta} =0$ of the QCD induced 
  potential $V(\theta)=\frac{1}{2} m^2_a f_a^2 \theta^2$  (see  \eqn{eq:Vapiexp})
and shift it   to 
\begin{equation}
\label{eq:nonzerovev}
\vev{\theta} = \frac{m^2_{*} \tan\delta_n}{n(m^2_a+m^2_{*})} \, .
\end{equation}
Taking for example operators that violate  PQ  by one unit ($n=1$) and   
assuming $ \lambda_1 \sim \tan\delta_1 \sim 1$, 
implies that  to satisfy $m_{*}^2 / m_a^2 \lesssim 10^{-10}$ 
the dimension of these operators should  be uncomfortably large: $d\geq 8,10,21$  respectively for  $f_a\sim  10^{8}, 10^{10}, 10^{15}\,$GeV.  

Of course, in QFT it is always possible to assume that there are no heavy states mediating PQ-violating interactions, 
so that operators of the type \eqn{eq:PQbreakMpl} do not arise. 
However, there is a general consensus that quantum gravity effects violate all global symmetries at some level. 
The standard argument is that particles carrying global charges can be swallowed by black holes, which subsequently 
may evaporate by emitting Hawking radiation~\cite{Hawking:1974sw}.\footnote{Local charges, such as 
the electric charge, cannot disappear because  Gauss law ensures that the electric field flux remains preserved when 
the charged particle falls into the black hole. Hence, charged black holes cannot evaporate entirely, but instead get stabilised 
as extremal black holes.}
One can then speculate  that non-perturbative quantum gravity formation of  ``virtual'' black holes  
would eventually result in an effective theory containing all types of 
operators compatible with the local gauge symmetries of the theory,
but violating global charge conservation, 
with  suppression  factors provided by appropriate powers of the cutoff scale $\mP$.
This latter feature is often justified  from the  requirement that quantum gravity effects should 
disappear when sending $\mP\to \infty$. 
However,  while  this is indeed  reasonable  for  perturbative quantum gravity   corrections, 
global charge violation is intrinsically a non-perturbative phenomenon,  and the assumption that 
the corresponding effects could be described only in terms of  operators suppressed by powers of  $\mP$
is at least questionable.  After all,  non-perturbative QCD  effects in the axion potential  are 
approximately described by a cosine term  $V(a) \approx -\Lambda_{\rm QCD}^4 \cos(a/f_a)$ 
with a positive power of $\Lambda_{\rm QCD}$, 
rather than by effective operators suppressed by inverse powers of $\Lambda_{\rm QCD}$.
Similarly,  one cannot exclude that  gravity   could give rise to operators 
like $\mP^3 (\Phi +\Phi^\dagger)$ or to other operators  not containing $\mP$ in 
the  denominator \cite{Abbott:1989jw,Coleman:1989zu,Kallosh:1995hi}.

 In the absence of  reliable ways to assess  the validity  of \eqn{eq:PQbreakMpl}  for describing quantum gravity effects, 
 other approaches have been pursued. Most noticeably, in   scenarios in which Einstein gravity is minimally coupled to the axion 
 field, non-conservation of global charges arises from  non-perturbative effects related  to   
wormholes that  can absorb the global charge and consequently  break the  symmetry. 
These effects are to some extent computable and have been studied for example in Refs.~\cite{Abbott:1989jw,Coleman:1989zu,Kallosh:1995hi} 
and, more recently, also in Ref.~\cite{Alonso:2017avz}.
These studies indicate that  in this setup global symmetries remain intact at any finite order in a perturbative expansion in $1/\mP$ 
so that  power-suppressed operators are not generated~\cite{Alonso:2017avz}. 
In fact, non-perturbative wormhole effects do generate additional cosine terms $\sim \cos(a/f-\delta_1)$ similar to the last term 
 in \eqn{eq:PQbreakMpl}. However, they come with an exponential  suppression factor $e^{-S_{\rm wh}}$ of the wormhole action. 
Then, if  $S_{\rm wh}$ is sufficiently large, as it appears possible in many cases~\cite{Kallosh:1995hi,Alonso:2017avz},  
the axion solution to the strong CP problem would not be endangered. 

The conclusion of this discussion is that, in a complete model, 
 it would be highly desirable  that the PQ symmetry could arise  as an accidental symmetry
 as a consequence of fundamental principles, like for example  gauge and Lorentz invariance.
 These principles  should not only guarantee that the symmetry remains perturbatively unbroken at the 
 renormalisable level, but they  should also  impede  writing down PQ violating effective operators of 
 dimension $d \leq 10$.  This,  unless some other mechanism can guarantee an appropriate strong 
 suppression of  the coefficients of any such higher dimensional  operator. 
In this respect,  in order to guarantee the quality of the PQ symmetry different mechanisms
have been put forth. 
They can be conceptually divided in three different classes:  
\begin{itemize}
\item \emph{Low $f_a$.} For $f_a \lesssim 10^3$ GeV, 
only $d>5$ is required in order 
not to generate a too large $\vev{\theta}$ from \eqn{eq:nonzerovev}. 
From this point of view the original WW axion model would have been 
perfectly natural, since in absence of SM-singlet fields 
the first gauge invariant PQ breaking operator
that one can write in the scalar potential is $(H_u H_d)^3$, which is $d=6$. 
Over the years, the increasing lower bounds on $f_a$
let the PQ quality problem to emerge. 
In `super-heavy' axion models, reviewed in \sect{sec:heavy}, one can evade 
the astrophysical constraints on $f_a \gtrsim 10^{8}$ GeV, 
e.g.~by modifying the QCD relation between $m_a$ and $f_a$. 
These models feature an axion decay constant of the order of $10^{4 \div 5}$ GeV, thus improving 
a lot on the PQ quality problem. 

\item \emph{Gauge protection}. 
New local symmetries can lead to an accidental PQ symmetry protected 
from higher-order PQ breaking operators, up to some fixed order.  
Various mechanisms have been proposed, based on 
discrete gauge symmetries~\cite{Dias:2002gg,Carpenter:2009zs,Harigaya:2013vja,Dias:2014osa,Harigaya:2015soa}, 
 Abelian~\cite{Barr:1992qq,Holman:1992us,Fukuda:2017ylt,Duerr:2017amf,Bonnefoy:2018ibr}
and non-Abelian~\cite{DiLuzio:2017tjx,Lee:2018yak}  gauge symmetries,   
composite dynamics 
\cite{Randall:1992ut,Dobrescu:1996jp,Redi:2016esr,Lillard:2018fdt,Gavela:2018paw}  
as well as extra-dimensional setups~\cite{Choi:2003wr,Cox:2019rro}.

\item \emph{Small coupling}. There is the possibility that the overall coupling $\lambda_n$ in \Eqn{eq:PQbreakMpl} 
is extremely tiny, even though effective operators are generated at relatively  low-dimension 
\cite{Cheung:2010hk}. 
Another possibility that was already mentioned above, 
is that  PQ breaking terms generated by quantum gravity come with a suppression  factor 
$e^{-S_{\rm wh}}$ which is  exponentially small for a sizeable wormhole action.

\end{itemize}
A more detailed account of all these mechanisms will be given in \sect{sec:protectingPQ}. 
All in all, we believe that 
theoretical efforts to search for a compelling mechanism to generate accidentally a PQ symmetry  
of the required high quality  keep being of primary importance. 
Meanwhile, this `incompleteness' of a theory, that is otherwise quite elegant  and particularly  rich of 
phenomenological implications, should not  discourage in any way experimental axion searches. 
Although theoretical UV completions of axion models and PQ protection mechanisms, for their very 
nature, tend to remain confined at high energy, thus challenging experimental tests, with the large 
number of ongoing and planned axion search experiments we are now entering  a data driven era.  
If the axion will be discovered, probing its fine properties with good accuracy could hopefully provide 
the clues needed to unveil the fundamental origin of the PQ symmetry.




\section{Axion cosmology}
\label{sec:section2}

The main goal of this Section is to review the mechanisms through which
axions contribute to the present DM energy
density. In~\sect{sec:friedmann} we briefly recall some basic
equations for cosmology and for early Universe thermodynamics, and we
establish the notations.  
The temperature dependence of the axion potential and axion mass 
is reviewed in \sect{sec:QCDsusceptibility}.  
In \sect{sec:misalignment} we
describe the mechanism of cold axions production from the misalignment 
mechanism. 
The contribution to the axion relic density from the
decay of topological defects is reviewed in \sect{sec:defects}.
\sect{sec:isocurvature} addresses the issue of isocurvature
fluctuations in the axion field and of the related bounds that can be
derived on the scale of inflation.  \sect{sec:axionmassbounds}
collects some bounds on the axion mass that can be obtained (mainly)
from cosmological considerations.  
In \sect{sec:benchmark}, we consider the important problem  of identifying  for
which mass range the axion can account for the totality of the DM.
In \sect{sec:hotaxions} we address
some issues related with the existence of a thermal population of
relativistic axions.  Finally, the role of axion-related
substructures, in the form of axion miniclusters and axion stars, is
reviewed \sect{sec:substructures}.

\subsection{Basics of cosmology and thermodynamics in the early Universe}
\label{sec:friedmann}

Observations at scales larger than 100 Mpc support the evidence that
the Universe is spatially homogeneous and isotropic.  Our starting
point is the description of such a Universe in terms of Einstein
equations
\beq
	\mathcal G^{\mu\nu} = 8\pi G_N \mathcal T^{\mu\nu}\,,
	\label{eq:einsteinequation}
\eeq
where the Einstein tensor $\mathcal G^{\mu\nu}$ describes the geometry
of the space-time and is defined in terms of (derivatives) of the
metric tensor $g_{\mu\nu}$ through the affine connection and its
derivatives.  The stress-energy tensor $\mathcal T^{\mu\nu}$ describes
the energy content of the Universe, and $G_N$ is Newton constant.  In
the following we rewrite it as $G_N = 1/\mP^2$ that defines the Planck
mass with a value $\mP = 1.221\times10^{19}\,$GeV.  The indices
$\mu,\nu$ run over the four space-time dimensions $\(t,x,y,z\)$
where $t$ is  {\it cosmic} time (i.e.~the time measured by a
physical clock at rest in the comoving frame) which, for any given
value, slices space-time into a three-dimensional homogeneous and
isotropic space manifold $\mathcal M_3$ with a constant curvature that
can be positive, negative or zero, corresponding respectively to an
open, closed or flat Universe. Observations and theoretical
considerations support the last possibility (flat Universe) for which
the line element is
\beq \mathrm{d}s^2 \equiv g^{\mu \nu} dx_\mu d x_\nu =\mathrm{d}t^2 -
R^2(t)\,\(\mathrm{d}x^2 +\mathrm{d}y^2+\mathrm{d}z^2\)\,,
	\label{eq:line_element}
\eeq
where $R = R(t)$ is the cosmic scale factor.
Eq.~\eqref{eq:line_element} defines the
Friedmann-Lema\^{i}tre-Robertson-Walker (FLRW) metric.  Consistency
with the symmetries of the metric (homogeneity and isotropy) requires
$\mathcal{T}_{\mu\nu}$ to be diagonal and with equal spatial
components.  A simple realisation is a perfect fluid at rest in the
comoving frame with time dependent energy density $\rho=\rho(t)$ and
pressure $P=P(t)$ for which the stress-energy tensor is
$\mathcal{T}^\mu_\nu ={\rm diag}\(\rho, -P,-P,-P\)$.
The stress energy in the Universe is conveniently described by the 
simple equation of state $P=w \rho$ so that a
FLRW Universe remains completely characterised by the 
total energy density and by the fractional size  and $w$-values 
of its components.

From the FLRW metric \eqn{eq:line_element} the affine connection is
computed, then the Einstein tensor, and inserting  it into Einstein's
field equations (\ref{eq:einsteinequation}) with 
the stress-energy tensor for the perfect fluid on the right-hand-side, yields
\begin{align}
H^2 \equiv \(\frac{\dot R}{R}\)^2 &= \frac{8\pi}{3\mP^2}\rho\,,\label{eq:friedmann1}\\
\dot H + H^2 = \frac{\ddot R}{R} &=
-\frac{4\pi}{3\mP^2}\,\(\rho + 3P\)\,, \label{eq:friedmann2} 
\end{align}
where the first equation is obtained from the 0-0 component and the
second from the spatial $j$-$j$ component after making use of
\eqn{eq:friedmann1} while, due to the symmetries of the system, all
the other components vanish.  $H \equiv \dot R /R$ with the dot
meaning  derivation with respect to cosmic time, expresses the rate of
the change in the scale factor, and is called the Hubble rate.
Combining the two Friedmann equations leads to the conservation law in
an expanding Universe:
\beq \dot\rho + 3H (P+\rho) = 0\,, \ \quad {\rm or\ equivalently} \
\quad \frac{\partial}{\partial t} (\rho R^3) = - P\, \frac{\partial
}{\partial t} (R^3).
	\label{eq:continuity_equation}
\eeq
The second form, whose physical meaning is that the change in
energy in a comoving volume is equal to minus the pressure times the
change in volume, is the first law of thermodynamics.  This 
form is particularly useful for deducing the scaling of the energy
density in matter and radiation with the expansion.  Matter is
pressureless ($w=0$) and immediately we obtain $\rho_{m} \propto R^{-3}$.
For radiation $P_{\rm rad} = \rho_{\rm rad} /3$ ($w=1/3$) and~\eqn{eq:continuity_equation}
can be recast as $\partial (\rho_{\rm rad} R^4)/\partial t = 0$. Hence
$\rho_{\rm rad} \propto R^{-4}$ where the additional factor of
$R^{-1}$ accounts for the redshift of radiation wavelength.
Finally,  from \eqn{eq:friedmann2} we see that an accelerated expansion 
($\ddot R/R > 0$) requires $\rho +3P < 0$  that is, given that the energy density 
is always positive, acceleration requires a negative pressure $P<- \rho/3$. The simplest possibility 
is $P_\Lambda=-\rho_\Lambda$ ($w=-1$) in which case from the first relation in \eqn{eq:continuity_equation}
we immediately obtain  $ \rho_\Lambda = {\rm const.}$ 
which corresponds to a Universe dominated by vacuum energy.


The FLRW metric~\eqn{eq:line_element} together with the Friedmann
equation (\ref{eq:friedmann1}) and \eqn{eq:continuity_equation}
provides the framework in which the so-called standard cosmological,
or $\Lambda$CDM, model is defined.  According to this model, the
present Universe is dominantly filled with a mysterious vacuum energy
component $\rho_\Lambda$ which accounts for about $2/3$ of the 
total energy density. The remaining  $1/3$ corresponds to a matter 
component, which is again dominated by an unknown type of DM. 
Known particle species as
baryons, electrons and (depending on their masses) neutrinos,
contribute a subleading amount $\rho_b \sim  0.2\, \rho_{\rm DM}$ while the
photon contribution is even smaller
$\rho_{\rm rad} \ll \rho_\Lambda, \rho_{\rm DM},\rho_b$.
%
However, in the early Universe, at temperatures above $\sim 1\,$eV,
 the Universe was dominated by radiation, $\rho \simeq \rho_{\rm rad}$.  
 The regime of radiation domination is
the most relevant one for the topics
that will be developed in the forthcoming Sections. 

%


Let us now review some basic notions of thermodynamics in an expanding
Universe.  For particles in kinetic equilibrium with a thermal bath at
temperature $T$, the phase space occupancy is given by the
distributions $f(E) = \[\exp\((E-\mu)/T\)\pm
1\]^{-1}$
with $+1$ for fermions and $-1$ for bosons, while $\mu$ is the
chemical potential of the particle species.  If a certain number of
species $p=1,2,3,\dots$ are in {\it chemical equilibrium}, then
$\sum_p \mu_p=0$.  In the relativistic limit $(T\gg m)$ the
contribution to the energy density of the thermal bath of a particle
 $p$ with $g_p$ internal degrees of freedom is
\beq
	\rho_p = \frac{\pi^2}{30}\,\eta_p \,g_p\,T_p^4\,,
	\label{eq:define_rhoj}
\eeq
where the statistical factor $\eta_p= 1 \;(7/8)$ for bosons
(fermions), and $T_p$ is the temperature that characterises the
$p$-particle distribution.  If a  species is kinetically decoupled from the
thermal bath, then $T_p$ might differ from $T$.\footnote{In the SM this
  occurs only for the neutrinos, and below $T\sim 1\,$MeV.}  In the
early Universe, a very important fiducial quantity is the total
entropy in a comoving volume $S= R^3 (\rho+P)/T$, and this is because
entropy it is conserved, or approximately conserved, in most phases of the
cosmological evolution.\footnote{This expression for $S$ holds
  whenever all the particle chemical potentials are negligible
  $|\mu|\ll T$, which is generally true to an excellent
  approximation.}  For relativistic species $P_p = \rho_p/3$ so that
the contribution to the entropy density $s=S/R^3$ of a relativistic
particle is
\beq
	s_p = \frac{2 \pi^2}{45}\,\eta_p \,g_p\,T_p^3\,.
	\label{eq:define_sj}
\eeq
It is customary to write the energy and entropy densities of
relativistic species in a compact form as
\begin{equation}
\label{eq:rho_and_s} 
\rho_{\rm rad} = \frac{\pi^2}{30} g_*(T) T^4\,, \qquad  
s_{\rm rad} = \frac{2 \pi^2}{45} g_S(T) T^3\,,   
\end{equation}
where the effective number of energy $(g_*)$ and entropy $(g_S)$
relativistic degrees of freedom are defined as
\begin{equation}
\label{eq:geff_gSeff}
g_*(T) =\sum_p \eta_p\,g_p\,\(\frac{T_p}{T}\)^4, \qquad 
g_S(T) =\sum_p \eta_p\,g_p\,\(\frac{T_p}{T}\)^3\,. 
\end{equation}
These expressions account for the possibility that for some species
$T_p\neq T$, in which case $g_S \neq g_*$, while they are otherwise
equal.  Since we will mainly deal with the Universe history during the
radiation dominated era ($\rho \simeq \rho_{\rm rad}$) in the
following we will drop the subscript $_{\rm rad}$ and simply use
$\rho$ and $s$.  When $g_S$ remains constant during the Universe
evolution, entropy conservation $dS \sim d\(T^3 R^3\)
= 0$
implies $T \propto 1/R \propto t^{-1/2}$.  Finally, from the Friedmann
equation (\ref{eq:friedmann1}) and from the expression for the total
energy density in the radiation dominated era \eqn{eq:rho_and_s}, one
obtains the following expression for the Hubble parameter:
\begin{equation}
\label{eq:Hubble}
H(T) = \(\frac{4 \pi^3 g_*(T)}{45}\)^{1/2}  \frac{T^2}{\mP} \simeq 1.66 \, g_*^{1/2} \frac{T^2}{\mP}\,. 
\end{equation}
In the early Universe, for $T>m_t$ all the SM degrees of freedom are
in the relativistic regime, and $g_* = 106.75$.  Other relevant values
are $g_*(m_b > T >m_c) = 75.75$, $g_*(T >T_{C}) = 61.75$ ($T_{C}$ 
is the QCD critical temperature),
$g_*(T >m_\pi) = 17.25$,
$g_*(T >m_e) = 10.75$. At temperatures  below the electron mass $g_*$ and $g_S$ 
differ, because  neutrinos are kinetically decoupled 
from the thermal bath and are  not reheated by $e^+e^-$ annihilation.  
The neutrino   temperature is in fact lower by a factor  $T_\nu/T_\gamma = (4/11)^{1/3}$ so that      
from  \eqn{eq:geff_gSeff} we obtain $g_*(T <m_e) \simeq 3.36$ and   $g_S(T <m_e) \simeq 3.91$.

\subsection{The  axion  potential and the axion mass at finite temperature}
\label{sec:QCDsusceptibility}

The dependence of the axion potential and  mass on the temperature of the  early Universe thermal bath 
is very important because it concurs  to determine  the present relic abundance of the axions generated 
through the misalignment mechanism. Although this is not the only mechanism of production 
of cold relic  axions, it is the most model independent and  possibly a  quite  relevant one.
However, deriving  a reliable expression for $V(a,T)$ is not an easy task. 
The chiral Lagrangian approach that in \sect{sec:AxionEFT} proved to be a powerful tool 
to determine axion properties at zero temperature cannot be used, because  convergence 
of the perturbative expansion deteriorates as the temperature approaches the critical 
temperature $T_C \simeq160\,$MeV where QCD starts deconfining, and for  $T\simeq T_C$ 
the chiral approach is clearly inadequate to describe QCD. 
However, it is expected that at larger temperatures  $T\gg T_C $ QCD becomes     
perturbative, in which  case other computational  approaches, as for example those 
based on   the so-called dilute instanton gas approximation (DIGA)~\cite{Callan:1977gz}
can represent a sensible approach. 
The result of computing  the  axion potential  around the background 
of a dilute instanton gas~\cite{Gross:1980br}  gives  a potential of the form 
\begin{equation}
\label{eq:VaT}
V(a,T)\big|_{T>T_C} = \chi(T) \[1- \cos\(\frac{a}{f_a}\) \]\,.
\end{equation}
The  finite temperature QCD topological susceptibility $\chi(T)$   is related 
to the temperature dependent axion mass as 
\begin{equation}
\label{eq:chiT}
\chi(T) =  f_a^2\, m^2_a(T) \,.
\end{equation}
The DIGA predicts a power-law dependence of the topological susceptibility 
on temperature with a rather large exponent 
$\chi(T) \propto \chi(0)\,T^{-8}$.  However, the reliability of the DIGA estimate has been 
questioned because  $\chi(T)$ exhibits an exponential dependence on quantum corrections
that  are difficult to  estimate 
perturbatively.\footnote{The range of validity of the DIGA and the corresponding theoretical uncertainty on $\chi(T)$ and 
on the axion relic abundance was given in Ref.~\cite{Dine:2017swf}. However, some 
lattice simulations (see e.g.~\cite{Bonati:2015vqz}) 
indicate that for temperatures in the range $[T_C , 4 \, T_C]$ 
$\chi(T)$ differs strongly from the DIGA 
both in the size and in the temperature dependence. 
Further lattice studies are needed in order to settle this issue.}
 To tackle this problem different techniques have been used, as for example a 
 semi-analytical approach based 
on the   interacting instanton  liquid model~\cite{Wantz:2009mi,Wantz:2009it}  
or non-perturbative  QCD lattice methods. 
In particular several groups have carried out dedicated QCD 
simulations on the lattice  in ranges of temperature relevant  
for the axion mass~\cite{Bonati:2015vqz,
Borsanyi:2016ksw,Petreczky:2016vrs,Bonati:2018blm,Burger:2018fvb}. 
Above some threshold temperature of the order of several 100 MeV, 
a power-law dependence close  to the DIGA prediction is found, 
however, sizeable deviations in general  appear at lower temperatures.
For definiteness we will   assume in the following that well above  $T_C$ 
the temperature dependence of the axion mass can be   parametrised as 
\beq
	m_a(T)  \simeq \beta\, m_a \left(\frac{T_C}{T}\right)^{\gamma}\,,
	 \label{eq:QCDaxion_mass}
\eeq
where $m_a$ is the zero-temperature axion mass,   the exponent has a value  $\gamma \approx 4$, 
 $T_C \approx 160\,$MeV (see e.g. ~\cite{Bazavov:2011nk}) while the parameter $\beta$, that depends 
 on the number of  quark flavours active at the temperature $T$ and on other 
details of QCD physics,  is found in the DIGA to be  of the order of a few$\ \times \ 10^{-2}$~\cite{Gross:1980br}.
Expanding  \eqn{eq:VaT} around $a/f_a =0$ gives 
\beq
	V(a, T)\big|_{T>T_C} \approx \frac{1}{2}\,m_a^2(T)\,a^2 \[1 + b_2(T) \frac{a^2}{f_a^2}+ \dots \]  \,,
	\label{eq:axion_potential_expansion}
\eeq
where  the coefficient of the quadratic term in the expansion is identified with the square of the 
axion mass, while  terms  higher than quadratic, that correspond to higher momenta of the topological 
charge distribution  $(\alpha_s/8\pi) G \tilde G$, describe  self-interactions of the axion field. In particular, 
 it is found in lattice simulations that the fourth order coefficient 
  has a value $b_2 \approx - \frac{1}{12}$, 
 as it would be expected  from the cosine potential (that is from  the DIGA). 
 Being this coefficient negative, the interaction between axions is attractive.

\subsection{Axion misalignment mechanism}
\label{sec:misalignment}

The  effective periodic potential $V(a, T)$ in \eqn{eq:VaT}  is responsible for 
the generation of a cosmological population of cold axions, whose
abundance was  first computed in Refs.~\cite{Abbott:1982af,Dine:1982ah,Preskill:1982cy}. 
In the so-called \emph{misalignment mechanism}, 
the present energy density stored in the zero modes\footnote{The 
axion field can be expanded in terms of a linear superposition 
of eigenmodes with definite co-moving wavevector ${\bf k}$ 
$a({\bf x},t) = \int d^3k \, a({\bf k},t) e^{i {\bf k} \cdot {\bf x}}$, 
and the axion zero mode refers to the component $a({\bf k} = 0,t)$.} of the axion field 
can be obtained by solving the equation of motion in an expanding Universe
\beq
	\ddot a + 3H\dot a - \frac{1}{R^2(t)}\nabla^2 a +  V'(a, T) = 0\,,
	\label{eq:equation_motion}
\eeq
where the Hubble rate $H$ has been defined in \eqn{eq:friedmann1}, 
$\nabla^2$ is the spatial Laplacian operator and $V'$ is the first derivative of the potential 
with respect to the axion field $a$. Although a proper
treatment would consist in solving Eq.~\eqref{eq:equation_motion}
numerically for a given effective potential, for illustrative purposes
and to better understand the physics involved let us consider the 
expansion given in Eq.~\eqref{eq:axion_potential_expansion}  truncated to the quadratic term. In terms of 
the axion angle $\theta(x) = a(x)/f_a$  Eq.~\eqref{eq:equation_motion}  then reads
\beq
	\ddot \theta + 3H\dot \theta - \frac{1}{R^2(t)}\nabla^2 \theta + m_a^2(t)\theta = 0\,,
	\label{eq:equation_motion1}
\eeq
where $m_a(t) \equiv m_a(T(t))$. In this approximation, the Fourier
transform of the axion field follows a similar expression as in
\eqn{eq:equation_motion}, with the replacement $\nabla^2 \to -k^2$ in
terms of the mode $k^2 \equiv {\bf k} \cdot {\bf k}$ 
for each field decomposition.  Hubble expansion
implies that the wavelength of each mode $k$ gets stretched as
$\lambda(t)=\frac{2\pi}{k} R(t)$. There are two qualitatively
different regimes in the evolution of a mode, depending on whether at
a given time its wavelength is larger than the horizon $\lambda(t)>t$,
or is inside the horizon $\lambda(t)<t$.  Recalling that $t = 1/(2H)$
super-horizon modes are then defined by the condition $k \lesssim R(t)H(t)$,
while for $k \gtrsim R(t)H(t)$ the mode is inside the horizon.  For
super-horizon modes the contribution from the spatial gradient
in~\eqn{eq:equation_motion1} can be neglected. If the mass term is
also negligible, the solution for the mode goes to a constant as
$t^{-1/2}$ (see \eqn{eq:initial_condition_theta} below). 
For modes well inside the horizon, and for an
adiabatic expansion, the solution to \eqn{eq:equation_motion1}
conserves the number of axions in each mode. The number density of
these higher frequency modes is suppressed with respect to the
contribution from super-horizon modes so we neglect it. Additional
details on higher modes can be found in Refs.~\cite{Linde:1985yf,
  Seckel:1985, Lyth:1990, Linde:1990yj, Turner:1991, Lyth:1992,
  Lyth:1992yy}.

When analysing Eq.~\eqref{eq:equation_motion1}, we see that the Hubble
drag effectively damps the evolution of the axion field as long as the
Hubble rate is significantly larger than the mass term. As the
Universe cools and the expansion rate slows down, the mass term
eventually begins to contribute to the evolution of the axion field
once the condition
\beq
	m_a(t_{\rm osc}) \approx 3H(t_{\rm osc})\,,
	\label{eq:condition_oscillations}
\eeq
has been fulfilled.\footnote{The factor three appearing in
  \eqn{eq:condition_oscillations} serves as a crude approximation to
  estimate the value of $t_{\rm osc}$. In general, this factor should
  be matched to a numerical solution.}
Eq.~\eqref{eq:condition_oscillations} defines implicitly the time
$t_{\rm osc}$ at which the axion mass becomes important,
and from this moment on, the axion field begins to oscillate in the
quadratic potential with oscillations that are damped by the expansion
rate.\footnote{The intuitive meaning of the condition
  \eqn{eq:condition_oscillations} is the following: in radiation
  domination $H(t)=1/(2t)$, hence the condition can be rewritten as
  $m_a(t_{\rm osc}) t_{\rm osc} \approx 1$.  In this regime, the
  angular frequency of the oscillations
is $\omega \approx m_a(t)$. 
Therefore, requiring that the  condition is satisfied, amounts to  
require that the Universe is sufficiently old  to host a 
sizeable  fraction of one oscillation period.}

For super-horizon modes $k\lesssim RH$ ($\nabla^2 \theta \simeq 0$) and as long as $t \ll t_{\rm osc}$
($m_a(t) \simeq 0$),~\eqn{eq:equation_motion1} reduces to
$\ddot \theta + (3/2t)\dot \theta = 0$. Defining the initial
conditions at the PQ symmetry breaking time as
$\theta(t_{\rm PQ}) \equiv \theta_{\rm PQ}$ and
$\dot\theta(t_{\rm PQ}) \equiv \dot\theta_{\rm PQ}$\footnote{As we will see below, 
the PQ symmetry can get broken before or after inflation. 
Only in the latter case the label PQ can consistently refer to quantities 
like temperature, Hubble rate, etc.~at the PQ breaking time. We 
will leave understood that in the case the symmetry is broken 
 during inflation, the same label 
PQ rather refers to a suitable early time after the end of inflation.\label{footnoteTPQ}}
 the solution
to~\eqn{eq:equation_motion1} reads
\begin{align}
        \theta(t) &= \theta_{\rm PQ} + \frac{\dot\theta_{\rm PQ}}{H_{\rm PQ}}\,\(1-\frac{R_{\rm PQ}}{R(t)}\)\,,
        \label{eq:initial_condition_theta}\\
        \dot\theta(t) &= \dot \theta_{\rm PQ}\,\(\frac{R_{\rm PQ}}{R(t)}\)^3\,,
        \label{eq:initial_condition_dottheta}
\end{align}
where $H_{\rm PQ} = H(t_{\rm PQ})$ and $R_{\rm PQ} = R(t_{\rm PQ})$. 
At  the time when the axion mass becomes relevant, defined   by~\eqn{eq:condition_oscillations} as $t_{\rm osc}$,  we have 
\beq
\label{eq:Tosc-fa}
        \frac{R_{\rm PQ}}{R(t_{\rm osc})} =   \frac{T_{\rm osc}}{T_{\rm PQ}}
        \approx \frac{T_{\rm osc}}{f_a} \lesssim 10^{-8}\,,
\eeq
where we have used $T_{\rm osc} \equiv T(t_{\rm osc}) \approx 1\,$GeV
and $T_{\rm PQ} \equiv T(t_{\rm PQ}) \simeq f_a \gsim 10^{8}\,$GeV.
The ratio of the scale factors in parenthesis in
\eqn{eq:initial_condition_theta} is thus a transient that rapidly
decays after the PQ phase transition.  We define the initial
conditions at the onset of oscillations
$\theta_i \equiv \theta(t_{\rm osc})  $ and
$\dot\theta_i\equiv \dot\theta(t_{\rm osc})$ as
\begin{align}
\label{eq:theta-i}
        \theta_i &= \theta_{\rm PQ} + \frac{\dot\theta_{\rm PQ}}{H_{\rm PQ}}\,, \\
\label{eq:dottheta-i}
        \dot\theta_i &=\dot \theta_{\rm PQ}  \(\frac{H(t_{\rm osc})}{H_{\rm PQ}}\)^{3/2}\,, 
\end{align}
where in the second equation we have used 
$R\propto 1/T$ and $H\propto T^2/\mP$. 
$\theta_{i}$ in \eqn{eq:theta-i} is known in the
literature as the {\it initial misalignment angle}. The effects of the
initial velocity $\dot\theta_i$ would be important if at $t_{\rm osc}$
the kinetic energy stored in the axion field $\dot a^2/2$ exceeds the
potential barrier $2 m_a^2 f^2_a$, 
see \eqn{eq:VaT}, 
that is if 
$\dot\theta_{i} > 2 m_a(t_{\rm osc})$. 
Using~\eqn{eq:condition_oscillations} we can
rewrite \eqn{eq:dottheta-i} as
\beq
\label{eq:dottheta-limit0}
\dot\theta_i = m_a(t_{\rm osc}) \, \frac{\dot \theta_{\rm PQ}}{H_{\rm
    PQ}} \frac{T_{\rm osc}}{3 T_{\rm PQ}}\,, \eeq
so that  we obtain the condition
\beq
\label{eq:dottheta-limit}
 \frac{\dot \theta_{\rm PQ}}{H_{\rm PQ}}  > \frac{6 \ T_{\rm PQ}}{T_{\rm osc}}\gsim 
 10^{8} \,.
  \eeq
  where for the numerical value we have used \eqn{eq:Tosc-fa}. We can
  conclude that unless the axion velocity at the PQ breaking scale is
  at least eight orders of magnitude larger than the expansion rate of the Universe, 
  the results for axion DM are indistinguishable from
  what is obtained by setting, as it is usually done, $\dot\theta_i=0$.
  Some models that assume that at $t_{\rm osc}$ the axion kinetic
  energy dominates the potential energy are reviewed in
  \sect{sec:thetadot}.

\begin{figure}[t!]
\begin{center}
	\includegraphics[width=0.7\linewidth]{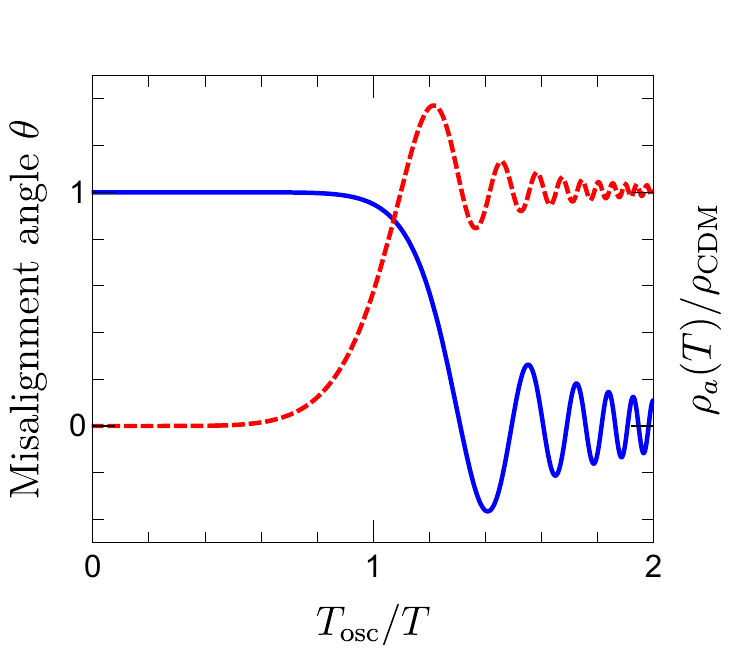}
	\caption{The axion angle $\theta$ (blue solid line) and the
          energy density of axions in units of the present CDM energy
          density (red dashed line), as a function of the inverse
          plasma temperature $T_{\rm osc}/T$. We set the initial
          misalignment angle $\theta_i = 1$.}
	\label{fig:evolution}
\end{center}
\end{figure}

 Regarding the misalignment angle $\theta_i$, the possible initial
  values depend on the cosmological history of the axion field. To
  better see this, let us consider the following two conditions:
\begin{itemize} 
	\item [{\bf a})] The PQ symmetry is spontaneously broken during inflation;
	\item[{\bf b})] The PQ symmetry is never restored after its spontaneous breaking occurs.
\end{itemize}
%
Condition {\bf a}) is realised whenever the axion energy scale is
larger than the Hubble rate at the end of inflation, $f_a > H_I$,
while condition {\bf b}) is realised whenever the PQ scale is larger
than the maximum temperature reached in the post-inflationary
Universe.
Broadly speaking, one of these two possible scenarios occurs:
\begin{itemize}

\item {\it Pre-inflationary} scenario. If both {\bf a}) and {\bf b}) are
  satisfied, inflation selects one patch of the Universe within which
  the spontaneous breaking of the PQ symmetry leads to a homogeneous
  value of the initial misalignment angle $\theta_i$. In this
  scenario, topological defects are inflated away and do not
  contribute to the axion energy density. However, other bounds that
  come from isocurvature modes~\cite{Crotty:2003rz, Beltran:2005xd,
    Beltran:2006sq} severely constrain this scenario, which
  require a relatively low-energy scale of inflation to be viable.

\item {\it Post-inflationary} scenario. If at least one of the conditions {\bf
    a}) or {\bf b}) is violated, the PQ symmetry breaks with 
    $\theta_i$ taking different values in patches that are
  initially out of causal contact, but that today populate the volume
  enclosed by our Hubble horizon. In this scenario, isocurvature
  fluctuations in the PQ field randomise $\theta$, with no preferred
  value in the power spectrum. Moreover, modes with $k \gtrsim RH$ decay
  because of the gradient term in the equation of motion. The proper
  treatment in this scenario is to solve numerically the equation of
  motion of the PQ field in an expanding Universe, in order to capture
  all features coming from the misalignment mechanism, including the contribution from
  topological defects, see \sect{sec:defects}. Here, we discuss how to
  approximate the computation of the relic density of cold axions,
  neglecting for the moment the important contribution from axionic
  strings and domain walls. In this scenario, the initial misalignment
  angle $\theta_i$ within a Hubble patch takes all possible values on the
  unit circle. The strategy to compute the
  present abundance relies on computing
  Eq.~\eqref{eq:equation_motion1} many times, each time drawing the
  initial value $\theta_i$ from a uniform distribution over the unit
  circle. For a quadratic potential, this is the same as assuming for
  the initial condition the value
\beq
	\theta_i \equiv \sqrt{\langle\theta_i^2\rangle} = \frac{\pi}{\sqrt{3}}\simeq 1.81\,,
	\label{eq:initialcondition_unbroken}
\eeq
where the angle brackets represent the value of the initial condition
averaged over $\[-\pi, \pi\)$.
For the periodic potential that defines the QCD axion  
this result is modified due to the presence of
non-harmonic
terms~\cite{Turner:1985si,Lyth:1991ub,Lyth:1992,Strobl:1994wk,Bae:2008ue,Visinelli:2009zm}
which can be parametrised in terms of a function $F(\theta_i)$~\cite{Lyth:1991ub} that
accounts for the anharmonic corrections in the axion potential, so
that
\beq
	\langle \theta_i^2\rangle = \frac{1}{2\pi}\int_{-\pi}^\pi \mathrm{d}\theta_i\,F(\theta_i)\theta_i^2,
	\label{eq:theta_avg}
\eeq
where $F(\theta_i)\to 1$ for small $\theta_{i}$ and increases monotonically  with $\theta_{i}$.
%
Including anharmonicities, numerically one  obtains  
$\theta_i \approx 2.15$~\cite{diCortona:2015ldu}, 
which is slightly larger than the value in \eqn{eq:initialcondition_unbroken}.
%
Thus, in this scenario the initial condition is  fixed by the averaging of $\theta_i$ over the Hubble
patch, which leaves the zero temperature  value of the axion mass as the sole
unknown for estimating the contribution to axion cold DM (CDM) from 
the misalignment mechanism.  (It should be kept in mind, however, that  in this case 
topological defects form after inflation and remain
within the horizon, contributing to the axion energy density, 
and that this contribution still has to be assessed
with sufficient confidence and precision,  see the discussion 
in \sect{sec:defects} below.) 
%
%
%
 

\end{itemize}
The evolution of the axion field can be described in terms of its
effective equation of state $w_a$ given by  
\beq
	w_a \equiv \left\langle\frac{\frac{1}{2}\dot a^2 - V(a, T)}{\frac{1}{2}\dot a^2 + V(a, T)}\right\rangle\,, 
	\label{eq:equation_of_state}
\eeq
where the angle brackets stand for a temporal average over times larger than the 
oscillation period.
The dependence of the axion energy density on the scale factor
ranges from a dark energy-like solution,
 $w_a = -1$, valid at 
$t \lesssim t_{\rm osc}$
(corresponding to the axion field frozen) 
to a CDM-like solution, $w_a = 0$, 
    for $t \gtrsim t_{\rm osc}$ when, due to the oscillating behaviour, 
    the average potential energy equals the average kinetic energy. 
The number of axions in a comoving volume is then frozen
from about $T_{\rm osc}$ on, with a number density of axions at the
onset of oscillations given by
\beq
	n_a(T_{\rm osc}) = \frac{b}{2}m_a(T_{\rm osc})\,f_a^2\,\langle\theta_i^2\rangle\,.
	\label{eq:numberdensity00}
\eeq
Here, $b$ is a factor of order one~\cite{Turner:1985si} that captures
the uncertainties derived from using the approximation in
Eq.~\eqref{eq:condition_oscillations} to estimate the number density
$n_a(T_{\rm osc})$, instead of numerically solving
Eq.~\eqref{eq:equation_motion}. Assuming that entropy remains 
conserved within a comoving volume since the axion field  starts
oscillating, the ratio of the axion number density to the entropy
density is then conserved to any temperature $T < T_{\rm osc}$,
\beq
	n_a(T) = n_a(T_{\rm osc})\,\frac{s(T)}{s(T_{\rm osc})}\,,
	\label{eq:numberdensity0}
\eeq
where $s(T)\propto T^{3}$ is given in \eqn{eq:geff_gSeff}. Neglecting the
contribution from the topological defects
the  energy density of axions as a function of the temperature, at $T\ll T_{C}$,  then is
\beq
	\rho_a^{\rm mis}(T) = m_a\,n_a(T_{\rm osc})\,\frac{g_S(T)}{g_S(T_{\rm osc})}\,\(\frac{T}{T_{\rm osc}}\)^3
	= \frac{b\beta}{2} \, m^{2}_{a} f^{2}_{a}  
	\,\langle\theta_i^2\rangle \, \frac{g_S(T)}{g_S(T_{\rm osc})}\,\frac{T^{3}T_{C}^{\gamma}}{T_{\rm osc}^{3+\gamma}}\,,
	\label{eq:VRMaxions}
\eeq
where in the second equality we have used  \eqn{eq:numberdensity00} 
and  \eqn{eq:QCDaxion_mass}.
$T_{\rm osc}$  appearing in this equation can be evaluated using  
the condition for the start of oscillations
\eqn{eq:condition_oscillations} together with the mass-temperature
relation in \eqn{eq:QCDaxion_mass}  and the Hubble-temperature relation
$H(T) = 1.66\, g_*^{1/2} \; T^2/\mP$ in \eqn{eq:Hubble}, obtaining
\beq
\label{eq:Tosc1}
T_{\rm osc}^{2+\gamma} \simeq \(\frac{\beta\, T_C^\gamma \, \mP}{4.98 \, g_*(T_{\rm osc})^{1/2}}\)\; m_a\,.
\eeq
As a reference, with $g_*(T>T_{C}) = 61.75$,  $\gamma=4$, $T_C=160\,$MeV  and  
$\beta \approx 0.026$~\cite{Turner:1985si}
one obtains $T_{\rm osc} \approx 800 \(\frac{m_{a}}{1\,\mu{\rm eV}}\)^{1/6}\,$MeV. 
Plugging now  \eqn{eq:Tosc1} into  \eqn{eq:VRMaxions} we obtain

\begin{equation}
\label{eq:rhoa_analytic}
	\rho_a^{\rm mis}(T) = 
	\frac{b\beta}{2} \, m^{2}_{a} f^{2}_{a}  
	\,\langle\theta_i^2\rangle \, \frac{g_S(T)}{g_S(T_{\rm osc})}\,
\( \frac{4.98 \, g_*(T_{\rm osc})^{1/2}}{\beta\, \mP\, m_{a}}\)^{\frac{3+\gamma}{2+\gamma}} T_{C}^{-\frac{\gamma}{2+\gamma}}
\,  T^{3}\,.
\end{equation}
In Fig.~\ref{fig:evolution} we show the evolution of $\theta(T)$ (blue solid line) and of the energy density of cold 
axions in units of the present CDM energy density 
$ \rho_a^{\rm mis}(T) /\rho_{\rm CDM}$ 
(red dashed line) as a function of the inverse plasma temperature $T_{\rm osc}/T$.
To draw the picture, we have set the initial misalignment angle $\theta_i = 1$, 
a vanishing  initial   axion energy density   when  the axion 
field is frozen to the configuration $\theta = \theta_i$ at  $T \gtrsim T_{\rm osc}$,  and 
the value of the axion mass has been chosen  so that the present value of  
$\rho_a^{\rm mis} $ saturates the  measured CDM energy density.
 Finally, the present  axion energy density from the misalignment mechanism 
 normalised to the  critical energy density  
  $\rho_{\rm crit} = 3\mP^2H_0^2/(8\pi) \approx 8.06 \times 10^{-11}\,h^2\,{\rm \,eV}^4$, 
  where  $h$  is defined in terms of the  Hubble constant as  $H_0 = 100\,h\,$km/s/Mpc, can be obtained from  
 \eqn{eq:rhoa_analytic} as: 
 \beq
	\Omega_a^{\rm mis} 
	h^2
 \approx  0.12\,  
	\(\frac{
	28
	\; \mu\text{eV}}{m_a}\)^{\frac{7}{6}}
	= 0.12\,
	\,\(\frac{f_a}{
	2.0
	\times 10^{11}{\rm \; GeV}}\)^{\frac{7}{6}} \,, 
	\label{eq:standarddensity}
\eeq
 where we have used $\langle \theta_i^2\rangle = (2.15)^{2}$, 
 $T_0 = 2.73 \, {\rm K} \simeq 2.35 \times 10^{-4}\,$eV and   $g_{S}(T_{0})=3.91$
 for the present photon temperature and effective number of entropic degrees of freedom, 
  $m_{a}f_{a}\simeq 5.7 \times 10^{15}\,$eV$^{2}$ from \eqn{eq:axionmassfa},
  $g_S(T\gtrsim T_{C}) = g_*(T\gtrsim T_{C}) = 61.75$,  $\gamma=4$, $T_C=160\,$MeV,   
$\beta = 0.026$ and $b=1$, and we have put in evidence a factor of the measured CDM abundance 
$\Omega_{\rm CDM} \approx 0.12$.  
Although the value of the axion mass that saturates the CDM density  obtained via this simple 
analytical estimate falls in the ballpark of the results of dedicated numerical studies, 
the  derivation  outlined above neglects several  effects,  and thus  the simple form of 
\eqn{eq:standarddensity}  should be  taken only as an illustrative  approximation. 
More reliable estimates  require describing with a better  accuracy the  dependence of the axion mass on 
the temperature,  including the effects of  the  axion potential anharmonic terms,  and taking  
properly into account  the details of the quark-hadron phase transition~\cite{DeGrand:1985uq,Bae:2008ue,Kim:2018knr}. 
%
%
%
Finally, \eqn{eq:standarddensity} has been obtained assuming a standard cosmology
and canonical properties for the axion.  
In \sect{sec:modifiedcosmo} we will challenge some of these assumptions,  
and we will argue that in non-standard scenarios different ranges of  mass   
values for CDM  axions are possible. \\

\subsection{Topological defects and their contribution to axion cold dark matter}
\label{sec:defects}

In post-inflationary scenarios  
the  production of cold axions from misalignment discussed in the previous section is not the only CDM 
production mechanism.
Additional contributions are possible because of the existence of topological defects, 
specifically strings and domain walls (DW) associated with the axion field
(for a review on topological defects of cosmological relevance see Ref.~\cite{Vilenkin:2000jqa}). 
 If the PQ symmetry is broken 
before inflation, topological defects related to the breaking of the PQ symmetry  (axionic strings) 
are inflated away, while those related to $N_{\rm DW}>1$  do not form.  
Topological defects are therefore relevant  only in the post inflationary scenario, 
a condition that will be assumed throughout this section.

Axion strings are formed through the Kibble mechanism~\cite{Kibble:1976, Kibble:1982dd} at the time of the spontaneous
breaking of the $U(1)$ PQ symmetry.
At the end of  the phase transition,  the phase $ a(x)/v_a$ of the PQ field acquires  
uncorrelated values between $0$ and $2 \pi$ in different  causally disconnected Hubble 
patches. Therefore, when these patches reenter the horizon,  the value of the axion field will have random 
fluctuations within every Hubble patch, and unavoidably there will be loops  around which $a(x)$ 
wraps all the values in the  domain $[0, 2\pi v_a]$. 
Any closed loop over which the phase changes by $ 2\pi v_a$ must necessarily contain a point in which 
the axion field  assumes all the values between 0 and $2\pi v_a$, i.e.~it is singular.
To avoid that the loop could be contracted in a topologically trivial manner, such points should form an infinite or a closed  string,  
 called axion string, where the axion is singular, while along any path encircling the string its value 
 changes by $2 \pi v_a$. 
Single string configurations are stable, however strings within string networks are not, because  
in this case strings can interact among each other, crossing strings can form loops that decay, 
radiating low momentum axions.
 %
%
Estimating the contribution of the string decay to the  abundance of cold axions is a complex problem, 
not yet completely solved, that  will be briefly address  in~\sect{sec:cosmicstrings}.

As the Universe expands, once the temperature lowers to $ \sim 1 $ GeV non perturbative QCD effects become important 
and the axion acquires a potential which provides an explicit breaking of the PQ symmetry. 
This  potential has periodicity $2 \pi f_a$ with $f_a=v_a/\NDW$, see \eqn{eq:DWdef}, 
and hence it is characterised by $\NDW$ equivalent minima, which are related by a  $ \mathbb{Z}_{N_{{\rm DW}}} $ discrete
 symmetry.  The axion relaxes to one of these  minima thus  
breaking spontaneously  $\mathbb{Z}_{N_{{\rm DW}}} $.   
In each  casually disconnected Hubble patch
a different minimum is randomly  chosen, so that   
after a few Hubble times, when several patches will have reentered the horizon,    
DWs  form as  field configurations that interpolate in space between neighbouring vacua.
%
The DW tension (energy per unit area)  for a QCD cosine potential is $\sigma =8 m_a f^2_a$, 
and the contribution to the Universe energy density from the  energy stored in 
such configurations largely exceeds the critical energy density. This  
constitutes a serious  issue, known as 
the axion DW problem, which  is  briefly addressed   in~\sect{sec:CosmoDW}. 
Here we  anticipate that the problem does not subsist in the case of $ N_{{\rm DW}} =1$, 
since in this case the network of  strings and DWs is unstable. 
Also for the case $ N_{{\rm DW}} >1$ there are ways to circumvent the DW problem, 
which  are rewieved  in \sect{sec:AxionsDW}.
%


\subsubsection{Cosmic axion strings}
\label{sec:cosmicstrings}

As predicted by the Kibble mechanism~\cite{Kibble:1976, Kibble:1980mv,Kibble:1982dd}, a network of global strings originates whenever a global $U(1)$ symmetry is spontaneously broken.
In the case of the PQ symmetry, the Kibble mechanism produces {\it axionic} strings, which are vortex-like defects that form as soon as the symmetry is spontaneously broken. 
As we shall see below, the string population consists in long strings and in string loops. 
The last, formed by the continuous intersection of long intercommuting strings, contribute to the majority of the radiated axions. 
%
As was mentioned above,  in models with $N_{\rm DW}=1$ the network of axionic topological defects is unstable. 
This is because in this case each string is attached to a single DW. 
This DW forms because, wrapping in space around the string, 
at some point the value of the axion field must change abruptly  by $2\pi v_a$. 
Since the system evolves to minimize the energy stored in the DW, two strings attached to the same 
wall are  pulled together until they eventually annihilate, so that the string and DW energy 
is released in  radiation of  
low-momentum axions~\cite{Vilenkin:1981kz,Davis:1986xc}. 
For $\NDW > 1$ instead, each string is attached to more than one DW that pull in 
different directions, reaching a stable equilibrium that prevents annihilation. 
%
%
Let us focus on the $N_{\rm DW}=1$ case.
The spontaneous decay of the axionic string network contributes to the present abundance of cold axions~\cite{Vilenkin:1981kz, Vilenkin:1982ks, Vilenkin:1986ku, Davis:1986xc,Harari:1987ht, Davis:1989nj, Lyth:1992aa}. 
Although all numerical  simulations~\cite{Battye:1993jv, Battye:1994au, Yamaguchi:1998gx, Yamaguchi:1999dy, Hiramatsu:2010yu, Hiramatsu:2010yn, Hiramatsu:2012sc, Hiramatsu:2012gg, Fleury:2015aca,Klaer:2017qhr, Klaer:2017ond, Gorghetto:2018myk, Vaquero:2018tib, Buschmann:2019icd,Gorghetto:2019hee} agree in confirming this qualitative result,  
 no quantitative agreement  has been reached yet regarding the 
 efficiency of cold axion production from  topological defects. 
  The main challenge being the limitations in numerical 
simulations of the evolution of the string-DW network, which requires that final results are extrapolated 
through several orders of magnitude in the ratio of  relevant dimensional quantities. 
The  uncertainty in estimating the overall cosmological abundance 
of cold axions has  an obvious impact on axion DM searches, and indeed 
it constitutes  one of the most important  open problems in axion physics.

\subsubsection{String populations}   
\smallskip
The population of strings can be  divided into two  subgroups: {\it long strings} and {\it string loops}. \\ [-10pt]

\noindent
{\it Long strings}  are string-like defects of size comparable to the horizon scale~\cite{Zeldovich:1978wj, Preskill:1979zi, 
Guth:1979bh, Vilenkin:1981kz, Vilenkin:1982ks,Davis:1986xc}.
The string energy per unit length, called string tension $\mu$, of an isolated string gets contributions 
from the string core where the divergence related to the singularity is cutoff by the heavy degree of freedom, 
that is by the radial mode $\varrho_a$ sitting on the top of its potential. For a standard potential 
as $V(\Phi)$ in~\eqn{eq:VPhiKSVZ} one has $\varrho_a^2  \sim f_a^2$ at the potential maximum, and 
the string core can be defined as a region of size $r_c \sim f_a^{-1}$ from the center. 
Thus, in integrating the volume energy density $r_c$ provides a short-distance cutoff. 
There is, however, also a  large-distance logarithmic divergence, that is cutoff 
by a length  of the order of the Hubble size   $t \sim H^{-1}$. 
For an single string configuration the string tension then is 
%
%
\beq
	\mu(t) \simeq \pi f_a^2\log\left( \eta f_a t\right) 
	\label{eq:mass_length}
\eeq
where $\eta$ is an $\mathcal{O}(1)$ factor, possibly dependent on time,  which takes into account the 
fact that the string can have non-trivial shape features on length scales  smaller  than the Hubble length.

In the real case one has to consider a full system of axion strings that can interact.
Already in earlier studies it was claimed that a system of interacting strings should evolve towards 
a particular configuration that is independent of the  initial 
conditions at early times~\cite{Kibble:1976, Kibble:1980mv,Vilenkin:1981kz}. This represents  a crucial 
property  to allow model-independent estimates of the cold axion contribution from topological defects.  
The existence of a unique asymptotic configuration can be justified qualitatively in a simple way.
To minimize the energy,  bended strings tend to straighten  emitting radiation, closed strings of 
size smaller than the Hubble radius tend to shrink further, long strings crossing each other  
recombine in new configurations of reduced length. 
The rate of these processes depends on the  overall length of strings within the Hubble volume
$\ell_{\rm tot}(t)$.
When this quantity  is large the rate of the length reducing processes is also large. 
However, when $\ell_{\rm tot}(t)$ falls below some critical value, these processes become inefficient,  
and $\ell_{\rm tot}(t)$  tends to grow because new strings enter the horizon. 
Thus equilibrium should be  reached for a particular value of the density of strings.  
The results of different simulations agree in confirming such a behaviour and, in particular, 
it is found that, regardless of the initial conditions,   at late times the number of strings for 
Hubble patch $\xi(t\to\infty)$ converges towards a value of  order of a few   
\cite{Yamaguchi:1998gx,Yamaguchi:1999yp,Hiramatsu:2010yu,Fleury:2015aca,Klaer:2017qhr}.
Quite noticeably,  other properties of the network, including the spectrum of axions emitted, also converge.

Given an overall length  of strings  $\ell_{\rm tot}(L)$ within a large volume $L^3$, 
the average number of strings of length $t\sim H^{-1}$ per Hubble volume $\sim t^{3}$  
is given  in terms of the length density  $\ell_{\rm tot}(L)/L^3$ as
\begin{equation}
\label{eq:xi_strings}
\xi(t) = \frac{\ell_{\rm tot}(L)}{L^3}\, t^2\,.
\end{equation}
For a string network, the  average separation between strings is then  $t/\sqrt{\xi}$, and this distance 
provides a cutoff for the large distance log divergence.  One can then define an effective 
string tension as 
\begin{equation}
\label{eq:mueff}
\mu_{\rm eff} (t) = \gamma(t)\, \pi f^2_a \,\log\(\frac{\eta f_a t}{\sqrt{\xi(t)}}\)  
\end{equation}
where $\gamma(t)$ is an effective boost factor,  typically of $\mathcal{O}(1)$, that accounts for the kinetic energy 
associated with the string network configuration.
%
In terms of the effective string tension the  energy density in strings then is: 
\beq
    \rho_{\rm long} = \xi(t) \frac{\mu_{\rm eff}(t)}{t^2} \,.
    \label{eq:axionicstring_simple}
\eeq
Hence, to the extent the theoretical expectation for the effective string tension \eqn{eq:mueff} is 
a good approximation (and numerical simulations indicate that this is the case)
the energy density of the string network at a particular time, once the equilibrium regime is reached,  
remains basically determined  from the density of strings. 
Various levels of sophistications have been used to describe the dynamics of   string networks. In the original formulation 
of the so-called one-scale model~\cite{Albrecht:1989mk,Bennett:1987vf}, the evolution of the network of  strings 
does not depend on   the string velocity. A more accurate description is obtained by  accounting for the effects   
of the velocity of the centre of mass of the string~\cite{Martins:1996jp, Martins:2000cs, Martins:2003vd}.
More sophisticated treatments also include  the effects of friction due to the interaction of the  strings with the 
surrounding matter and radiation~\cite{Vilenkin:1984ib, Martins:1995tg}. 
The mean-squared velocity of the long string gas $\langle v^2\rangle$ enters the equation of state $w_{\rm str}$ that describes the evolution of the energy density, 
namely~\cite{Vachaspati:1986cc}
\beq
	w_{\rm str} = \frac{2\langle v^2\rangle - 1}{3}\,.
	\label{eq:w_string}
\eeq
For ultra-relativistic strings $\langle v^2\rangle \approx 1$, and a free network of long strings evolves as radiation, while in the case of slow-moving strings $\langle v^2\rangle \approx 0$ and $w_{\rm str} = -1/3$. In general, we expect the equation of state for the string gas to vary within the range $-1/3 \leq w_{\rm str} \leq 1/3$. Numerical simulations of  string networks in a radiation-dominated cosmology  hint to a mean velocity  $\langle v^2\rangle^{1/2} \approx 0.5$, see
 for example Table~1 in Ref.~\cite{Martins:2018dqg}. 

\smallskip
\noindent
{\it String loops.} A population of  closed loops of  sub-Hubble size $\ell$ and energy density $\rho_{\rm loop}$ forms from the continuous intersection of long  strings.  Segments  of high curvature in  long strings can also split off forming isolated loops.
Loops vibrate and shrink while releasing energy into a spectrum of radiated axions. The power loss of the network into radiation can be described through the dissipation of the energy   $E_{\rm loop} = \mu_{\rm eff}(t) \ell$ of a closed loop 
of length $\ell$ into axions as~\cite{Battye:1993jv, Battye:1994qa, Martins:1995tg},
\beq
	\frac{dE_{\rm loop}}{dt} = \kappa_s\,\mu_{\rm eff}(t)\,,
	\label{eq:decayrateloop}
\eeq
where $\kappa_s \approx 0.15$ is a dimensionless quantity~\cite{Vachaspati:1986cc, Sakellariadou:1990ne, Sakellariadou:1991sd} computed at a fixed value of the string velocity $\langle v^2\rangle^{1/2} \approx 0.5$. 
%
Although in principle the length of the loops formed could be ranging at any size, numerical simulations~\cite{Battye:1994au, Vanchurin:2005pa, Martins:2005es, Ringeval:2005kr, Olum:2006ix, BlancoPillado:2011dq, Blanco-Pillado:2017oxo} show that the initial length of the large loop at its formation tracks the time of formation as
\beq
	\ell(t_I) =\alpha_{\rm loop} t_I\,,
	\label{eq:define_alphaloop}
\eeq
where $\alpha_{\rm loop}$ is an approximately constant loop size parameter which gives the fraction of the horizon size at which loops predominantly form. The loop spectrum is described in terms of a loop formation rate $r_\ell(\ell_I, t_I)$, which is related to the rate of the energy density dissipated into axions by~\cite{Kibble:1984hp, Cui:2008bd}
\beq
	\frac{d\rho_{\rm loop}}{dt} = \int_0^{\infty} d\ell\, \mu_{\rm eff}(t) \,\ell\, r_\ell(\ell, t)\,.
	\label{eq:loop_conversion1}
\eeq
Eventually, a loop of initial size $\ell_I$ shrinks and disappears by a time $t_F$ defined implicitly as $\ell(t_F, \ell_I) = 0$. 
%
\smallskip

The total energy density of the string network  receives contributions from both string 
populations $\rho_{\rm str} = \rho_{\rm long} + \rho_{\rm loop}$. 
%
Numerical simulations indicate that  roughly 80\% of the string length per Hubble
patch is contained in long strings, while the remaining 20\% is equally distributed 
in sub-Hubble loops of  different lengths, ranging from the horizon scale $\sim H^{-1}$
to the  size of the string core $\sim f_a^{-1}$~\cite{Gorghetto:2018myk}. Note that the ratio 4:1 in favour of long strings 
justifies the approximation of using the effective tension defined for the long string in \eqn{eq:mueff}  also for the string loops. 

\subsubsection{Spectrum of radiated axions}
\label{sec:string_radiated_axions}

In order to extract the energy density of the radiated axions from the simulation of the evolution of an axionic string network, the evolution of the energy density in the strings $\rho_{\rm str}$ is compared to what is obtained for a gas of free strings of energy density $\rho_{\rm free}$ that do not radiate~\cite{Gorghetto:2018myk}. The evolution of a free, non-intercommuting string gas is described by the kinetic equation
\beq
	\frac{d\rho_{\rm free}}{dt} + 3H\(1+w_{\rm str}\)\rho_{\rm free} = 0\,,
	\label{eq:freestrings}
\eeq
where $w_{\rm str}$ is given in Eq.~\eqref{eq:w_string}. The solution to Eq.~\eqref{eq:freestrings} is $\rho_{\rm free} \propto a^{-3(1+w_{\rm str})}$ which, in the case of a radiation-dominated cosmology and fixing the proportionality constant to match the initial energy density of the string network in Eq.~\eqref{eq:axionicstring_simple}, gives
\beq
	\rho_{\rm free}(t) = \xi \frac{\mu_{\rm eff}(t)}{t_{\rm PQ}^2}\(\frac{t}{t_{\rm PQ}}\)^{-\frac{3}{2}(1+w_{\rm str})}\,.
\eeq
When the axion emission is included, Eq.~\eqref{eq:freestrings} is modified to include the effective energy lost in the emission of axions per unit time $\Gamma_{{\rm str} \to a}$, defined as the difference in the rates at which the energy densities in the free string gas and the string network change,
\beq
	\Gamma_{{\rm str} \to a}(t) \equiv \dot\rho_{\rm free}(t) - \dot\rho_{\rm str}(t) = H(1 - 3w_{\rm str})\rho_{\rm str} - \frac{\rho_{\rm str}}{\mu_{\rm eff}}\,\frac{d\mu_{\rm eff}}{dt}\,.
	\label{eq:globalstring}
\eeq
Once the power radiated by the string network $\Gamma_{{\rm str} \to a}$ has been obtained, the energy density of the radiated axions follows the evolution $\rho_a^{\rm str} + 4H\rho_a^{\rm str} = \Gamma_{{\rm str} \to a}$; the axion energy density from the decay of the string network contributes to an additional source on top of the abundance expressed in Eq.~\eqref{eq:VRMaxions} obtained from the 
misalignment mechanism. 
The number density of axions associated to this process is~\cite{Gorghetto:2018myk}
\beq
	n_a^{\rm str} = \int^t\,dt'\frac{\Gamma_{{\rm str} \to a}(t')}{H(t')}\(\frac{R(t')}{R(t)}\)^3\,\int\frac{dk}{k}\,F(k) = \frac{\(1-3w_{\rm str}\)\pi f_a^2}{t^{3/2}}\int^t\,dt'\,\frac{\ln\left(f_a t'\right)}{(t')^{1/2}}\,\int\frac{dk}{k}\,F(k)\,,
	\label{eq:numberdensityaxions}
\eeq
where the spectral energy density $F(k) \propto k^{-q}$ has been explored in different regimes in the literature depending on the value of the spectral index $q$. A choice $q>1$ assumes that axions are radiated away on a timescale comparable to the Hubble time, and describes a power spectrum ranging over all modes from $k \approx 1/\ell(t_I) \approx H(t_I)/\alpha_{\rm loop}$ to infinity~\cite{Battye:1993jv, Battye:1994qa}. On the other hand, assuming that strings efficiently shrink emitting all of their energy at once leads to a flat power spectrum per logarithmic interval with an infrared cutoff at the wave mode $k \approx H$ and an ultraviolet cutoff at $k = f_a$, with a harder spectral index $q=1$~\cite{Harari:1987ht, Hagmann:1990mj, Chang:1998tb}. Demanding that the spectrum is normalised over the given interval results in
\beq
	F(k) = \begin{cases}
	\frac{q-1}{\alpha_{\rm loop}^{q-1}}\,\(\frac{k}{H}\)^{-q}, & \hbox{for $q > 1$}\,,\\
	\frac{1}{\ln \(f_a/H\)}\,\frac{H}{k}, & \hbox{for $q = 1$}\,.
	\end{cases}
	\label{eq_axionspectrum}
\eeq
The integration of Eq.~\eqref{eq:numberdensityaxions} with the spectrum in Eq.~\eqref{eq_axionspectrum} and $q> 1$ finally leads to
\beq
	n_a^{\rm str} \approx \frac{\(1-3w_{\rm str}\)\pi f_a^2}{2t_{\rm osc}}\,\times \begin{cases}
	\alpha_{\rm loop}\frac{q-1}{q}\,\ln\left(f_a t_{\rm osc}\right), & \hbox{for $q > 1$}\,,\\
	1, & \hbox{for $q = 1$}\,.
	\end{cases}
	\label{eq:numberdensityaxions0}
\eeq
This expression shows that most of the axions are radiated by loops right before $t_{\rm osc}$, when DWs dissipate the network. The computation matches the estimation for the decay of the string network at the time of DW formation~\cite{Davis:1986xc, Battye:1993jv}. The number density of axions at time $t_{\rm osc}$ from Eq.~\eqref{eq:numberdensityaxions0} results from a steeply falling integrand function of $t$, so that the dominant contribution comes from loops originating nearly instantaneously at values $t_I \sim t_{\rm osc}$~\cite{Davis:1986xc, Battye:1993jv, Battye:1994au}. The use of the harder spectrum with $q=1$ generically leads to results for the string contributions which are smaller by a factor $\log(f_a t_{\rm osc}) \approx 70$.

\subsubsection{Cosmological domain walls}
\label{sec:CosmoDW}

Cosmological DWs emerge with the spontaneous breaking of discrete symmetries~\cite{Vilenkin:1984ib, Vilenkin:1982ks,Vilenkin:2000jqa}. 
When the PQ symmetry is explicitly broken by nonperturbative QCD effects, the axion acquires a periodic potential with  $N_{\rm DW}$ equivalent minima.
Thus, the original $U(1)$ symmetry, which guarantees the equivalence of all values of $\theta$, is broken into a 
discrete $\mathbb{Z}_{N_{\rm DW}}$ symmetry.
Uncorrelated patches will choose different ground states randomly selecting one of the $N_{\rm DW}$ equivalent vacua. 
Axion DWs appear at the boundaries of physical regions that are in different minima, providing a smooth interpolation between the two vacua. 
Such configurations have a thickness $\delta_{\rm wall} \approx 1/m_a(t_{\rm osc})$ and an energy per unit area $\sigma_{\rm wall} \approx 8m_a^2(t_{\rm osc})v_a$ for the cosine potential~\cite{diCortona:2015ldu}.
Notice that there are axion DWs even in the case of $N_{\rm DW}=1$, which have an interpolating field configuration starting and ending in the same vacuum, but winding around the bottom of the Mexican hat potential once~\cite{Sikivie:2006ni}.
%
At the time when the axion acquires a mass, axion strings become the edge of DWs, forming a string-wall network~\cite{Vilenkin:1984ib,Vilenkin:2000jqa} whose dynamics depends on the value of $N_{\rm DW}$. 
In the case of $N_{\rm DW}=1$, strings disrupt the DWs soon after their formation by separating them into smaller 
pieces~\cite{Vilenkin:1982ks, Barr:1986hs} and dissipating the network whose  decay 
produces a spectrum of radiated axions which also includes a cold axion component that adds up  to 
the axion CDM produced by misalignment~\cite{Hiramatsu:2012gg, Gorghetto:2018myk, Vaquero:2018tib, Buschmann:2019icd}.

In contrast, in theories with $N_{\rm DW} > 1$  the DWs are generally stable,  and eventually come to dominate the energy density 
of the  Universe, largely overshooting any acceptable value. However, some solutions to this problem  
also exist, and are discussed in  \sect{sec:AxionsDW}.

\subsection{Axion isocurvature fluctuations}
\label{sec:isocurvature}

This topic is relevant in the pre-inflationary scenario. We consider an
early inflationary period described by the quasi-de Sitter space-time
metric, with a nearly-constant Hubble rate $H_I$. In this setting, the
axion field is a massless spectator. Inflationary models generically
predict the appearance of primordial scalar and tensor fluctuations,
which redshift to super-horizon scales to later evolve into primordial
curvature perturbations as well as primordial gravitational waves,
leaving an imprint in the CMB  radiation anisotropy and on the large-scale
structure~\cite{Mukhanov:1981xt, Guth:1982ec, Hawking:1982cz,
  Starobinsky:1982ee, Bardeen:1983qw, Steinhardt:1984jj}.  CMB
features distinct peaks both in its angular power spectrum of
temperature (TT) as well as in the temperature-polarisation
cross-power spectrum (TE), with the first peak in TT  located at
$\ell = 220.6 \pm 0.6$ at 68\%~CL by the  {\it
  Planck} collaboration survey~\cite{Akrami:2018vks}. Other information
extracted from the CMB includes the polarisation angular power
spectrum (EE).

Adiabatic curvature perturbations generated during inflation show a
nearly scale-invariant dimensionless power spectrum of primordial
curvature perturbations with amplitude $\PkO$, with a mild dependence
on the co-moving wavenumber $k$ which is parametrised by a scalar
spectral index $n_S$ around an arbitrary reference scale
$k_0$\footnote{The reference scale $k_0$ is also called the ``pivot''
  scale in cosmology, and refers to the comoving scale at which
  measurements are taken. For example, the \textit{Planck} mission
  often refers to the pivotal scale
  $k_0 = 0.05\,$Mpc$^{-1}$~\cite{Aghanim:2018eyx}. For isocurvature
  and tensor modes, the constraints are quoted at three different
  scales: $k_{\rm low} = 0.002\,$Mpc$^{-1}$,
  $k_{\rm mid} = 0.050\,$Mpc$^{-1}$, and
  $k_{\rm high} = 0.100\,$Mpc$^{-1}$~\cite{Akrami:2018odb}.}
as~\cite{Kosowsky:1995aa, Leach:2002dw, Liddle:2003as} \beq \Pk \equiv
\PkO\,\left(\frac{k}{k_0}\right)^{n_S-1}\,.
	\label{curvature_perturbations}
\eeq
If the axion field $a$ originates during inflation, it also
inherits  quantum fluctuations with the typical standard deviation
$\sigma_a$ of a massless scalar field in the accelerated expansion,
see e.g.~Ref.~\cite{Riotto:2002yw},
\beq
	\sigma_a = \sqrt{\langle a^2\rangle} \simeq \frac{H_I}{2\pi}\,,
\eeq
and with a corresponding standard deviation for the distribution of
the axion angle $\sigma_\theta = \sigma_a/f_a$. Since these 
fluctuations are independent of the
quantum fluctuations of the inflaton field, they are of the
isocurvature type. More in detail, the fluctuations in the axion field
do not perturb the total energy density during inflation, but change
the value of the axion number density with respect to entropy density,
$\delta \(n_a/s\)\neq
0$.
Axion isocurvature fluctuations convert into curvature perturbations
at the time at which the axion mass becomes relevant, around the QCD
phase transition~\cite{Crotty:2003rz, Beltran:2005xd, Beltran:2006sq},
due to the feedback of isocurvature into curvature
modes~\cite{Bozza:2002fp, Bozza:2002ad}. Since the axion field behaves
as CDM at recombination, these axion perturbations imprint into the
temperature and polarisation fluctuations in the CMB and, because of
their isocurvature nature, they are completely uncorrelated with the
adiabatic curvature perturbations.

The magnitude of the axion isocurvature perturbations $\Delta^2_a(k) = \Delta^2(k_0) (k/k_0)^{n_I-1}$  
where $n_I$ the isocurvature spectral index,   is given by the  relative fluctuation in the  axion energy 
density~\cite{Kobayashi:2013nva}. 
In terms   of the initial  angle $\theta_i$  the amplitude  can be written as
\beq
	\Delta_a (k) = 
	\frac{\delta \Omega_{\rm CDM}}{\Omega_{\rm CDM}} = \mathcal{F}^a_{\rm CDM} 
	\,\frac{\delta \ln \Omega_a}{\delta \theta_i}\,\sigma_\theta \simeq 	\mathcal{F}^a_{\rm CDM} \,
		\frac{H_I}{\pi \theta_i f_a } \,, 
	\label{eq:amplitude_iso}
\eeq
where $\mathcal{F}^a_{\rm CDM}  = \Omega_a/\Omega_{\rm CDM}$ is the relative contribution of 
axions to CDM, we have assumed $\delta \Omega_{\rm CDM} = \delta \Omega_a$   and  the last step 
holds in  the small $\theta_i$ regime in which anharmonic corrections can be neglected and 
 $\Omega_a \propto \theta_i^2$  (see for  example \eqn{eq:rhoa_analytic}).
%
%
Measurements of the anisotropies in the CMB can be used to constrain
$\Delta^2_a(k) $,  since data show that primordial fluctuations
are predominantly adiabatic with little space left for fluctuations of
the isocurvature type. 
%
The \textit{Planck} mission constrains the fraction of uncorrelated 
isocurvature fluctuations by placing an upper bound  on the quantity 
\beq
\label{eq:beta} 
	\beta(k) \equiv \frac{\Delta^2_a(k)}{\Pk + \Delta^2_a(k)}\,,
\eeq
where the spectra are computed at a pivot scale $k$. Here, we
have used the measurements of the CMB temperature and polarisation
anisotropies from the \textit{Planck} satellite, jointly with the
BICEP2/Keck Array (BK15) for the combined \textrm{TT, TE, EE + lowE +
  lensing + BK15} dataset at 68\% CL~\cite{Ade:2013zuv,
  Planck:2013jfk, Barkats:2013jfa, Ade:2015tva, Ade:2015lrj,
  Ade:2018gkx, Aghanim:2018eyx, Akrami:2018odb}. We list  the values
used in this Review in Table~\ref{tab:planckdata}, where data in the
form $A\pm B$ are reported at 68\% CL, while upper limits are reported
at 95\% CL.
Isocurvature bounds place a stringent constraint on the scale of
inflation, so that  models of axion DM  in the pre-inflationary 
scenario require a relatively low Hubble rate $H_I$
compared to the scale $H_I \lesssim 10^{13}\,$GeV which has currently
been probed using CMB tensor modes. Using the  results in Table~\ref{tab:planckdata},  
from Eqs.~(\ref{eq:amplitude_iso}) and (\ref{eq:beta})   the  following constraint 
on the inflationary scale can be derived: 
\beq
	H_I \lesssim  
	  \frac{0.9 \times 10^7 }{\mathcal{F}^a_{\rm CDM}}\;
	\(\frac{\theta_i}{\pi} \;\frac{f_a}{10^{11}{\rm \,GeV}} \)\,{\rm GeV}\,. 
	\label{eq:isocurvature}
\eeq

The bound from  isocurvature  perturbations can, however, be evaded in many ways.
The simplest one is to appeal to the post-inflationary scenario,  since in this case there is
no massless Goldstone boson during inflation,  although, for $N_{\rm DW} > 1$, one then 
has to  deal with  the DW problem, see \sect{sec:CosmoDW}.
However, other possibilities are viable also  within scenarios in which   the PQ symmetry 
is already broken during inflation. 
For example, in Ref.~\cite{Linde:1991km}  it was argued that  that the isocurvature bound holds 
only  under the assumption that $f_a$  does not change during the last stages of inflation or 
after it, and   it was argued that this assumption is not valid in general.  
Axion models embedded  into a  {\it hybrid inflation}  scenario
in which the background value  of the axion field   slowly decreases, rolling down 
to its equilibrium value $\vev{|\Phi |} = v_a/\sqrt{2} \propto f_a$  were  in fact  
shown to  be compatible with an inflationary scale several orders of magnitude larger 
than the reference value in \eqn{eq:isocurvature}, 
while still keeping   $f_a \sim   10^{11}\,$GeV  ($m_a\sim 6\cdot 10^{-5}\,$eV)
and without the need to tune  $\theta_i$ to particularly small values.  
Ref.~\cite{Folkerts:2013tua} considered instead   the possibility of a  non-minimal derivative  coupling of the axion field  
to gravity. While the new coupling does not affect the density of  axion CDM, it 
can effectively suppress  the isocurvature perturbations during inflation.
Another solution studied in Ref.~\cite{Jeong:2013xta}  relies on the observation 
that  if the axion acquires a sufficiently large mass already during inflation, its quantum fluctuations 
at super-horizon  scales would become significantly suppressed, thereby  relaxing the constraint 
on the inflation scale. 
This can be achieved  by assuming  a phase in the  early Universe in which 
QCD becomes strong   thus rendering the axion massive. 
Such a  scenario was in fact first proposed in Refs.~\cite{Dvali:1995ce,Banks:1996ea} 
in the attempt of suppressing the cosmological abundance of axions from the
misalignment mechanism, a task  that, as is outlined  in \sect{sec:preI} below,
eventually turned out to be rather difficult to accomplish.   
However, an early  strong QCD phase 
yielding an axion mass larger than the Hubble scale over a certain period
of inflation was reconfirmed in Ref.~\cite{Choi:2015zra} 
as a viable mechanism to  suppress  isocurvature  axion fluctuations.
Finally,   the issue would of course disappear within inflationary models 
 that could naturally respect  the isocurvature constraint on the inflationary
Hubble scale in~\eqn{eq:isocurvature}, and  indeed this was shown to be possible 
in low-scale models of hybrid inflation~\cite{Schmitz:2018nhb}.

\begin{table}
	\footnotesize
	\begin{center}
	\renewcommand{\arraystretch}{1.4}
	\begin{tabular}{|@{\hspace{1 cm}}l|@{\hspace{0.5 cm}} l|@{\hspace{0.5 cm}} l|}
	\hline
	\textbf{Parameter}	& \textbf{Prior} & \textbf{Reference}\\
	\hline\hline
	$\Omega_{\rm CDM}h^2$ & $0.1200 \pm 0.0012$& Base $\Lambda$CDM, \textit{Planck} TT, TE, EE + lowE + lensing~\cite{Aghanim:2018eyx}\\
	$\ln\(10^{10}\Delta^2_\mathcal{R}\)$	&	$3.044 \pm 0.014$ & Base $\Lambda$CDM, \textit{Planck} TT, TE, EE + lowE + lensing~\cite{Aghanim:2018eyx}\\
		$\beta$	&	$ < 0.038$ at 95\% CL & CDI $n_{\rm II} = 1$, \textit{Planck} TT, TE, EE + lowE + lensing~\cite{Akrami:2018odb}. \\
	$n_S$	&	$0.9649 \pm 0.0042$ & $\Lambda$CDM+$r$, \textit{Planck} TT, TE, EE + lowE + lensing~\cite{Aghanim:2018eyx}\\
	$r $		&	$ < 0.056$ at 95\% CL & $\Lambda$CDM+$r$, \textit{Planck} TT, TE, EE + lowE + lensing+BK15~\cite{Akrami:2018odb}\\
	\hline
	\end{tabular}
	\end{center}
	\caption{Cosmological parameters 
          from the joint  {\it Planck} and 
          BICEP2/Keck Array (BK15) analysis of the combined  \textrm{TT, TE, EE + lowE + lensing + BK15} 
          dataset,  from Refs.~\cite{Aghanim:2018eyx,Akrami:2018odb}.
          The  value of $\Delta^2_\mathcal{R}$  and the constraint on the primordial isocurvature fraction $\beta$ 
           defined in \eqn{eq:beta}   are at the default pivot scale $k_0 = 0.05\,$Mpc$^{-1}$.
          The limit on the tensor to scalar ratio $r$ is at the pivot scale $k_0 = 0.002\,$Mpc$^{-1}$.          
          }
	\label{tab:planckdata}
\end{table}




\subsection{Cosmological bounds on the axion mass}
\label{sec:axionmassbounds}

Before addressing the upper bounds that can be set on the axion 
mass from  early Universe and cosmological consideration, 
let us mention  a generic (non cosmological) argument
that is sometimes invoked to claim a {\it lower} limit on the axion mass. 
This argument is based on the belief that an axion  decay constant $f_a$ above 
the Planck scale is  incompatible with an effective QFT description 
so that,  according to  \eqn{eq:axionmassfa}, $f_a \lesssim \mP$  
would yield $m_a \gtrsim 5\times 10^{-13}\,$eV.  
Note, however, that even this seemingly model-independent bound can be circumvented
in a particular axion construction that exploits the  Kim-Nilles-Peloso  mechanism~\cite{Kim:2004rp}. 
This mechanism  requires the existence of at least two axions, and  can produce an effective 
super-Planckian  axion scale, although the original fundamental scale is sub-Planckian~\cite{Kim:2004rp}.
In any case this lower limit would not  compete with  the lower limit on the axion 
mass  $m_a \gsim 2\times 10^{-11}\,$eV that can be obtained from considerations of 
the super-radiance phenomenon (a topic that  will be touched on in \sect{sec:Astro_bounds_axion_gravity}) 
and that as long as axion self interactions are sufficiently feeble, is truly model independent.

Moving now to review the cosmological  arguments that allow to constrain from above 
the value of the axion mass, it should be mentioned 
from the start that  altogether these  bounds are  much 
less constraining than the  limits that can be inferred from astrophysical processes (discussed in   
 \sect{sec:Astro_bounds}),  by translating the bounds on the different axion 
 couplings  $\gag,\gae,\gaN$ into lower  bounds on the axion decay constant $f_a$ and 
 in turn on upper limits on $m_a$.   Although the upper limits derived in this way  
 are admittedly model dependent, and a precise number cannot  be given without specifying  
 the theoretical setup,  the indication that  $m_a \lesssim 0.1\,$eV  is  rather solid. 
 In spite of this, the cosmological bounds are well worth mentioning because they rely on 
 completely independent arguments.\\

{\it Hot dark matter.}\  
As will be discussed in more detail in~\sect{sec:hotaxions}, at temperatures below the QCD phase transition  
axions interact with pions  $(a\,\pi \leftrightarrow \pi\, \pi)$  and nucleons  $(a\, N \leftrightarrow \pi\, N)$ 
with a strength that increases  as $m_a$ is increased. 
Then, for sufficiently large values of  $m_a$  a thermal population of axions 
arises, which would constitute a Hot DM (HDM) component.
However, HDM abundance is severely constrained by data from cosmological surveys.
Using  7-year data from the Wilkinson Microwave Anisotropy Probe  (WMAP-7)   
together with  Sloan Digital Sky Survey (SDSS) observations  and 
the Hubble constant from Hubble Space Telescope (HST) 
Ref.~\cite{Hannestad:2010yi} set the bound $m_a\lsim 0.72\,$eV. 
A  bound on $m_a$ using only CMB data  was reported in~\cite{Archidiacono:2013cha}.
More recently,    using Planck 2015 results the bound  was  strengthen 
to  $m_a\lsim 0.53\,$eV~\cite{DiValentino:2015wba}.
A study of the potential of future cosmological surveys  for further constraining the axion mass
 has been presented in Ref.~\cite{Archidiacono:2015mda}. \\

{\it  Baryon to photon ratio from Big Bang Nucleosynthesis (BBN) and from CMB.}\ 
Axions decay to two photons with a lifetime inversely  proportional to the fifth power of the mass,
\beq
	\tau_{a} = \frac{64\pi}{g_{a\gamma}^2\,m_a^3} \sim 10^{24} 
	 \(\frac{{\rm eV}}{m_a}\)^5 \,  {\rm s}. 
	\label{eq:axion_lifetime}
\eeq
For $m_a \gsim 20\,$eV  the axion lifetime drops below the age of the Universe,  
hence, for a sufficiently large mass, the decay happens early enough and  the HDM bound 
no longer applies. However,  if axions decay not too early, non-thermal photons injected in 
the bath would  produce a spectral distortion in the CMB, and this  can be used to derive 
restrictive constraints~\cite{Masso:1997ru}. However,  
if decays occur early enough,  photons will have the time to thermalise
and the CMB would not be affected. 
Nevertheless, the additional entropy injected into the bath   increases  the number of photons  
relative to that of baryons, decreasing  the ratio $\eta_b=n_b/n_\gamma$. 
Nowadays we know with  good precision the value of $\eta_b$,  which is  inferred independently  
from BBN and from CMB data, and there is a beautiful agreement between the two results. 
Since the relevant physics processes responsible for the primordial abundances of  light elements and for 
the CMB occur in different cosmological eras that are  characterised by  temperatures that differ by about 
six orders of magnitude,   the amount of entropy that can be injected in the thermal bath 
between  $T_{\rm BBN}\lesssim 1\,$MeV  and $T_{\rm CMB}\lesssim 1\,$eV  is tightly constrained.  
This argument was used in Ref.~\cite{ Cadamuro:2010cz} to exclude $m_a\lsim 300\,$keV.
A later study that used Planck  data, together with new  inferences of primordial
element abundances, pushed the limit up to $m_a\gsim 1\,$MeV~\cite{Millea:2015qra}, 
while an  assessment of the robustness of the bounds with respect to possible deviations 
in the standard cosmology  was recently presented in  Ref.~\cite{Depta:2020wmr}.
 These bounds  are nicely complemented by constraints from beam dump and reactor experiments 
 which exclude  axion (and ALP) masses from MeV up to several GeVs~\cite{Riordan:1987aw,
 Bjorken:1988as,Blumlein:1990ay,Blumlein:1991xh,Mimasu:2014nea,Jaeckel:2015jla,
 Dobrich:2015jyk, Brivio:2017ije,Bauer:2017ris,Dolan:2017osp,Dobrich:2019dxc,Gavela:2019cmq}.
 Altogether, these considerations allow to identify $m_a \lsim 0.53\,$eV as the 
cosmologically allowed  axion mass  window.

%
%


\subsection{Benchmark axion mass region for  $\Omega_a \simeq \Omega_{\rm CDM}$}
\label{sec:benchmark}


The present axion relic abundance $\Omega_a$ depends on multiple  
production mechanisms that have been reviewed in the previous sections. 
Here, we discuss the benchmark axion DM mass regions both in 
the post- and pre-inflationary PQ breaking scenarios, 
including also the constraints coming from various cosmological considerations. 

\subsubsection{Post-inflationary scenario}
\label{sec:postI}

In post inflationary scenarios  we would expect that 
the  contribution to CDM from axions could be assessed with 
particular precision, given that estimates of the density of axions 
produced  from the misalignment mechanism do not suffer from uncertainties 
related to particular choices of the initial conditions. 
Unfortunately this is not completely true. Other  important sources of uncertainties 
are in fact present and hamper, for example, precise  determinations of the  value of $m_a$ for which axions can account 
for the whole of CDM, a quantity to which we will refer as  {\it the  DM axion mass}.
These uncertainties are  primarily   related to 
the serious  difficulties that one  encounters in trying to assess  
the contribution to the population of 
cold axions from the decay of topological defects, 
but additional uncertainties are also  associated with  evaluations of  
the  dependence on temperature of the topological  susceptibility $\chi(T)$, see \eqn{eq:chiT}. 
As a consequence,  although it is not uncommon to find in the literature values of the DM axion mass  
with relatively small errors,  differences between  different estimates can easily be several  times larger 
than the quoted errors.  This is simply due to the fact that the theoretical 
uncertainties mentioned above are not accounted for.  To be conservative, one can take  as a reasonable range 
for the DM axion mass   $m_a \approx \(10 - 100\){\rm   \, \mu eV}$,  and likely  with an even larger upper 
value if $N_{\rm DW} > 1$.
It  is clear, however, that to  to guide experimental searches  for axion DM  any improvement in controlling 
the  major sources of uncertainties is highly desirable. To this aim, several groups are carrying out extended 
simulations  of the network of axionic strings in order to improve in the understanding of their effect. 
In this respect, models with $N_{\rm DW} = 1$ are more promising because  they do not require 
additional assumptions about the dissipation of the domain wall network, and for this 
reason most studies focus on this case.
Hiramatsu {\it et
  al.}~\cite{Hiramatsu:2010yu} simulate the evolution of the axionic
string network using an efficient identification scheme of global
strings in order to assess the energy spectrum and the axion decay
constant, which is found to be
$f_a \lesssim 3 \times 10^{11}{\rm \,GeV}$ or
$m_a \gtrsim 20{\rm \,\mu eV}$, while the string stretching parameter
appearing in Eq.~\eqref{eq:axionicstring_simple} is found to be
$\xi = 0.87 \pm 0.14$. Klaer and Moore~\cite{Klaer:2017qhr} have used
the expertise in Refs.~\cite{Fleury:2015aca, Fleury:2016xrz} and
developed a numerical scheme to handle on the lattice the large hierarchy 
$\log\(f_a t_{\rm osc}\) \sim 70$  between the scale of the string core $f_a \sim r_c^{-1}$  and the 
string separation scale   at the time when the axion acquires a mass of order 
$t_{\rm osc} \sim H^{-1}_\o$. 
Using their technique along with the QCD
susceptibility obtained in Ref.~\cite{Borsanyi:2016ksw}, Klaer and
Moore find that the total axion production when strings are included
is somewhat less efficient than in the angle-averaged misalignment
case. They quote a value of the axion mass
$m_a = \(26.2 \pm 3.4\){\rm  \, \mu eV}$~\cite{Klaer:2017ond}.
Gorghetto {\it et al.}~\cite{Gorghetto:2018myk,Gorghetto:2019hee} find an axion spectrum
peaked at the energy of the order of the string core scale that, 
 when extrapolated to the physical parameter region,
would lead to a negligible number density of relic axions from strings.
However, they also showed that the presence of small logarithmic 
corrections to the spectrum shape could completely alter such a 
conclusion, and their ongoing studies in this direction indicate that this might
indeed be the case.\footnote{G. Villadoro, private communication.}
 Buschmann {\it et al.}~\cite{Buschmann:2019icd} have
performed high-resolution simulations of the evolution of the PQ field
starting at the epoch before the PQ phase transition and ending at
matter-radiation equality, about the time at which axion miniclusters
collapse. The value of the axion mass obtained from the simulations is
$m_a = \(25.2 \pm 11.0\){\rm
  \, \mu eV}$.
Both results in Refs.~\cite{Gorghetto:2018myk, Buschmann:2019icd} find
a logarithmic deviation to the number of strings per Hubble patch from
the scaling regime. Kawasaki {\it et al.}~\cite{Kawasaki:2014sqa}
estimate the abundance of CDM axions both from the misalignment mechanism and
the decay of topological defects, obtaining the mass range
$m_a = \(115 \pm 25\)
{\rm \,\mu eV}$
for the models with the domain wall number $N_{\rm DW} = 1$. The
results for the present axion energy density as a function of the
axion mass obtained in Refs.~\cite{Klaer:2017ond, Buschmann:2019icd}
are consistent with each other, while the results in
Ref.~\cite{Kawasaki:2014sqa} are larger by a factor of order ten. Such
a discrepancy could be partially due to the explicit separation of the
misalignment and string decay production mechanisms enforced in
Ref.~\cite{Kawasaki:2014sqa}, which leads to an over-counting of the
axion energy density~\cite{Klaer:2017ond}. The simulation by Vaquero
{\it et al.}~\cite{Vaquero:2018tib} differs from that in
Ref.~\cite{Buschmann:2019icd} for a number of details, including the
use of the ``fat'' string scheme~\cite{Moore:2001px}, initial
condition, the measurement of the CDM energy density, the length and
the details of the evolution to the matter-radiation equality. These
results strongly depend on the choice of the temperature dependence of
the QCD topological susceptibility. Refs.~\cite{Hiramatsu:2010yu,
  Kawasaki:2014sqa, Buschmann:2019icd} have derived their results
using a susceptibility with an index corresponding to $2\gamma = 6.68$
in Eq.~\eqref{eq:QCDaxion_mass}, using the parametrisation in the
Interacting Instanton Liquid Model~\cite{Wantz:2009it}. This exponent is milder than what obtained
in Ref.~\cite{Borsanyi:2016ksw} for which, at
temperatures $T \gg T_C$,  $2\gamma = 8.2$. 
The bottom line is that, although the misalignment is the most model independent and  
  best understood of all the axion CDM production mechanisms, 
no consensus has been achieved yet about the share of axion CDM ascribable 
to the misalignment mechanism 
with respect to the  much less  understood contributions  from topological defects.
This reflects in a large uncertainty in determining the real value for the
axion mass that can fully  saturate the  DM density.

\subsubsection{Pre-inflationary scenario}
\label{sec:preI}

\begin{figure}[t!]
\begin{center}
	\includegraphics[width=0.7\linewidth]{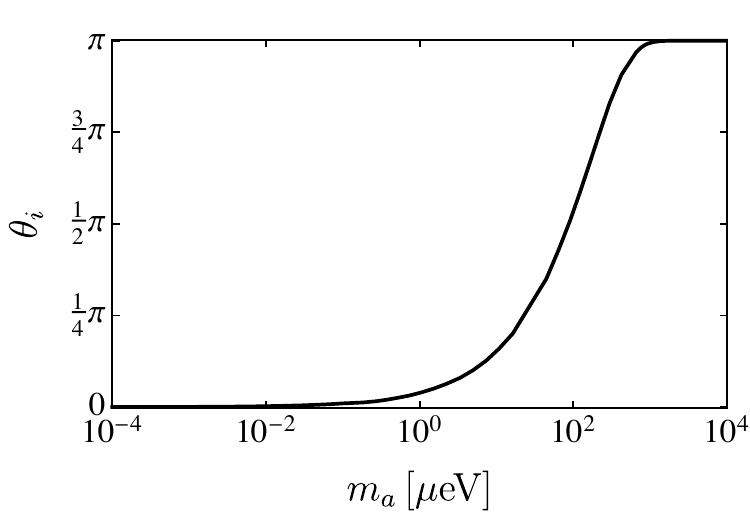}
	\caption{The relation between the DM axion mass and the initial misalignment angle in the 
	pre-inflationary scenario.}
	\label{fig:bound_standardFIXED}
\end{center}
\end{figure}

This scenario is realised whenever the PQ symmetry is spontaneously
broken during inflation, $H_I < f_a$, and it is not restored
afterwards~\cite{Dine:1982ah}. In this scenario, the axion field is
homogeneous through various Hubble patches, with a unique value of
$\theta_i$ characterising the whole observable Universe. As shown in
Fig.~\ref{fig:bound_standardFIXED}, each value of $\theta_i$ is related to
a unique value of the DM axion mass. 
A global fit of this
scenario~\cite{Hoof:2018ieb} performed using the DarkBit
module~\cite{Workgroup:2017lvb} of the GAMBIT numerical
code~\cite{Athron:2017ard} yields the DM axion mass range
$0.12{\rm \,\mu eV} \leq m_a \leq 0.15{\rm \,meV}$ at the 95\%
equal-tailed confidence interval of the marginalised posterior
distribution accounting for the QCD axion (both the KSVZ and the DFSZ
models) and taking into account results from various observations and
experiments in the likelihood including the light-shining-through-wall
experiments, helioscopes, cavity searches, distortions of gamma-ray
spectra, supernovae, horizontal branch stars and the hint from the
cooling of white dwarfs. An important assumption that impacts on the
result is the choice for the prior on $\theta_i$, which in
Ref.~\cite{Hoof:2018ieb} is assumed to be uniform over the interval
$\[-\pi, \pi\)$.

Small initial values of $\theta_i$ might also occur naturally,
i.e.~without any fine tuning, in low-scale inflation models in which
inflation lasts sufficiently long~\cite{Graham:2018jyp,
  Guth:2018hsa}. 
  If $H_I \lesssim \Lambda_{\rm QCD}$ the axion
acquires a mass already during inflation, the $\theta_i$-distribution
flows towards the CP conserving minimum and, for long durations of
inflation, stabilises around sufficiently small $\theta_i$ values.  As
a result the QCD axion can naturally give the DM abundance
for axion masses well below the classical window, down to
$m_a \approx 10^{-12}\,$eV~\cite{Graham:2018jyp,Guth:2018hsa}.

Another way  to suppress the value of the axion misalignment angle  is by 
assuming a period in the early Universe during which the QCD coupling constant
takes a value larger than the present one, allowing 
the colour group to become strong for a certain period,  during which therefore 
$m_{a}\gsim H$~\cite{Dvali:1995ce}
(see also~\cite{Banks:1996ea}).   The axion field   is then dynamically 
driven to its minimum at early times. Attempts to realise this scenario have 
generally relied on supersymmetric models in which  the 
 gauge coupling constant is related to the expectation value of  
 some moduli field, like the the dilaton,  that initially sits 
 away from its true minimum and  later adjusts to its present value.  
 Ref.~\cite{Choi:1996fs}  questioned this possibility and   concluded that,  
  under generic conditions,  an early phase of stronger QCD 
 is not useful for raising the cosmological upper bound of the axion scale, 
although, as was mentioned in~\sect{sec:isocurvature},  it can still be effective for  
suppressing axion isocurvature fluctuations~\cite{Jeong:2013xta,Choi:2015zra}. 
 The negative  conclusion of  Ref.~\cite{Choi:1996fs} can however be circumvented, and  
 a viable realisation in which  $f_a \sim \mathcal{O}(10^{16}\!-\!10^{17})\,$GeV  
 can be obtained has been recently given  in Ref.~\cite{Co:2018phi}.

Going in the opposite direction, values of the initial misalignment angle that are
close to the hilltop of the potential $\theta_i \simeq \pi$ could
drive the axion energy scale to values $f_a \lesssim 10^{12}\,$GeV
($m_a \gtrsim 10\,\mu$eV), because of the relevance of the
non-harmonic terms in the axion potential. With the choice
$\pi - \theta_i \simeq 10^{-3}$, one obtains
$f_a \approx 10^{10}\,$GeV. It is possible that the axion field has
been driven to such a value of $\theta_i$ during
inflation. This can be again obtained by enhancing the QCD 
confinement scale  so that the axion  acquires a potential 
during inflation~\cite{Co:2018mho}, or by assuming that the 
  inflation scale is  below the QCD scale~\cite{Takahashi:2019pqf}.

\subsubsection{Summary of cosmological  bounds}

\begin{figure}[t!]
\begin{center}
	\includegraphics[width=0.75\linewidth]{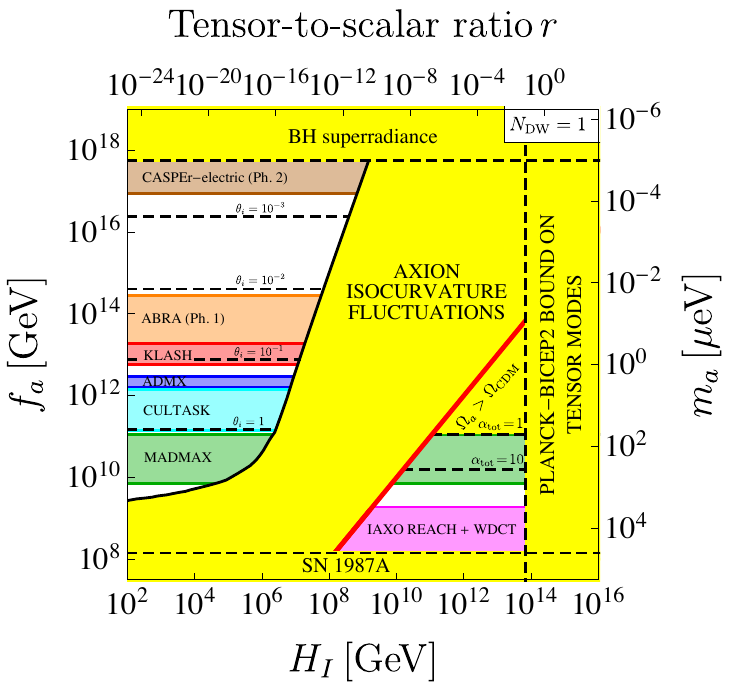}
	\caption{Region of axion parameter space where the axion
          constitutes the totality of the DM observed. The
          axion mass scale on the right corresponds to
          Eq.~\eqref{eq:axionmass} for the case $N_{\rm DW} = 1$. If
          the PQ symmetry breaks during inflation and the axion
          spectates inflation ($f_a \gtrsim H_I$, pre-inflationary
          scenario), axion isocurvature perturbations constrain the
          parameter space to the region on the top left, which is
          marked by the values of $\theta_i$ necessary to achieve the
          observed CDM density for a given value of $f_a$. If the PQ
          symmetry breaks after inflation ($f_a < H_I$, post-inflationary
          scenario), the axion is the CDM particle only for a specific
          value of $f_a$, which takes into account the contributions
          from the decay of topological defects $\alpha_{\rm
            tot}$.
          The lower bound on $f_a$ results from astrophysical
          considerations~\cite{RAFFELT1986402, Raffelt:2006cw,
            Viaux:2013lha, Giannotti:2017hny}, the upper bound on
          $f_a$ relies on the non-detection in LIGO of gravitational
          waves associated with the super-radiance phenomenon from
          stellar-mass black holes~\cite{Arvanitaki:2010sy,
            Arvanitaki:2014wva}, the upper bound on $H_I$ comes from
          the non-observation of tensor modes in the
          CMB~\cite{Ade:2015xua, Akrami:2018odb,
            Aghanim:2018eyx}. The coloured transparent bands indicate
          future reaches of planned or ongoing experiments covering
          the allowed regions of the parameter space: CASPEr-Electric Phase 2 (bronze), ABRACADABRA
          (ABRA Ph.1, orange), KLASH (red), ADMX (blue), CULTASK (Cyan), MADMAX (green), and IAXO
          (magenta).}
	\label{fig:bound_standard}
\end{center}
\end{figure}

The various constraints on the parameter space of the QCD axion can be
summarised as in Fig.~\ref{fig:bound_standard}, in which we focus on 
the KSVZ model (with $E/N = 0$ and $N_{\rm DW} = 1$), 
and we show the axion energy scale $f_a$ (left Y
axis) as a function of the Hubble rate at the end of inflation $H_I$
(bottom X axis). The parameter space of the DM axion depends on four
quantities, namely the axion energy scale $f_a$, the initial
misalignment angle $\theta_i$, the Hubble rate during inflation $H_I$,
and the contribution from topological defects to the total axion
energy density $\alpha_{\rm tot} \equiv \rho_a/\rho_a^{\rm mis}$. For
the QCD axion, the axion mass $m_a$ is related to $f_a$ as in
Eq.~\eqref{eq:axionmassfa}, see the different grids used on the Y
axes. For single-field inflation, the tensor-to-scalar ratio $r$ is
proportional to $H_I^2$, as expressed in the top X axis. The parameter
space is bound to the right by the non-detection of primordial
gravitational waves by the $Planck$-BICEP2 joint analysis~\cite{Ade:2015tva},
which set an upper limit on the scale of inflation $H_I$. The axion
energy scale $f_a$ is bound from below by various astrophysical
considerations, for example the duration of the neutrino burst from the supernova (SN) 1987A, see Fig.~\ref{fig_KSVZ_astro_bounds} and \sect{sec:Astro_bounds_nEDM}. This translates into an upper bound on the mass of the QCD axion. The phenomenon of superradiance excludes the portion of the axion mass given in
Eq.~\eqref{eq:BH_superradiance_fpq} 
(see \sect{sec:Astro_bounds_axion_gravity}). The solid red line marks
the watershed $f_a = H_I$ and separates the region where the axion is
present during inflation (top-left region, pre-inflationary scenario) from
the region where the axion field originates after inflation 
(bottom-right region, post-inflationary scenario). This line has to be thought as a
qualitative bound between the two scenarios considered, since the
exact details depend on the inflationary model, the
preheating-reheating scenarios, and axion particle physics.

For $f_a \gtrsim H_I$ (pre-inflationary scenario), we show the values
$\theta_i \in \{1, 10^{-1}, 10^{-2}, 10^{-3}\}$. The parameter space in this
scenario is bound to the right by the non-observation of axion
isocurvature fluctuations in the CMB by the \textit{Planck}
mission~\cite{Ade:2015xua, Akrami:2018odb, Aghanim:2018eyx}, subject
to the requirement that axions constitute the DM. The change
in the slope corresponds to the effect of the non-harmonic terms in
the axion potential when
$\theta_i \gtrsim \mathcal{O}(1)$~\cite{Visinelli:2009zm}. The
possible presence of axion isocurvatures in the spectrum of the CMB
relies on the fact that the PQ symmetry has never been restored after
the end of inflation. For example, in the unified 
Standard Model--Axion--seesaw--Higgs portal inflation (SMASH)
model of Refs.~\cite{Ballesteros:2016euj,Ballesteros:2016xej,Ballesteros:2019tvf},
although the axion energy scale can be as large as
$f_a \sim 4\times10^{16}\,$GeV, the PQ symmetry is restored immediately
after the end of inflation and isocurvature modes are absent. Caveats
that allow to evade the bound from isocurvature fluctuations include
the presence of more than one axion~\cite{Kitajima:2014xla}, or the
identification of the inflaton with the radial component of the PQ
field~\cite{Fairbairn:2014zta}, see \sect{sec:axion_inflation}.

Note, that in single-field inflation models in which the axion
constitutes the DM, the pre-inflationary scenario requires a value
of the tensor-to-scalar ratio $r$ which is well below the projected
sensitivity of forthcoming cosmological surveys, $r\sim 10^{-3}$,
which is forecast for example by the next-generation ground-based
cosmic microwave background experiment
CMB-S4~\cite{Abazajian:2016yjj}.\footnote{If the DM is in the
  form of QCD axions, a joint analysis of axion direct detection
  experiments and future CMB-S4 experiments is able to probe the range
  $2.5 \times 10^6 \lesssim H_I /{\rm GeV} \lesssim 4 \times
  10^9$~\cite{Abazajian:2016yjj},
  which is otherwise not accessible through CMB tensor modes
  alone.} For example, an explicit realisation of an inflation model
with an extremely low value of the tensor-to-scalar ratio
$r \sim 10^{-13}$ such that the axion is the DM has been
recently presented in Ref.~\cite{Tenkanen:2019xzn}.

For $f_a \lesssim H_I$ (post-inflationary scenario), the axion is present
during inflation as the Goldstone boson of the PQ field. The initial
value of the misalignment angle is averaged out over many Hubble
patches. We have used $\langle \theta_i^2\rangle \approx (2.15)^2$ as
in Ref.~\cite{diCortona:2015ldu}, and we show results for two cases:
1) $\alpha_{\rm tot} = 1$ (no contribution from the decay of
topological defects), and 2) $\alpha_{\rm tot} = 10$, where the
parameter $\alpha_{\rm tot} \equiv \rho_a / \rho_a^{\rm mis}$
parametrises the ratio of the present energy density in axions
$\rho_a$ to that coming from misalignment only $\rho_a^{\rm mis}$. We
have also reported the forecasts for the sensitivities of various
experiments, either planned or already running, including the Cosmic
Axion Spin Precession Experiment (CASPEr) in its Phase 2~\cite{Budker:2013hfa}, A
Broadband/Resonant Approach to Cosmic Axion Detection with an
Amplifying B-field Ring Apparatus (ABRACADABRA) in its Phase 1 ($B_{\rm max} = 5\,$T and Volume$=1 {\rm \,m^3}$) and in the resonant configuration~\cite{Kahn:2016aff, Ouellet:2018beu}, the KLoe magnet for Axion SearcH (KLASH)~\cite{Alesini:2017ifp, Alesini:2019nzq}, the Axion Dark Matter Experiment (ADMX)~\cite{Duffy:2006aa, Asztalos:2009yp,
  Asztalos:2011bm, Stern:2016bbw, Braine:2019fqb}, the region labelled CULTASK 
  which combines the expected sensitivity of the Center for Axion and Precision Physics (CAPP) haloscope in both configurations CAPP-12TB and CAPP-25T~\cite{Semertzidis:2019gkj, Lee:2020cfj} (see also Ref.~\cite{Kim:2020kfo}), the MAgnetized Disc and Mirror Axion eXperiment
(MADMAX)~\cite{TheMADMAXWorkingGroup:2016hpc}, and the International
Axion Observatory (IAXO)~\cite{Vogel:2013bta, Vogel:2015yka,
  Armengaud:2019uso}. Notice that we have chosen the bounds from
laboratory searches by setting the ratio $E/N = 0$ as in the original
KSVZ model. 
Additional details
on the experimental setups for axion helioscopes are given in
\sect{sec:Helioscopes}, for axion haloscopes in \sect{sec:Haloscopes},
and for laboratory searches in \sect{sec:lab_searches}. 



\subsection{QCD axions as dark radiation}
\label{sec:hotaxions}

So far, we have discussed the population  of non-thermal axions that
stems from the classical evolution of the axion field,  which can represent an 
important contribution  to the  total amount of CDM. 
A thermal  population of axion is also expected from more conventional production 
mechanisms like particle scatterings, and for the typical 
sub-eV range of axion masses it will represent  a  dark radiation component.

The energy density in relativistic particles  at the epoch of matter-radiation decoupling
is conventionally described in terms of an effective number of neutrino species 
$N_{\rm eff}$ as 
\beq \rho_\r \equiv
\[1 + \frac{7}{8}\(\frac{T_\nu}{T_\gamma}\)^{4}\,N_{\rm eff}\]\,\rho_\gamma\,, 
	\label{eq:Neff}
\eeq
where $\rho_\gamma$ is  the energy density of the CMB photons, and  
$T_\nu/T_\gamma=(4/11)^{1/3}$ is the ratio of  neutrino to  photon temperature.
If there are no new  relativistic particles  besides the  three SM neutrinos the prediction is  
$N_{\rm eff}= N_{\rm eff}^{\rm SM} = 3.045$~\cite{deSalas:2016ztq}.\footnote{See
  also Ref.~\cite{Mangano:2005cc} where the value
  $N_{\rm eff}^{\rm SM} = 3.046$ was obtained,  and 
  Refs.~\cite{Dodelson:1992km, Hannestad:1995rs, Dolgov:1997mb} for
  earlier works.} 
%
A new relativistic particle  species,  such as thermal axions, 
would then appear as a modification in the number of effective neutrinos
$N_{\rm eff}$:  
\beq \rho_\r  =
\rho_\gamma + \rho_\nu + \rho_a =
\[1 + \frac{7}{8}\(\frac{T_\nu}{T_\gamma}\)^{4}\,N^{\rm SM}_{\rm eff}
+\frac{1}{2} 
\(\frac{T_a}{T_\gamma}\)^{4}\]\,\rho_\gamma\,,
	\label{eq:darkradiationrho}
\eeq
where $\rho_\nu$ and $\rho_a$ are respectively  the energy density of neutrinos and axions and 
$T_a$ is the temperature of the  thermal population of axions, characterised by one bosonic degree of freedom.  
Confronting \eqn{eq:Neff} 
and \eqn{eq:darkradiationrho} we see that the  contribution of thermal axions would appear as
 an excess in effective  number of neutrinos
\begin{equation}
\label{eq:Deltanu}
	\Delta N_{\rm eff} \equiv  N_{\rm eff} -N^{\rm SM}_{\rm eff} 
	= \frac{4}{7}\(\frac{T_a}{T_\nu}\)^4\,.
\end{equation}
Several groups have provided  estimates of $N_{\rm eff}$ 
by using cosmological data (see e.g. Refs.~\cite{Hannestad:2010yi,
  Archidiacono:2013fha, Lesgourgues:2014zoa} and   
  Refs.~\cite{Hannestad:2005df,Melchiorri:2007cd,Hannestad:2010yi,Archidiacono:2013cha,
  DiValentino:2015wba,DiValentino:2015zta,Ferreira:2018vjj}  
  in particular for the contribution of thermal axions). 
The current best-fit  value   
$N_{\rm eff} = 2.99 \pm 0.17$ at 68\% CL  comes from
combining  \textit{Planck} 2018 TT, TE, EE+lowE+lensing datasets plus Baryon Acoustic 
Oscillations (BAO) data~\cite{Aghanim:2018eyx}. This determination is 
fully consistent with the SM value, hence it  constrains any additional 
contribution from dark radiation. 

To evaluate quantitatively the possible effects of a thermal axion population on $N_{\rm eff} $
we need to estimate the ratio $T_a/T_\nu$.  Let us denote with $\Gamma_a$ the rate of reactions
that keep the axions in thermal equilibrium
\begin{equation}
\label{eq:Gamma_a}
\Gamma_a = \sum_i n_i \langle v \sigma_i \rangle \,, 
\end{equation}
where  for simplicity  we  consider only two-body  processes 
with cross sections 
$\sigma_i = \sigma (p_i a \leftrightarrow p_j p_k)$ 
with $p_i,p_j,p_k$ particle species in thermal equilibrium,  
$n_i$ the number density of $p_i$,  $v\approx 1$  the velocity
of scatterers assumed to be relativistic, and the brackets 
denote a thermal average. Axions decouple from the thermal bath 
when the  reaction rate $\Gamma_a$ falls below  the Hubble expansion rate,
and after decoupling  they maintain a thermal distribution which remains 
unaffected  by other phenomena occurring in the plasma. 
Let us consider a decoupling temperature $T_d$  for which  
the number of entropy degrees of freedom, including the axion,  
is $g_S(T_d)  +1$ 
and  all the particles share the same temperature. 
At a temperature $T_2$ well below  1\,MeV only the 
photon ($g_\gamma=2$),   the three SM neutrinos $\(\sum_i g_{\nu_i} = 6\)$ and the axion ($g_a =1$) 
are  relativistic,
with respective temperatures $T_\gamma= T_2$, $T_\nu = \(\frac{4}{11}\)^{1/3} T_\gamma$ and $T_a$.
 Including statistical factors, entropy conservation  gives
\begin{equation}
\label{eq:entropy_axion}
\[g_S(T_d) +1\] \(T_d \, R_d\)^3 = 
\[2 
\(\frac{T_\gamma}{T_a}\)^3+
\frac{7}{8} 6 
\(\frac{T_\nu}{T_a}\)^3 + 1 \] \(T_a R_2\)^3 \,,
\end{equation}
where $R_d$ and $R_2$ are the cosmological scale factors  at the respective temperatures $T_d$ and $T_2$. 
Since axions do not get reheated by subsequent  particle annihilation, their temperature 
simply scales as $T\propto R^{-1}$ so that $T_d R_d =T_a R_2$. \Eqn{eq:entropy_axion}
then gives
\begin{equation}
\label{eq:Ta}
\frac{T_a}{T_\nu} = \left\{\frac{2}{g_S(T_d)} \[ \(\frac{T_\gamma}{T_\nu}\)^3 +\frac{21}{8}\]\right\}^{1/3}
=\(\frac{43}{4\,g_S(T_d)}\)^{1/3} \,,
  \qquad \Delta N_{\rm eff} \simeq 0.027 \(\frac{106.75}{g_S(T_d)}\)^{4/3}\,,
\end{equation}
where in the second equation we have normalised $g_S$ 
to  the total number of SM degrees of freedom  
$g_S(T>m_t) = 106.75$.  Finally, in terms of the present number density of 
CMB photons $n_\gamma \simeq  411\,$cm$^{-3}$ 
the number  density of thermal axions is easily obtained as $n_a \simeq (43/22)\,n_\gamma/ g_S(T_d)$. 

 In the early Universe, there are various processes  involving  different types of  particles which  can produce a 
thermal  population of axions~\cite{Turner:1986tb,Masso:2002np,Graf:2010tv, Salvio:2013iaa}. 
Interactions with the gluons exist for any type of axion and  are model independent.
Coloured fermions interacting with the axion are also a necessary ingredient 
of any axion model, but in this case there is a difference 
if these states are exotic and heavy, as in  KSVZ models 
($m_\Q \propto v_a$)  or if they are instead SM quarks and much lighter, 
as in  DFSZ models  ($m_q \propto v \ll v_a$). 
 Also, the axion  mixes with the neutral  pion and this gives rise to an axion-pion interaction  
 (see the discussion in \sect{sec:axionpion}). 
 This interaction  can be  particularly important when,  at temperatures $m_\pi \lsim T \lsim T_C$,  
the abundance of pions in the plasma is of the order of the photon abundance. 
Let us discuss these thermalisation channels in more  detail. \\

\noindent
\emph{Axion-gluon coupling.}\ 
The relevant processes  for axion thermalisation  involving  the axion coupling to the gluons 
are  (i)~$a \, q\leftrightarrow  g\,  q$
and $a\,  \bar q\leftrightarrow  g \, \bar q$, (ii) ~$a\,  g \leftrightarrow  q\,  \bar q$,  
  (iii)~$a\,  g \leftrightarrow g\,  g$, all of which have a cross section of order 
\begin{equation}
\label{eq:sigma_ag}
\sigma_{ag}  \simeq \frac{\alpha_s^3}{8\pi^2} \frac{1}{f_a^2}\,.
\end{equation}
At temperature well above $m_t$  all the six quark flavours populate 
the thermal bath, then, including the statistical factor $3/4$  for the fermion number densities,  
 for reaction (i) with quark and antiquarks we have  $n_q + n_{\bar q} = 54\, n_{\rm eq}$, while 
 (ii) and (iii) contribute a factor $2 n_g = 32\, n_{\rm eq}$,  where 
 $n_{\rm eq} = \frac{\xi(3)}{\pi^2} T^3$ is the equilibrium 
distribution for one bosonic degree of freedom, $\zeta$ is the Riemann zeta 
function and $\zeta(3) \approx 1.2$.
From this and by using the cross section~\eqn{eq:sigma_ag} 
we can estimate the total scattering rate $\Gamma_a$~\eqn{eq:Gamma_a}.
Recalling the expression for $H$ given in   \eqn{eq:Hubble}  the decoupling condition  $\Gamma_a \lsim H$  
 yields  
\begin{equation}
\label{eq:Td_ag}
T_d \simeq 12.5  \frac{\sqrt{g_*(T)}}{\alpha_s^3} \frac{f_a^2}{\mP} \simeq
 4.0\cdot 10^{11} \(\frac{f_a}{10^{12}\,{\rm GeV}}\)^2\,{\rm GeV}, 
\end{equation}
where in the second relation we have used  $g_*(T) = 106.75+1$ and $\alpha_s(T) \simeq 0.03$ 
which are the values appropriate for temperatures of the order $10^{11}\div 10^{12}\,$GeV.
From \eqn{eq:Td_ag} we obtain   $\Delta N_{\rm eff} \simeq 0.027$ and
$n_a \simeq 7.5\,$cm$^{-3}$ for   the  present abundance of thermal  axions.
Let us note, however, that since above the PQ breaking scale there is no axion, this result is only  valid for   
$T_d < v_a =2 N f_a$,  that is for $f_a \lsim 4 N \cdot 10^{12}\,$GeV. 
Also, if the  scale of inflation and the reheating temperature are  
below $T_d$, this thermal population gets inflated away.  It is then important
to consider other processes that can be effective for producing a thermal axion 
population   at lower temperatures.  \\
 
\noindent
\emph{Axion-quark coupling.}\ 
The Compton-like scattering process $q\,  g \leftrightarrow  q\,a$ and its CP conjugate, 
which involve the axion coupling to coloured states, can be important  as long as 
$T\gsim m_q$ when the coloured fermions  are relativistic 
(for  $T < m_q$,  $\Gamma_a$  is   Boltzmann suppressed by $n_q$).
Since the axion coupling to the KSVZ quarks 
is parametrically larger than the coupling to SM quarks by a factor  $\sim v_a/v$ we 
concentrate on this case. The cross section is 
\begin{equation}
\label{eq:sigma_aQ}
\sigma_{a\Q} \simeq \alpha_s \(\frac{m_\Q}{v_a}\)^2 \frac{1}{T^2}\,, 
\end{equation}
and using $n_\Q + n_{\bar \Q} = 9 \, n_{\rm eq}$ the decoupling condition yields
\begin{equation}
\label{eq:Td_aQ}
T_d \simeq 3 \cdot 10^8 \(\frac{m_\Q}{10^8\, {\rm GeV}}\)^2 \(\frac{10^{12}\,{\rm GeV}}{v_a}\)^2\, {\rm GeV},
\end{equation}
where we have used $\alpha_s \approx 0.05$ valid for temperatures around  $T\approx 10^8\,$GeV. 
The resulting values of $\Delta N_{\rm eff} $ and $n_a$ are only slightly smaller than in the previous case
if entropy injection from $\Q$ decays into SM particles is taken into account (see \sect{sec:KSVZ-like}
for a discussion on this point). 
Note however, that the requirement $T_d>m_\Q$ implies  that this result holds only for 
 $m_\Q \gsim  3\cdot 10^7 \(\frac{v_a}{10^{12}\,{\rm GeV}}\)^2$. \\

 \noindent
\emph{Axion-pion and axion-nucleon couplings.}\ 
There are also processes that can  keep axions in thermal equilibrium 
after the quark-hadron phase transition, and that could be  important 
because  in this case it is  unlikely that the corresponding population of  thermal axions could  
be wiped out by inflation. The  most model independent  mechanisms 
are  pion-axion conversion  $\pi\, \pi \leftrightarrow \pi\, a$, whose interaction term is 
described by \eqn{eq:apicoupdef}, and  scatterings involving nucleons   $ N\, \pi \leftrightarrow N  a$,
with $N=n,p$.
The corresponding  cross sections were computed in Ref.~\cite{Chang:1993gm}.
The rates of these two processes have a different behaviour with the temperature, and in particular 
scattering off nucleons  becomes subdominant below $T\lesssim 200\,$MeV because of the 
exponential suppression in the number density of protons and neutrons.  
For the pion channel    Ref.~\cite{Hannestad:2005df} gives the following expression based 
on dimensional considerations and with numerically fitted coefficients:
\begin{equation}
\label{eq:Gamma_api}
\Gamma_{a\pi} \simeq 0.215\; C_{a\pi}^2 \, \frac{T^5}{f_a^2f_\pi^2}\,   h\(\frac{m_\pi}{T}\)\,, 
\end{equation}
with $C_{a\pi}$  given in \eqn{eq:Capidef}, and $h(x)$, normalised as $h(0)=1$
a rapidly decreasing function  of its argument.  To give an example of the values of $f_a$ for which axion thermalisation 
can occur after the QCD phase transition, let us fix $T_d = T_C \simeq 160\,$MeV,  in which case we have 
$g_* = 17.25$ and   $h\(\frac{m_\pi}{T_C}\) \simeq 0.8$,  
and let us consider the hadronic axion with  $C_{a\pi}\simeq 0.12$, see \eqn{eq:Capi}.  
This  decoupling temperature is obtained for  $f_a \simeq 3 \times 10^7\,$GeV  which, 
as we will see in \sect{sec:Astro_bounds}, is in conflict  with various   astrophysical bounds.  
Phenomenologically acceptable  values of $f_a$ would imply  higher decoupling temperatures
$T_d\gg T_C$   for which, however,   nucleons and pions are deconfined into the 
fundamental QCD degrees of freedom and the previous analysis breaks down. \\

We have seen that when axion thermalisation occurs above the electroweak phase 
transition,  one can expect  a contribution $\Delta N_{\rm eff} \sim 0.027$.
Other processes not considered here, as  the Primakoff process  $\gamma \, q \leftrightarrow  a \, q$
that  could be enhanced by a large value of the  axion-photon coupling, or   reactions  involving the 
SM quarks~\cite{Salvio:2013iaa,Ferreira:2018vjj} or leptons~\cite{DEramo:2018vss},  for which  large 
enhancements of the model-dependent couplings can be obtained  within some type of construction
(see \sect{sec:axion_landscape_beyond_benchmarks}), could contribute to thermalise the axions at  lower 
temperatures,  implying  larger contributions to  $N_{\rm eff} $. In particular, 
the  sensitivity of future cosmological measurements to  axion couplings to all the SM degrees 
of freedom  has been assessed  in~\cite{Baumann:2016wac}. 
A hot axion component generated in this way has 
been also invoked~\cite{DEramo:2018vss}  to alleviate the existing discrepancies  between  early and the late Universe 
determination of the Hubble constant $H_0$ (see Ref.~\cite{Verde:2019ivm} for a recent review). 
Projected sensitivities of  forthcoming   experimental determinations  of   $N_{\rm eff} $ suggest that   
axion  contributions to the radiation density at the level of  a 
few percent 
might  still be detectable.  In particular, conservative configurations of the next generation of 
ground-based CMB experiments, CMB-S4~\cite{Abazajian:2016yjj}, can reach a sufficient   accuracy  
in the measurement of $N_{\rm eff}  $ to test the minimal contribution of  
axions (or of any other light particle with zero spin thermalised above the electroweak phase transition)
at the $1\sigma$ level. 
$N_{\rm eff}  $  is indeed a unique measurement in cosmology. The overall importance of precise determinations
of this observable cannot be understated and, as we have seen, it   can provide fundamental 
information also on axion phenomenology.


\subsection{Axion miniclusters and axion stars}
\label{sec:substructures}

An important feature of the post-inflationary scenario is that at the time when the axion 
acquires a mass, around $T\sim T_\o$  (see \sect{sec:misalignment})  the  value of the initial misalignment angle $\theta_i$ changes 
by $\mathcal{O}(1)$  from one causal patch to the next.  Accordingly, the density of cold 
axions produced by the misalignment mechanism is characterised  by  sizeable 
inhomogeneities  $\delta\rho_a/\rho_a \sim  \mathcal{O}(1)$. 
The free streaming length of the misalignment population of cold axions  is too short 
to erase  these inhomogeneities  before the time $t_\e$ of matter-radiation equality, so that 
at $T\sim T_\e$  the density perturbations  decouple from the Hubble expansion and 
start growing by gravitational instability,  rapidly forming gravitationally bound objects, called axion 
miniclusters~\cite{Hogan:1988mp, Kolb:1993zz,Kolb:1993hw, Kolb:1994fi}.
The scale of minicluster masses is set by the total mass in axions within one  Hubble volume of 
radius  $ 2 R_\o\sim  H_\o^{-1}= \sqrt{\frac{3 \mP^2}{8\pi\rho_\o}} \,$ at the time  when the 
axion mass becomes relevant, since after the onset of oscillations the number of cold axions per 
comoving volume remains conserved. At $T_\o$ the Universe energy density is radiation dominated so that  
$\rho_\o \simeq \rho_\r = \frac{\pi^2}{30} g_* T^4_\o$.  The energy  enclosed in a Hubble volume then is 
\begin{equation}
\label{eq:Mosc}
M_\r(T_\o) = \frac{4 \pi R_\o^3}{3} \, \rho_\r = \frac{3}{32 \pi} \sqrt{\frac{5}{\pi g_*}}\, \frac{\mP^3}{T^2_\o}\,.
\end{equation}
As the Universe expands, the energy in radiation gets redshifted, and at the  matter-radiation 
equality temperature $T_\e$  it provides an estimate of the gravitationally bound minicluster mass
 $M_{\rm MC} \approx M_\r(T_\e)$, that is
\begin{equation}
\label{eq:MMC}
M_{\rm MC} \approx  M_\r(T_\o) \frac{T_\e}{T_\o} \simeq 1.3 \times 10^{46} \;\(\frac{800\, \mathrm{MeV}}{T_\o}\)^3\frac{T_\e}{0.8\, \mathrm{eV}}\, \mathrm{GeV} 
\approx 10^{-11} \, M_\odot\,,
\end{equation}
where we have used $g_*(T_\o) = 61.75$, and 
$M_{\odot}\approx 2\cdot 10^{30}\,\mathrm{kg} = 1.3\cdot 10^{57}\,\mathrm{GeV} $ represents one solar mass.  
The typical radius of the overdensities when they decouple from the Hubble flow  at  $T_\e$ is also easily  
estimated. From  $R_\o \approx 0.1\,$km as is obtained from the expression given 
above \eqn{eq:Mosc}  and using entropy conservation $(R_\e T_\e)^3 g_S(T_\e)= (R_\o T_\o)^3 g_S(T_\o)$  gives
\begin{equation}
\label{eq:RMC}
 R_{\rm MC}\approx R_\e = R_\o \frac{T_\o}{T_\e} \(\frac{g_S(T_\o)}{g_S(T_\e)}\)^{1/3} \approx 2.5 \cdot 10^8 \, \mathrm{km}\,,
\end{equation}
where we have used $g_S(T_\e)\simeq 3.91$. 
The numbers  in Eqs.~(\ref{eq:MMC}) and (\ref{eq:RMC}) are of course  indicative 
 and can easily vary by  a couple of orders of magnitude.  For example it was recently argued in 
Ref.~\cite{Enander:2017ogx} that  the characteristic size of the density fluctuations 
is smaller than the Hubble horizon at $T_{\rm osc}$, implying that   
typical miniclusters are both lighter and smaller with respect to the quoted numbers.\footnote{Even 
larger variations in  mass and size are  obtained when considering non standard cosmological scenarios, 
as was done for example in Refs.~\cite{Nelson:2018via,Visinelli:2018wza}.} 
%
The dynamics of the collapse of non-linear density fluctuations that leads
to the formation of axion miniclusters has been recently assessed
through a semi-analytical Press-Schechter
approach~\cite{Fairbairn:2017sil, Fairbairn:2017dmf, Enander:2017ogx}
or by numerical simulations~\cite{Vaquero:2018tib, Buschmann:2019icd,Eggemeier:2019khm}.

Axion miniclusters would lead to important detection features. First,
their typical density $\rho_{\rm MC} \simeq 3  M_{\rm MC}/(4 \pi R_{\rm MC}^3) \approx  0.2\times 10^6\,$GeV cm$^{-3}$
is about a factor $10^6$ larger than the local DM density, so that 
in an encounter with the Earth the rate of conversion of axions into photons in dedicated resonant
cavities or haloscopes (see \sect{sec:Haloscopes}), would be accordingly 
temporarily  enhanced~\cite{Sikivie:2006ni}. The expected signal would
be time-dependent due to the revolution and rotation motions of the
Earth, which leads to a detectable annual
modulation~\cite{Drukier:1986tm} and possibly even to a detectable
diurnal modulation~\cite{Knirck:2018knd}. Since the cavity would have
to be tuned to the correct frequency in order to capture the signal, a
broadband axion cavity resonator has been proposed to exploit this
possibility, see Ref.~\cite{Irastorza:2018dyq}. Second, the presence
of axion miniclusters could be assessed through picolensing of
individual clusters~\cite{Kolb:1995bu}, or microlensing of a halo
formed of axion miniclusters that hierarchically
merge~\cite{Fairbairn:2017dmf, Fairbairn:2017sil}. This latter
possibility could also be used to assess the fraction of cold axions
that are bound into clumped objects, a fraction that can also be
accessed through numerical simulations. Third, axion minicluster could
be disrupted by the gravitational field of a nearby star encounter, or
by the mean galactic gravitational field, leading to tidal streams of
axions that would enhance the local CDM density by about one order of
magnitude~\cite{Tinyakov:2015cgg}.

As we have seen in \sect{sec:misalignment} when $m(t_\o) \simeq 3 H(t_\o)$ a 
non-relativistic and cold  population of axions  ($\vev{p_a} \ll m_a \ll T_\o $)
 is created. 
  Using as reference values  $T_\o \sim 1\,$GeV and $f_a \sim 10^{12}\,$GeV
their number density  can be estimated as   $n_a (T) = m_a(T_\o) f^2_a \frac{s(T)}{s(T_\o)} 
\simeq  \frac{3\,f_a^2}{\mP T_\o} T^3 \sim  10^5\, T^3$.
This corresponds to a huge value  of the initial phase-space density 
$\mathcal{N}_a \sim n_a/\vev{p_a}^3 \gg 10^5 (T_\o/m_a)^3 \approx 10^{47}$
which, given that effects of collisions are completely negligible, remains enormous at 
all subsequent times during the Universe expansion.
Around the time of matter radiation equality gravitational instabilities grow 
and bound systems can form.
As we have seen in \sect{sec:QCDsusceptibility} axions  have an attractive self-interaction and, although  
extremely tiny $\sim (m_a/f_a)^4$, it can still produce significant effects due to the large phase space density,
causing in particular a relaxation of  gravitationally bound axions clouds. If the relaxation is efficient, 
an {\it axion star} could form  within a time scale compatible with the age of the Universe. 
The  possibility of formation of  axion stars was first studied in Ref.~\cite{Tkachev:1991ka}.
Axion stars can be modelled as   oscillation-like solutions of the
Klein-Gordon equation associated with the axion potential in
Eq.~\eqref{eq:Vapiexp} and coupled to the Einstein or Poisson equation
to account for the feedback into the gravitational field. Contrarily
to axion miniclusters, the mass of the axion star is not fixed once
setting the mass of the QCD axion. The typical mass of the axion star
is fixed once a formation mechanism is imposed~\cite{Schive:2014dra,
  Eggemeier:2019jsu}. Axion stars are described by a real
pseudo-scalar field that oscillates with time, with a frequency that
is related to the mass of the axion. This configuration is different
from what is obtained from a self-gravitating condensate made of a
{\it complex} boson field, which is known in the literature as a
`boson star'~\cite{Kaup:1968zz, Ruffini:1969qy}. Both configurations
possess black hole-like solutions for a null field and, in the case of
the complex scalar field of mass $m_\phi$, the boson star cannot grow
to masses larger than the critical mass
$M_* = 0.633/(Gm_\phi)$~\cite{Breit:1983nr}. If the axion potential
were quadratic as in Eq.~\eqref{eq:axion_potential_expansion}, the
axion star would also possess a slightly smaller critical mass
$M_* = 0.607/(Gm_a)$~\cite{UrenaLopez:2002gx}, see also
Ref.~\cite{Helfer:2016ljl}.

The formation of axion stars might proceed either by gravitational
cooling out of the virialised minicluster or during a process of
violent relaxation~\cite{Seidel:1993zk, Guzman:2006yc, Levkov:2018kau,
  Eggemeier:2019jsu}, leading to a solution in the weak gravity
regime~\cite{Chavanis:2011zi, Chavanis:2011zm}, with the mass and the
radius of the axion star being related by~\cite{Ruffini:1969qy,Membrado:1989ke,Chavanis:2011zm}
\beq
	R_{\rm as} = \frac{9.9}{Gm_a^2\,M_{\rm as}}\,.
\eeq
This solution is known in the literature as the ``dilute'' regime,
since the average energy density inside the axion star is much smaller
than the energy scale $(m^2_\pi f^2_\pi)^{1/4}$ 
at which the axion potential
saturates. As the density of the star increases, self-interactions
become more relevant and destabilise the equilibrium when the mass of
the star approaches the critical value~\cite{Helfer:2016ljl,Levkov:2016rkk, Visinelli:2017ooc}
%
%
\beq
	M_{\rm as, crit} \approx 18.4 \, \mP\,\frac{f_a}{m_a}\,.
\eeq
This result is valid for an axion field moving in the potential described in Eq.~\eqref{eq:VaChPT} 
with $m_u/m_d = 0.48$.

\section{Astrophysical signatures and bounds}
\label{sec:Astro_bounds}

\begin{figure}[b!]
	\centering
	\includegraphics[width=0.45\linewidth]{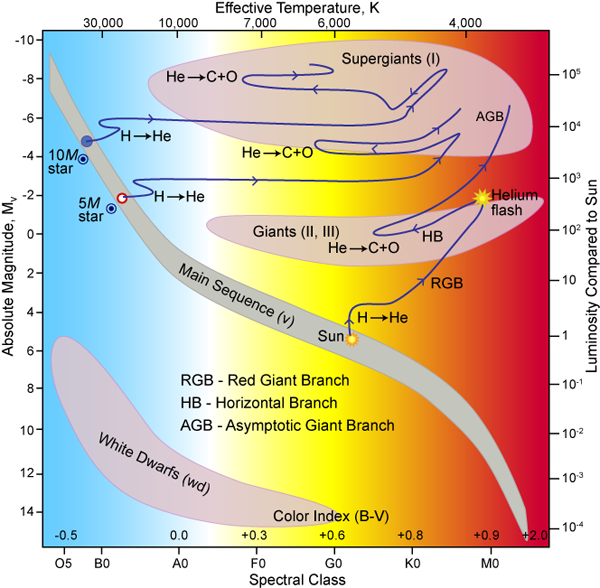}
	\hspace{1.cm}
	\includegraphics[width=0.45\linewidth]{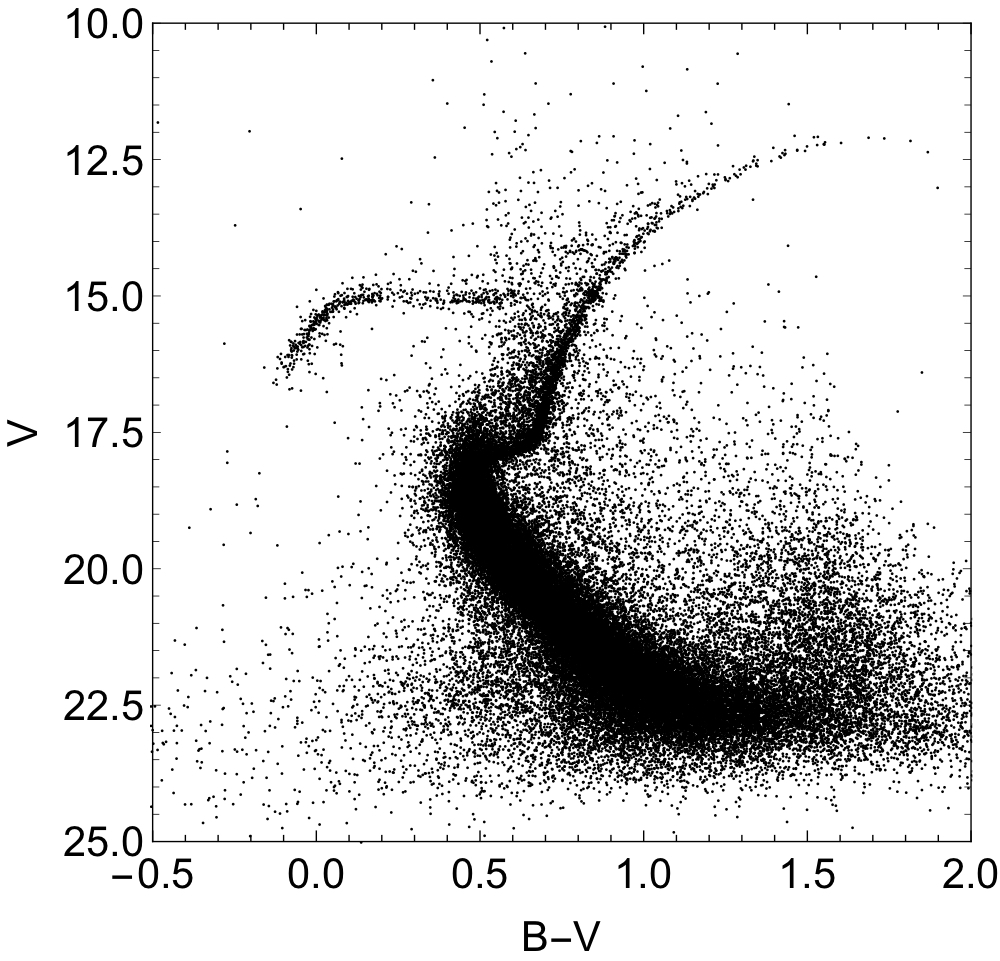}
	\caption{
	In the left panel, the theoretical Hertzsprung Russell (HR) or Colour Magnitude Diagram (CMD), showing the evolution of luminosity and surface temperature of stars with different initial masses.
	To the right, the observational CMD of the M5 globular cluster, which shows the luminosity and surface temperature of stars at a fixed time (isochrones). More massive stars evolve more rapidly and are found in more advanced stages. The density of stars in different regions of the observational CMD reflects the duration of the corresponding evolutionary stage. 
	The luminosity (energy emitted per unit time) is conventionally measured in magnitude. In the figures we show the magnitude in the visual (V) band.
	The surface temperature is show as the B-V colour, that is the difference between the blue (B) and visible (V) brightness.
	Note that, for historical reasons, the temperature increases towards the left of the diagram. So, stars with higher surface temperature (blue) are found to the left.
	See text for more details.	
	The figure to the left is reproduced (with permission) from \url{https://physics.aps.org/articles/v2/69}.
	}
	\label{fig_HR_diagram}
\end{figure}
%

Astrophysics, and stellar evolution in particular, offer powerful methods to probe the axion couplings to SM fields~\cite{Raffelt:1990yz,Raffelt:1996wa,Raffelt:2006cw}. 
The observational properties of stars are conveniently shown in the Hertzsprung Russell (HR) or Colour Magnitude Diagram (CMD), which shows the stellar luminosity (or magnitude) versus the surface temperature (expressed through the colour index).

No matter their initial mass, stars spend most of their life burning H into He in their core (main sequence).
The post main sequence evolution depends on the stellar initial mass.
A schematic picture  of this evolution is shown in the left panel of \fig{fig_HR_diagram}.
After the hydrogen in the core is exhausted, stars of roughly the same mass as our Sun 
enter the subgiant phase, burning hydrogen in a thick shell. It follows the Red Giant Branch (RGB) stage, with hydrogen burning in a thin shell surrounding an inert  He core. 
The evolution in the RGB continues until the temperature in the core is high enough to ignite the He in the core (He-flash). 
Afterwards, the star moves to the Horizontal Branch (HB) region of the diagram. 
Such low mass stars never reach the conditions (temperature and density) required to ignite heavier elements and end up as carbon-oxygen White Dwarfs (WDs). 

Stars a few times the mass of the Sun do not undergo a He-flash and ignite helium soon after the end of the main sequence stage,  transitioning to the cold (red) region of the HR diagram. 
The evolution during the He-burning stage may show a peculiar journey to the bluer region of the diagram and back, called the blue loop (see, e.g., the $5\,M_{\odot}$ track in the left panel of \fig{fig_HR_diagram}).
Stars with an initial mass larger than about $8M_{\odot}$ do not become WD but undergo a core collapse, giving rise to a type II Supernova (SN) explosion and leaving a compact 
Neutron Star (NS) or, if very massive, a black hole.

The diagram in \fig{fig_HR_diagram} is theoretical. It shows the evolutionary tracks of individual stars. 
Observationally, one extracts colour and magnitude of individual stars (at a fixed time) and shows the results in a diagram similar to the one shown in the right panel of \fig{fig_HR_diagram}. 
From the  stellar population it is possible to reconstruct the evolutionary times of each stage (the longer the evolutionary time, the larger the stellar population corresponding to that phase), which can then be compared with the theoretical predictions extracted from numerical stellar evolution codes.

The method presents evident  difficulties related to statistics (particularly for fast evolutionary stages), stellar contamination, interstellar absorption of the stellar light, etc. 
Nevertheless, numerical simulations reproduce with a remarkable level of agreement  the observed CMD of particular stellar populations and allow to set stringent bounds on new physics. 
The emission of axions (or other light particles) from stars might, in fact,  impact their expected evolution and spoil the agreement with observations. 

The aim of this section is to provide an updated summary of the bounds on axions derived from   stellar astrophysics considerations. 
In addition, we will briefly present the results of the axion interpretation of some observations of anomalous stellar evolution that have been reported in the last two decades (see, e.g., references~\cite{Giannotti:2017hny,Hoof:2018ieb,DiVecchia:2019ejf} for more detailed discussions). 
Our general approach will be to present first all the results in a model independent way. 
The impact on the axion benchmark models (KSVZ and DFSZ-type) will also be discussed at the end of the section.

\subsection{Axion-photon coupling}
\label{sec:Astro_bounds_gag}

\begin{figure}[t]
	\includegraphics[width=0.3\linewidth]{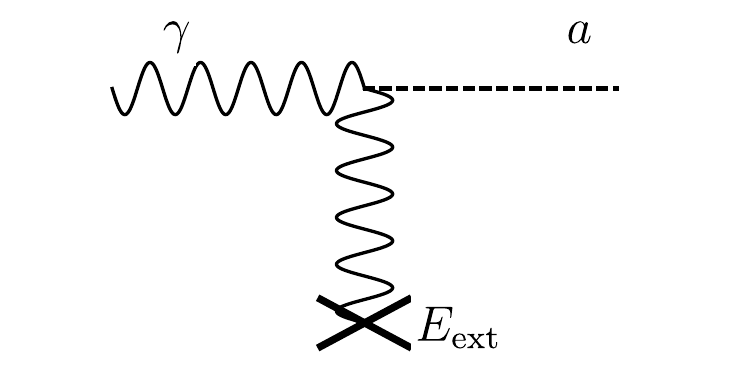}
	\includegraphics[width=0.3\linewidth]{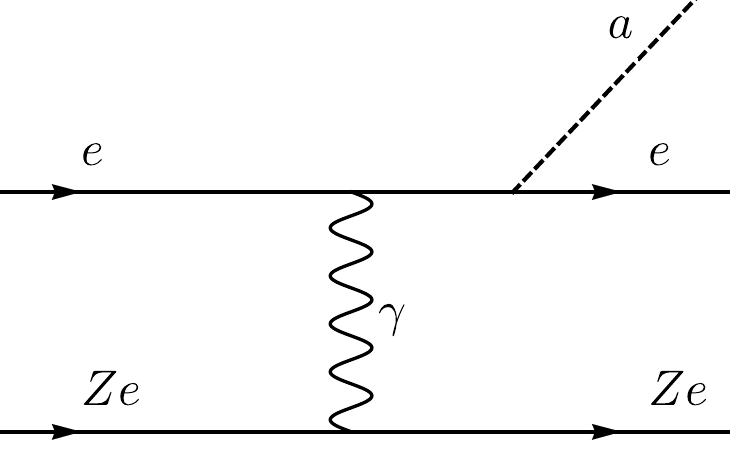}
	\hspace{0.8cm}
	\includegraphics[width=0.3\linewidth]{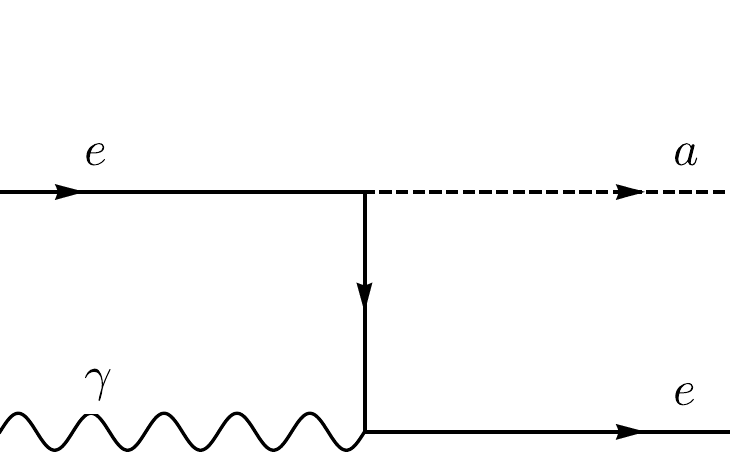}
	\caption{From left to right: axion Primakoff processes in an external electric field; axion bremsstrahlung process; and Compton processes. In the case of the bremsstrahlung process, $Ze$ represents either an ion or an electron.  
	}
	\label{fig_Feyn_Primakoff}
\end{figure}

In the contest of stellar evolution, the most relevant process induced by the axion-photon coupling, $g_{a\gamma}$ (\sect{sec:axionphoton}) is the Primakoff process (\fig{fig_Feyn_Primakoff}), 
consisting in the conversion of thermal photons in the electrostatic field of electrons and nuclei
\begin{align}
\gamma+ Ze\to a+ Ze \,.
\end{align}
Neglecting degeneracy effects and the plasma frequency (a good assumption in plasma conditions when the Primakoff process is the dominating axion production mechanism), it is possible to provide a semi-analitical expression for the energy-loss rate per unit mass in axions~\cite{Friedland:2012hj}:
\begin{align}
\label{eq:Primakoff_approx}
\varepsilon_P\simeq 2.8\times 10^{-31} Z(\xi^{2}) \left( \frac{g_{a\gamma}}{\rm GeV^{-1}} \right)^{2}
\frac{T^7}{\rho}\, {\rm erg\,g^{-1}\,s^{-1}}\,,
\end{align}
where $T$ and $\rho$ are in K and in g cm$^{-3}$ respectively.
The coefficient $Z(\xi^{2})$ is a function of $\xi^{2}\equiv(\kappa_{S}/2T)^{2}$, with $\kappa_{S}$ being the Debye-Huckel screening wavenumber.
It can be explicitly  expressed as an integral over the photon distribution (see Eq.~(4.79) in Ref.~\cite{Raffelt:1990yz}).
Ref.~\cite{Friedland:2012hj} proposed the analytical parametrisation
\begin{equation}
\label{eq:alternativefit}
{\textstyle Z(\xi^{2})\simeq 
\left(\frac{1.037\xi^{2}}{1.01+\xi^{2}/5.4}+\frac{1.037\xi^{2}}{44+0.628\xi^{2}}\right)
\ln\left(3.85+\frac{3.99}{\xi^{2}}\right)}\,,
\end{equation}
which is better than 2\% over the entire range of $\xi$.
 In general, $Z(\xi^{2})$ is ${\cal O}(1)$ for relevant stellar conditions. For example, in the core of the Sun, $\xi^{2}\sim12$ and $Z\sim6$ and in the core of a low-mass He burning star, $\xi^{2}\sim2.5$ and $Z\sim3$ \cite{Raffelt:1990yz}, while in a  $10 M_{\odot}$  He burning star, $\xi^{2}\sim0.1$ and $Z\sim0.4$~\cite{Friedland:2012hj}.
\begin{figure}[t]
	\centering
	\includegraphics[width=0.6\linewidth]{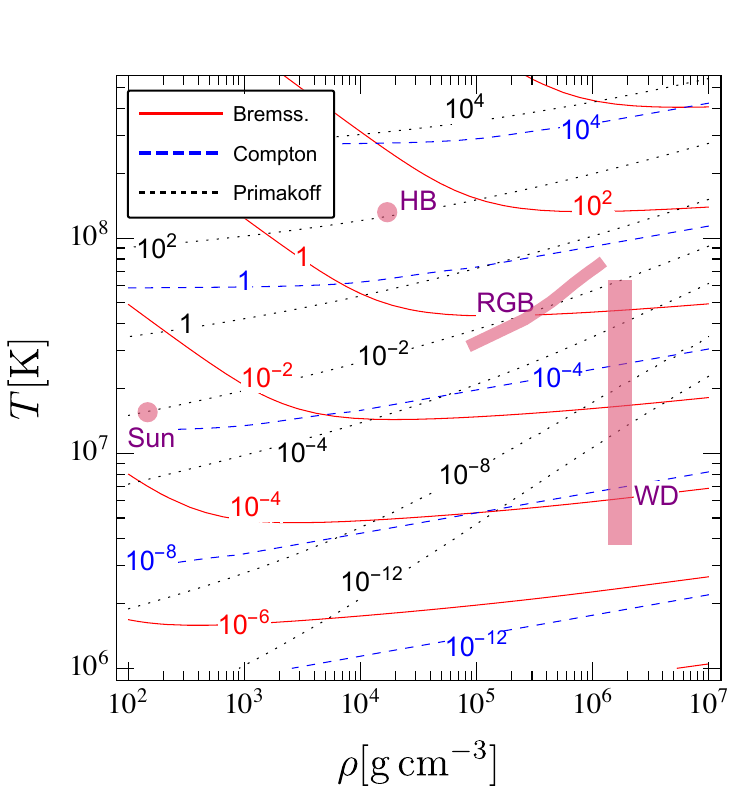}
	\caption{Contours of the axion energy-loss rates per unit mass, $\varepsilon_a$, in erg$\,$g$^{-1}$s$^{-1}$, for a pure He plasma. Different lines represent different channels, as shown in the legend. 
		The Primakoff process is calculated for $ g_{a\gamma}=0.65\times 10^{-10} $GeV$ ^{-1} $, corresponding to the bound from HB stars~\cite{Ayala:2014pea,Straniero:2015nvc}.
		The Bremsstrahlung and Compton processes are calculated for $ g_{ae}=4.3\times 10^{-13} $, corresponding to the RGB bound from M5~\cite{Viaux:2013lha}.		
The onset of the degeneracy region is visible in the bending of the bremsstrahlung contours. 
		The central temperature and density of the Sun~\cite{Vinyoles:2016djt}, RGB stars, HB stars and WDs are also shown, for reference.  
		In the case of HB and RGB, these are the results of a numerical simulation of a 0.8 $ M_{\odot} $ model as obtained with the FuNS code~\cite{Straniero:2019dtm}.
		The WD region is estimated using a polytropic model of WDs with mass from $ 0.6 $ to $ 0.7 M_{\odot}$,  as discussed in Ref.~\cite{Raffelt:1996wa}, and spans luminosities in the range between $ 0.5\times 10^{-4} $ and $ 0.5 L_{\odot}$.
		Except for the WD case, the thickness of the lines has no significance. 
	}
	\label{fig_bremsstrahlung_compton}
\end{figure}

As shown in \fig{fig_bremsstrahlung_compton}, the Primakoff process has a steep dependence on the stellar temperature, which controls the number of thermal photons, but is suppressed at high density because of the effects of a large plasma frequency and of the reduction of electron targets~\cite{Raffelt:1987yu} (in such conditions, \eqn{eq:Primakoff_approx} ceases to be valid).
Hence, this process is strongly suppressed in the degenerate core of WDs and RGB stars. 
Indeed, the strongest bounds on the axion-photon coupling are derived from the analysis of stars with a low density and high temperature core. 
In the following, we present the relevant stellar arguments used to constrain this coupling. \\

\begin{figure}[t]
	\begin{center}
		\includegraphics[width=0.45\linewidth]{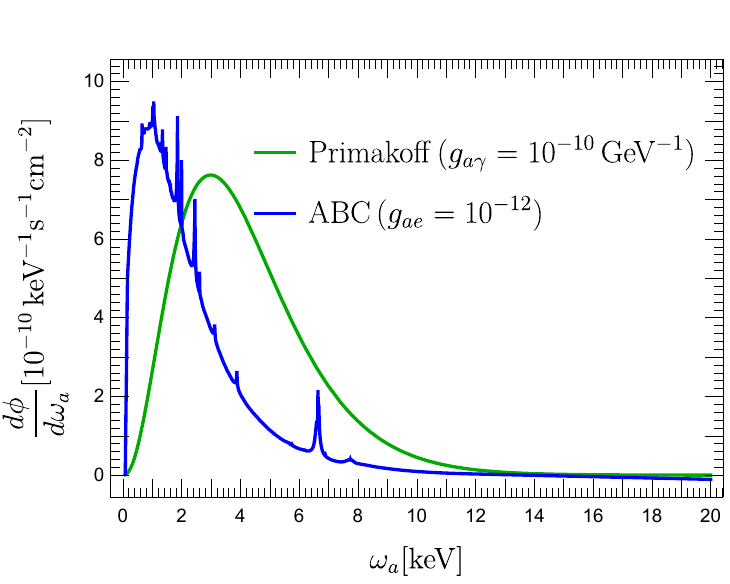}
	\hspace{0.2 cm}
		\includegraphics[width=0.45\linewidth]{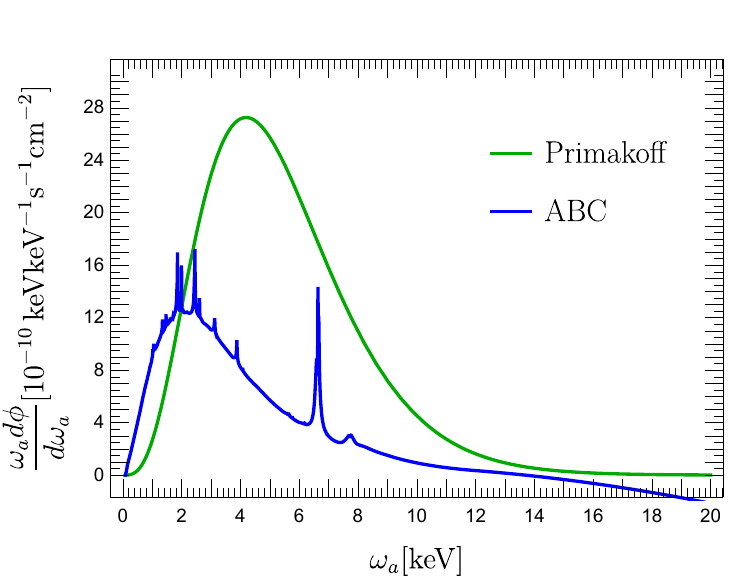}
		\caption{Solar axion spectrum. 
To the left shows the axion number and to the right the axion energy.
The couplings are the same in both graphs.}
		\label{fig_solar_flux}
	\end{center}
\end{figure}

\emph{The Sun.}
Given its low density and, more importantly, its proximity, the Sun provides a good environment  to test the axion-photon coupling.
Moreover, as we shall see, the Sun is an important source for axions to be detected in terrestrial experiments (see \sect{sec:Helioscopes}).
The (number) spectrum of axions produced in the Sun is shown in the left panel in \fig{fig_solar_flux}.
Interestingly, the axion spectra produced by processes induced by the axion-photon and the axion-electron couplings are similar, if we consider couplings of the order of  the current bounds (cf.~\sect{sec:Astro_bounds_gae} for the astrophysical bounds on the axion-electron coupling). 
%
%
This is a very impactful result for experimental searches, as we discuss in \sect{sec:Helioscopes}.
For considerations about stellar cooling, however, the relevant quantity is the axion luminosity, not the axion number, and this is dominated by Primakoff axions since they are, in average, more energetic (cf.~right panel of \fig{fig_solar_flux}).

Strong bounds on exotic cooling processes in the Sun can be set from helioseismological considerations~\cite{Schlattl:1998fz,Vinyoles:2015aba}. 
The current bound is  
$g_{a\gamma}\leq 4.1\times 10^{-10}\,{\rm GeV^{-1}}$ at $3\,\sigma$~\cite{Vinyoles:2015aba}, which corresponds to   
$g_{a\gamma}\leq 2.7\times 10^{-10}\,{\rm GeV^{-1}}$ at $2\,\sigma$.
%
A somewhat weaker bound, $g_{a\gamma}\leq 7\times 10^{-10}\,{\rm GeV^{-1}}$, was inferred in Ref.~\cite{Gondolo:2008dd} from the axion impact on solar neutrinos. \\

\emph{R-parameter and HB stars.} 
The major problem with the Sun as a source for axions is the relatively low temperature of its core.
This limits strongly the Primakoff emission rate. 
Stars in the  HB stage, which follows the RGB phase, have a low core density of about $ 10^{4} $\,g$ \, $cm$ ^{-3} $ and a high temperature (see \fig{fig_bremsstrahlung_compton}), providing excellent conditions to produce axions through the Primakoff process. 

To assess the effects of axions on the evolution of HB stars, it is convenient to introduce the R-parameter,  defined as the ratio of the number of stars in the HB and in the upper portion of the RGB: $ \mathcal{R}=N_{\rm HB}/N_{\rm RGB} $.
In the presence of axions, this parameter is expected to be
\begin{align}
\label{eq:R-parameter}
\mathcal{R}=\mathcal{R}_0(Y)-
F_{a\gamma} \left( \frac{g_{a\gamma}}{10^{-10}{\rm GeV}^{-1}} \right) -
F_{ae} \left( \frac{g_{ae}}{10^{-12}} \right)\,,
\end{align}
where $\mathcal{R}_0(Y)$ is a function of the helium abundance ($Y$) in the GC and the $F$ are some positive-defined functions of the axion couplings.\footnote{It is easy to infer that $F_{a\gamma}$ must be positive. A finite axion-photon coupling would contribute to the stellar energy-loss, particularly in the HB stage, since Primakoff is suppressed in the degenerate plasma typical of the RGB core. Thus, a large $g_{a\gamma}$ would shorten the life of  HB stars and, consequently, their expected numbers in a cluster. The argument for the positivity of $F_{ae}$ goes as follows. 
An efficient energy-loss channel, such as the one induced by a large coupling of axions to electrons, would delay the He-ignition in the RGB core 
(He-flash), allowing the core to grow more. This would produce HB stars with more massive cores, which would evolve more rapidly.  
Thus the number of stars in  the HB would be reduced, and the value of the $\mathcal{R}$-parameter lowered. 
Direct effects of  $g_{ae}$ on HB stars energy losses, as for example through Compton emission, would contribute to 
speedup  the evolution  reducing further $\mathcal{R}$.   
}
For completeness, we are including the contribution from processes induced by the axion coupling to photons, $g_{a\gamma}$, as well as the axion coupling to electrons, $g_{ae}$, which will be reviewed in more detail in \sect{sec:Astro_bounds_gag}.
The positivity condition of the $F$ insures that the axion emission can only lower the value of the $R-$parameter and that there could be a degeneracy between the effects of the axion couplings to electrons and photons. 
Although at the present a full numerical study that includes axions coupled to both electrons and photons does not exist, reference~\cite{Giannotti:2015kwo} provided approximate analytical expressions for the F in \eqn{eq:R-parameter}: 
%
\begin{align}
\label{eq:R-parameter_functions}
&\mathcal{R}_0(Y)=0.02+7.33\, Y\,;\\
&F_{a\gamma}(x)=0.095 \sqrt{21.86+21.08 \,x}\,;  \\
&F_{ae}(x)=0.53\,x^2+0.039
\left( \sqrt{1.23^2+100 \,x^2} -1.23-4.36\,x^{3/2}\right) . 
\end{align}

Neglecting the axion-electron coupling and adopting the value $Y=0.2535$ for the helium abundance~\cite{Izotov:2013waa}, reference~\cite{Ayala:2014pea} derived the upper bound on the axion-photon coupling
\begin{equation} 
g_{a\gamma}<0.66\times 10^{-10}\,\ \textrm{GeV}^{-1} \qquad (95 \% \,\ \textrm{CL})  \, .
\label{eq:g_agamma_HB_2sigma}
\end{equation}
This is known as the HB bound,
 since by neglecting the axion-electron coupling we are essentially ignoring the RGB evolution.
In addition to the result in \eqn{eq:g_agamma_HB_2sigma}, the analysis in~\cite{Ayala:2014pea} inferred an $ R $ parameter somewhat larger than observed, indicating a 2$ \,\sigma $ preference for a small, non-vanishing axion-photon coupling, later confirmed in~\cite{Straniero:2015nvc} using an updated value of the He abundance~\cite{Izotov:2014fga,Aver:2015iza}.
The result,
\begin{align}
g_{a\gamma}=(0.29\pm 0.18)\times 10^{-10}\,{\rm GeV}^{-1}  \qquad (68 \% \,\ \textrm{CL}) \, ,
\label{eq:g_agamma_HB_1sigma}
\end{align}
is known as the HB hint.\\

\emph{Massive Stars.} 
Further insights on the axion-photon coupling can be extracted from the analysis of intermediate mass stars,  $ M\sim 8-12 M_{\odot} $~\cite{Friedland:2012hj,Carosi:2013rla}.
As discussed at the beginning of this section, the (core) He burning stage of these stars is characterized by a migration towards the blue (hotter) region of the CMD and back.
This journey is known as the \textit{blue loop}. 
The existence of the loop is corroborated by many astronomical observations. 
In particular, this stage is essential to account for the observed Cepheid stars (see, e.g., \cite{kippenhahn}).
The disappearance of the  blue loop stage
in the luminosity ranges where Cepheid stars are observed
is forbidden~\cite{Carosi:2013rla}.
A nonvanishing axion-photon coupling would reduce the time a star spends in the blue loop stage and, consequently, the number of \emph{blue} versus \emph{red} stars of a given luminosity. 
According to the analysis in~\cite{Friedland:2012hj}, based on numerical simulations of solar metallicity stars in the $ 8-12 M_{\odot} $ mass range, a coupling larger than  
$ \approx 0.8\times 10^{-10} {\rm GeV}^{-1} $ 
would cause the complete disappearance of the blue loop.
The result is comparable to the globular cluster bound.
Somewhat lower values of $g_{a\gamma} $ might help explaining the observed deficiency of blue with respect to red supergiants discussed, e.g., in~\cite{McQuinn:2011bb}.
The numerous uncertainties in the microphysics and in the numerical description of the blue loop stage have not permitted a more quantitative assessment of this possibility~\cite{Giannotti:2015dwa}.

More recently, the analysis of SN progenitors has also indicated a preference for additional cooling, in the form of axions or, possibly, other light particles~\cite{Straniero:2019dtm}.
Surveys show that in many cases the SN type II progenitors are red supergiants with a certain maximal (surface) luminosity. To stay below this luminosity, stars would need to be relatively light, contrary to observations. 
Standard modifications to the stellar codes, e.g., adding rotations, overshooting, etc.,  do not help but rather worsen the agreement with the observations. 
The addition of a novel cooling channel, however, might help reconciling the simulations with the observations.
Because of the more efficient cooling, the  development of the 
envelop would freeze at lower luminosities, allowing for more massive stars to end up with the required surface luminosity~\cite{Straniero:2019dtm}. 
In the case of axions, the hint is to rather large couplings to both electrons and photons, close to the current HB and RGB bounds.
However, the data sample is still too sparse to draw definitive conclusions and the identification of reliable axion couplings is  prohibitive. 
The situation will largely improve with the data from the Large Synoptic Survey Telescope (LSST)~\cite{0912.0201,Drlica-Wagner:2019xan}, which will likely identify a large number of SN progenitors (see Section~3 in Ref.~\cite{Drlica-Wagner:2019xan}).

\subsection{Axion-electron coupling}
\label{sec:Astro_bounds_gae}
The axion-electron coupling, $g_{ae}$ (\sect{sec:axionelectron}), induces several processes relevant for stellar evolution (see Ref.~\cite{Raffelt:1996wa} for a comprehensive presentation). 
The most important  are the Atomic recombination and de-excitation, the electron and ion Bremsstrahlung, and the Compton process, collectively known as the ABC processes.\footnote{Another astrophysical process discussed in the literature is the electron-positron annihilation, $ e^{+} e^{-} \to \gamma + a $~\cite{Pantziris:1986dc}, which plays, however, a less significant role in stellar evolution.}
The atomic recombination and deexcitation processes are an important contribution to the solar axion spectrum~\cite{Redondo:2013wwa}  (see \sect{sec:Helioscopes}) but can be ignored in  numerical simulations of stellar evolution.

At high densities, particularly in electron degeneracy conditions, the  most efficient axion production mechanism is the electron/ion bremsstrahlung process
\begin{align}
e +Ze\to  e + Ze +a\,,
\end{align}  
shown in the central  panel of \fig{fig_Feyn_Primakoff}.
The axion energy-loss rates per unit mass for the case of a pure He plasma is shown in \fig{fig_bremsstrahlung_compton}.
As clear from the figure, at high density and relatively low temperature, when electrons become degenerate, the bremsstrahlung rate 
has a very mild dependence on the density.
On the other hand, in nondegenerate conditions the rate 
depends linearly on the density.
In both cases, there is also a dependence on the stellar chemical composition.
Explicit expressions for the energy-loss rates per unit mass in the degenerate (d) and nondegenerate (nd) limits  are provided in Ref.~\cite{Raffelt:1994ry}. Approximately, 
\begin{align}
& \varepsilon_{\rm ND}
\simeq 47\, g_{ae}^{2} T^{2.5}\frac{\rho}{\mu_e}
\sum\frac{X_j Z_j}{A_j}\left(Z_j+\frac{1}{\sqrt{2}}\right) {\rm erg\,g^{-1}\,s^{-1}}\,,\\
& \varepsilon_{\rm D}
\simeq 8.6 \times 10^{-7} F\, g_{ae}^{2} T^{4}\left(\sum\frac{X_j Z_j^2}{A_j}\right)  {\rm erg\,g^{-1}\,s^{-1}}\,,
\label{eq_bremsstrahlung_degenerate}
\end{align}
where $T$ and $\rho$ are in K and in g cm$^{-3}$ respectively, $\mu_e=\left( \sum X_j Z_j/A_j \right)^{-1}$ is the mean molecular weight per electron, $ X_j $ is the relative mass density of the j-th ion, and $ Z_j,~A_j $ its charge and mass number respectively.\footnote{In the typical plasma conditions where the bremsstrahlung is relevant, one finds $Z_j/A_j\approx 1/2$. So, the rate has a dependence on the chemical composition of the plasma and increases in the case of high $Z$. 
In particular, the rate is larger in a CO WD core than in the core of a RGB star, composed mostly of He.}
The mild density dependence of the degenerate rate is accounted for by the dimensionless function $F$.
An explicit expression for this function can be found in~\cite{Raffelt:1994ry} (see also section 3.5 of Ref.~\cite{Raffelt:1996wa} for a pedagogical presentation).
Numerically, it is of order 1 for the stellar plasma conditions, $\rho \sim 10^{5}-10^{6}$ and $T\sim 10^{7}-10^{8}$, of interest for our discussion here, when the degenerate bremsstrahlung process dominates. 
The intermediate regime between degenerate and nondegenerate conditions in \fig{fig_bremsstrahlung_compton} is calculated as 
$\varepsilon_B=( 1/\varepsilon_{\rm B}^{\rm (d)}+ 1/\varepsilon_{\rm B}^{\rm (nd)})^{-1}$, 
following the prescription in Ref.~\cite{Raffelt:1994ry}.

The Compton process
\begin{align}
\gamma +e \to a + e
\end{align}  
 (right panel of \fig{fig_Feyn_Primakoff}) accounts for the production of axions from the scattering of thermal photons on electrons.
The Compton axion emission rate is a steep function of the temperature
\begin{align}
\varepsilon_{\rm C}\simeq 2.7\times 10^{-22} g_{ae}^2 \frac{1}{\mu_e}\left( \frac{n_{e}^{\rm eff}}{n_e} \right)\,T^6 \,{\rm erg\,g^{-1}\,s^{-1}} \,,
\end{align}
where $n_e$ is the number density of electrons while $n_{e}^{\rm eff}$ is the effective number density of electron targets.
At high densities, degeneracy effects reduce $n_{e}^{\rm eff}$, suppressing the Compton rate (cf.~\fig{fig_bremsstrahlung_compton}).
The Compton process can effectively dominate over the bremsstrahlung only at low density and high temperature. 

Below, we review the bounds on the axion-electron coupling derived by the most relevant stellar systems. \\

\emph{White Dwarfs.} 
The strongest bounds on the axion-electron coupling are inferred from observations of stars with a dense core, where the bremsstrahlung is very effective.
These conditions are realised in WD and RGB stars.
As discussed above, the WD phase is the last stage of the evolution of a low mass star, after the nuclear energy sources are exhausted. 
Hence, the evolution of a WD is essentially a cooling process, governed by photon radiation 
and neutrino emission, 
with the possible addition of novel energy-loss channels, e.g. in axions.

There are at least two ways to test the cooling of WDs and, consequently, exotic cooling theories.
First, one may study the WD Luminosity Function (WDLF), representing the distribution of WDs versus luminosity.
While cooling, the WD luminosity decreases.
Thus, the efficiency of the cooling reflects in the shape of the WDLF. 
Additionally, one can measure the secular drift of the oscillation period, $ \dot{P}/P $, of WD variables, which is practically proportional to the cooling rate $ \dot T/T $.

%
\begin{table}[t]
	\begin{center}
		\begin{tabular}{ l c  c c c  c c}
			Star  				&  $ P $(s)&  $ \dot{P}_{\rm obs} $(s/s) 		&  $ \dot{P}_{\rm th} $(s/s) 			& $g_{ae}^{(\rm best)}$   	&  $g_{ae}^{\rm (max)}(2\sigma)$ \\ \hline
			G117 - B15A 			&  215 	& $ (4.2 \pm 0.7)\times 10^{-15} $	&  $ (1.25 \pm 0.09)\times 10^{-15} $		&  $4.9\times 10^{-13}$ 	& $ 6.0\times 10^{-13} $  \\
			R548 		 	 	&  213 	&  $ (3.3 \pm 1.1)\times 10^{-15} $ 	&  $ (1.1 \pm 0.09)\times 10^{-15} $ 		&  $4.8 \times 10^{-13} $ 	&  $6.8 \times 10^{-13} $ \\
			PG 1351+489			&  489 	& $ (2.0 \pm 0.9)\times 10^{-13} $ 	& $ (0.81 \pm 0.5)\times 10^{-13} $ 		& $ 2.1 \times 10^{-13} $ 	& $ 3.8 \times 10^{-13} $ \\
			L 19-2	(113) 		&  113 	&  $ (3.0\pm 0.6)\times 10^{-15} $ 	&  $ (1.42 \pm 0.85)\times 10^{-15} $		&  $5.1 \times 10^{-13} $ 	&  $7.7 \times 10^{-13} $\\
			L 19-2 	(192) 		&  192 	&  $ (3.0\pm 0.6)\times 10^{-15} $ 	&  $ (2.41\pm 1.45) \times 10^{-15} $		&  $ 2.5 \times 10^{-13} $ 	&  $6.1  \times 10^{-13} $ \\ 
			\hline
		\end{tabular}
		\caption{Hints, $g_{ae}^{(\rm best)}$, and bounds, $g_{ae}^{\rm (max)}$, on the axion-electron coupling from WD 
		variable stars~\cite{Giannotti:2016hnk,Corsico:2019nmr}. $P$ is the period of the variable star and $\dot P$ its time derivative. We report the measured (observed) values and the theoretical predictions. 
		}
		\label{tab:Pdot}
	\end{center}
\end{table}

Let us begin with the WDLF. 
Current numerical analyses suggest the bound $g_{ae}\lesssim 2.8\times 10^{-13}$~\cite{Corsico:2019nmr}.
However, this result does not come with a credible confidence level because of the large theoretical and observational uncertainties (see, e.g., discussion in~\cite{Bertolami:2014wua}).
Moreover, the analyses show, fairly consistently  (though not universally) an anomalously large energy-loss.
In particular~\cite{Bertolami:2014wua}, using data from the Sloan Digital Sky Survey (SDSS) and the SuperCOSMOS Sky Survey (SCSS), showed that the axion coupling $ g_{ae}\simeq 1.4\times 10^{-13} $ is favoured with respect to the standard model case at about 2$ \sigma $ confidence level.\footnote{The additional energy can also be accounted for by hidden photons~\cite{Giannotti:2015kwo,Chang:2016qfl} but not by anomalous neutrino electromagnetic form factors~\cite{Bertolami:2014noa}.}
A more recent analysis of the data in Ref.~\cite{Bertolami:2014wua}, found~\cite{Giannotti:2017hny}
\begin{align}
g_{ae}=1.5^{+0.6}_{-0.9}\times 10^{-13}\qquad (95 \% \,\ \textrm{CL}) \,.
\end{align}
These results were confirmed in a later study~\cite{Isern:2018uce}, which attempted to reduce some systematic uncertainties, particularly those due to the star formation rate, by studying the WDLF of the thin and thick disc, and of the halo. 
A considerable improvement is expected from the next generation of astrophysical observations. 
Data from the GAIA satellite have already increased the catalog of WDs by an order of magnitude with respect to SDSS \cite{1805.01227,1807.03315}.
The Large Synoptic Survey Telescope (LSST) is expected to detect even  fainter WDs, ultimately increasing the census of WDs to tens of millions~\cite{0912.0201,Drlica-Wagner:2019xan}.

An independent method to study the cooling of WDs is the analysis of the period change of the WD variables. 
Unfortunately, the period changes very slowly, $\dot P/P\approx 10^{-18}\,{\rm s}^{-1}$ in most measured cases (see Table~\ref{tab:Pdot}), 
and an accurate assessment  of this change requires decades of accurate data taking. 
Therefore, although there are many known WD variables, $\dot P/P$ has been measured only for a handful of them (see Ref.~\cite{Corsico:2019nmr} for an update review of this subject).
Interestingly, in all cases the observed period change rate is always larger than the expected one, $ \dot{P}_{\rm obs} > \dot{P}_{\rm th} $, hinting at an unexpected cooling channel.
Such result could be attributed to an axion coupled to electrons with the couplings shown in Table~\ref{tab:Pdot}.
The highest level of discrepancy is observed in G117 - B15A but that may be due to some assumptions about the trapping mode and should perhaps be reconsidered~\cite{Bertolami:2014wua}.
The combined analysis of all other WD variables in the table gives a fairly good fit,
$\chi^2_{\rm min}/ $d.o.f.$ =1.1 $, for $ g_{ae}=2.9\times 10^{-13} $ and favours the axion (or ALP) interpretation at $2\,\sigma $. \\

\emph{Red Giants.} 
Another strong bound on the axion-electron coupling is inferred from the luminosity of the tip of the RGB in Globular Clusters (GC).
After the hydrogen in the core of a main sequence star is exhausted, the stellar core contracts and the star enters the RGB phase. During the RGB evolutions, the star expands, its surface cools down, and its luminosity increases (see \fig{fig_HR_diagram}). 
For sufficiently low mass stars, as  those populating a GC, the electrons in the stellar core eventually become degenerate.
Meanwhile, the core continues to contract and heat up, while its mass grows by the H-shell burn.
The process continues until the core reaches the conditions necessary to ignite He. 
At this time, known as the He-flash, the star reaches the point of highest luminosity in the CMD, known as the RGB tip.
The luminosity of the RGB tip is an excellent observable to probe the cooling of the star during the RGB phase, being the He ignition extremely sensitive to the temperature.
%
Thus, any additional cooling, in the form of axions or other light, weakly interacting particles, can be effectively constrained by observations of the luminosity of the RGB tip.
In the case of axions, the most efficient production mechanisms during the RGB production are the electron bremsstrahlung and the Compton processes.
The two clusters studied so far,\footnote{We point out, however, that very recently, Ref.~\cite{Diaz:2019kim} considered 50 GC to find the upper bound $g_{ae}=2.5\times 10^{-13}$ on the axion-electron coupling. The result is slightly more stringent than the one we discuss in this section.
}
M5~\cite{Viaux:2013lha} and M3~\cite{Straniero:2018fbv},
indicate fairly consistent, though not identical, results for the axion-electron coupling.\footnote{In particular, the analysis of M5~\cite{Viaux:2013lha} suggests a stronger hint to a non-vanishing axion-electron  coupling
\begin{align}
\begin{array}{ll}
g_{ae}=1.88^{+1.19}_{-1.17}\times 10^{-13}\,, & {\rm at~}  68\% {\rm ~CL} \\
g_{ae}\leq 4.3\times 10^{-13}\,,  &{\rm at~} 95\% {\rm ~CL}\,,
\end{array} 
\end{align}
while the observations of M3 are rather consistent with expectations ($g_{ae}^{\rm best}=0.05\times 10^{-13}$) and suggest a somewhat stronger bound $g_{ae}\leq 2.6\times 10^{-13}$ at 95\% CL~\cite{Straniero:2018fbv}.
Reference~\cite{Serenelli_2017} attributes the disagreement between theory and observations in~\cite{Viaux:2013lha}, at least partially, to the convention used for the screening of the nuclear reaction rates.  
} 
%
The combined analysis indicates the bound
\begin{subequations}
\begin{align}
& g_{ae}^{\rm (best)}= 1.4\times 10^{-13}\,,  \\
& g_{ae}\leq 3.1\times 10^{-13}\,, \qquad {\rm at~} 95\% {\rm ~CL}
\end{align}
\end{subequations}
%
%
\begin{figure}[t]
	\centering
	\includegraphics[width=0.7\linewidth]{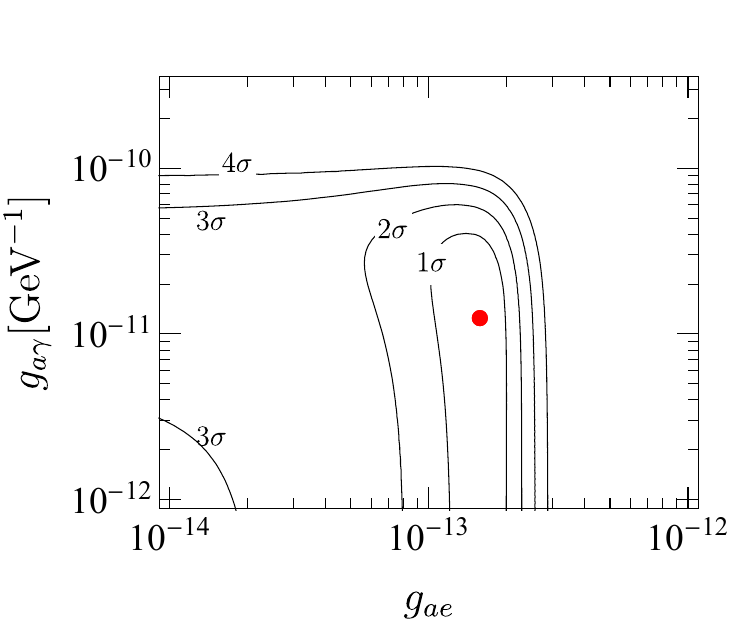}
	\caption{Stellar hints on general Axion Like Particles interacting with electrons and photons~\cite{DiVecchia:2019ejf}. 
	The hints are derived from the global analysis of WD pulsation, the WD luminosity function, RGB and HB stars. 
	The best fit parameters are indicated with the red dot.
}
	\label{fig_gae_gag_parameter_space}
\end{figure}

Several uncertainties, including the cluster morphology and distance as well as uncertainties in the nuclear reaction rates, affect the exact determination of the RGB bound on the axion-electron coupling~\cite{Viaux:2013hca,Viaux:2013lha,Straniero:2018fbv}.
The use of multi-band photometry of multiple globular clusters would provide a substantial improvement (see, e.g.,~\cite{Straniero:2018fbv,Diaz:2019kim}).
The error in the globular cluster distance, currently the largest observational uncertainty, will be reduced considerably (perhaps by as much as a factor of 10) with the release of the GAIA 
data relevant for GCs, expected in 2022~\cite{Gaia}. 

The combination of  hints from the WDLF, the WD pulsation, and RGB stars gives the 1$ \,\sigma $ preferred interval
\begin{equation}
g_{ae}=1.6^{+0.29}_{-0.34} \times 10^{-13},
\end{equation}
with $ \chi^2_{\rm min}/ $d.o.f.$ =14.9/15=1.0$, and favours the axion (or ALP) solution at slightly 
more than $3\,\sigma $~\cite{Giannotti:2017hny}.

Combining the results from the
WDLF, the WD pulsation, and RGB stars, discussed in this section, 
with the analysis of the $R$ parameter (\sect{sec:Astro_bounds_gag}), 
one finds the hinted regions presented in \fig{fig_gae_gag_parameter_space}~\cite{DiVecchia:2019ejf}.
The analysis shows a preference for an axion coupling to electrons at the level of 3$\,\sigma$.
On the other hand, the coupling to photons is compatible with zero at 1$\,\sigma$.
Notice, however, that the last conclusion cannot be drawn in the case of specific axion models, such as DFSZ I and II, which predict well-defined relations among couplings. \\

\subsection{Axion-nucleon coupling}
\label{sec:Astro_bounds_gaN}

Finite axion-nuclei interactions allow for further ways axions may impact the evolution of stars. 
Non-thermal processes, such as nuclear transitions in stars (particularly, the Sun) with the emission of axions, provide an interesting channel to produce a possibly detectable axion flux. However, their impact on stellar evolution is minimal. 
Thermal processes turn out to be quite more relevant, in this respect. 

\begin{figure}[h]
	\centering
	\includegraphics[width=0.3\linewidth]{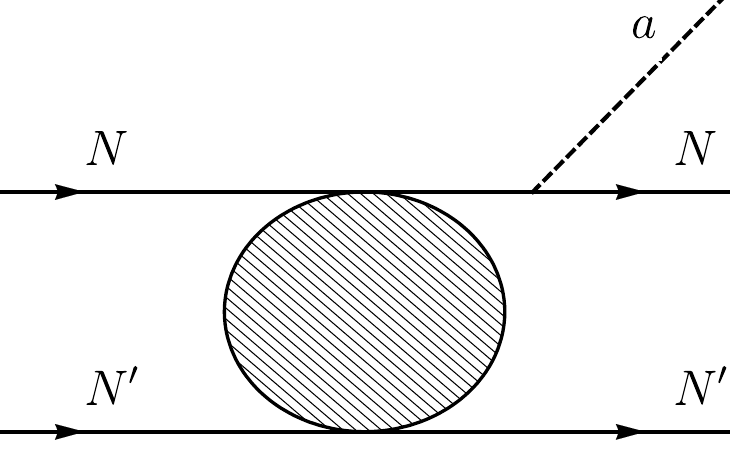}
	\hspace{1.5cm}
	\includegraphics[width=0.3\linewidth]{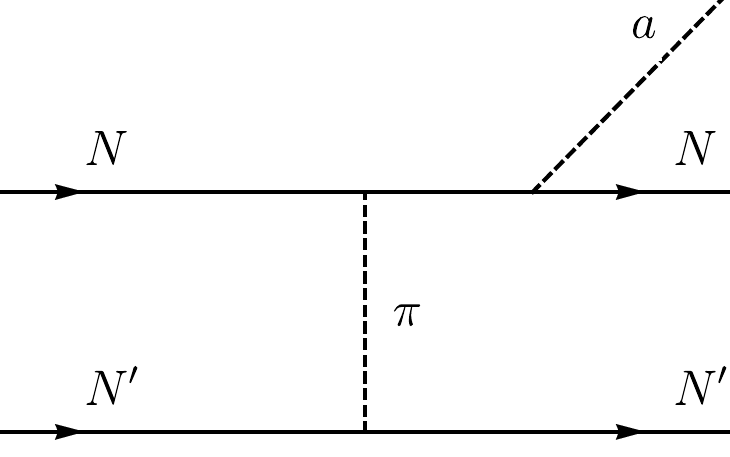}
	\caption{
	Axion nuclear bremsstrahlung (one of the possible diagrams). $N,N^{\prime}$ represent either a proton or a neutron. 
	To the right is the Feynman diagram corresponding to the OPE approximation. 
	}
	\label{fig_Feyn_NuclearBremsstrahlung}
\end{figure}

The most relevant thermal process involving the axion-nucleon coupling is the nucleon bremsstrahlung
\begin{align}
& N+N^{\prime}\to N+N^{\prime}+a\,,
\end{align}
with $N,N^{\prime}=n,p$, where $n$ represents  a neutron and $p$ a proton.
The Feynman diagram for these processes is shown in \fig{fig_Feyn_NuclearBremsstrahlung}.
One of the major difficulties in dealing with the nucleon bremsstrahlung is the description of the nuclear interaction, shown as a blob in figure~\ref{fig_Feyn_NuclearBremsstrahlung}.
A substantial simplification, known as the One Pion Exchange (OPE) approximation, is to assume that the interaction is mediated by the exchange of a single pion, as shown in the right panel in figure~\ref{fig_Feyn_NuclearBremsstrahlung}.
The OPE framework is not always justified (see, e.g., Ref.~\cite{Raffelt:1990yz,Hanhart:2000ae}) but it does provide a starting point for more accurate computations (cf.~Ref.~\cite{Carenza:2019pxu} for a recent review of the role of OPE and its corrections).
Nevertheless, it is evident from the figure that the pion mass in the propagator is going to suppress the emission rate unless the temperature is such that the typical momentum  exchanged in the collision, which is of the order of the nucleon momentum $q_N\sim (3m_NT)^{1/2}$, is larger than the pion mass. 
This demands $T\gtrsim 10$MeV, a temperature typical of the core of Supernovae (SNe) and NS.
Therefore, only SNe and NS may (and, indeed, do) provide an environment to test the axion nucleon bremsstrahlung. 

Approximate emission rates for the $nn$ scattering (the $pp$ scattering is similar) in the limit of nondegenerate and degenerate nuclei are given below~\cite{Raffelt:1996wa}
\begin{subequations}\label{eq_nuclear_bremsstrahlung_approx}
\begin{align}
& \varepsilon_{{\rm ND}}\approx 2.0\times 10^{38} g_{an}^2 \rho_{14}\,T_{30}^{3.5}\, {\rm erg}\, {\rm g}^{-1} {\rm s}^{-1} \,,\\
& \varepsilon_{{\rm D}}\approx 4.7\times 10^{39} g_{an}^2 \rho_{14}^{-2/3}\,T_{30}^{6}\, {\rm erg}\, {\rm g}^{-1} {\rm s}^{-1} \,,
\end{align}
\end{subequations}
%
where $T_{30}=T/30$MeV and $\rho_{14}=\rho/10^{14}{\rm g} \, {\rm cm}^{-3}$.
Notice that eqs.~\eqref{eq_nuclear_bremsstrahlung_approx} are only a crude approximation of the emission rate, calculated in the OPE approximation and ignoring the pion mass and medium effects. 
However, they do show the steeper temperature dependence of the degenerate emission rate and the stronger density dependence in the nondegenerate limit. 

Below we report the recent bounds on the axion couplings to nuclei from SN and NS. 
However, we emphasise that, at the time of writing, there is  an intense effort to
provide a more reliable description of the nuclear processes that produce axions at very high density,\footnote{Many-body effects can be quite large in a high density medium. 
The inclusion of such effects in the description of axion production processes in SN and NS has a long history. 
Discussions can be found in Ref.~\cite{Keil:1996ju,Carenza:2019pxu,Raffelt:1990yz}.  
Very recently, Ref.~\cite{2020arXiv200304903B} reevaluated such medium effects,
showing a significant dependence of the axion couplings to nucleons upon the environment density. 
These latest effects are not included in the results presented in this review, which reports
the latest  bounds on the axion-nucleon couplings available in the literature at the
 time of writing. 
}
and such bounds are often reconsidered and reassessed. \\

\begin{figure}[t]
	\centering
	\includegraphics[width=0.7\linewidth]{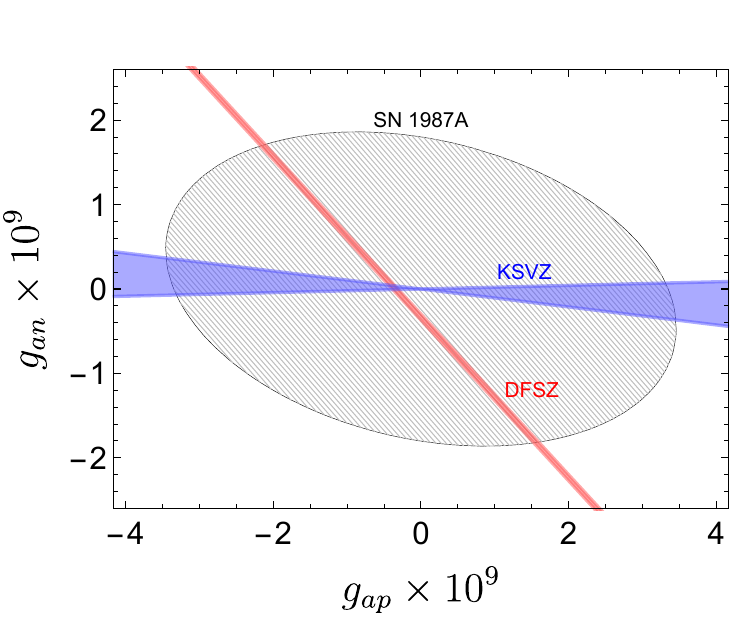}
	\caption{SN 1987A bound on the axion-nucleon couplings.
		In hatched grey, the region allowed by the SN 1987A bound derived  in~\cite{Carenza:2019pxu}. 
		The parameter space for KSVZ and DFSZ axions is superimposed. 
		The width of the lines represent the current uncertainties according to~\cite{diCortona:2015ldu}.
		}
	\label{fig_SN_bound}
\end{figure}

\emph{SN 1987A.}
The most well known argument to constrain the axion interaction with protons and neutrons is the one based on the observed neutrino signal from SN 1987A~\cite{Turner:1987by,Burrows:1988ah,Raffelt:1987yt,Raffelt:1990yz}.
The signal duration depends on the efficiency of the cooling and is compatible with the assumption that SN neutrinos carry about 99\% of the  energy released in the explosion.
For a light, weakly interacting particle, a bound can be extracted from the requirement that it does not contribute more than neutrinos, about $2\times 10^{52} $ erg s$^{-1}  $, to the cooling of the young SN, with a typical core conditions of $T\sim 30$ MeV and $\rho\sim 10^{14}$g$\,$cm$^{-3}$. 
The most recent analysis for the axion case~\cite{Carenza:2019pxu} derived the bound\footnote{\eqn{eq:gan_gap_SN_bound} shows a surprisingly subdominant contribution of the proton scattering to the emission rate, quite more accentuated than what reported in previous analyses~\cite{Raffelt:1990yz,Giannotti:2017hny}.
The reason is that, besides being less abundant than neutrons, protons are nondegenerate while neutrons are partially degenerate and the emission rate in SN conditions is more efficient for degenerate nuclei, as evident from Eqs.~\eqref{eq_nuclear_bremsstrahlung_approx}. } 
\begin{equation} 
g_{an}^2+ 0.29\, g_{ap}^2 + 0.27\, g_{an}\,g_{ap}\lesssim 3.25 \times 10^{-18} \,,
\label{eq:gan_gap_SN_bound}
\end{equation}
shown in \fig{fig_SN_bound}. 
%

Strongly interacting axions may be trapped in the SN core. 
In this case the emission is reduced, as they thermalise and are effectively emitted from an \emph{axiosphere}, similarly to what happens to neutrinos. 
The most recent analysis  found  that this condition is satisfied for $g_{an}=g_{ap}\gtrsim 10^{-7}$.
However, even trapped axions may extract more energy than neutrinos from the young SN and couplings all the way to $g_{an}=g_{ap}\approx 10^{-4}$ should be probably  excluded~\cite{Carenza:2019pxu}. \\

\emph{Neutron Stars.}
Observations of the cooling of NS also provide information about the axion-nucleon coupling~\cite{Keller:2012yr,Sedrakian:2015krq,Hamaguchi:2018oqw,Beznogov:2018fda,Sedrakian:2018kdm}. 
In particular, the unexpectedly rapid cooling of the NS in CAS A was attributed to the presence of axions with coupling to neutrons~\cite{Leinson:2014ioa}
\begin{align}
\label{eq:NS_CASA_hint}
g_{an}\simeq 4\times 10^{-10}\,.
\end{align}
However, the anomalous rapid cooling may also be originated in the  phase transition of the neutron  condensate  into a multicomponent  state~\cite{Leinson:2014cja}.
More recently, the data have been explained assuming a neutron triplet superfluid transition occurring  at the present time, $t\sim 320$ years, and that proton superconductivity is operating at $t\ll 320$ years~\cite{Hamaguchi:2018oqw}.
The neutron triplet superfluid transition accelerates the neutrino emission through  the breaking and reformation of neutron Cooper pairs.
Under these assumptions the data can be fitted well, leaving little room for additional axion cooling.
Quantitatively,
\begin{align}
\label{eq:NS_CAS_A}
g_{ap}^2+1.6\, g_{an}^2\leq 1.1\times 10^{-18}\,.
\end{align}
An even stronger bound, though only on the axion-neutron coupling, 
\begin{align}
\label{eq:NS_J1731}
g_{an}\leq 2.8\times 10^{-10}\,,
\end{align}
was inferred from observations of the NS in HESS J1731-347~\cite{Beznogov:2018fda}.
A considerable less stringent result, 
\begin{align}
\label{eq:NS_Sedrakian}
g_{an}\lesssim (2.5-3.2)\times 10^{-9}, 
\end{align}
was derived more recently in Ref.~\cite{Sedrakian:2018kdm},
the range depending on the adopted value for $\tan\beta$.\footnote{Notice that Ref.~\cite{Sedrakian:2018kdm} uses the opposite for $\tan \beta$ and so what they call $\cos \beta$ is our $\sin \beta$ and viceversa.} 
Interestingly, this latest analysis accounted also for a possible axion emission 
by electron bremsstrahlung in the neutron star crust,
which cannot \emph{a priori} be ignored for non-hadronic axion models, such as the DFSZ.
The resulting rate is, however, in most cases subdominant with respect to the processes induced by nuclear couplings 
and becomes relevant only when the axion-neutron coupling is very small.




\subsection{Axion coupling to the neutron EDM}
\label{sec:Astro_bounds_nEDM}

As discussed in \sect{sec:summaryaxioncoupl} a fundamental consequence of  QCD axion models is that axions couple to the neutron EDM, effectively driving it to zero and solving the strong CP problem.
The axion-neutron EDM vertex can be parametrised with the coupling $g_d$ defined through the Lagrangian term (cf. \eqn{eq:Laint1})
%
\begin{align}
\label{eq:neutron_EDM}
\mathcal L_d =-
\frac{i}{2}g_d\, a\, \bar{n}\,\sigma_{\mu\nu}\gamma_{5}n\, F^{\mu\nu}\,.
\end{align}
This coupling induces the process $n+\gamma\to n+a$, allowing the production of axions that, in turn, contribute to the SN cooling. 
As discussed, observations of the SN 1987A neutrino burst limit the amount of possible exotic cooling rate to be less than the neutrino's.
A rough estimate gives~\cite{Graham:2013gfa}
\begin{align}
\label{eq:SN_bound_neutron_EDM}
g_{d}\leq 4 \times 10^{-9}\,{\rm GeV}^{-2}\,.
\end{align}

We emphasise  that the interaction in~\eqref{eq:neutron_EDM} is generic to any QCD axion model and does not demand any other assumption besides the solution of the strong CP problem.
Therefore, the bound~\eqref{eq:SN_bound_neutron_EDM} is, effectively, a bound on the PQ constant or, equivalently, on the axion mass.\footnote{Here, we are assuming the standard relation, \eqn{eq:axionmass}, between axion mass and decay constant.
We discuss mechanisms to modify this relation in \sect{sec:heavy}.\label{footnote:ma_fa}}
If we express the coupling in terms of the axion mass, 
$g_d\approx 6\times 10^{-10} (m_a/{\rm eV}){\rm GeV}^{-2}$,
the SN bound implies $f_a\gtrsim 9\times 10^{5}$ GeV or, equivalently, $m_a\lesssim 7\,$eV.

\subsection{Axion CP-odd couplings}
Stellar evolution provides also strong bounds on the axion CP odd couplings, discussed in \sect{sec:CPvaxioncoupl}.
The axion scalar couplings to electrons $g^{S}_{ae}$ can be constrained in globular cluster stars, where such particles can be produced through Compton scattering or bremsstrahlung~\cite{Raffelt:2012sp}.
The strongest bound is derived by the luminosity of the tip of the RGB.
A semiquantitative argument, based on the assumption that any novel emission rate would spoil observations unless $\varepsilon\lesssim 10\,{\rm erg\,s^{-1}g^{-1}}$, gives the rather restrictive bound~\cite{Hardy:2016kme}
\begin{align}
\label{eq:RGB_scalar_electron_coupling}
g^{S}_{ae}\leq  0.7 \times 10^{-15}\,.
\end{align}
HB stars provide a slightly less restrictive bound, $g^{S}_{ae}\leq 3 \times 10^{-15}$.

The scalar coupling to nuclei is likewise constrained in RGB stars~\cite{Hardy:2016kme}
\begin{align}
\label{eq:RGB_scalar_nucleon_coupling}
g^{S}_{aN}\leq  1.1 \times 10^{-12}\,,
\end{align}
while HB stars provide the less restrictive bound $g^{S}_{aN}\leq 6 \times 10^{-12}$.

These astrophysical bounds are the dominant constraints on the coupling to nuclei for masses above 1 eV or so. 
However, for lower masses the experimental bounds on $5^{{\rm th}}$ force are much stronger (see Fig.~1 in Ref.~\cite{Raffelt:2012sp}).

%

\subsection{Axion coupling to gravity and black hole superradiance}
\label{sec:Astro_bounds_axion_gravity}

In some cases, astrophysical considerations can provide insights  on the couplings of axions to gravity, 
without assuming any interaction with standard model fields.
Such considerations are, therefore, completely model-independent. 
The case of black holes (BH) discussed below is particularly interesting since, just like in the case of the other bounds discussed in this section, there is no assumption that axions are initially present, i.e. there is no requirement for axions to be the DM. 

Axions form gravitational bound states around black holes whenever their Compton length is of the order of the black holes radii. 
The phenomenon of \emph{superradiance}~\cite{Penrose:1969pc} 
then guarantees that the axion occupation numbers grow exponentially, providing a way to extract very efficiently energy and angular momentum from the black hole~\cite{Arvanitaki:2010sy,Arvanitaki:2014wva}.
The rate at which the angular momentum is extracted depends on the black hole mass and so the presence of axions could be inferred by observations of black hole masses and spins. 
Current observations exclude the region~\cite{Arvanitaki:2014wva,Cardoso:2018tly} 
\begin{align}
\label{eq:BH_superradiance_fpq}
6\times 10^{17}\,{\rm GeV}\leq f_a\leq 10^{19}\,{\rm GeV} \,,
\end{align}
corresponding to the mass region (Cf. footnote \ref{footnote:ma_fa}) $6\times 10^{-13}{\rm eV}\leq m_a\leq 10^{-11}{\rm eV}$.\footnote{The results reported are the ones in the most recent analysis, Ref.~\cite{Cardoso:2018tly}.}
We underline that superradiance can start from a quantum mechanical fluctuation and does not require the prior existence of an axion population.

The condition for the BH superradiance relies on the assumption that the axion self interaction is small, which is why the bounds concern such large values of $f_a$.
For sufficiently large  couplings, the axion cloud could collapse in what is known as a \emph{bosenova},\footnote{Axion couplings to other fields, e.g. photons, would also induce an effective axion self-interaction. However, such couplings would need to be too large to have  any significant effect. For example, the value of $g_{a\gamma}$ required to induce a self-coupling as large as what expected from the axion potential in \eqn{eq:Vapiexp}, for masses $\sim 10^{-12}$eV, is several orders of magnitude larger that the value excluded by the HB bound. This justifies plotting $E/N$ up to very large values, as we do in~\sect{sec:Experiments}. 
} 
producing periodic bursts which should be observable by Advanced LIGO and VIRGO~\cite{Arvanitaki:2014wva}.

Other observational signatures of BH superradiance are the gravitational waves produced in the transition of axions between gravitational levels or from axions annihilation to gravitons~\cite{Arvanitaki:2014wva,Arvanitaki:2016qwi}.

\subsection{Summary of astrophysical bounds}
\label{sec:Astro_bounds_summary}

A summary of all  bounds on the axion couplings from stellar evolution is shown in Table~\ref{tab:Astro_bounds} where, 
whenever possible, we have reported the bounds at $2\,\sigma$ and the hints at $1\sigma$.
\begin{table}[t]
	\begin{center}
		\begin{tabular}{ l l l }
			Star  				&  Hint ($1\,\sigma$)										&  Bound  ($2\,\sigma$) 						 		\\ \hline
			\vspace{0.2cm}
			Sun 					&   -- 												&  $g_{a\gamma}\leq 2.7\times 10^{-10}\,{\rm GeV^{-1}}$ 		  	\\
			\vspace{0.2cm}
			WDLF 				&  $g_{ae}=1.5^{+0.3}_{-0.5}\times 10^{-13}$ 					&  $g_{ae}\leq 2.1 \times 10^{-13}$ 		  	\\
			\vspace{0.2cm}
			WDV 		 	 	&  $g_{ae}=2.9^{+0.6}_{-0.9}\times 10^{-13}$ 		 			&  $g_{ae}\leq 4.1 \times 10^{-13}$ 	 		 \\
			\vspace{0.2cm}
			RGB Tip				&  $g_{ae}=1.4^{+0.9}_{-1.3} \times 10^{-13}$ 		(M3+M5)				&  $g_{ae}\leq 3.1 \times 10^{-13}$ 	(M3+M5)	 	\\
			\vspace{0.2cm}
				 				&  --													&  $g^{S}_{ae}\leq 0.7 \times 10^{-15}\,;~~ g_{aN}^{S}\leq 1.1 \times 10^{-12}$ 		 	\\		
			\vspace{0.2cm}
			HB 					&  $g_{a\gamma}=(0.3\pm 0.2)\times 10^{-10}\,{\rm GeV}^{-1}$ 	&  $g_{a\gamma}\leq 0.65 \times 	10^{-10}\,{\rm GeV}^{-1}$ 		\\
			\vspace{0.2cm}
			  					&   		--											&  $g^{S}_{ae}\leq 3 \times 10^{-15}\,;~~ g_{aN}^{S}\leq 6\times 10^{-12}$ 		 	\\		
			\vspace{0.2cm}
			SN 1987A 			&  -- 	&  $ g_{an}^2+ 0.29\, g_{ap}^2 + 0.27\, g_{an}\,g_{ap}\lesssim 3.25 \times 10^{-18}  $ 			 \\ 
			\vspace{0.2cm}
			 					&   --	&  $ g_{d}\lesssim 4 \times 10^{-9}\,{\rm GeV}^{-2} $ 	($\Rightarrow f_a\gtrsim 9\times 10^{5}$ GeV)		 \\ 
			\vspace{0.2cm}
			NS in CAS A 			&  -- 	&  $ g_{ap}^2+1.6\, g_{an}^2\lesssim 1.1\times 10^{-18} $ 			 \\ 
			\vspace{0.2cm}
			NS in HESS J1731-347 	&  -- 	&  $g_{an}\leq 2.8\times 10^{-10}$			 \\ 
			\vspace{0.2cm}
			Black Holes 	&  -- 		&  $f_a\leq 6\times 10^{17}\,{\rm GeV~ or} f_a\geq 10^{19}\,{\rm GeV}$			 \\  
\hline
\end{tabular} 
		\caption{Summary of stellar hints and bounds on axions.	
		The hints are all at 1$\sigma$ and the bounds at 2$\sigma$, except for the case of SN 1987A and NS in CAS A, for which a confidence level was not provided. 
		We have not reported the hint from the NS in CAS A~\cite{Leinson:2014ioa} since it is in tension with the more recent bound in~\cite{Hamaguchi:2018oqw}. }
		\label{tab:Astro_bounds}
	\end{center}
\end{table}
The bounds were derived without assuming any model dependence and are therefore quite general.

Particularly  interesting among the astrophysical  considerations are the bounds from BH superradiance and the SN 1987A bound on the neutron EDM, since they are the only ones that provide a bound on the axion mass rather than on the couplings.\footnote{Astrophysical considerations are, of course, affected by many uncertainties. In particular, the bound on the neutron EDM is based on a simple estimate and should be probably revised. Additionally, we are assuming a standard relation between the axion mass and decay constant (Cf. footnote \ref{footnote:ma_fa}).} 
Combined, they constrain the axion mass in the range $2\times 10^{-11} {\rm eV}\leq m_a \leq 7 {\rm eV}$.

\begin{figure}[t]
	\centering
	\includegraphics[width=0.7\linewidth]{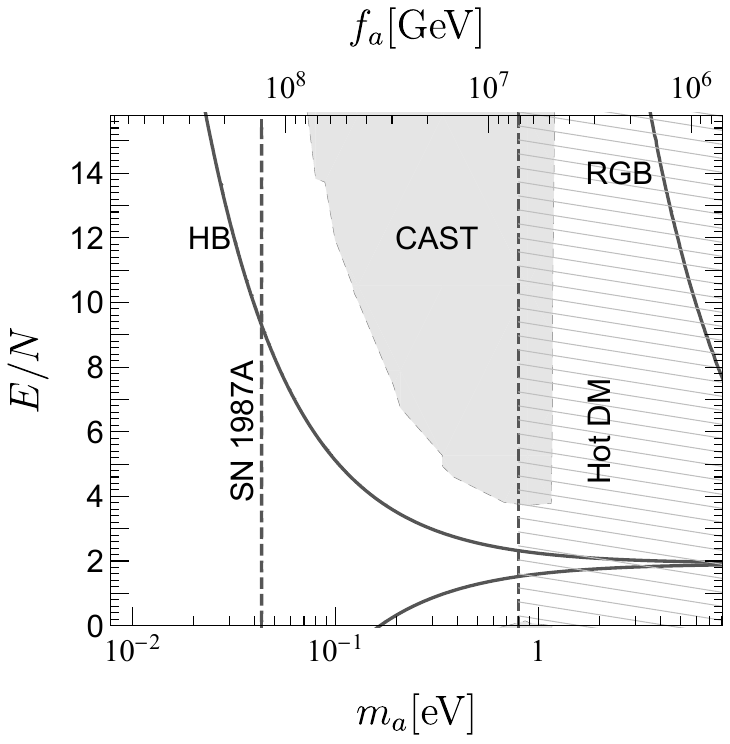}
	\caption{Astrophysical bounds on the hadronic axions (at $2\,\sigma$).
	The region to the right of the curves is excluded.
	The bound from SN 1987A is shown with dashed lines since the bound is less robust than the others~\cite{Carenza:2019pxu,Fischer:2016cyd} and its statistical significance is not well defined.	
	The area probed by CAST is shown in light grey.	
	The lightly hatched region is excluded by the hot DM bound~\cite{Archidiacono:2013cha} 
(see \sect{sec:hotaxions}).
}.
	\label{fig_KSVZ_astro_bounds}
\end{figure}

In the case of the other bounds, specific model-dependent  relations connect the different couplings, as well as the axion mass. 
The astrophysical bounds for hadronic axion models are shown in \fig{fig_KSVZ_astro_bounds}.
The models are parametrised in terms of the axion mass and $E/N$.
We superimpose also the region excluded by the CERN Axion Solar Telescope (CAST) for reference. 
The astrophysical hinted region is not shown since hadronic axions cannot fit very well the required parameters~\cite{Giannotti:2017hny}.

Hadronic axions are naturally electrophobic. 	
For axion masses between $10^{-9}$ and 1 eV, and with  $E/N$ between 0 and 15, the parameter $C_{ae}$, defined in \eqn{eq:gagammagaf}, which parameterise the interaction with electrons, is always confined in the range $\approx 5-15\times 10^{-3}$.
Consequently, the RGB bound is always suppressed with respect to the HB and the hot DM bounds for such axions. 

\begin{figure}[t]
	\centering
	\includegraphics[width=0.48\linewidth]{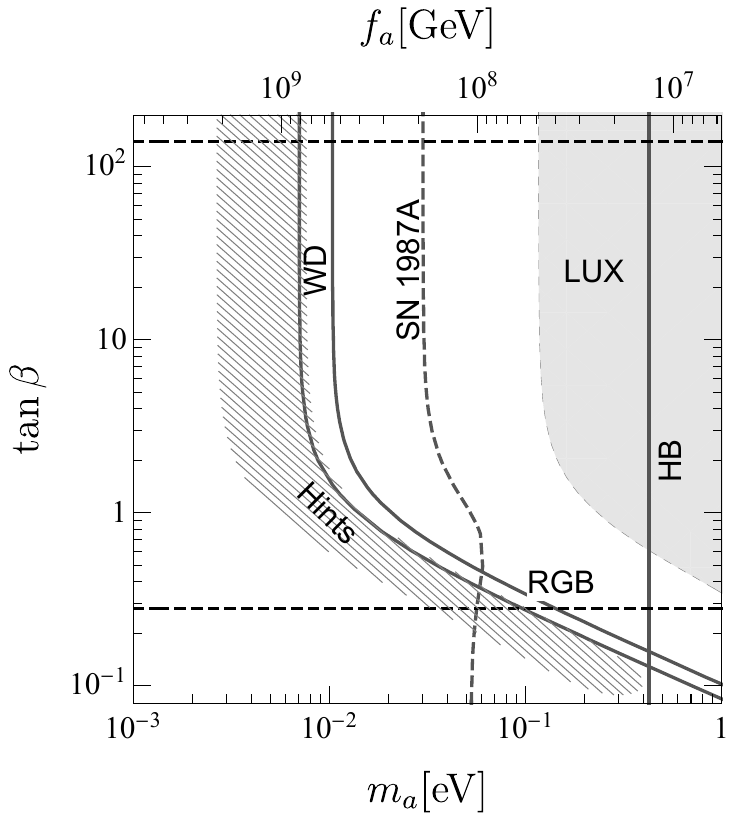}
	\hspace{0.1 cm}
	\includegraphics[width=0.48\linewidth]{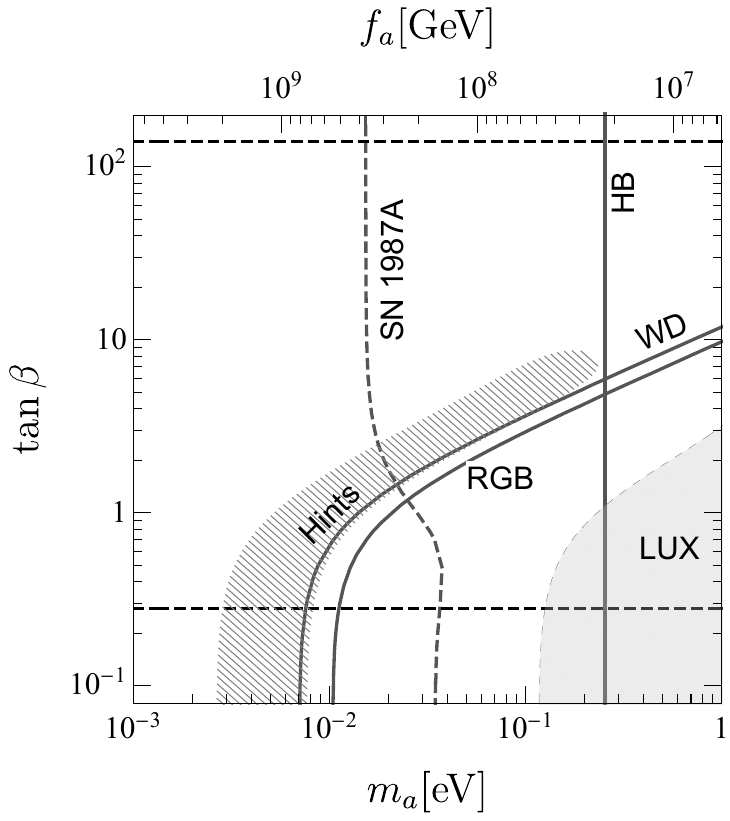}
	\caption{Astrophysical bounds on the DFSZ axions.
	The left panel represents the DFSZ I model.
	The right panel the DFSZ II.
	The region to the right of the curves are excluded by astrophysical considerations.
	The WD curve refers to the WD luminosity function bound (refer to the text for more information).
	The hatched region refers to the combined hints from the WD luminosity function, the WD pulsation, HB and RGB stars.
	The bound from SN 1987A is shown with dashed lines since the bound is less robust than the others~\cite{Carenza:2019pxu}.		
	}
	\label{fig_DFSZ_astro_bounds}
\end{figure}

The analogous plot for DFSZ axions is shown in \fig{fig_DFSZ_astro_bounds}.
Again, we superimpose the region experimentally excluded, in this case by the Large Underground Xenon (LUX) experiment (CAST does not probe these models).
In this case, the stellar anomalous observations can be fitted quite well~\cite{Giannotti:2017hny}.
The hinted region is, however, quite difficult to explore experimentally and probably only the International AXion Observatory (IAXO) will be able to access parts of it in the near future~\cite{Armengaud:2019uso}. 

%


%

\clearpage
%
%
%
 


\section{Experimental Searches}
\label{sec:Experiments}

In the interest of keeping this review  self-consistent, we briefly describe the status of the axion experimental searches. 
More complete reviews can be found in Refs.~\cite{Kim:2008hd,Graham:2015ouw,Irastorza:2018dyq,Sikivie:2020zpn}.


\subsection{Solar axions and helioscopes}
 \label{sec:Helioscopes}

The Sun is a natural source of axions. 
These are produced through Primakoff  (\sect{sec:Astro_bounds_gag}) and ABC (\sect{sec:Astro_bounds_gae}) processes:
\begin{align}
\label{eq:solar_flux_1}
\frac{dN_a}{dt d\omega}
=\left( \frac{g_{a\gamma}}{{\rm GeV}^{-1}} \right)^{2}\,n_{a\gamma}(\omega) +g_{ae}^2\,n_{ae}(\omega) \,,
\end{align}
where $n_{a\gamma}$ represents the Primakoff contributions while $n_{ae}$ gets contribution from atomic recombination and deexcitation, bremsstrahlung, and Compton processes, 
$n_{ae}=n_{ae}^{A}+n_{ae}^{B}+n_{ae}^{C}$.
Good analytical approximations for these coefficients exist for all but the atomic recombination and deexcitation processes (see, e.g., Ref.~\cite{Andriamonje:2007ew} and \cite{Barth:2013sma}):
\begin{subequations}\label{eq:solar_flux_approx}
\begin{align}
&n_{a\gamma}(\omega)\approx
1.69\times 10^{58}
e^{-0.829 \,\omega}\omega^{2.45}
\,{\rm keV}^{-1}{\rm s}^{-1}\,; \\
& n_{ae}^{B}\approx 7.4\times 10^{62}
\frac{\omega\,e^{-0.77\,\omega}}{1+0.667\,\omega^{1.278}}
\,{\rm keV}^{-1}{\rm s}^{-1}\,; \\
& n_{ae}^{C}\approx 3.7\times 10^{60}
\omega^{2.987}\,e^{-0.776\,\omega}
\,{\rm keV}^{-1}{\rm s}^{-1} \,.
\end{align}
\end{subequations}
The total number of axions emitted by the Sun per second is,
%
\begin{align}\label{eq:solar_spectrum_integrated}
\frac{dN_a}{dt}
=1.1\times 10^{39} 
\left[ 
\left( 
\frac{g_{a\gamma}}{10^{-10}{\rm GeV}^{-1}} 
\right)^2  
+ 0.7\,
\left( 
\frac{g_{ae}}{10^{-12}} 
\right)^{2} 
\right] \,{\rm s}^{-1}  \,.
\end{align}
%
Evidently, the axion flux gets a similar contribution from Primakoff and ABC axions for couplings of phenomenological interest (cf.~\fig{fig_solar_flux}).
Notice, however, that the ABC flux is peaked at slightly lower energies than the Primakoff, a fact that allows to distinguish between the two fluxes (and so infer some information about the underlying axion model) if enough data is collected in an helioscope experiment~\cite{Jaeckel:2018mbn}.\footnote{For example, the number of axions produced in the energy bin $1\leq \omega/{\rm keV}\leq 2$ is 12.5\% of the total, in the case of hadronic axions (no coupling with electrons) but it moves quickly to $\sim 30\%$ if $g_{ae}/g_{a\gamma}>5\times 10^{-2}\,{\rm GeV}$, which is typical for DFSZ axions. }

The weight of the two contributions in \eqn{eq:solar_spectrum_integrated} depends on the specific axion model.
In terms of 
$g_{e 12}=g_{ae}/10^{-12}$
and  $g_{\gamma 10}=g_{a\gamma}/10^{-10}{\rm GeV}^{-1}$, we have
\begin{subequations}
\label{eq:ge12gg10}
\begin{align}
& {\rm KSVZ}:  ~ ~\,~ g_{e12}/g_{\gamma 10} \approx 0\,; \\
& {\rm DFSZ~I}:  ~~g_{e12}/g_{\gamma 10}  = 20 \sin^{2}\beta\,;  \\
& {\rm DFSZ~II}:  ~g_{e12}/g_{\gamma 10} = 12 \cos^{2}\beta\,.
\end{align}
\end{subequations}
%

%

%
%

Experiments that aim at detecting the solar axion flux are known as axion helioscopes. 
The most notable example is the Sikivie helioscope~\cite{Sikivie:1983ip},
or simply helioscope, 
which adopts a strong laboratory magnetic field for the coherent conversion of solar axions into X-ray photons. 
However, the solar axion flux may be detected through other means, for example through the Primakoff-Bragg conversion~\cite{Paschos:1993yf} or the axio-electric effect~\cite{Derbin:2011gg,Arisaka:2012pb}.

The Sikivie helioscope makes use of the axion coherent conversion in a transverse magnetic field $ B $. 
The conversion probability is 
\begin{align}
\label{eq:Helioscope_agamma_probability}
P_{a\to \gamma}=\left( \dfrac{g_{a\gamma}\, B\, L}{2}\right)^2 \dfrac{{\rm sin}^2(qL/2)}{(qL/2)^2}\,,
\end{align}
where $ q=q_\gamma-q_a $ is the momentum transfer provided by the magnetic field and $L$ is the length of the magnet.
Coherence is ensured whenever $qL\ll 1$.
In this condition, the probability does not depend on the axion mass and energy.
In vacuum,  the relativistic approximation ($\omega_a\gg m_a$) gives  $ q\simeq m_a^2/2\omega $.
A small mass, therefore, ensures coherence on macroscopic scales.

Whenever the coherence condition is verified, the conversion probability scales as $(g_{a\gamma}BL)^2$, rapidly increasing with the magnetic field and the size of the magnet. 
When coherence is lost, the sensitivity is reduced proportionally to $m_a^{-2}$.\footnote{Given enough data (and a large enough axion mass), the spectral distortion induced by the loss of coherence may allow to pin down the axion mass~\cite{Dafni:2018tvj}.}
Since for QCD axion models $g_{a\gamma}^2\propto m^2_a$ the sensitivity line follows exactly the axion model line in the $g_{a\gamma}-m_a$ plane  once coherence is lost at large masses. 

To regain sensitivity, the coherence can be restored using a buffer gas in the magnet beam pipes~\cite{vanBibber:1988ge}.
In this case the momentum transfer is $ q\simeq (m_a^2-m_\gamma^2)/2\omega $, where $m_\gamma$ is the effective photon mass in the gas. 
Tuning the effective photon mass to the axion mass allows to effectively regain coherence.

The CERN Solar Axion Experiment (CAST)~\cite{Zioutas:1998cc}, a 3-rd generation and currently the most advanced running axion helioscope,\footnote{CAST is expected to stop operating in 2020.} reported the bound $g_{a\gamma}\leq 0.66\times 10^{-10}\,{\rm GeV}^{-1}$ for masses $m_a\leq 20\,{\rm meV}$, while reaching the $m_a\sim$eV range at high masses~\cite{Anastassopoulos:2017ftl}.
 Going beyond that mass may be less interesting  because of the hot DM bound~\cite{Hannestad:2005df,Archidiacono:2013cha} (see \sect{sec:hotaxions}).
 
\begin{figure}[t]
	\centering
	\includegraphics[width=0.7\linewidth]{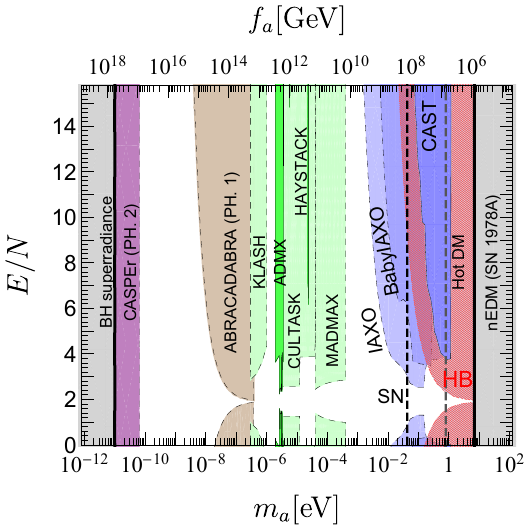}
	\caption{
	Hadronic axion parameter space. Interactions with electrons are neglected. 	
	Experimental bounds are shown with solid lines while projected sensitivities with dashed lines.
	The Helioscope lines refer to the latest results from CAST~\cite{Anastassopoulos:2017ftl} and to the expected sensitivity of BabyIAXO and IAXO~\cite{Armengaud:2019uso}.
	The sensitivity of the haloscope experiments is calculated assuming that axions comprise the totality of the cold dark matter in the Universe. 
	In green are the cavity experiments. Darker colour corresponds to actual data while in lighter colour we show the sensitivity of proposed  experiments. 
	The region labelled CULTASK shows the combined expected sensitivity of CAPP-12TB and CAPP-25T (Cf. Table~\ref{tab:summary_probes}). 
The sensitivity of ABRACADABRA refers to phase 1 ($B_{\rm max}=5$T and Volume=1$\,m^{3}$~\cite{Kahn:2016aff}) for the resonant case. 
CASPER refers to CASPER electric, phase 2, and indicates the most optimistic scenario compatible with the QCD uncertainty in the calculation of the nEDM (Cf.~\eqn{eq:Cangamma}).  }
	\label{fig_KSVZ_parameter_space}
\end{figure}

\begin{figure}[t]
	\centering
	\includegraphics[width=0.7\linewidth]{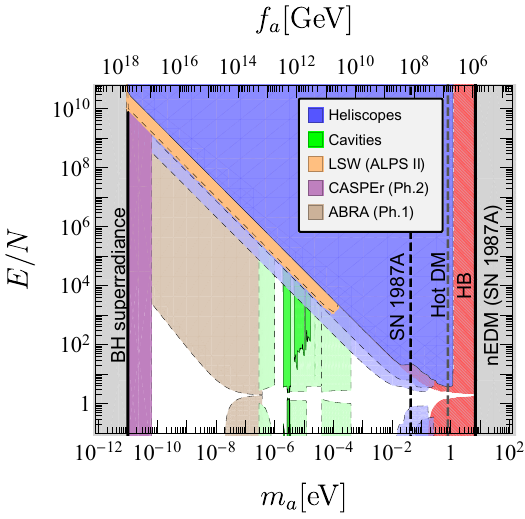}
	\caption{Phenomenological and experimental status for the  hadronic axion. 
	See caption of \fig{fig_KSVZ_parameter_space} for details.   }
	\label{fig_hadronic_parameter_space_grandAngolo}
\end{figure}

\begin{figure}[t]
	\centering
	\includegraphics[width=0.48\linewidth]{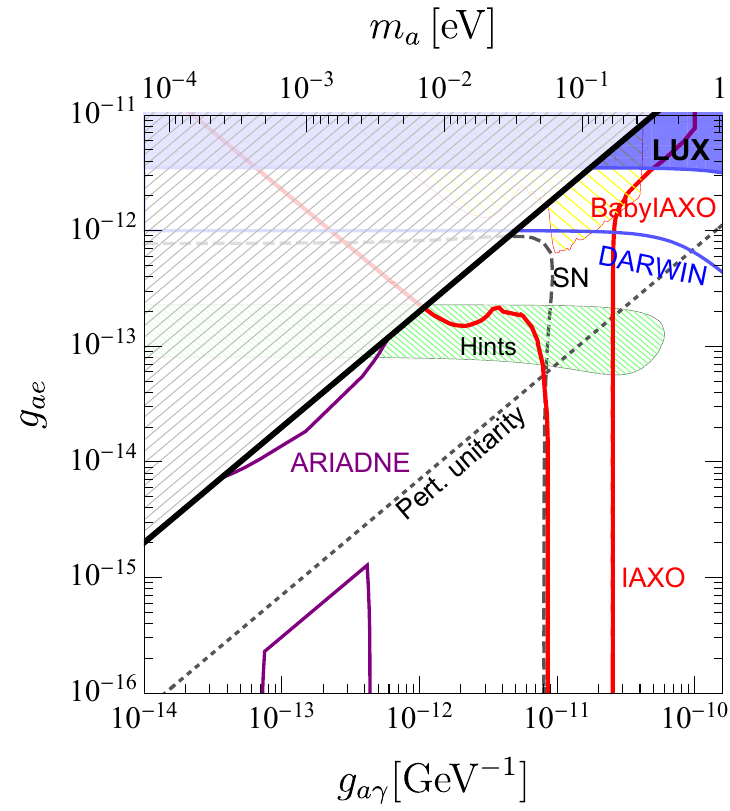}
	\hspace{0.4 cm}
	\includegraphics[width=0.48\linewidth]{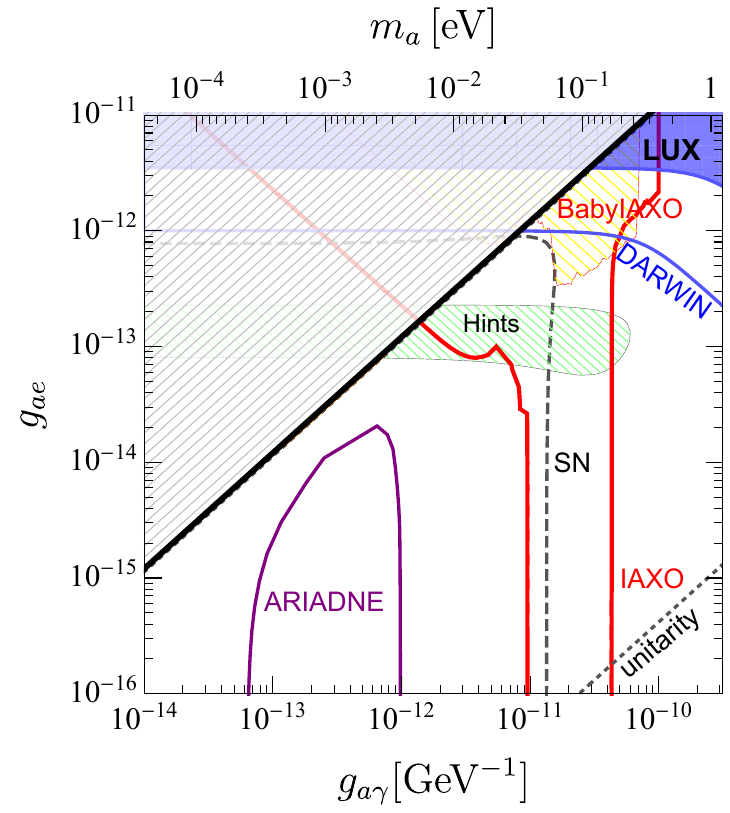}
	\caption{DFSZ axion I (left) and II (right) parameter space.
	The hatched region in the upper left corner is not accessible to such axions models. 
	The green region represents the stellar evolution hints from HB, RGB and WD stars, as discussed in the text.
	The SN 1987A bound is shown with a dashed line to reflect the higher level of uncertainty with respect to other stellar bounds.}
	\label{fig_DFSZ_gae_gag}
\end{figure}

The proposed International Axion Observatory (IAXO), a 4-th generation axion helioscope~\cite{Irastorza:2011gs}, is expected to increase the sensitivity by a factor of $\gtrsim 10^4$, probing the axion-photon coupling down to $g_{a\gamma}\sim $a few $10^{-12}\,{\rm GeV^{-1}}$ at low mass and exploring significant regions of the KSVZ and DFSZ axion models~\cite{Armengaud:2019uso}. 
A scaled down (and significantly  less expensive) version of IAXO, called BabyIAXO, will likely start operations in the mid of the current decade, in DESY.
Though considerably  less powerful than its brother IAXO, BabyIAXO~\cite{Armengaud:2019uso} will still be sensitive to  DFSZ axion models, unique in this respect among the experiments probing the axion mass region above a few meV.
The sensitivity of IAXO and BabyIAXO to hadronic and DFSZ axion models is shown in Figs.\ref{fig_KSVZ_parameter_space}, \ref{fig_hadronic_parameter_space_grandAngolo}, and  \ref{fig_DFSZ_gae_gag}. 

There are other technologies to detect solar axions.
One explored option is to exploit the coherent enhancement of the axion conversion into photons, via Primakoff effect, when the solar axion beam satisfies the Bragg condition of scattering with a crystal plane~\cite{Avignone:1997th}.
However, the sensitivity of these experiments is not competitive with the Sikivie helioscope or with the astrophysical bounds. 

Large Weakly Interacting Massive Particle (WIMP) detectors such as LUX and XENON100 have the capability to detect solar axions through the axio-electric effect and do, in fact, provide bounds on the axion-electron coupling.
Specifically, 
the XENON100 collaboration~\citep{Aprile:2014eoa} reported $g_{ae} < 7.7\times 10^{-12}$ (90 \% CL), LUX~\citep{Akerib:2017uem}, $g_{ae} < 3.5\times 10^{-12}$, and 
PandaX-II~\citep{Fu:2017lfc}, $ g_{ae}<4\times 10^{-12} $.
These searches are particularly interesting since they permit the exploration of the axion-electron coupling in a very wide mass region.
However, probing the parameter space allowed by the cooling of  WDs and RGB may prove extremely challenging even for axio-electric helioscopes of the next generation~\cite{Irastorza:2018dyq}.

\subsection{Helioscopes sensitivity  to $\gag$ and  $\gae$}
\label{sec:app_helioscope_potential}

Here we provide simple, approximate expressions to extract the helioscope sensitivity to general axion models, accounting for the fact that axions can be produced in the Sun through processes induced by their couplings to electrons and to photons. 
By helioscope here we mean any instrument that can detect solar axions. 
In particular, we consider the standard Sikivie helioscope, which uses an external magnetic field to convert axions into photons, and the axionelectric helioscope, which operates through the axion interaction with electrons.

In the Sikivie helioscope, solar axions are converted into photons in a transverse magnetic field $ B $. 
The conversion probability is given in \eqn{eq:Helioscope_agamma_probability}, where we remind that $ q $ is the momentum transfer provided by the magnetic field.
In vacuum, $ q\simeq m_a^2/2\omega $.
Whenever $qL\ll 1$, the probability does not depend on the energy. 

The expected number of events in the Sikivie helioscope is
\begin{align}
N_\gamma-N_b=\dfrac{S\Delta t}{4\pi D_{\odot}^2}\int \frac{dN_a}{dt d\omega} P_{a\gamma}\epsilon\, d\omega
\end{align}
where $ N_b $ is the total background, $D_{\odot} \simeq 1.5\times 10^{11}\,$m is the distance to the Sun, $ S $ is the detector total area, $ \Delta t $ the exposure time, and $ \epsilon $ a parameter that measures the detection efficiency.
In general, the integral should be restricted to some $ \omega $ region.
We assume that these threshold values are accounted for by $ \epsilon $.

Be $ \bar{g}_{a\gamma} $ the bound on the axion-photon coupling in the case of $ g_{ae}=0 $.
In the general case, we find
\begin{align}
g_{a\gamma}^2\int (g_{a\gamma}^2 n_{\gamma}+g_{ae}^2 n_e)\tilde{P}_{a\gamma}\epsilon \, d\omega 
\leq \bar{g}_{a\gamma}^{4}\int n_{\gamma} \tilde{P}_{a\gamma}\epsilon\, d\omega 
\end{align}
where $ g_{a\gamma} $ is given in units of GeV$ ^{-1} $ and $ \tilde{P}_{a\gamma} $ is the oscillation probability divided by $ g_{a\gamma} $.
In the case of small axion mass, the factor $ q L$ in the expression for the probability is small and $ P_{a\gamma} $ does not depend on the axion energy. 
We can assume that the factor $ \epsilon $ is also roughly constant in the energy interval relevant for solar axions, if the experimental cuts in $ \omega $ are performed far from the regions where the two fluxes are large.\footnote{This is not always the case. For example, in~\cite{Barth:2013sma} the analysis is restricted to the energy range between 0.8 and 6.8 keV.}
In this case, the integrals can be performed\footnote{We integrate between $\omega=100 $eV and $\omega =12$ keV. Below 100 eV the emission rate is less clear and unaccounted processes may contribute~\cite{Irastorza:2018dyq}.} and one finds the very simple relation
\begin{align}
\label{eq:Sikivie_helioscopes_gag_gae}
g_{\gamma 10}^{2}\left( g_{\gamma 10}^{2} +0.7 g_{e 12}^{2}\right) \lesssim \bar{g}_{\gamma 10}^{4}\,,
\end{align}
where, we remind, $g_{\gamma 10}=g_{a\gamma}/10^{-10}{\rm GeV}^{-1}$ and $g_{e12}=g_{ae}/10^{-12}$.
At large masses, when coherence is lost, one cannot assume $P_{a\gamma}$ constant anymore. 
Assuming a $\omega^{2}$ dependence for $P_{a\gamma}\epsilon $ we find that the coefficient 0.7 should be replaced with 0.3.
Indeed, in this case Primakoff has a greater weight since it is peaked at higher energy. 
In this case, however, a gas can be used to restore coherence.

We  now turn to the helioscopes based on the axio-electric effect.
For nonrelativistic electrons and ultrarelativistic axions, the cross section for the axio-electric effect 
is proportional to the photoelectric cross section~\cite{Dimopoulos:1986mi,Pospelov:2008jk,Derevianko:2010kz}  
\begin{align}
\sigma_{ae}(\omega)=\dfrac{g_{ae}^{2}}{8\pi \alpha}\left( \dfrac{\omega}{m_e}\right)^{2}\sigma_{ph} (\omega)
\simeq 2.1\times 10^{-29}\,g_{e12}^2\,\omega_{{\rm keV}}^2\,\sigma_{ph} (\omega)\,,
\end{align}
where $\omega _{{\rm keV}}=\omega/ $keV. 
The pronounced peak at low energy ($\omega_a\sim 1\,$keV), characteristic of the photoelectric cross section, which would favour the ABC over the Primakoff flux, is compensated by the $ \omega^2 $ term in the case of the axio-electric cross section. 
Therefore, in general the Primakoff flux is not negligible and should be included in the estimates of the experimental potential. 
For example, setting $ g_{ae}=10^{-12} $ and $ g_{a\gamma} =10^{-10}$GeV$ ^{-1} $, 
we find $ g_{a\gamma}^2\int \sigma_{ae} n_\gamma d\omega \simeq 2 g_{ae}^2\int \sigma_{ae} n_e d\omega  $.
Proceeding similarly to what done in the case of the Sikivie helioscopes, we set $ \bar{g}_{ae} $ to the experimental bound for a purely ABC spectrum and find the very simple relation for the axion couplings:
\begin{align}
g_{ae}^2\int (g_{a\gamma}^2 n_{\gamma}+g_{ae}^2 n_e)\tilde{\sigma}_{ae}\epsilon \, d\omega 
\leq \bar{g}_{ae}^{4}\int n_{\gamma} \tilde{\sigma}_{ae}\epsilon\, d\omega \,,
\end{align}
where $ \tilde{\sigma}_{ae} $ is the axio-electric cross section for $ g_{ae}=1 $. 
If we assume that $ \epsilon $ does not have a strong dependence on $ \omega $, the integrals can be performed analytically and we find:
\begin{align}
\label{eq:axioelectric_helioscopes_gag_gae}
g_{e 12}^{2}\left( 2g_{\gamma 10}^{2} +  g_{e 12}^{2}\right) \lesssim \bar{g}_{e 12}^{4} \,.
\end{align}

\subsection{Haloscopes and DM axions}
 \label{sec:Haloscopes}

The local DM density is estimated to be about $0.45 \,{\rm GeV/cm^{3}}$.
If the axion paradigm is correct and axions do make up the totality of the DM matter, we should expect (locally) about 
$4.5\times 10^{14} ({\rm \mu eV}/m_a){\rm cm}^{-3}$ non-relativistic axions, with energy $\omega_a\sim m_a(1+O(10^{-6}))$,
where the correction to the energy derives from the estimated axion velocity distribution ($v\sim 10^{-3})$.
Several experimental techniques have been developed  to detect such a huge number of nonrelativistic axions. 
Collectively, such experiments are known as \emph{axion haloscopes}.
As discussed in \sect{sec:section2}, the exact axion mass required for the axion to account for the totality of the DM in the Universe is unknown.
Although, historically, there has been a preference for the mass region around  a few $\mu$eV, theoretical uncertainties about the initial conditions (\sect{sec:misalignment}) and the production mechanisms,  particularly the contribution from cosmological defects (\sect{sec:defects}), make the pinning of the exact mass very uncertain.
Moreover, in \sect{sec:extendingma} we present mechanisms that can shift the relevant mass region from $\lesssim$ neV all the way up to 10 meV or so, in the region accessible to the next generation of axion helioscopes (see \sect{sec:Helioscopes}). 
Designing  experiments that can probe this entire mass region is, therefore, extraordinary important to detect axions, if they really are a significant DM component.

The conventional haloscope technique is the resonant cavity haloscope~\cite{Sikivie:1983ip}, 
which employs the Primakoff conversion of DM axions in a strong magnetic field that permeates a resonant microwave cavity.
The conversion is resonant if the axion energy $\omega_a= m_a(1+O(10^{-6}))$ matches a cavity mode.
Cavity experiments are well suited to search for axions in the $\mu$eV mass range, where they have reached extremely high sensitivities.
In particular, the Axion Dark Matter eXperiment (ADMX), which is the most mature axion haloscope, 
has already reached into the KSVZ and DFSZ parameter space (under the assumption that axions are the totality of the DM) for masses $m_a\sim 3\mu$eV~\cite{Du:2018uak,Braine:2019fqb}.
A drawback is the slow mass-scanning time, which scales quadratically with the desired signal to noise level, $\Delta t\propto (S/N)^2$.\footnote{The rate at which a mass range can be scanned is controlled  by the Dicke radiometer equation~\cite{Dicke:1946}:
\begin{equation}
\label{equ:dicke}
S/N = \frac{P_a}{P_N}\sqrt{\Delta\nu t} = \frac{P_a}{T_S}\sqrt{\frac{t}{\Delta \nu}} \,,
\end{equation}  
where $P_N = \Delta\nu T_S$ is the thermal noise power, $\Delta\nu$ is the bandwidth and $T_S$ is the system noise temperature (physical + receiver noise) (cf.~e.g.~\cite{Carosi:2013rla,Irastorza:2018dyq}). 
}
Furthermore, the cavity size has to match with great accuracy the axion Compton wavelength.
Therefore, scanning higher masses requires smaller cavities with the consequent loss 
of sensitivity at fixed scanning time (the power of the signal scales linearly with the volume).\footnote{A recent promising way-out consists in exploring higher order resonant modes~\cite{Kim:2019asb}.}

Nevertheless, an intense program to probe the higher mass range is currently in place.
A series of ultra low temperature cavity experiments at IBS/CAPP 
(CULTASK) 
promise the exploration of  the mass region between ADMX and MADMAX (see below)~\cite{Semertzidis:2019gkj}.
CAPP-8TB is designed to search for axions with mass 6.62 to 7.04 $ \mu $eV, 
with enough sensitivity to detect DFSZ or KSVZ axions~\cite{Lee:2019mfy}.
Operations of the more ambitious CAPP-12TB and CAPP-25T could begin in the early 2020s~\cite{Semertzidis:2019gkj}.
Combined, they are 
expected to explore the axion mass region from 
$\sim$ 3 to $\sim $ 40 $\mu$eV, down to the KSVZ model ($E/N=0$), 
with sensitivity 
 to DFSZ I couplings for masses up to  $\sim 10\,\mu$eV
(Cf. \fig{fig_gag_parameter_space} and Table~\ref{tab:summary_probes}).
%
Meanwhile, the HAYSTACK experiment is probing masses about an order of magnitude larger than ADMX.
It recently reported the first results for a scan in the mass range $23.5-24 \mu$eV, with sensitivity down to $g_{a\gamma}=2\times 10^{14}\,{\rm GeV^{-1}}$~\cite{Brubaker:2016ktl}.
At considerably  higher masses, ORGAN~\cite{McAllister:2017lkb} plans to probe the mass region $\sim 60-210\mu$eV. 

A quite different haloscope concept, MADMAX~\cite{Brun:2019lyf} uses movable booster dielectric discs to enhance  the photon signal and, as shown in \fig{fig_KSVZ_parameter_space} and~\ref{fig_hadronic_parameter_space_grandAngolo}, the projected sensitivity ranges from $\sim 50$ to a few 100~$\mu$eV~\cite{TheMADMAXWorkingGroup:2016hpc}.
A similar range of masses could be probed with tunable axion plasma haloscopes~\cite{Lawson:2019brd}, which employ the axion coupling to plasmons. 
The resonance condition is induced by the matching of the axion mass with the plasma frequency and is, therefore, completely uncorrelated with the size of the experiment. 
Finally, the meV mass range might also become accessible using topological antiferromagnets~\cite{Marsh:2018dlj}.
The current study predicts enough sensitivity to probe, perhaps in its second stage, DFSZ axions in the $1-3$meV mass range. 

Differently from the above experiments, the QUAX (QUaerere AXion)  experiment~\cite{Barbieri:2016vwg} 
aims at detecting axion DM via the axion coupling to electrons.
While the Earth moves through the cold axion halo, 
the coupling to the electron spins would induce spin flips in a magnetised material placed inside a static magnetic field.
This effect can be detected  using Nuclear Magnetic Resonance (NMR) techniques. 
The experiment aims at the mass range $m_a\sim 100\, \mu$eV, similar to the range of MADMAX. 
Recent experimental data~\cite{Crescini:2018qrz} excluded the range of couplings $g_{ae}>4.9\times 10^{-10}$ at 95\% CL, for an axion mass of 58$\,\mu$eV.
This first bound is not yet comparable with the stellar constraints discussed in~\sect{sec:Astro_bounds_gae}, and is still several orders of magnitude larger than what predicted in DFSZ axion models.

An intriguing proposal to test the axion DM paradigm in a wide mass range, $m_a \sim 0.2-40\,\mu$eV, is through the detection of radio signals from the axion conversion into photons in NS magnetospheres~\cite{Hook:2018iia,Edwards:2019tzf,Leroy:2019ghm}.
The NS magnetosphere hosts a very intense magnetic field and a variable plasma frequency. 
If DM axions do exist, they would convert in such field at the radial distance where the plasma frequency matches the axion mass, and produce an observable flux.
The detection potential of current and future radiotelescopes is discussed in~\cite{Hook:2018iia} and expected to reach enough sensitivity to probe the DFSZ axion parameter region.  
Another suggestion for probing cold dark matter axions  with forthcoming radio telescopes 
such as the Square Kilometre Array (SKA)~\cite{Bacon:2018dui} was put forth in~\cite{Caputo:2018ljp}
and further elaborated in~\cite{Caputo:2018vmy}. 
The proposed strategy is that of detecting axion decay into photons at radio frequencies monitoring astrophysical 
targets such as dwarf spheroidal galaxies, the Galactic Centre and halo, and galaxy clusters.
Depending on the environment and on the mass of the axion, a stimulated  enhancement  of the decay rate   
may amplify the photon flux by serval orders of magnitude, bringing the signal  within the reach 
of next-generation radio telescopes. 

Probing lower masses is also quite challenging, requiring larger cavities and magnets.
The KLASH (KLoe magnet for Axion SearcH) experiment aims at the mass region $\simeq 0.3-1\,\mu$eV, just below the ADMX range.
According to the preliminary study, KLASH has the potential to probe axion-photon couplings close to the DFSZ benchmarks  in the given mass range~\cite{Alesini:2019nzq}.
Experiments that aim at exploring even lower masses adopt different techniques 
to avoid the problem, inherent  in all cavity searches, of matching the axion wavelength with extremely large cavity sizes. 
ABRACADABRA (A Broadband/Resonant Approach to Cosmic Axion Detection with an Amplifying B-field Ring Apparatus) 
uses a toroidal magnet and a pickup loop to detect the variable magnetic flux induced by the oscillating current produced by DM axions in the static (lab) magnetic field.
The experiment can operate as a broadband or as a resonant experiment by using an untuned or a tuned magnetometer respectively.
The aim is to probe a wide mass region below $10^{-8}\,$eV, with sensitivity to DFSZ axions for masses in the range $0.1-10\,$neV (cf.~Figs.~\ref{fig_KSVZ_parameter_space} and~\ref{fig_hadronic_parameter_space_grandAngolo}).
In the first data release, masses between $3.1\times 10^{-10}\,$eV and $8.3\times 10^{-9}$ eV were explored, with slightly less sensitivity to the axion-photon coupling than CAST~\cite{Ouellet:2018beu}.

Another ingenuous idea to probe low masses is to exploit the axion coupling to the neutron EDM, \eqn{eq:neutron_EDM}~\cite{Graham:2013gfa}.
The oscillating axion field generates an oscillating nEDM that can be detected using NMR techniques.
The Cosmic Axion Spin Precession Experiment (CASPEr), in its Electric version (CASPEr-Electric), employs a ferroelectric crystal which possess a large, permanent internal electric field with which the axion field interacts.
In phase 2, this experiment is expected to reach the sensitivity necessary to probe QCD axions~\cite{JacksonKimball:2017elr}, as shown in purple in \fig{fig_KSVZ_parameter_space} and~\ref{fig_hadronic_parameter_space_grandAngolo}.

CASPEr Electric is particularly  interesting from the modelling  point of view, since it would effectively probe the axion-gluon coupling which, differently  to the other couplings, is  model independent.
A consequence is that, in the case of a signal induced by a QCD axion, it may be possible to infer the (local) axion DM fraction. 
In fact, the value of $g_d$ can be inferred, within the QCD uncertainties, from the value of the axion mass (identifiable since it sets the oscillation frequency) and the signal strength is a function only of $g_d$ and of the (local) axion abundance. 
This possibility is unique among the axion haloscope experiments. 

\subsection{Searches for axions produced in the laboratory }
\label{sec:lab_searches}

Pure laboratory searches are also emerging as powerful options to search for axions.
Such methods avoid uncertainties related to the use of  natural sources for the axion flux.   
However, the sensitivities of pure laboratory experiments are currently far from 
the benchmark QCD axion regions and aim at testing, 
more generically, the ALPs parameter space. 

One of the most mature experimental  strategies to search for ALPs is the Light Shining Through a Wall (LSW)~\cite{VanBibber:1987rq}.
A powerful photon source, e.g. a laser beam, is used to produce axions, which are then reconverted into photons after crossing a \emph{wall} opaque to light but not to axions. In both cases, the conversion is induced by a strong laboratory magnetic field. 
The Any Light Particle Search (ALPS) experiment has probed the mass region $m_a\lesssim 100\,\mu$eV, yet only for couplings not competitive with CAST.
Its updated version ALPS II (data-taking expected starting from 2021) will surpass CAST and probe unexplored parameter space.
However, the sensitivity is expected to be still far from the DFSZ and KSVZ regions and may be of interest for QCD axions only in the case of very photophilic models (see \fig{fig_hadronic_parameter_space_grandAngolo}).
Currently, the strongest existing limit on the axion-photon coupling using this technique is 
$g_{a\gamma}<3.5\times 10^{-8}\,{\rm GeV^{-1}}$ (95\% CL), for 
$m_a\leq$ 0.3 meV, achieved by the OSQAR 
experiment \cite{Ballou:2015cka}.
Similarly, polarisation experiments use a laser beam in a strong magnetic field to search for axions or other light particles coupled to photons.
The PVLAS experiment has been exploring this possibility for over a decade. 
The most recent reported constraints on the axion-photon coupling are comparable to  the OSQAR bound but extend to slightly higher mass~\cite{DellaValle:2015xxa}.
Some improvement is expected with the Vacuum Magnetic Birefringence experiment (VMB@CERN)~\cite{LOI2018}.

Long range monopole-monopole and dipole-monopole interactions 
(cf.~\sect{sec:CPvaxioncoupl}), induced by CP-odd couplings, provide other possibilities to search for axions~\cite{Moody:1984ba}. 
The CP-odd axion couplings expected in the SM are tiny, as evident from \eqn{eq:thetaeffSM}. 
However, additional contributions are allowed in theories beyond the SM and the current experimental bounds allow for considerably larger values.
Long range forces are severely constrained by precision measurements of Newton’s law and tests of the equivalence principle (see, e.g., \cite{Irastorza:2018dyq} and references therein).
A better strategy for axion detection consists in using NMR techniques to detect the axion field sourced by a macroscopic object.
This program will be carried out by the ARIADNE experiment~\cite{Arvanitaki:2014dfa}.
Interestingly, in the most optimistic scenario (largest allowed CP odd couplings), ARIADNE is expected to have enough sensitivity to probe the $g_{aN}^{S} g_{an}$ combination of couplings down to values expected for the  DFSZ axion~\cite{Arvanitaki:2014dfa,Geraci:2017bmq}.
The forecasted  sensitivity under these assumptions is shown in \fig{fig_DFSZ_gae_gag}.
Standard KSVZ axions are not accessible to ARIADNE, since in that case the coupling to neutrons is vanishingly small. 

Somewhat similarly, QUAX-$g_{p}g_{s}$ probes the $g^{S}_{aN} g_{ae}$ combination.
However, even in the most optimistic case, the expected sensitivity is still far from the coupling region expected in the case of KSVZ or DFSZ axions~\cite{Crescini:2017uxs}.  

%

%
%

\subsection{Summary of experimental constraints} 
\begin{figure}[th]
	\centering
	\includegraphics[width=0.85\linewidth]{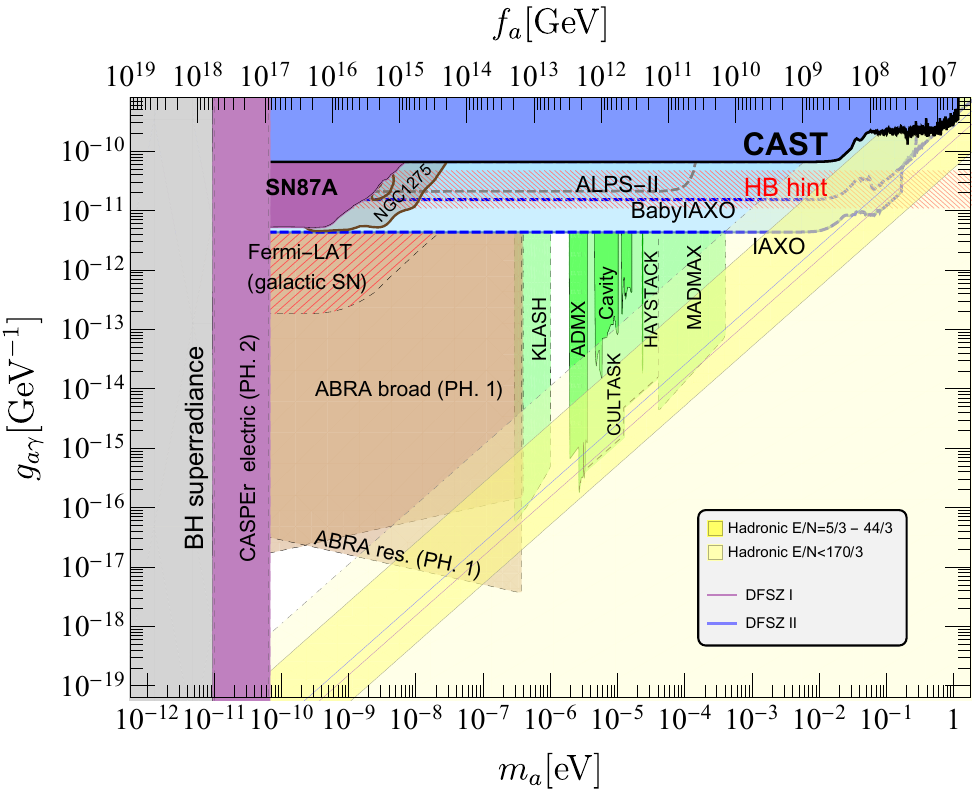}
	\caption{Phenomenological summary of the axion-photon  interactions. 
	We show also the region accessible to CASPEr electric in phase II, when it will be able to probe the model independent axion coupling to gluons. 
	The region presents the most optimistic scenario compatible with the QCD uncertainty in the calculation of the nEDM (Cf.~\eqn{eq:Cangamma}). 
	The region expected for hadronic axions for certain ranges of $E/N$ is shown in yellow. The relevance of these particular ranges for $E/N$ is discussed in \sect{sec:axion_landscape_beyond_benchmarks}. For completeness, we also show the position of the DFSZ I and DFSZ II axions. However, in the case of the helioscopes the figure does not take into account the possible contribution of $g_{ae}$ to the axion production. Refer to \fig{fig_DFSZ_gae_gag} for a more comprehensive analysis of the DFSZ axion models. }
	\label{fig_gag_parameter_space}
\end{figure}
\begin{figure}[t]
	\centering
	\includegraphics[width=0.7\linewidth]{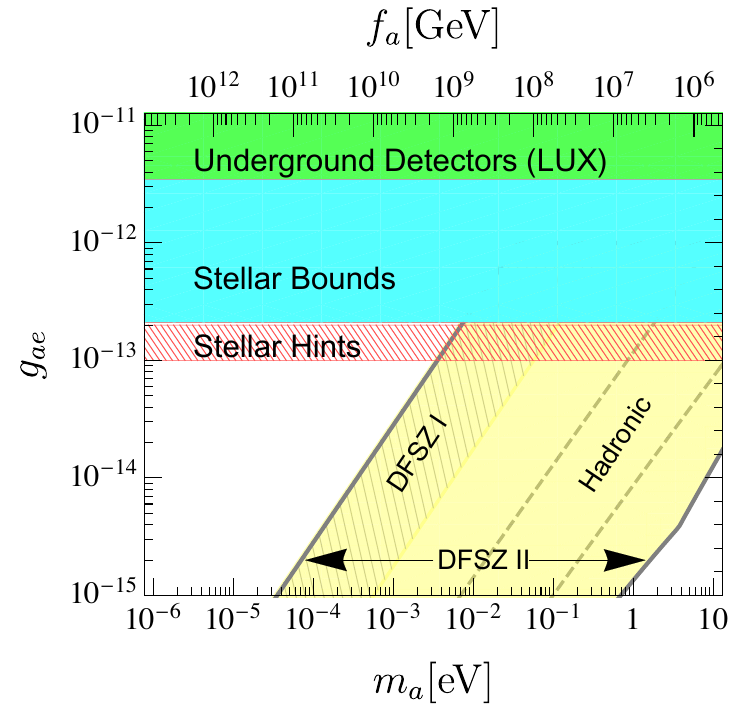}
	\caption{Phenomenological summary of the axion-electron interactions.
	The hadronic region is estimated for $E/N$ between 5/3 and 44/3 (see \sect{sec:axion_landscape_beyond_benchmarks}).
	Notice the changing in the slope of the DFSZ II line at high mass. 
	This happens when tan $\beta$ is such that the tree level coupling of the DFSZ II axion to electrons is subdominant with respect to  the 1-loop contribution. }
	\label{fig_gae_parameter_space}
\end{figure}
\begin{figure}[t]
	\centering
	\includegraphics[width=0.48\linewidth]{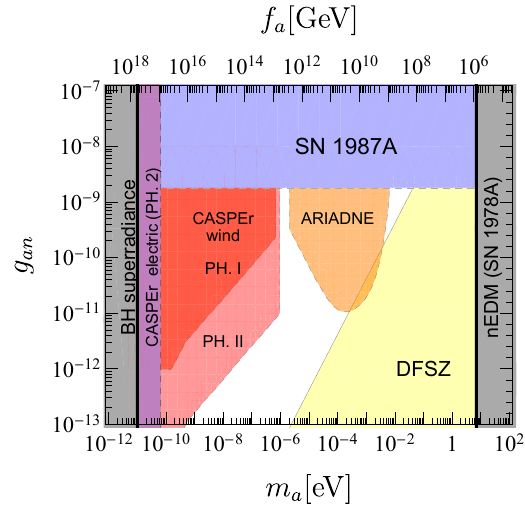}
	\hspace{0.cm}
	\includegraphics[width=0.48\linewidth]{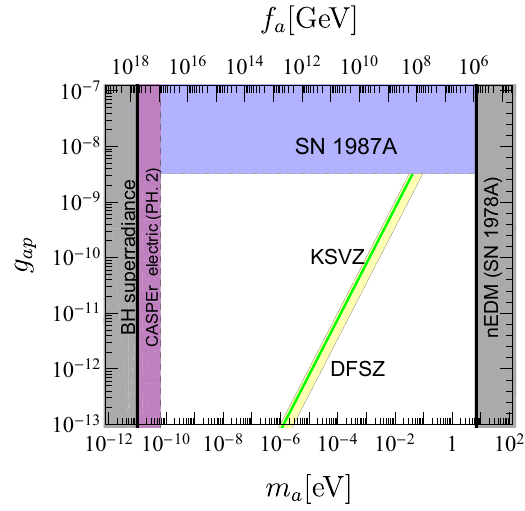}
	\caption{Phenomenological summary of the axion-nucleons interactions. 
	In the case of the ARIADNE experiment, which measures the product $g^S_{aN}g_{an}$, the sensitivity is estimated for the most optimistic case of the scalar coupling, 
	$g^S_{aN}=10^{-12}{\rm GeV}/f_a$ (see Ref.~\cite{Arvanitaki:2014dfa}).
	Notice that the axion-proton coupling remains largely unexplored.	The hadronic axion coupling to neutrons is zero, within the errors~\cite{diCortona:2015ldu}. Hence the absence of a KSVZ axion line in the left panel. }
	\label{fig_gaN_parameter_space}
\end{figure}

In this section we summarise the experimental and astrophysical bounds on the individual axion couplings.
Table~\ref{tab:summary_probes} provides a quick reference to the major probes for each coupling. 
More details can be found in \fig{fig_gag_parameter_space}, for what concerns the axion-photon coupling, 
\fig{fig_gae_parameter_space} for the axion-electron coupling,
and \fig{fig_gaN_parameter_space} for the axion couplings to protons and neutrons.
Notice that, in all cases, we are assuming that the axion solves the strong CP problem. 
Hence, we show the  CASPERr electric potential in all cases since the experiment probes the model independent coupling to QCD gluons. 
However, we are not assuming any specific model and, hence, we are allowing for the couplings to be uncorrelated from the mass. 

In the figures, we are showing the parameter space for DFSZ axions (I and II) and for hadronic axions within a specific range of $E/N$ (cf. \sect{sec:KSVZ-like} and, in particular, \fig{fig:KSVZDFSZbands}).
The reader should refer to \sect{sec:axion_landscape_beyond_benchmarks} for a discussion of axion models beyond these benchmarks.
\begin{table}[!h]
	\begin{center}
\begin{tabular}{|c|l|l|l|}
	\hline
	Coupling & Source & Probes & Notes  \\ \hline
	\multirow{20}{*}{$ g_{a\gamma} $} 
		& \multirow{4}{*}{Astro} 
			& Sun & $ g_{a\gamma}\leq 2.7\times 10^{-10} $GeV$ ^{-1} $ for $ m_a $ up to a few keV\\	
	 	&	& HB-stars &  $ g_{a\gamma}\leq 0.65 \times 10^{-10} $GeV$ ^{-1} $ for	$ m_a $ up to a few 10 keV\\ 
		&	& SN 1987A &  $ g_{a\gamma}\lesssim 6\times 10^{-9} $GeV$ ^{-1} $ for	$ m_a\lesssim $ 100 MeV \\  
		&	&  &  $ g_{a\gamma}\lesssim 5.3\times 10^{-12} $GeV$ ^{-1} $ for	$ m_a\lesssim  4.4\times 10^{-10}$ eV \\  	\cline{2-3}
	 	& \multirow{9}{*}{Cosmo} 
	 		& ADMX & $ m_a\sim 2-3.5 \mu$eV. DFSZ for $ m_a\sim 3 \mu$eV   \\
		&	& HAYSTACK & $ g_{a\gamma}\sim (2-3)\times 10^{-10}$GeV$ ^{-1} $ for $ m_a\sim (23-24) \mu$eV \\	
		&	& MADMAX & DFSZ for $ m_a\sim 0.04-0.4 $meV (\emph{expected})\\	
		&	& CULTASK & DFSZ for $ m_a\sim 3-12 \mu$eV (\emph{expected}, CAPP 12-TB)  \\	
		&	&  		    & KSVZ $(E/N=0)$ for $ m_a\sim 3-40 \mu$eV (\emph{expected}, CAPP 25-T)  \\	
		&	& KLASH & $g_{a\gamma}\sim 3\times 10^{-16}$GeV$^{-1} $ for $ m_a\sim 0.3-1 \mu$eV (\emph{expected, Ph. 3}) \\	
		&	& ABRACADABRA & $ m_a\sim 2.5\times 10^{-15} - 4\times 10^{-7} $eV (\emph{expected}). \\	
		&	&   		& DFSZ for $ m_a\sim 40-400 $neV (\emph{expected}, ABRA res, \emph{Ph. 1}) \\	
		&	& Radio astronomy&  DFSZ for $ m_a\sim 0.2-20 \mu$eV (\emph{expected}) \\	\cline{2-3}
	 	& \multirow{5}{*}{Sun} 
		& CAST & $ g_{a\gamma} =0.66 \times 10^{-10}$GeV$^{-1} $ for $ m_a\lesssim 20 $meV\\
		&	& BabyIAXO & $ g_{a\gamma} =0.15 \times 10^{-10}$GeV$^{-1} $ for $ m_a\lesssim 10 $meV  (\emph{expected}) \\	
		&	&   &  DFSZ for $ m_a\sim 60 - 200 $meV  (\emph{expected}) \\	
		&	& IAXO & $ g_{a\gamma} =4.35 \times 10^{-12}$GeV$^{-1} $ for $ m_a\lesssim 10 $meV  (\emph{expected}) \\	
		&	&   &  DFSZ for $ m_a\gtrsim 8 $meV  (\emph{expected}) \\	\cline{2-3}
	 	& \multirow{2}{*}{Lab} 
		& PVLAS & $ 10^{-7} {\rm GeV^{-1}}\lesssim g_{a\gamma} \lesssim 10^{-6}$GeV$ ^{-1} $ for $ m_a\sim 0.5 - 10$meV \\
		&	& OSQAR & $ g_{a\gamma} \simeq 4\times 10^{-8}$GeV$ ^{-1} $ for $ m_a\lesssim 0.4 $meV  \\		
		&	& ALPS II & $ g_{a\gamma} \simeq 2\times 10^{-11}$GeV$ ^{-1} $ for $ m_a\lesssim 60 \mu$eV (\emph{expected}) \\		\hline
	\multirow{7}{*}{$ g_{ae} $} 
		& \multirow{2}{*}{Astro} 
		& RGB-stars & $g_{ae}\leq 3.1 \times 10^{-13}$ for $ m_a $ up to a few 10 keV  \\	
		&	& WDs &  $g_{ae}\leq 2.1 \times 10^{-13}$ for $ m_a $ up to a few keV \\		\cline{2-3}
		& \multirow{5}{*}{Sun} 
		&	LUX  & $ g_{ae}\leq 3.5 \times 10^{-12} $\\
		&	& XENON100 & $ g_{ae}\leq 7.7 \times 10^{-12} $ \\	
		&	& PandaX-II & $ g_{ae}\leq 4 \times 10^{-12} $ \\	
		&	&LZ  & $ g_{ae}\leq 1.5 \times 10^{-12} $ \\
		&	&DARWIN  & $ g_{ae}\leq 1 \times 10^{-12} $\\ 	\hline
	\multirow{2}{*}{$ g_{a\gamma}g_{ae} $} 
		& \multirow{2}{*}{Sun} 
		&	helioscopes  & Depends on explicit values of $ g_{a\gamma} $ and $ g_{ae} $. \\
		&  &  (CAST, LUX,...)  &  Can be extracted from \eqn{eq:Sikivie_helioscopes_gag_gae} and \eqn{eq:axioelectric_helioscopes_gag_gae}  \\ \hline
	\multirow{4}{*}{$ g_{an} $} 
		& \multirow{1}{*}{Astro} 
		&	SN 1987A, NS  &   $g_{an}\leq 2.8\times 10^{-10}$ (from NS in HESS J1731-347 )	 \\ \cline{2-3}
		& \multirow{4}{*}{Lab} 
		&	ARIADNE  & measures $g^S_{aN}g_{an}$ down to DFSZ for $ m_a\sim 0.25-4 $meV \\
		&   &	 		  &  (\emph{expected for most optimistic choice $g^S_{aN}=10^{-12}{\rm GeV}/f_a$}~\cite{Arvanitaki:2014dfa}).\\
		&  &	CASPEr wind &  From $ m_a \simeq 3.6\times 10^{-12} $eV ($ g_{an}\simeq 1.1 \times 10^{-14} $) and \\ 
		&  &	 &  $ m_a \simeq 9.5\times 10^{-7} $eV ($ g_{an}\simeq 5.1 \times 10^{-12} $) (\emph{expected, Ph. 2}). \\ 
		\hline
	\multirow{1}{*}{$ g_{ap} $} 
		& \multirow{1}{*}{Astro} 
		&	SN 1987A  & $ g_{ap}\lesssim 3.3\times 10^{-9} $ for $ g_{an}=0 $ (hadronic axions).    \\ \hline
	\multirow{2}{*}{$ g_{d} $} 
		& \multirow{1}{*}{Astro} 
		&	SN 1987A  & $ g_{d}\leq 3 \times 10^{-9}\,{\rm GeV}^{-2} $  \\ \cline{2-3}
		& \multirow{1}{*}{Lab} 
		&	CASPEr electric & QCD axion for $ \log \left( \frac{m_a}{{\rm eV}}\right) \simeq -11^{+0.8}_{-0.9}$ (\emph{expected, Ph. 2})\\ \hline
\end{tabular}
		\caption{
Summary of astrophysical and experimental probes on the axion couplings. 
		For proposed experiments, in the notes column we indicate the expected mass range and potential, as extracted from the original literature.
		Whenever necessary, we also indicate the phase (Ph.) during which such results are expected.  	
		The haloscope potential reported assumes that the totality of the DM component in the Universe is made of axions.
		Whenever an experiment can probe DFSZ axions, we have indicated the mass range using the sentence ``DFSZ for $m_a...$".		
		In the case of CASPEr electric, we indicate the mass range in which the experiment may be sensitive to QCD axions. 
		The error indicates the QCD uncertainty.}
\label{tab:summary_probes}
\end{center}
\end{table}

As evident from the figures, the axion-photon coupling is by far the most studied. 
Besides the experiments and bounds discussed in this section, we have added to the figure also the bound from 
the search for spectral irregularities in the gamma ray spectrum of NGC~1275~\cite{TheFermi-LAT:2016zue}, 
from the non-observation of gamma rays from SN~1987A~\cite{Payez:2014xsa} and the Fermi Lat potential in the case of a future galactic SN~\cite{Meyer:2016wrm}. 
Of course, such considerations apply to axions coupled to photons much more strongly than what expected in the benchmark axion models.

The axion-electron coupling is quite more difficult to probe, experimentally.
The most efficient way is through helioscopes based on the axio-electric effect, discussed in \sect{sec:Haloscopes}.
However, such experiments are still quite inefficient in the region below the astrophysical bounds. 
Given the very weak dependence of their potential on volume and observation time, this situation is not likely to change in the near future~\cite{Irastorza:2018dyq}.

The strongest astrophysical  constraints on the axion-nucleon coupling are derived from the cooling of NS and from SN 1987A (cf. \sect{sec:Astro_bounds_gaN}).
Experimentally, the couplings can be probed with NMR techniques.
The experimental potential depends on the spin content of the nuclei adopted in the experiment. 
Our estimates for  \fig{fig_gaN_parameter_space} are based on the single-particle Schmidt model~\cite{Schmidt},
which predicts that only neutrons contribute to the nuclear spin of $^3$He (adopted in ARIADNE) and $^{129}$Xe (adopted in CASPEr wind).\footnote{cf. Table 1 and 4 of Ref.~\cite{Kimball:2014dna}.}
Thus, assuming the Schmidt model, neither  ARIADNE nor CASPEr wind can probe the axion-proton coupling, implying that they would be blind to, e.g., KSVZ axions.
%
%
This result is not completely correct. 
The single-particle Schmidt model does not always reproduce the experimental results well~\cite{Kimball:2014dna}.
In particular, in the case of $^3$He there is evidence of a small proton contribution to the nuclear spin~\cite{Kimball:2014dna}. 
Moreover, recently CASPEr-ZULF-Comagnetometer used a mixture of $^{13}$C and $^{1}$H~\cite{Wu:2019exd}, which allows to access the axion-proton coupling. 
At any rate, the axion-proton coupling is significantly less probed experimentally and the current bounds on $g_{ap}$ are not competitive with 
the astrophysical constraints. 

In this section, we have focussed the attention to a subset of well known laboratory axion experiments, and more in general on 
experimental axion probes that either have already produced data, or  that have passed some crucial step 
towards their realisation, as for example having  published a conceptual design report or having 
obtained funding to develop a  R\&D  phase. In no way we have aimed at a complete review  
(a more comprehensive  account of the experimental panorama can be found in Ref.~\cite{Irastorza:2018dyq,Sikivie:2020zpn}).  
For example,  recent results obtained by using superconducting LC circuit techniques at ADMX~\cite{Crisosto:2019fcj} 
or from the pathfinding run at the Western Australia haloscope ORGAN~\cite{McAllister:2017lkb} have not 
been included. Most importantly, let us stress that the panorama  of different 
proposals, that in many cases put forth  originally new  detection techniques, is much wider than what can  
be guessed  from our brief review.  To give a taste of the number of existing experimental 
projects and ideas, and to highlight  the intellectual dynamism that permeates the  community interested  in axion searches,  
let us mention proposals for new heliospcopes (TASTE)~\cite{Anastassopoulos:2017kag}, 
or for variant  haloscopes   (ORPHEUS) \cite{Rybka:2014cya},   (RADES)~\cite{Melcon:2018dba,Melcon:2020xvj}, also 
including multilayers optical techniques~\cite{Baryakhtar:2018doz}  
or  the possibility of broadband axion searches (BEAST)~\cite{McAllister:2018ndu}, 
new LSW based experiments (STAX/NEXT)~\cite{Capparelli:2015mxa,Spagnolo:2016zjj,Ferretti:2016aut}, 
detection techniques based on laser spectroscopy (AXIOMA)~\cite{Santamaria_2015,Braggio:2016saq},
methods based on precision magnetometry~\cite{Arvanitaki:2014dfa},
and on antiferromagnetically doped topological insulators
(TOORAD )~\cite{Marsh:2018dlj}, or even more exotic ideas, like that of a global network of optical 
magnetometers sensitive to nuclear- and electron-spin couplings, that could reveal  transient  
astrophysical events as  the passage through Earth  of  axion stars (GNOME)~\cite{Pustelny:2013rza}.




\clearpage

\section{The axion landscape beyond benchmarks}
\label{sec:axion_landscape_beyond_benchmarks}

Shortly after the KSVZ~\cite{Kim:1979if,Shifman:1979if} and DFSZ~\cite{Zhitnitsky:1980tq,Dine:1981rt}
invisible axion models had been  constructed,  several variants started to appear in the literature. 
They  were often motivated by the attempt of  making more natural the preferred values of  the axion decay constant 
$10^9\,$GeV $\lesssim f_a\lesssim$  $10^{12}\,$GeV which does not match neither the electroweak  nor the GUT scale, 
by constructing the axion as a composite state of a new non-abelian gauge  
interaction~\cite{Farhi:1980xs,Kim:1984pt,Kaplan:1985dv,Randall:1992ut} that becomes 
strong at the required scale.\footnote{Composite axions will be the subject of \sect{sec:CompositeAxions}.}
At the same time, these constructions  implied different properties for the axion, as for example 
sizeable differences  in the strength  of the axion coupling to the photon with respect to the KSVZ 
and DFSZ models.
Thus, the idea that the \emph{axion window}, i.e.~the region in the $(m_a, g_{a\gamma})$ 
plane predicted by 
QCD axion models, could be sizably larger than what  suggested  by  canonical KSVZ/DFSZ   
constructions, was already contemplated as a likely possibility by some  early phenomenological 
studies~\cite{Kaplan:1985dv,Randall:1992ut,Cheng:1995fd}. 
Recent years have witnessed a proliferation of proposals for axion experiments~\cite{Graham:2015ouw,Irastorza:2018dyq} (see also \sect{sec:Experiments}) and, although at present all ongoing attempts to detect 
the axion  rely on the axion-photon coupling, a certain number of new proposals, often based on 
cutting-edge experimental techniques, are also sensitive to the axion couplings to nucleons and electrons. 
In many cases  the projected sensitivity of new proposals does not reach the level required to test 
the KSVZ and DFSZ models, 
and this is especially true for  first generation experiments which, at least in their initial operational stages,  
mainly represent a proof-of-concept for the effectiveness of new axion search techniques. For these experiments 
it is often understood that the  explorable regions would at best test the possible existence of 
ALPs, that is particles that share some of the properties of the QCD axion, 
but that are not requested to solve the strong CP problem and, as a consequence, can exist 
within a much wider parameter space region.  

The  aim of this section is to carry out a thorough review of past and recent axion models, focussing 
on their predictions for the axion couplings to photons $\gag$ (\sect{sec:gag}), 
to electrons $\gae$ (\sect{sec:gae}), to protons $\gap$ and 
neutrons $\gan$ (\sect{sec:gaN}), and their constraints from astrophysics and direct searches (\sect{sec:Enhanced_couplings_and_astrophysics}).
The highlight of this section is to specify under which conditions the values of these couplings can be 
either enhanced or suppressed with respect to their canonical values, 
and in which cases new features,  like for example 
flavour changing axion couplings, can emerge  (\sect{sec:gaFCNC}). 
To classify the large number of possibilities, it turns 
out to be convenient to maintain a distinction between KSVZ-like and DFSZ-like scenarios. 
 KSVZ-like  refers  to the class of models in which the  SM fermions and the electroweak Higgs fields 
 do not carry PQ charges, the axion-electron coupling vanishes at leading order,  and the 
leading contribution to the axion-nucleon and  axion-pion couplings  only depends on the anomalous QCD term $({a}/{f_a}) G \tilde G$.
 DFSZ-like  instead  broadly refers to  the class of models in which the SM particles are charged  
under the PQ symmetry so that the couplings between the axion and the SM fermions acquire a 
contribution proportional  to the quarks and leptons PQ charges.

In \sect{sec:extendingma} we will  address the issue of the range of  axion masses $m_a$ for which 
the axion can account for the whole cosmological density of DM. 
We will identify  different variants in both the particle physics and cosmological models 
that allow to extend the canonical mass range  towards smaller or larger values of $m_a$. 

Finally, in \sect{sec:heavy}, we will consider mechanisms that allow to modify drastically the
$m_a$-$f_a$ relation while maintaining  the solution of the
strong CP problem, yielding $m_a \gtrsim 100$ keV axions which, due to their large mass, 
 can  avoid most astrophysical constraints.  \\

\subsection{Enhancing/suppressing  $g_{a\gamma}$}
\label{sec:gag}

In \Eqn{eq:gagamma2} we have given the general expression for the axion coupling to photons.
Including NLO corrections, which were computed in Ref.~\cite{diCortona:2015ldu}, and 
using $m_u/m_d \simeq 0.48(3)$,  one obtains the expression given in \Eqn{eq:Cagamma}
that we repeat  below for convenience:
\beq 
\label{eq:gag}
g_{a\gamma} = \frac{\alpha}{2\pi f_a} 
\[ \frac{E}{N} - 1.92(4) \] .
\eeq
We recall that  $E$ and $N$ are respectively the coefficients of the PQ electromagnetic and colour 
anomalies whose    general expression is given in \eqn{eq:EN}. 
%
%
The original KSVZ model discussed in \sect{sec:KSVZ} adopted the simplest choice in which the 
QCD anomaly is induced by heavy exotic coloured fermions $\Q$, singlets 
under $SU(2)_L\times U(1)_Y$, so that  $E/N=0$ and $\gag$ is then determined solely by the 
model-independent numerical factor ``$-1.92$''  originating from the $a/f_a G\tilde G$ term. 
This, however, implies that  $\Q$-baryon number, associated with the non-anomalous $U(1)$ symmetry $\Q\to e^{i\beta}\Q$,
is exactly conserved (being protected by the gauge symmetry).  Hence, after confinement, the $\Q$'s would 
give rise to stable fractionally charged hadrons which have not been observed~\cite{DiLuzio:2016sbl}. 
Thus, charged 
$\Q$'s that could mix with the light quarks and decay, and that would give rise to $E/N\neq 0$, 
should be considered phenomenologically preferred.  
 For the DFSZ models discussed in Section~\ref{sec:DFSZ}  
the SM quarks and leptons charges imply either $E/N= 8/3$ or $2/3$, depending on whether the 
leptons are coupled to the $d$-type (DFSZ-I) or to the $u$-type (DFSZ-II) Higgs. 
From \eqn {eq:gag} we see that 
large  enhancements of  the axion-photon coupling require $E/N \gg 2$, while the axion would 
approximately decouple from the photon if $E/N \approx 2$. Below we 
discuss several cases in which these possibilities can be realised.

\subsubsection{KSVZ-like scenarios}
\label{sec:KSVZ-like}

The model dependence of the axion-photon coupling and the possibilities 
to arrange for large enhancements or for approximate decoupling, was first 
discussed by Kaplan in Ref.~\cite{Kaplan:1985dv}, where a variant of Kim's  composite  
axion model~\cite{Kim:1984pt} (see also \cite{Choi:1985cb}) was put forth. 
Since composite axions will be the subject of \sect{sec:CompositeAxions},  here we only provide 
 the essential ingredients to explain the main features of Kaplan's  model. 
A gauge group factor $SU(\mN)$ 
of metacolour (or axicolour) that becomes strongly interacting at a large scale $\Lambda \sim f_a$ is added 
to the SM gauge group $SU(3)_c\times SU(2)_L\times U(1)_Y$, together with  two multiplets of 
exotic Dirac  fermions $\psi$ and $\xi$ that are $SU(2)_L$ singlets and transform 
under  $SU(\mN)\times SU(3)_c\times U(1)_Y$ as 
\begin{equation}
\label{eq:kaplanfermions}
\psi \sim (\mN,3)_{y_1} , \qquad 
\xi \sim (\mN,1)_{y_2} \,,
\end{equation}
where $y_1,y_2$ denote hypercharges.
The exotic quarks have no mass term so that there is a $U(1)^4$ global symmetry for the kinetic term
corresponding  to $U(1)_{B_\psi}\times U(1)_{B_\xi}\times U(1)_{\tilde A} \times U(1)_A$. 
The first two factors are  vector-like and ensure  $\psi$ and $\xi$ baryon number conservation. 
The remaining two axial symmetries get spontaneously broken by the condensates 
$\langle \bar \psi_L \psi_R\rangle=\langle\bar\xi_L \xi_R\rangle\neq 0$
resulting in two NGBs that acquire masses due to the anomaly 
of the corresponding currents with the strongly interacting gauge groups. 
The current $ \tilde J_\mu^5  = \frac{1}{2}  \(\bar \psi \gamma_\mu\gamma_5 \psi +
 \bar\xi \gamma_\mu\gamma_5 \xi\)$ has an anomaly with  $SU(\mN)$  hence the corresponding 
NGB acquires a large mass of order $f_a$. 
In contrast,  the current $J_\mu^5 =\frac{1}{\sqrt{24} }\( \bar \psi \gamma_\mu\gamma_5 \psi -
3 \bar\xi \gamma_\mu\gamma_5 \xi\)$  has only  colour and  electromagnetic anomalies:
\begin{equation}
\label{eq:kaplananomaly}
\partial^\mu J_\mu^5 =\frac{g^2_s}{16\pi^2}\frac{\mN}{2 \sqrt{24} } G^a_{\mu\nu}\tilde G^{a\mu\nu}
+ \frac{e^2}{16\pi^2}\frac{3 \mN (y^2_2-y^2_1)}{\sqrt{24} } F_{\mu\nu}\tilde F^{\mu\nu}\,,
\end{equation}
so that the corresponding NGB, which is identified with the axion, acquires a tiny mass $m_a \sim m_\pi f_\pi (N/f_a)$.
The axion coupling to photons  involves the anomaly coefficient ratio $E/N = 6 (y_1^2-y_2^2)$ 
and is thus proportional to $g_{a\gamma} \propto 6 (y_1^2-y_2^2) -1.92$.    
%
As remarked in Ref.~\cite{Kaplan:1985dv}, by choosing $(y_1,y_2)=(0,1)$ one obtains 
an enhancement in $\gag$ of a factor of ten with respect to the DFSZ model with $E/N=8/3$, 
while  $(y_1,y_2)=(2/3,1/3)$ give $ E/N-1.92=0.08$  which amounts to  a suppression of a factor of ten. 
Although this composite axion model clearly relies on a different construction with respect to 
KSVZ models with fundamental scalars, it can still be classified as being of the KSVZ-type 
since the SM states do not carry PQ charges and at the leading order $\gae = 0$. 
Models in which the axion does not couple directly to the leptons are also known 
as \emph{hadronic} axion models.
Note that similar constructions for composite axions  always require at least two sets of 
massless exotic fermions   in order to 
be able to generate a light axion that can solve the QCD $\theta$ problem, since 
a single set would give rise to just one heavy axion for which the dominant potential term would 
drive $\theta_\mN \to 0$ leaving $\theta$ at a generic $O(1)$ value (unless an accurate alignment 
between the two angles is arranged for, see \sect{sec:heavy}). 
Taken as a mechanism  to enhance or suppress $\gag$,
composite axion models are not particularly economical since, as we will see,  
KSVZ models involving two sets of exotic fermions allow for much larger enhancements 
of $\gag$  as well as for more accurate  cancellations between the two contributions to the coupling.  
%
Clearly in the first KSVZ construction the choice of a  $SU(2)_L\times U(1)_Y$ singlet representation for 
the heavy vector-like quarks  $\Q$  represented the simplest possibility. Colour representations with 
nontrivial  electroweak quantum numbers, as in the Kaplan model, would work as well, 
with the difference that 
at least some of the heavy quarks  would be electrically charged yielding $E/N\neq 0$.   
While in principle any electroweak representation is equally good for implementing the PQ solution, not all are 
phenomenologically allowed, 
and only a few are in fact phenomenologically preferred. 
A systematic  attempt  of classifying the viable  representations 
$R_\Q = (\cal{C}_\Q,\cal{I}_\Q,\cal{Y}_\Q)$  under the $SU(3)_c\times SU(2)_L\times U(1)_Y$  gauge group
was carried out in Refs.~\cite{DiLuzio:2016sbl,DiLuzio:2017pfr}.    
Four possible phenomenological criteria to classify the representations as 
{\it phenomenologically preferred} were identified:
 \begin{itemize}  \itemsep -2pt 
 \item[$(i)$]  $R_\Q$  should not induce Landau poles (LP)  below a scale $\Lambda_\mathrm{LP}$ of the order of the Planck scale; 
  \item[$(ii)$]  $R_\Q$  should not  give rise to  cosmologically dangerous strongly interacting relics; 
\item[$(iii)$]  $R_\Q$'s yielding  domain wall number  $\NDW=1$ are preferred by cosmology;
\item[$(iv)$]  Improving gauge coupling unification can be an added value. 
\end{itemize} 
These four criteria have rather different discriminating power. 
Gauge coupling unification $(iv)$  is a desirable feature for any
particle physics model. However, improved   
 unification for some $R_\Q$ might simply occur as an accident because of the many
different representations that one can consider, as well as from  the freedom 
in choosing  the relevant mass scale $m_\Q$ between a few TeV (from exotic quark searches at the LHC) 
and $ \Lambda_{\rm GUT}$.
Besides this, from the theoretical point of view envisaging a GUT completion of KSVZ axion
models in which only a fragment $R_\Q$ of a complete GUT multiplet receives a mass
$m_\Q \lsim v_a\ll \Lambda_{\rm GUT}$, as is necessary to assist gauge coupling unification, 
while all the other fragments
acquire GUT-scale masses, is not straightforward, and it appears
especially challenging in all the cases in which the PQ symmetry commutes
with the GUT gauge group (see  \sect{sec:axionGUTs} and in particular footnote \ref{foot:unificaxion}). 
Difficulties in constructing explicit realisations  of $R_\Q$ assisted 
unification  thus  suggest that discriminating the $R_\Q$ on the basis of  the last property $(iv)$ 
might be  less meaningful than what one would initially guess. 
Axion models with $\NDW>1$ face the so called  cosmological DW problem~\cite{Sikivie:1982qv}
(see \sect{sec:CosmoDW}) 
while models with   $\NDW=1$  are cosmologically safe~\cite{Vilenkin:1982ks,Barr:1986hs},
hence motivating the criterion $(iii)$ above. 
However, one can envisage various solutions 
of the DW problem  
based either on cosmology or model-building 
(see \sect{sec:AxionsDW}).
Due to these considerations, in Refs.~\cite{DiLuzio:2016sbl,DiLuzio:2017pfr} 
criteria $(iii)$ ($\NDW=1$) and $(iv)$ 
(improved gauge coupling unification) 
were only considered as desirable features of a KSVZ axion model, 
but not sufficiently discriminating to be used as criteria for 
identifying possibly pathological $R_\Q$.
The first two conditions are instead quite selective for discriminating among 
different representations, as it will be reviewed in the following. \\

$(i)$ {\it Landau poles}.  Large representations can often induce LP in the hypercharge, weak, or
strong gauge couplings $g_1,\, g_2,\,g_3$ at some uncomfortably
low-energy scale $\Lambda_{\rm LP} <  \mP$.  
In  Refs.~\cite{DiLuzio:2016sbl,DiLuzio:2017pfr}, rather than  $\Lambda_{\rm LP} \sim \mP$ a more conservative choice 
 $\Lambda_{\rm LP} \sim 10^{18}\,${\rm GeV}  was made. This was justified by the fact that 
at energy scales approaching
$\mP$, gravitational corrections to the running of the gauge couplings
can become relevant, and explicit computations show that they go in
the direction of delaying the emergence of LP~\cite{Robinson:2005fj}. 
Therefore  a value of $\Lambda_{\rm LP}$ for which gravitational corrections 
are presumably negligible was chosen. 
The scale at which the LP arises was evaluated by using two-loops  
gauge beta functions\footnote{Accidental cancellations in the 
one-loop gauge beta function for higher-dimensional multiplets 
can lead indeed to erroneous estimates of the LP (see e.g.~\cite{DiLuzio:2015oha}).} 
and by  setting the threshold for including  the $R_\Q$ 
representations  in the running of the gauge couplings at $m_\Q = 5 \cdot 10^{11}$ GeV.
By  recalling that  $m_\Q = y_\Q  \NDW f_a/\sqrt{2}$ (see \sect{sec:KSVZ}), 
this value can be justified
in post-inflationary scenarios in terms of the 
cosmological limit on $f_a$  that follows from the requirement 
$\Omega_a \lesssim \Omega_{\rm DM}$, and in pre-inflationary scenarios 
by  demanding additionally that  the initial value of  the axion field  is not tuned 
to be much smaller than $f_a$.

To see what is the maximum value of the $\gag$ coupling allowed by the LP condition, 
one can start by considering a single representation
$R_\Q=  (\mathcal{C}_\Q,\mathcal{I}_\Q,\mathcal{Y}_\Q)$ and look for the maximum allowed values 
of the $E_\Q/N_\Q$ coupling factor. 
From \eqn{eq:NKSVZ}  this factor  can be written as
\begin{equation} 
\label{eq:EQNQ}
\frac{E}{N} = 
\frac{E_\Q}{N_\Q} = \frac{d({\cal{C}}_{\Q})}{T(\mathcal{C}_Q)} 
\left[\frac{1}{12} \(d({\cal{I}}_{\Q})^2 -1\) + {\cal{Y}}_{\Q}^2 \right] \,. 
\end{equation}
The minimum value of the denominator is $T(\mathcal{C}_Q)=\frac{1}{2}$ which corresponds 
to colour triplets. 
For a hyperchargless representation the LP condition in $g_2$ is saturated with a fermion quadruplet 
$ R_\Q(\mathcal{I}_\Q^{\rm max}) = (3,4,0)$ 
while for an $SU(2)_L$ singlet  the LP condition in $g_1$ is saturated by 
$R_\Q(\mathcal{Y}_\Q^{\rm max}) = (3,1,5/2)$, where 
the rational value $5/2$ is a simple approximation to the real valued result.
The maximum allowed hypercharge value $\mathcal{Y}_\Q = \frac{5}{2}$ in turn  
implies the condition $\mathcal{Y}_\Q \sqrt{d({\cal{I}}_{\Q})} \leq \frac{5}{2}$.  $E/N$ 
is then maximal for the value of $d({\cal{I}}_{\Q})$ that, within the allowed range $1\leq d({\cal{I}}_{\Q})\leq 4$, maximises the expression 
in square brackets in  \eqn{eq:EQNQ}  subject to this condition, that is, it maximises the function $d({\cal{I}}_{\Q})^2 -1 +75/d({\cal{I}}_{\Q})$.
The maximum is found for~\cite{DiLuzio:2017pfr} 
\begin{equation}
\label{eq:LPmaxEQNQ}
  R_\Q(\mathcal{Y}_\Q^{\rm max}) =(3,1,{5}/{2}) \quad \Rightarrow  \quad {E}/{N} = {75}/{2}. 
 \end{equation}
Larger values can be obtained by adding  the two representations   $R_\Q(\mathcal{Y}_\Q^{\rm max}) \oplus 
 R_\Q(\mathcal{I}_\Q^{\rm max}) $ to maximise the numerator in \eqn{eq:EQNQ} 
 (this saturates the LP limits for $g_1$ and $g_2$)  while the minimal possible value of the denominator 
 $\sum_\Q \mX_\Q  d({\cal{I}}_{\Q})  T(\mathcal{C}_\Q) = \pm 1/2$ can be retained by adding a 
 $SU(2)_L\times U(1)_Y$ singlet  in the adjoint of colour 
(with index $T(8) = 3$) and  with opposite sign of the PQ charge:  $(1+4)\times T(3)-T(8) = -1/2$. 
In this way one obtains the maximum value of $E/N$ compatible with the LP condition: 
\begin{equation}
\label{eq:LPmaxEN}
(3,1,{5}/{2}) \oplus (3,4,0) \ominus (8,1,0) \quad \Rightarrow  \quad {E}/{N} = - {135}/{2}. 
 \end{equation}
%
 (Here and below the symbol $\ominus$ will refer to an irreducible  fragment of a reducible representation 
  that has opposite sign of the PQ charge with respect to the other fragments, that is it couples  
 to the Hermitian conjugate of the PQ field  as  $y_\Q \bar \Q_L \Q_R \Phi^\dagger$ rather than as in   \Eqn{eq:LaKSVZ1}.) 
 Equivalent  possibilities can be obtained  with different representations $ (3,1,\mathcal {Y}_\Q) \oplus (3,4,\mathcal {Y}_{\Q'})$ 
  satisfying $\mathcal{Y}_\Q^2 + 4 \mathcal {Y}_{\Q'}^2= (5/2)^2$,  as for example  
  $(3,1,{3}/{2}) \oplus (3,4,1)$, etc. \\
%
 
$(ii)$ {\it Cosmologically dangerous relics}. 
The transformation properties of $R_\Q$ under the SM gauge group constrain 
the type of couplings of the $\Q$'s with SM particles, and in several cases this might result in 
stable or long-lived anomalously heavy `hadrons'. 
In post-inflationary scenarios this can represent a cosmological issue. In pre-inflationary 
scenarios instead the $\Q$'s would be diluted away by the exponential expansion 
and  would be harmless. 
Therefore, condition $(ii)$ restricts the viable $R_\Q$ only under the assumption 
that  $U(1)_{PQ}$ is broken after inflation and that $m_\Q < T_{\rm RH}$. 
In this case  the $\Q$'s will attain, via their gauge
interactions, a thermal distribution. 
This provides well defined 
initial conditions for their cosmological history, which then in principle 
depends only on their mass $m_\Q$ and representation $R_\Q$.

For some $R_\Q$, as for example the $SU(2)_L\times U(1)_Y$ singlets of the original 
KSVZ model~\cite{Kim:1979if,Shifman:1979if} as well as for the three representations in \eqn{eq:LPmaxEN}, 
$\Q$ decays into SM particles are forbidden.  Moreover 
the heavy quarks can only hadronise into fractionally charged hadrons 
 (see the Appendix in Ref.~\cite{DiLuzio:2017pfr}). These $\Q$-hadrons must then exist 
today as stable relics. Searches for fractionally charged particles in ordinary matter 
limit their abundance with respect to ordinary nucleons to
$n_\Q/n_b \lsim 10^{-20}$~\cite{Perl:2009zz} which  is orders of
magnitude below any reasonable estimate of the  $\Q$'s  cosmological 
relic abundance and of their expected concentration in bulk matter. 
The set of viable $R_\Q$'s is then restricted to the much smaller subset
that yields  integrally charged  or neutral colour singlet $\Q$-hadrons, in which case
the limits on $n_\Q/n_b$ are less tight. 
 
Whether this type of $\Q$-hadrons 
can be excluded crucially depends on  carrying out reliable estimates of 
their cosmological abundance  $\Omega_\Q$, which is, however,    
a non-trivial task. In these estimates one generally assumes a symmetric scenario $n_\Q=n_{\bar \Q}$ since 
any asymmetry would eventually quench $\Q\bar \Q$ annihilation resulting in stronger bounds.
At temperatures above the QCD phase transition the $\Q$'s annihilate as free quarks, and
the annihilation cross section can be computed perturbatively for any representation 
with reliable results. 
In this regime,  for all masses above a few TeV  one obtains  
$\Omega_\Q\gg  \Omega_{\rm DM} $, which would firmly exclude stable $\Q$'s if it were not 
for the fact that  after confinement,  $\Q$-hadrons  can restart annihilating.   
Obtaining reliable estimates of $\Omega_\Q$ in this non-perturbative regime is challenging.
A  large cross section typical of inclusive hadronic scattering
$\sigma_{\rm ann} \sim (m_\pi^2 v)^{-1}$ 
was assumed in Ref.~\cite{Dover:1979sn}, however, 
it was  remarked in Ref.~\cite{Nardi:1990ku} that the relevant process is exclusive containing  
no $\Q$'s in the final state,  resulting in a  cross section a few orders of magnitude smaller.  
Ref.~\cite{Arvanitaki:2005fa}
suggested that annihilation could be catalysed by the formation 
of quarkonia-like  bound states in the collision of  a $\Q$- and a $\bar \Q$-hadron.
Refs.~\cite{Kang:2006yd,Jacoby:2007nw}  reconsidered 
this mechanism arguing that  $\Omega_\Q$ could indeed be efficiently reduced to the level 
that energy density considerations would not be able to exclude stable relics with
$m_\Q\lesssim 5  \cdot 10^3$ TeV.
 Ref.~\cite{Kusakabe:2011hk} studied this mechanism more
quantitatively and remarked that  
the possible formation of $\Q\Q...$ bound states, besides 
$\Q\bar \Q$, would hinder annihilation rather than catalyse it.
Even if consensus on how to estimate reliably $\Omega_\Q$ has not been reached yet, 
and  $\Omega_\Q <\Omega_{\rm DM}$ remains an open possibility, 
all estimates suggest that   present 
concentrations would still be rather large, at least $10^{-8} \lsim n_\Q/n_b \lsim 10^{-6} $.
Although it can be questioned if  similar concentrations have to  be  expected also 
in  the Galactic disc~\cite{Dimopoulos:1989hk,Chuzhoy:2008zy}, 
  searches for anomalously heavy isotopes in terrestrial, lunar, and meteoritic materials, yield
limits on $n_\Q/n_b$ many orders of magnitude below these estimates~\cite{Burdin:2014xma}. 
Moreover, even  tiny amounts of heavy $\Q$'s in the interior of celestial bodies (stars, neutron stars, Earth) 
would produce all sorts of effects like instabilities~\cite{Hertzberg:2016jie}, collapses~\cite{Gould:1989gw}, 
anomalously large heat flows~\cite{Mack:2007xj}. Therefore,  unless  an extremely efficient mechanism exists 
that keeps $\Q$-hadrons completely separated from ordinary matter, cosmologically stable 
heavy relics of this type are excluded. This implies that if exotic $\Q$ exist, they must decay.

As  shown in  Ref.~\cite{DiLuzio:2017pfr},   for  each  $R_\Q$
that  allows for  integrally charged  or neutral colour singlet $\Q$-hadrons,  
it is always possible to construct gauge invariant operators of some dimension 
that can mediate their decay into SM particles.  
The issue is which range of lifetimes $\tau_\Q$ are cosmologically safe, and   
this can then be translated into an upper bound on the 
dimension of the relevant decay operators. The larger is the dimension, 
the longer is the lifetime, the more dangerous is the relic.  
Cosmological observations severely constrain  the allowed values of $\tau_\Q$.
For $\tau_\Q \sim (10^{-2} \div 10^{12})\,$s the decays of
superheavy quarks with $m_\Q\gg 1\,$TeV would affect
BBN~\cite{Kawasaki:2004qu,Jedamzik:2006xz,Jedamzik:2007qk,Kawasaki:2017bqm}.
Early energy release from  decays with lifetimes
$\sim (10^{6} \div 10^{12})\,$s is strongly constrained also by limits on
CMB spectral distortions~\cite{Hu:1993gc,Chluba:2011hw,Chluba:2013wsa}. $\Q$'s
decaying around the recombination era ($t_{\rm rec} \sim 10^{13}\,$s)
are tightly constrained by measurements of CMB anisotropies.  Decays
after recombination would give rise to free-streaming photons visible
in the diffuse gamma ray background \cite{Kribs:1996ac},  and 
Fermi LAT \cite{Ackermann:2012qk}  excludes $\tau_\Q\sim (10^{13}\div 10^{26})\,$s. 
Note that these last constraints
are also able to exclude lifetimes that are several order of magnitude
larger than the age of the Universe, $t_U\sim 4\cdot 10^{17}\,$s.
 The conclusion is that for the post-inflationary case  cosmologically stable heavy $\Q$'s with 
$m_\Q < T_{\rm{RH}}$  are strongly disfavoured and likely ruled out, while for   
unstable $\Q$'s  the  lifetime must satisfy  $\tau_\Q\lsim 10^{-2}\,$s.

\begin{figure}[t]
\begin{center}
\includegraphics[width=.65\textwidth]{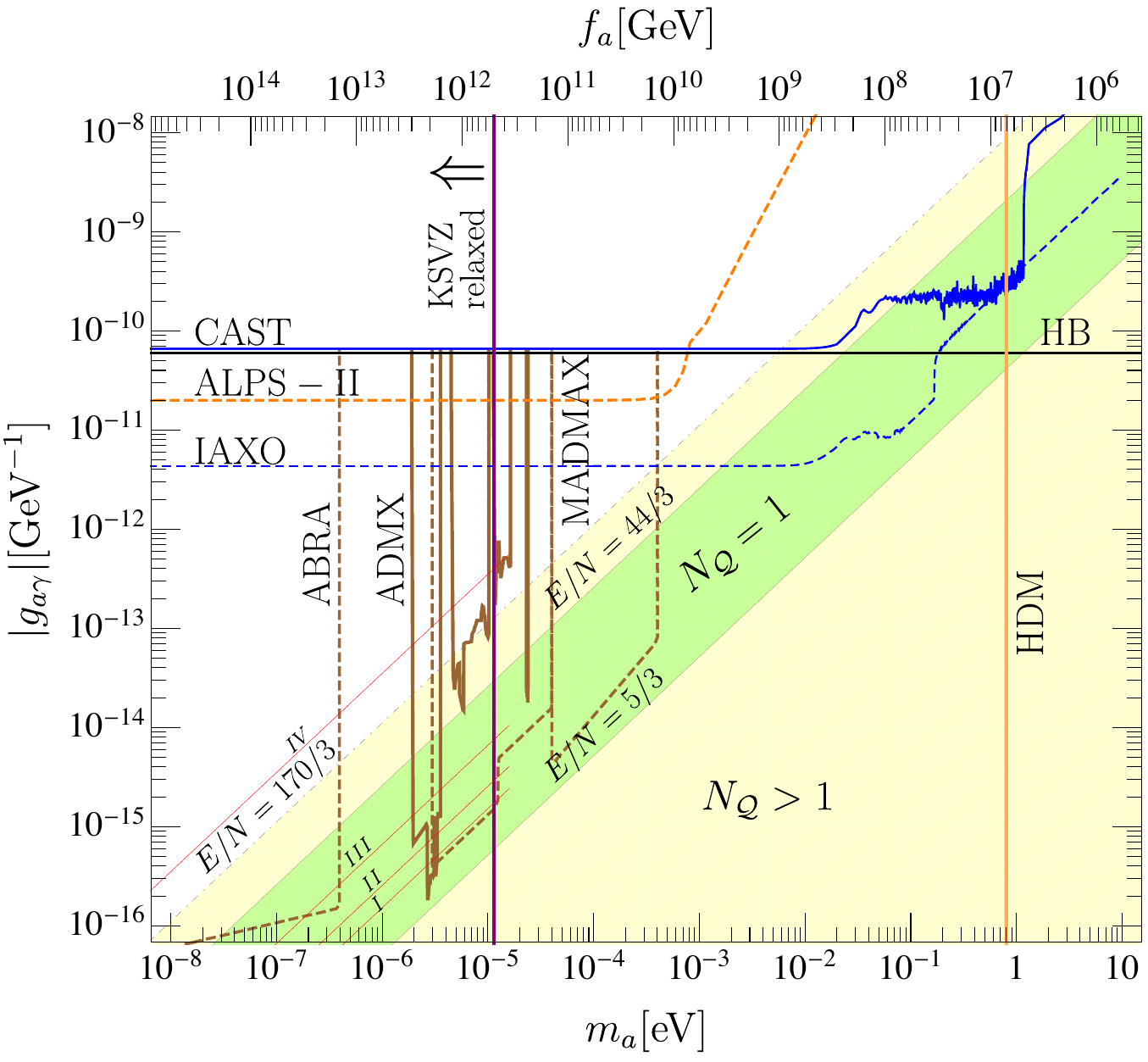}
\caption{\label{fig:KSVZDFSZbands}
The $g_{a\gamma}$-$m_a$ window for preferred 
axion models.  The two lines labelled $E/N=44/3$ and $5/3$  
encompass KSVZ models with a single $R_\Q$, while  
the region below $E/N=170/3$  allows for   
more  $R_\Q$'s satisfying both the conditions on the absence of LP and of stable relics.  
The red lines labelled from I to IV
(only partially drawn not to clutter the figure)  
 indicate where  the DFSZ-type of models lie (see \sect{sec:DFSZ-like}).
Current exclusion regions and expected
experimental sensitivities are 
delimited by  solid  and dashed lines  respectively.
To improve  readability, only some experiment names are indicated. 
More details about the experimental panorama can be found in  \fig{fig_gag_parameter_space}. 
On the left side of the vertical violet line, 
corresponding to a pre-inflationary PQ breaking scenario, 
the upper limit on $E/N$ for KSVZ models can get
relaxed (see text).}
\end{center}
\end{figure}

All $R_\Q$'s which allow for decays via renormalizable operators  
easily satisfy the above requirements,  while for higher-dimensional 
operators suppressed by $\mP$  and for   
 $m_\Q\gsim 800\,$TeV  operators of dimension not larger than  $d=5$  are needed~\cite{DiLuzio:2016sbl}.   
For $d=6$, even for the largest values compatible with post-inflationary scenarios 
$m_\Q \sim f^{\text{max}}_a\sim 10^{12}\,$GeV  decays occur dangerously close to
BBN.  Operators of $d=7$ and higher are always excluded.
Table~\ref{ViableRQ}  (adopted from Ref.~\cite{DiLuzio:2016sbl}) collects 
the representations that satisfy both  the LP and the cosmological constraints. 
The largest value $E/N=44/3$ is obtained for $R_8$, while the weakest coupling is obtained for 
$R_3$ giving $E/N-1.92 \sim - 0.25$. 
These two values  correspond to the two lines encompassing the green band in \fig{fig:KSVZDFSZbands}. 
Only the first two representations $R_1$ and $R_2$ have 
 $\NDW=1$, while  
only $R_3 = (3,2,1/6)$ is able to  improve considerably unification with
respect to the SM~\cite{Giudice:2012zp,DiLuzio:2016sbl}.

With multiple representations sizably larger values  of $E/N$ can be obtained.
Saturating the LP condition using only cosmologically safe representations yields~\cite{DiLuzio:2016sbl}
\begin{equation}
\label{eq:COSMOmaxEN}
(3,3,-{4}/{3}) \oplus (3,3,-1/3) \ominus (\bar 6,1,-1/3) \quad \Rightarrow  \quad {E}/{N} =  {170}/{3},  
 \end{equation}
which is only about 20\% smaller than the  value of $E/N=-135/2$   
in \eqn{eq:LPmaxEN} obtained by imposing  only the LP condition. 
This maximum value for the $\gag$ coupling in KSVZ models 
is depicted with a dot-dashed line in Fig.~\ref{fig:KSVZDFSZbands}. 
For  $f_{a} > 5 \times 10^{11}\,$GeV we forcedly have to assume a pre-inflationary scenario  
were the condition $m_\Q > T_{\rm{RH}}$ could be easily fulfilled.
In this case the limit  from cosmological considerations does not apply, however, 
the limit from the LP analysis still holds, so that the bound is only mildly relaxed.    
The corresponding region lies  on the left-hand side of the purple vertical line in \fig{fig:KSVZDFSZbands} 
labelled $f_{a} > 5 \times 10^{11}\,$GeV. 

Besides enhancing the axion-photon coupling, more $R_\Q$'s can also
weaken $g_{a\gamma}$, and even yield an approximate axion-photon decoupling. 
This  requires an ad hoc choice of  $R_\Q$'s, but no numerical fine tuning.  With two $R_\Q$'s there are
three such cases: $R_6 \oplus R_9$; $R_{10} \oplus R_{12}$ and $R_4 \oplus R_{13}$ which give respectively
$E_c/N_c = (23/12, 64/33,41/21) \approx (1.92,1.94,1.95)$ so that within theoretical errors 
even a complete decoupling is possible. In all these cases the axion could be eventually detected  
only via its (model-independent) coupling to the nucleons, given that  
the coupling to electrons is loop suppressed~\cite{Srednicki:1985xd}. 
Finally, with multiple representations additional possibilities featuring $\NDW=1$ open up 
as for example $R_{12} \ominus R_9$ or 
$R_{12} \ominus R_{10}$ for which  $N = T(8)-T(6)=3 - 5/2=1/2$
coincides with what is  obtained with a single  $SU(3)_c$ fundamental.

\begin{table}[t!] 
\renewcommand{\arraystretch}{1.2}
\centering
\begin{tabular}{@{\kern3em} c @{\kern3em}  c @{\kern3em} c@{\kern3em} c@{\kern3em} c@{\kern3em} c  @{\kern3em}}
\hline
 \hline
& $ R_\Q$ &  $\mathcal{O}_{\Q q }$ & $\phantom{\Big|} \Lambda^{\!R_\Q}_{\rm LP}$[GeV] & $E/N$ 
& $N_{\rm DW}$ \\ 
\hline
$d\leq4$ & $R_1$:$\,(3,1,-\tfrac{1}{3})$  & 
$\overline{\Q}_L d_R$ 
& $9.3 \cdot 10^{38} (g_1)$ & $2/3$ & $1$ \\ 
& $R_2$:$\,(3,1,+\tfrac{2}{3})$ & 
$\overline{\Q}_L u_R$
& $5.4 \cdot 10^{34} (g_1)$ & $8/3$ & $1$ \\ 
& $R_3$:$\,(3,2,+\tfrac{1}{6})$ & 
$\overline{\Q}_R q_L$
& $6.5 \cdot 10^{39} (g_1)$ & $5/3$ & $2$ \\ 
& $R_4$:$\,(3,2,-\tfrac{5}{6})$ & 
$\overline{\Q}_L d_R H^\dag$
& $4.3 \cdot 10^{27} (g_1)$ & $17/3$ & $2$ \\
& $R_5$:$\,(3,2,+\tfrac{7}{6})$ & 
$\overline{\Q}_L u_R H$
& $5.6 \cdot 10^{22} (g_1)$ & $29/3$ & $2$ \\
& $R_6$:$\,(3,3,-\tfrac{1}{3})$ & 
$\overline{\Q}_R q_L H^\dag$ 
& $5.1 \cdot 10^{30} (g_2)$ & $14/3$ & $3$ \\
& $R_7$:$\,(3,3,+\tfrac{2}{3})$ & 
$\overline{\Q}_R q_L H$ 
& $6.6 \cdot 10^{27} (g_2)$ & $20/3$ & $3$ \\ [3pt]
$d=5$ & $R_8$:$\,(3,3,-\tfrac{4}{3})$ & 
$\overline{\Q}_L d_R H^{\dag 2}$ & $3.5 \cdot 10^{18} (g_1)$ & $44/3$ & $3$ \\
& $R_9$:$\,(\bar 6,1,-\tfrac{1}{3})$ & 
$\overline{\Q}_L \sigma  d_R \cdot G $ & $2.3 \cdot 10^{37} (g_1)$ & $4/15$ & $5$ \\
& $R_{10}$:$\,(\bar 6,1,+\tfrac{2}{3})$ & 
$\overline{\Q}_L \sigma  u_R \cdot G $ & $5.1 \cdot 10^{30} (g_1)$ & $16/15$ & $5$ \\
& $R_{11}$:$\,(\bar 6,2,+\tfrac{1}{6})$ & 
$\overline{\Q}_R \sigma  q_L \cdot G $ & $7.3 \cdot 10^{38} (g_1)$ & $2/3$ & $10$ \\
& $R_{12}$:$\,(8,1,-1)$ & 
$\overline{\Q}_L \sigma  e_R \cdot G $ & $7.6 \cdot 10^{22} (g_1)$ & $8/3$ & $6$ \\
& $R_{13}$:$\,(8,2,-\tfrac{1}{2})$ & 
$\overline{\Q}_R \sigma  \ell_L \cdot G $ & $6.7 \cdot 10^{27} (g_1)$ & $4/3$ & $12$ \\
& $R_{14}$:$\,(15,1,-\tfrac{1}{3})$ & 
$\overline{\Q}_L \sigma  d_R \cdot G $ & $8.3 \cdot 10^{21} (g_3)$ & $1/6$ & $20$ \\
& $R_{15}$:$\,(15,1,+\tfrac{2}{3})$ & 
$\overline{\Q}_L \sigma  u_R \cdot G $ & $7.6 \cdot 10^{21} (g_3)$ & $2/3$ & $20$ \\
\hline
\hline
  \end{tabular}
  \caption{\label{ViableRQ} 
    $R_\Q$  allowing for $d\leq 4$ and $d=5$ decay operators 
    ($\sigma\cdot G \equiv \sigma_{\mu\nu} G^{\mu\nu}$) and yielding the first LP 
    above $10^{18}$ GeV
    in the gauge coupling given in parenthesis in the fourth column.
The anomaly contribution to $g_{a\gamma}$ is given in the fifth
column, and the DW number in the sixth one. Table adapted from Ref.~\cite{DiLuzio:2016sbl}} 
\end{table}
 
\subsubsection{DFSZ-like scenarios}
\label{sec:DFSZ-like}

As we have seen in  Section~\ref{sec:DFSZ} in DFSZ-type of models~\cite{Zhitnitsky:1980tq,Dine:1981rt} 
besides the SM-singlet field $\Phi$, two or more Higgs doublets $H_i$ carrying PQ charges
are introduced. The SM fermion content is not enlarged, but in general both quarks and leptons carry PQ
charges since they are coupled to the Higgs fields.  The electromagnetic and colour $U(1)_{PQ}$
anomalies then depend on the SM fermions gauge quantum numbers 
as well as on their model dependent PQ charge assignments.
Hence, several variants of DFSZ axion models are possible, some of
which have been discussed, for instance, in Refs.~\cite{Cheng:1995fd,Kim:1998va}.  
For most of these variants the axion-photon coupling remains within the KSVZ regions 
highlighted in Fig.~\ref{fig:KSVZDFSZbands}, and only in some specific cases the
KSVZ upper limit $E/N=170/3$ can be exceeded. We will now review under 
which conditions this can occur.

Let us  assume $n_H\geq 2$
Higgs doublets $H_i$ coupled to quarks and leptons
via Yukawa interactions, and to the singlet field $\Phi$ through scalar
potential terms.  The kinetic term for the scalars carries a
$U(1)^{n_H+1}$ rephasing symmetry that must be explicitly broken to
$U(1)_{PQ}\times U(1)_Y$ so that the anomalous PQ current 
is unambiguously defined, and to avoid additional Goldstone bosons with  
couplings too mildly suppressed just by the electroweak scale.  
Renormalizable non-Hermitian monomials  involving $H_i$ and
$\Phi$ are then required to provide an  explicit
breaking $U(1)^{n_H+1}\to U(1)_{\rm PQ}\times U(1)_Y$.
This implies that the PQ charges of all the Higgs doublets, and hence also of  
 the SM fermions, are interrelated and cannot be arbitrarily chosen.  
 To keep the discussion as general as possible let us use a notation in which 
 each fermion bilinear is coupled to a specific scalar doublet, so that 
 $\mX(\bar u_{L_j}  u_{R_j}) = - \mX_{H_{u_j}}$,   
  $\mX(\bar d_{L_j}  d_{R_j}) = - \mX_{H_{d_j}}$,   
 $\mX(\bar e_{L_j}  e_{R_j}) = - \mX_{H_{e_j}}$  
 and we have taken  the hypercharge of $H_{d_j,\, e_j}$  opposite to 
 that of $H_{u_j}$.
The  ratio of anomaly coefficients $E/N$ can then be written as 
\begin{equation}
\label{eq:EoNgeneralDFSZ}
\frac{E}{N} = 
 \frac{\sum_j \left( 
 \frac{4}{3}  \mX_{H_{u_j}} +
  \frac{1}{3}  \mX_{H_{d_j}} +
     \mX_{H_{e_j}}\right)}
     {\sum_j \left( \frac{1}{2}   \mX_{H_{u_j}} +
              \frac{1}{2}   \mX_{H_{d_j}}  \right)} 
              = \frac{2}{3} + 2\,\frac{\sum_j \left( \mX_{H_{u_ j}} + \mX_{H_{e_j}} \right)}{
 \sum_j \left( \mX_{H_{u_ j}} + \mX_{H_{d_j}}\right)} \, , 
\end{equation}
where the denominator, being proportional to the PQ-colour 
anomaly, cannot vanish.
The results for the two  DFSZ models discussed in  Section~\ref{sec:DFSZ} are recovered 
from \eqn{eq:EoNgeneralDFSZ} by  dropping the 
generation index and by setting  for DFSZ-I $H_e = H_d$ which yields $E/N=8/3$, 
and for DFSZ-II $H_e = \tilde H_u$  which yields $E/N=2/3$. 
In both cases the  axion-photon couplings,
that correspond to the oblique red lines labelled I and II in Fig.~\ref{fig:KSVZDFSZbands},
fall inside the $N_\mathcal{Q}=1$ KSVZ region.  

Let us now consider the so called DFSZ-III variant~\cite{Cheng:1995fd}
in which the leptons couple to a Higgs doublet $H_e$ different from $H_{u,d}$. 
 In order to enforce the breaking  $U(1)^4 = U(1)_e\times U(1)_u\times U(1)_d\times U(1)_\Phi\to
U(1)_{PQ}$,  in the scalar potential the three doublets  $H_e$, $H_u$ and $H_d$
must be appropriately coupled among them and/or to $\Phi^2$ or $\Phi$.
To find  which new values of $E/N$ are allowed in DFSZ-III 
we follow Ref.~\cite{DiLuzio:2017pfr}. Let us 
consider the mixed bilinear scalar monomials
$(H_e H_u)\,,(H_e^\dagger H_d),\, (H_u H_d)$.  It is easy to see 
that these bilinear terms alone yield for $E/N$ the same two possibilities 
obtained for DFSZ-I and II. Combining these bilinears among themselves or  
with their Hermitian conjugates to build quartic couplings two new possibilities arise: 
\begin{eqnarray}
\label{quadlinear} \nonumber
(H_eH_u)\cdot (H_u H_d) \quad  &\Longrightarrow& \quad  \mX_{H_e} =  -(2\mX_{H_u}+\mX_{H_d})\,,
                                                           \qquad E/N = -4/3 \,, \\ \nonumber
(H_e^\dagger H_d) \cdot  (H_u H_d)     \quad  &\Longrightarrow& \quad \mX_{H_e} =
                                                           \mX_{H_u}+2
                                                           \mX_{H_d}\,, \qquad\quad\ \;  E/N = 14/3\,. 
\end{eqnarray} 
The largest coupling is obtained for the first possibility giving  $\gag\propto E/N-1.92 \simeq -3.25$. 
We see that the largest axion-photon coupling allowed in DFSZ-III still  falls
within the $N_\mathcal{Q}=1$ band in Fig.~\ref{fig:KSVZDFSZbands}.\footnote{Note that
  the  charges of the DFSZ-III variants in  Ref.~\cite{Cheng:1995fd} do not allow to build PQ and gauge
  invariant mixed terms at the renormalizable level. 
  Consequently, $U(1)^4$ cannot get broken to a single $U(1)_{PQ}$.}

More possibilities in choosing the PQ charges become possible if
we allow for generation dependent PQ charge assignments, a scenario that 
was already  mentioned in Section~\ref{sec:IntroFlavourViolating} 
and that  will be studied  
more extensively in Section~\ref{sec:gaN}. 
The maximum freedom corresponds to the case in which there are three Higgs doublets for
each fermion species ($H_{e_1}, H_{e_2}, H_{e_3}$, etc.) so that the scalar rephasing 
symmetry is $U(1)^{n_H+1}$ with $n_H=9$, a  scenario that was labelled   DFSZ-IV in Ref.~\cite{DiLuzio:2017pfr}. 
Differently from DFSZ-I, II and III, this scenario should not be understood as 
giving rise to realistic models, since it might be easily plagued by various 
phenomenological issues. 
However, it turns out to be useful to consider this construction since bounding 
from above the maximum possible $E/N$ in DFSZ-IV automatically provides an upper bound
on $E/N$ for all cases with generation dependent PQ charges and
$2\leq n_H \leq 9$ Higgs doublets coupled to the SM fermions.
For the derivation of this upper bound we refer to Ref.~\cite{DiLuzio:2017pfr}, the result is that  
the maximum possible axion-photon coupling in DFSZ-IV models  corresponds to
\begin{equation}
\label{eq:DFSZ-IV}
  \left(E/N\right)^{\text{DFSZ-IV}}_\text{max}= 524/3 \,. 
\end{equation}
 Although the
value in \eqn{eq:DFSZ-IV} exceeds by a factor of three the maximum KSVZ
value $E/N=170/3$ plotted in \fig{fig:KSVZDFSZbands}, the construction through which $(E/N)_{\text{max}}$
was obtained in  Ref.~\cite{DiLuzio:2017pfr} is sufficiently cumbersome to 
suggest that the $N_\mathcal{Q}>1$
region in  Fig.~\ref{fig:KSVZDFSZbands} 
can still be considered as representative also
of most of DFSZ-IV type of models. 
In  DFSZ-IV  it is also possible to obtain axion-photon decoupling
ensuring  at the same time a  correct breaking of the  global
symmetries $U(1)^{9+1}\to U(1)_{\rm PQ}\times U(1)_Y$. One  example  is given by:
\beq
\mX_{H_{u_j}} =(2,4,8)\,\mX_\Phi\,, \qquad
\mX_{H_{d_j}} =(0,-2,-4)\,\mX_\Phi\,,  \qquad 
\mX_{H_{e_j}} =(-1,-3,-5)\,\mX_\Phi\,,
\eeq
for which  \eqn{eq:EoNgeneralDFSZ} yields $E/N=23/12 \approx 1.92$ and  hence $\gag \approx 0$ 
(within theoretical errors).   

The values of $E/N$ associated to the maximum and minimum of $\gag$ for the different classes of
models discussed so far are summarised in \Table{summaryEoN}.  Note that
  differently from the KSVZ models analysed in \sect{sec:KSVZ-like}, the limits on
  the axion-photon coupling in DFSZ models do not depend on the details of
  the cosmological evolution of the Universe, and therefore hold also within
  the region on the left of the violet vertical line 
  labelled $f_a > 5\times 10^{11}\,$GeV in Fig.~\ref{fig:KSVZDFSZbands}.
\begin{table}[t!] 
\renewcommand{\arraystretch}{1.2}
\centering
\begin{tabular}{@{} |l|c|c| @{}}
\hline
 &  $E/N(\gag^{\text{max}})$ & 
 $E/N(\gag^{\text{min}})$  \\ 
\hline
\hline
KSVZ ($N_Q=1$) &  ${44}/{3}$ &  ${5}/{3}$  \\ 
KSVZ ($N_Q>1$) &  ${170}/{3}$ &  ${23}/{12}$  \\ 
\hline
DFSZ-I-II ($n_H=2$) &  ${2}/{3}$ &  ${8}/{3}$  \\ 
DFSZ-III ($n_H=3$) &  $-{4}/{3}$ &  ${8}/{3}$  \\ 
DFSZ-IV ($n_H=9$) &  ${524}/{3}$ &  ${23}/{12}$   \\ 
\hline
  \end{tabular}
  \caption{\label{summaryEoN} 
    Values of $E/N$ corresponding to the maximum and minimum values of
    $\gag$ for different classes of models.  Only KSVZ models 
    that satisfy  the first two selection rules $(i)$ and $(ii)$ discussed 
    in \sect{sec:KSVZ-like}  are included.} 
\end{table}



\subsubsection{Enhancing $\gag$ above  KSVZ- and DFSZ-like scenarios} 
\label{sec:gag-clockwork}

The anomalous  triangle diagrams responsible for the axion-photon coupling
depend on the  gauge and global charges of the fermions running in the loops. 
$E/N$ can then be boosted either by increasing the contribution of the 
gauge charges (subject to LP constraints) or by increasing the value of 
the global charges (not subject to LP constraints). 
In the previous sections we have given some examples that exploit  the former possibility. 
More elaborated scenarios that rely on the second possibility 
allow to  enhance the axion coupling 
to photons by much larger (exponentially enhanced) factors, via 
mechanisms loosely inspired by  `clockwork'  type of 
constructions~\cite{Choi:2014rja,Choi:2015fiu,Kaplan:2015fuy,Giudice:2016yja}.
One of such possibilities for the QCD axion was put forth in  Ref.~\cite{Farina:2016tgd} 
(see also \cite{Higaki:2015jag,Higaki:2016yqk})
and is based on 
a KSVZ type of model.
Several singlet scalar fields $\Phi_k$ ($k=0,1,\dots, n$) are introduced, 
coupled among them through non-Hermitian monomials $\Phi_k^\dagger \Phi^3_{k+1}$ 
so that only a single global $U(1)$ remains unbroken, and a charge relation 
$\mX_{\Phi_k} = 3^{-k} \mX_{\Phi_0}$ is enforced.
A vector-like representation of   coloured quarks is coupled to  $\Phi_n$ 
while another pair of electromagnetically charged but colour neutral   leptons 
is coupled to a different scalar $\Phi_p$ with $p<n$. Hence  the anomaly coefficient ratio
$E/N$  acquires from the ratio between the PQ charges of the coloured/uncoloured fermions 
an enhancement factor  $3^{n-p}$.
Cosmological  constraints on  clockwork axions
have been analysed in Ref.~\cite{Higaki:2016jjh,Long:2018nsl}.


A different possibility was proposed in Ref.~\cite{DiLuzio:2017pfr}. 
This is based on a DFSZ type of construction and, for this reason,  
has the virtue of  exponentially enhance also the axion-electron coupling
(see \sect{sec:gae}).  
 Let us consider a DFSZ-like scenario with $2+1+(n-1)$ Higgs doublets.
The up and down  Higgs scalars are coupled to the PQ symmetry breaking singlet
through the   term 
$H_u H_d \Phi^2$ so that the  PQ charge for $H_u$ 
satisfies the relation $\mX_{H_u} = -2\mX_\Phi -  \mX_{H_d} $. 
Let us define $H_1=H_u$ and add $n-1$ additional scalar 
doublets  $H_k$ ($k=2,3,\dots,n$) with the same hypercharge of 
$H_1$ but whose couplings to the  SM fermions are forbidden 
by the PQ symmetry. The additional doublets are    
coupled among each other via non-Hermitian quadrilinear terms 
 $(H^\dagger_{k} H_{k-1})(H_{k-1} H_d)$, 
 so that  a single $U(1)$ 
 global symmetry survives,  while 
  their PQ charges satisfy $\mX_{H_k} = - 2^k \mX_\Phi - \mX_{H_d}$. 
 Finally, let us couple the lepton Higgs doublet $H_e$ as
$(H_e H_n)(H_n H_d)$ so that $\mX_{H_e}= 2^{n+1} \mX_\Phi + \mX_{H_d}$.  
Inserting in  \eqn{eq:EoNgeneralDFSZ}  the expressions for $\mX_{H_u}$ and $\mX_{H_e}$ 
we readily obtain:
\beq
\label{eq:ENclockwork}
\frac{E}{N} = \frac{8}{3} -2^{n+1}\,. 
\eeq
Since even in the most conservative case of  doublets
with electroweak scale masses,  $n$ can be as large as fifty before a LP is hit below the Planck scale,  
it is always possible to obtain exponentially large axion-photon couplings. 

Other possibilities for large enhancement of $\gag$ exist. Some examples  
that are qualitatively different from the ones sketched above can be found 
in Ref.~\cite{Agrawal:2017cmd,Daido:2018dmu}.


\subsection{Enhancing/suppressing  $g_{ae}$}
\label{sec:gae}

As discussed in \sect{sec:Astro_bounds_gae}, the axion-electron coupling $g_{ae}$ 
can activate different processes that  alter stellar evolution and that are 
particularly  significant   in the white dwarf and red giant evolutionary stages. 
The corresponding astrophysical bounds  have been reviewed  in \sect{sec:Astro_bounds_gae}.  
Although they are  remarkably strong,  translating  them into constraints  on fundamental  
parameters as the axion decay constant or the axion mass should be done with care, 
because the relation between  $g_{ae}$ and  $f_a$ (or $m_a$) depends crucially on the specific
model (cf.~\Eqn{eq:Cae} for its general expression). 
For example, in the KSVZ scenarios electrons do not carry a PQ charge, and hence 
they couple to the axion 
only through a  photon-electron loop generated by the axion-photon coupling,  
so that the resulting $g_{ae}$ is loop suppressed and tiny (see \eqn{eq:Caeradiative}).  
For this reason astrophysical limits on $g_{ae}$ provide only weak constraints on KSVZ models 
(cf.~also \fig{fig_KSVZ_astro_bounds}). 

\subsubsection{Enhancing $\gae$ in KSVZ-like scenarios}
\label{sec:gaeKSVZenhanced} 
Large enhancements with respect to  conventional KSVZ scenarios can be  straightforwardly obtained by 
recalling that while in  the original  model $E/N=0$, models that involve non-trivial heavy quarks representations like 
the ones discussed in \sect{sec:KSVZ-like},  can yield  particularly large $E/N$ factors (see \Table{ViableRQ}). In this case  
$g_{ae}$ gets enhanced by the same factor that enhances $g_{a\gamma}$.  
In another class of  KSVZ models  the one-loop induced axion-electron coupling can get an extra 
contribution from  loops involving new particles. 
This is what happens for example in type-I  seesaw scenarios for neutrino masses 
in which  the heavy RH  neutrinos $N_R$ obtain their mass $M_R$ 
from a coupling $N_R N_R \Phi$ to the PQ symmetry breaking scalar singlet~$\Phi$. 
These models are soundly  motivated by the fact that the seesaw and the PQ symmetry breaking  scales 
naturally fall in the same intermediate range  $M_R \sim f_a\sim 10^9 \div 10^{12}$ GeV 
so that an identification seems natural (see also \sect{sec:connection_neutrinos}). 
In this case, in order to acquire their mass, the RH neutrinos must carry a PQ charge, and 
their couplings to the axion give rise to new loops \cite{Shin:1987xc,Pilaftsis:1993af,Garcia-Cely:2017oco}  that can 
enhance the axion-electron coupling by  up to one or two orders of magnitude with respect to  
conventional hadronic axion models, see for example Refs.~\cite{Shin:1987xc,Dias:2014osa,Ballesteros:2016euj}. 




\subsubsection{Enhancing $\gae$ in DFSZ-like scenarios}
\label{sec:gaeDFSZenhanced} 
In DFSZ scenarios the axion-electron coupling arises at the tree level and hence the astrophysical limits
on $g_{ae}$ are more effective in providing tight constraints on the fundamental axion parameters. 
Clearly,  in the DFSZ variants 
reviewed in \sect{sec:DFSZ-like} in which $g_{a\gamma}$ gets enhanced  via an  $E/N$ value  
that is boosted  by a large PQ electron charge, also $g_{ae}$ is accordingly enhanced. 
However,  defining correctly the  coupling of electrons to the {\it physical} axion 
involves some subtleties.   Let us consider for example the DFSZ construction with $2+1+(n-1)$ 
Higgs doublets outlined in \sect{sec:gag-clockwork} that  produces an exponential 
enhancement in   $E/N$,  see \eqn{eq:ENclockwork}.
The physical axion is identified by imposing  the orthogonality condition between the 
PQ and hypercharge currents  
  $J^{\rm PQ}_{\mu}|_a   =\sum_i \mX_i v_i \partial_\mu a_i $ 
  and    $J^{\rm Y}_{\mu} |_a = \sum_i Y_i v_i \partial_\mu a_i $,  where the sum runs over all 
  the scalars  $\{\Phi, H_u, H_d, H_e, H_{k\geq 2}\}$, $a_i$ are the scalar (neutral) orbital modes, and $v_i$ are   
the corresponding   VEVs. 
Note that since $H_{u,d,e}$ need to pick-up a VEV to generate masses for the fermions, 
even in the case that all the $H_{k\geq 2}$ have positive mass squared terms, 
non-vanishing  $v_{k\geq 2}\neq 0$  unavoidably arise  because of the potential terms 
$(H_k^\dagger H_{k-1})(H_{k-1} H_d)$  in which they appear linearly.  
The size of these VEVs is controlled by the size of the quadrilinear coupling constant, 
hence it is natural to expect $v_{k} < v_u$ with a size that decreases  with increasing $k$.
To proceed, it is convenient to express the PQ charges $\mX_{H_d},\mX_{H_u}$, and $\mX_{H_k}$  in terms of 
$\mX_{H_e}$ and $\mX_\Phi$ by using the relations given above~\eqn{eq:ENclockwork}  as 
\begin{equation}
\label{eq:chargesHe}
\mX_{H_d} = \mX_{H_e} - 2^{n+1}\mX_\Phi,
\qquad
\mX_{H_k} = - \mX_{H_e} +(2^{n+1}-2^k)\mX_\Phi,
\end{equation}
where again we have defined $H_u=H_1$.
Recalling that $Y(H_e) = Y(H_d) = - Y(H_k) =\frac{1}{2}$ 
the orthogonality condition can be written  as: 
\begin{equation}
\label{eq:orthogonalityHk}
 \sum_i 2 Y_i  \mX_i v_i^2 
 = \mX_{H_e} v^2 
- \mX_\Phi \(2^{n+1} v_d^2 + \sum_{k=1}^{n} \(2^{n+1} -2^k\) v_k^2\) =0 \, ,
\end{equation}
where $v^2 = v^2_e+v^2_d+\sum_{k=1}^n v^2_k$ is the electroweak VEV.
The resulting  PQ  charge of the electron is:
\begin{equation}
\label{eq:Xeclockwork}
\mX_{H_e} = \frac{\mX_\Phi}{v^2}\(2^{n+1} v_d^2 + \sum_{k=1}^{n} \(2^{n+1} -2^k\) v_k^2\) \, ,
\end{equation}
%
and the axion-electron coupling can  be written as 
%
\beq 
g_{ae} = \frac{\mX_{H_e}}{2N} \frac{m_e}{f_a} 
=
- 2^{n+1}\, \frac{m_e}{6 f_a}\[\frac{v_d^2}{v^2} + \sum_{k=1}^{n} \(1-\frac{1}{2^{n+1-k}}\) \frac{v_k^2}{v^2}\]\, ,
\eeq
where  we have used $2 N= 3 (\mX_{H_u} + \mX_{H_d})= -6\mX_\Phi$.
For large $n$ the size of $\gae$ is bounded from below by 
\beq 
\label{eq:gae-perturb}
|g_{ae}|  
\gtrsim
2^{n+1} \(\frac{v_u^2}{v^2}+\frac{v_d^2}{v^2}\) \frac{m_e}{6f_a} \, ,
\eeq
where the inequality is saturated when all the $v_{k\geq 2}$ are 
negligible. Hence  the axion coupling to the electrons  is always exponentially 
enhanced.\footnote{Only for $v_e = v$ one could have $\gae \to 0$. However, the sum of the VEV 
ratios in \eqn{eq:gae-perturb}  is bounded by the perturbativity 
of the top Yukawa coupling to be $\gtrsim 0.1$.}

\noindent
\subsubsection{Suppressing $\gae$ in DFSZ-like scenarios:  the electrophobic axion}
\label{sec:gaeDFSZsuppressing} 

A more intriguing  possibility within DFSZ-type of models is that of  decoupling the axion from the electrons, 
 especially if this could be done consistently with 
the conditions  that enforce nucleophobia. In fact, in this case all the most stringent 
astrophysical limits would  evaporate. Such an axion would be appropriately defined as 
{\it astrophobic}~\cite{DiLuzio:2017ogq}.
One might think that in order to decouple the electron from the axion it would be sufficient 
to introduce a third (leptonic) Higgs doublet neutral under the PQ symmetry, so that the leptons 
would not carry PQ charges either. 
However, 
perhaps a bit unexpectedly, this can work only for a few specific values of the VEVs ratios
of the two hadronic Higgs~\cite{Bjorkeroth:2019jtx}. To see this let us consider  
a model with three-Higgs doublets, that is now convenient to define having the same hypercharge, 
wherein $ H_{1,2} $ couple to quarks 
while $ H_3 $ couples to the leptons.  
We need to require that the four $U(1)$ rephasing symmetries of the
kinetic term of the three scalar doublets $H_{1,2,3}$ and of the PQ symmetry breaking field $\Phi$ 
are broken by renormalizable potential terms so that   $U(1)^4 \to U(1)_Y\times U(1)_{PQ}$. 
This can be done either by coupling the leptonic Higgs
doublet $H_3$ to both hadronic Higgses ($H_{1,2}$), or by coupling one
of the two hadronic Higgses to the other two doublets:
\begin{equation}
\label{eq:H1H2H3}
    H_3^\dagger H_1 \Phi^m + H_3^\dagger H_2 \Phi^n 
    ~~\mathrm{or}~~  
    H_3^\dagger H_{1,2} \Phi^m + H_{2}^\dagger H_{1} \Phi^n . 
\end{equation}
For renormalizable operators one has, without loss of generality,
$m=1,2$ and $n=\pm1,\pm2$, with the convention that  negative values of $n$ mean
Hermitian conjugation $\Phi^{n} \equiv (\Phi^\dagger)^{|n|}$.  
This implies certain conditions for the  PQ charges of the scalar, for example for the first pair of 
operators we have
\begin{equation}
\label{eq:mn1}
    -\mX_3 +\mX_1 + m   =0,   \qquad
    -\mX_3 +\mX_2 + n   = 0  \,,
\end{equation}
where for simplicity we have set $\mX_\phi = 1$.  Similar relations hold if 
the second pair of operators in \Eqn{eq:H1H2H3} is chosen. 
Besides these two conditions,   orthogonality between the physical axion and the  Goldstone boson of hypercharge implies: 
\begin{equation}
\label{eq:3vevs}
    \mX_1 v^2_1 + \mX_2 v_2^2 + \mX_3 v_3^2   = 0 . 
\end{equation}
To see under which conditions the leptons can be decoupled from the axion, let us set $\mX_3 \to 0$.
We obtain 
\begin{equation}
\label{eq:goodbeta}
\tan^2\beta \equiv \frac{v^2_2}{v^2_1} = - \frac{\mX_1}{\mX_2} = - \frac{m}{n}\,, 
\end{equation}
so that with the constraints in \eqn{eq:mn1},  $\mX_3\approx 0$ can be consistently enforced 
only  for   $\tan^2\beta  \approx \frac{1}{2}, 1, 2$. 
It can be verified that in the other two cases in \Eqn{eq:H1H2H3}
$\tan\beta = 1$ is the only possibility. 
As it was remarked in Ref.~\cite{Bjorkeroth:2019jtx}, 
it is intriguing that the condition for nucleophobia~\eqn{eq:cpMcn} which,  as we have seen 
in \sect{sec:nucleophobic},  requires 
 $\tan^2\beta \approx \frac{m_d}{m_u} \approx 2$, is consistent with one of the $\tan\beta$ values 
that can enforce electrophobia or,  said in another  way,  it is  consistent with a suitable set of 
operators involving the three Higgs doublets, namely $ H_3^\dagger H_1 \Phi^2 + H_3^\dagger H_2 \Phi^{\dag}$.  
 




\subsection{Enhancing/suppressing   $g_{aN}$}
\label{sec:gaN}

The defining property of axions is that they couple to gluons via the anomalous term, and 
hence  they  should couple as well to the nucleons $N=n,p$.  
 In fact, from the model building point of view it is much more difficult to arrange  
for strong enhancements/suppressions  of  the axion-nucleon coupling $\gaN$  than it is for    $\gag$ and $\gae$.  
Nevertheless, a handful of  working mechanisms  has been put forth, and will be reviewed in this section. 
Of special interest is the possibility  of approximate axion-nucleon decoupling, that will be reviewed 
in \sect{sec:nucleophobic}, because in this case the tightest 
constraints on $f_a$ and $m_a$, inferred from astrophysical bounds 
on $\gaN$ by assuming a conventional axion model, see \sect{sec:Astro_bounds_gaN}, 
would be accordingly relaxed, and  regions in the $\gag-m_a$ plane that are generally 
considered  excluded, would become allowed  and thus worthwhile to be explored experimentally. 

\subsubsection{Enhancing $g_{aN}$}
\label{sec:gaNenhanced}
A possible way to enhance the axion coupling to nucleons is discussed in 
Ref.~\cite{Marques-Tavares:2018cwm}. The idea is to uncorrelate the 
axion-nucleon coupling from the axion mass by assigning $U(1)_{\rm PQ}$ 
charges to SM quarks such that the latter do not contribute to the QCD anomaly. 
Effective dimension five operators are introduced that  couple 
the axion multiplet $\Phi$ to the light quarks and to the corresponding Higgs. 
Upon PQ symmetry breaking these terms give rise to the quark Yukawa 
couplings which, however, also involve the axion through a term  
$e^{i a/f}$,  
where $f$ is the $U(1)$ symmetry breaking order parameter. 
As usual, by reabsorbing the axion via a quark field redefinition, derivative axion couplings 
are generated  from the kinetic terms, and eventually  an axion-nucleon interaction
\beq 
\label{eq:gaNenhanc}
\frac{\partial_{\mu}a}{f} \bar N \gamma^\mu \gamma_5 N \, 
\eeq
suppressed by the scale $f$ results.  The QCD anomaly of the PQ current is due instead to 
heavy vector-like fermions $\mathcal{Q}$  that  
couple to a scalar $\Phi_k$ whose PQ charge is related to the charge of $\Phi$ as 
$\mX_{k} = 3^{-k} \mX_\Phi$, as the result of a clockwork-type of potential 
$ \sum_j  \Phi^\dagger_j \Phi_{j+1}^3 +{\rm h.c.}$  (see~\sect{sec:gag-clockwork}). 
Hence the anomalous term reads  
\beq 
\label{eq:gagggaNenhanc}
\frac{a}{3^k f} \frac{\alpha_s}{8\pi} G \tilde G \, ,
\eeq
where the factor $3^k$ in the denominator is due to  the exponential suppression of the  heavy quarks PQ charges  
$\mX_\mathcal{Q}\sim 3^{-k}$.
The axion decay constant is thus defined as $f_a = 3^k f$, so that  the axion coupling to the nucleons in 
\eqn{eq:gaNenhanc}  is exponentially  enhanced with respect to the usual $1/f_a$ scaling.  

\bigskip

A different, and possibly simpler way to exponentially enhance $\gaN$  was  
 recently  proposed in \cite{LucLucaEN:inprep1}. 
 It relies on clockworking directly a set of  Higgs doublets in a DFSZ-type of construction. 
Let us take $n+1$ doublets with hypercharge $Y=-\frac{1}{2}$ and let us consider 
the following scalar terms: 
\begin{equation}
\label{eq:gaNclock}
H_0^\dagger H_1 \Phi^2, \qquad \sum_{k=2}^n \(H^\dagger_{k-1} H_k\)\(H^\dagger_{k-1} H_0\)\,.
\end{equation}
It is easy to verify the charge relation  
$\mX_k = \mX_0 - 2^k \mX_\Phi$ 
so that $\mX_n$ gets exponentially enhanced. 
Consider now the following generation dependent quark couplings: 
\begin{equation}
\label{eq:gaNexpH}
\(\bar u_{1L} u_{1R}  H_n + \bar d_{1L} d_{1R}  \tilde H_0 \) +
\(\bar u_{2L} u_{2R}  H_0 + \bar d_{2L} d_{2R}  \tilde H_n \) +
\(\bar u_{3L} u_{3R}  H_1 + \bar d_{3L} d_{3R}  \tilde H_0 \) \,, 
\end{equation}
where $\tilde H = i \sigma_2 H^*$.
All CKM mixings can be generated by adding for example $(\bar u_{2L} u_{1R}  H_1)+
(\bar u_{3L} u_{1R}  H_n)$ together with all the additional terms consistent with the charge
assignments implied by  these two terms plus the terms in ~\eqn{eq:gaNexpH}.
The anomalies of the first two generations that contain the Higgs multiplet $H_n$  with 
the  `large charge' cancel each other, so that the QCD (and QED) anomaly coefficient is determined 
only  by the  third generation:
\begin{equation}
\label{eq:expgaN_anomaly}
2N =\(\mX_{u_{3L}} - \mX_{u_{3R}} \)
+ \(\mX_{d_{3L}}-  \mX_{d_{3R}} \)
= \mX_{H_1} - \mX_{H_0} = - 2\mX_\Phi\,.
\end{equation}
As a result, the axion coupling to the light up-quark gets exponentially enhanced as
\begin{equation}
\label{eq:expPQup}
c^0_{u1} = \frac{
\mX_{u1L}-\mX_{u1R}
}{2 N} \approx 2^{n-1}\,, 
\end{equation}
which in turn gives rise to an exponentially enhanced $\gaN$, see~\eqns{eq:cpmatrixelem1}{eq:cnmatrixelem1},
while the axion-photon coupling $\gag$ remains of standard size.




\subsubsection{Suppressing  $g_{aN}$: the nucleophobic axion}
\label{sec:nucleophobic}

%

Arranging for a  strong suppression of the axion-nucleon coupling $g_{aN}$
is difficult. In KSVZ models in which the SM fermions do not carry  PQ charges
this is not possible, because  $g_{aN}$ remains determined by a model-independent 
contribution that is induced  by the axion coupling to the gluon fields. 
In DFSZ models the pathways to   enforce $\gaN\approx 0$
must comply with  tight theoretical constraints.
For example, a necessary (but not sufficient) condition 
is that the PQ-colour anomaly 
is only determined by the light quarks, that is, either the two heavier generations
are  not charged under PQ, or they have cancelling anomalies \cite{DiLuzio:2017ogq}. 
This unavoidably requires generation dependent axion couplings to the  quarks,
which in turn implies that nucleophobic  axions are mediators of flavour changing interactions,
see \sect{sec:gaFCNC}.
An early study in the direction of suppressing $\gaN$ was presented in  Ref.~\cite{Krauss:1987ud}, 
where some phenomenological aspects connected to nucleophobic axions  
were investigated, with the notably exception of  SN1987A related limits, 
given that  the Supernova  explosion occurred only a few months before the 
completion of that paper. 
In another early reference~\cite{Hindmarsh:1997ac}  a class of models in which the axion 
couples to a single quark was studied, and it was found that when the coupling is only to the light up-quark 
it was possible to engineer  for a strong  suppression of $\gaN$. Clearly this 
is in agreement with the condition  that to obtain $\gaN\approx 0$ only  the light 
quark generation is allowed to contribute to the PQ anomaly. 
More recently, a dedicated study of the various possibilities for constructing nucleophobic 
axion models was  presented  in Ref.~\cite{DiLuzio:2017ogq}. We  will now review the main 
results  following this reference. 

To identify which conditions must be satisfied to suppress both axion couplings to protons and neutrons
it is convenient to recast   \eqn{eq:cpmatrixelem1} and \eqn{eq:cnmatrixelem1}
into the following  isospin conserving and isospin violating linear combinations: 
\begin{align}
\label{eq:cpPcn}
C_{ap} + C_{an} &= \left( c^0_u + c^0_d -1 \right) (\Delta u + \Delta d) -2 C_{a,{\rm sea}} \, , \\
\label{eq:cpMcn}
C_{ap} - C_{an} &= \left( c^0_u - c^0_d  - f_{ud} \right) (\Delta u - \Delta d) \, ,
\end{align}
 where we have defined $f_{ud} = \frac{m_d - m_u}{m_d + m_u}\approx \frac{1}{3}$.
The last term in \eqn{eq:cpPcn} accounts for the contribution 
to the nucleon couplings of the sea quarks, that has been neglected in the treatment 
of \sect{sec:axionnucleon} (but included in the full expressions for the couplings \eqns{eq:Cap}{eq:Can}) and 
that cancels out in \eqn{eq:cpMcn}. 
The leading contribution comes from the strange quark  $ C_{a,{\rm sea}} = 0.038\, c^0_s +\dots$, and the 
overall value of the correction is less than $10\%$  of the contribution of  the  valence quarks.\footnote{Unless otherwise noticed,  
it is understood  that axion-nucleon decoupling  will refer to a suppression of 
$\approx 10\%$ level compared to conventional KSVZ-like couplings.}
%
The (approximate) vanishing of  \eqns{eq:cpPcn}{eq:cpMcn}
can then be taken as the defining condition for the nucleophobic axion.
Note, also, that since the axion-pion coupling  $C_{a\pi}$ is proportional to $C_{ap}-C_{an}$, see \eqn{eq:Capidef}, 
nucleophobic axions are  also pionphobic.
In variant DFSZ models with two Higgs doublets $H_{1,2}$ and non-universal PQ charge 
assignment \cite{DiLuzio:2017ogq} 
both conditions $ C_{ap} \pm C_{an} \approx 0$ can be realised~\cite{DiLuzio:2017ogq}. 
To see this, let us focus on the first generation Yukawa terms
\begin{equation}
    \bar q_{1L} u_{1R} H_1 + \bar q_{1L} d_{1R} \tilde H_2 .  
\label{eq:firstgen}
\end{equation}
%
Neglecting the effects of flavour mixing, which are assumed to be small throughout this 
Section,\footnote{%
  In the presence of flavour mixing, $c^0_{u,d} \to c^0_{u,d} + \Delta c^0_{u,d}$,  
  where $\Delta c^0_{u,d}$ involves quark mass diagonalisation matrices.  
These effects will be discussed in \sect{sec:gaFCNC}, see also 
 Ref.~\cite{DiLuzio:2017ogq}.} 
 the axion couplings to the light quark fields
 are (cf.~\Eqn{eq:c0fPQcharges} for their UV origin in terms of PQ charges)
\begin{align}
    c^0_u &= \frac{1}{2N}
 \left(\mX_{q_1} - \mX_{u_1}\right)
    =\frac{\mX_1}{2N} , \\
    c^0_d &= \frac{1}{2N} 
    \left(\mX_{q_1} - \mX_{d_1}\right)  
    = -\frac{\mX_2}{2N}    .
\end{align} 
Here $\mX_{u_1}=\mX(u_1)$, etc.~denote the PQ charges of the fermion fields
while $\mX_{1,2}=\mX(H_{1,2})$. The coefficient of the PQ colour anomaly is then
\begin{equation} 
    2N = \sum_{i=1}^3
    \left(  2 \mX_{q_{i}} -\mX_{u_{i}} - \mX_{d_{i}}\right) ,
\end{equation}
while the contribution to the colour anomaly from light quarks only can be written as:
\begin{equation}
    2N_\ell = 2 \mX_{q_1}  - \mX_{u_1} - \mX_{d_1}  = \mX_1-\mX_2 .
\end{equation}
Hence, the first condition for ensuring approximate nucleophobia reads
(cf.~\eqn{eq:cpPcn})
\begin{equation}
\label{eq:cuPcd}
    c^0_u + c^0_d = \frac{N_\ell}{N} = 1.
\end{equation}
This neatly shows that only models in which the colour anomaly is determined solely by
the light $u,d$ quarks 
have a chance to be
nucleophobic.  This  can be realised in two ways: 
\begin{itemize}
\item[$(i)$] either the contributions of the two heavier generations vanish identically  ($N_2=N_3=0$); 
\item[$(ii)$] or  they  cancel each other  ($N_2 = - N_3$ and $N_\ell = N_1$). 
\end{itemize}
Assuming only  two Higgs doublets, the first possibility  can be easily realised by taking one 
Higgs doublet  (say $H_1$) with a vanishing PQ

The first possibility was for example realised in a scenario in which the axion couples only to the up quark~\cite{Hindmarsh:1997ac}
(see also \cite{Saikawa:2019lng})   so that $2N = 2N_\ell$ is a straightforward result. 
%
Alternative  possibilities  in which the PQ charges for the two heavier generations  are non-trivial were 
found in Ref.~\cite{Bjorkeroth:2018ipq}.  In the attempt of looking for connections between  PQ symmetries  
and the quark flavour puzzle, with the aim of building predictive scenario for the quark masses and mixings,  
the authors of this last reference studied  which PQ symmetries could  force the vanishing 
of a certain number of  entries in the Yukawa matrices, in such a way  that a maximal parameter 
reduction consistent with observations is enforced.\footnote{In fact the requirement of no massless quarks, no vanishing mixings and one CP violating phase in the CKM matrix
implies that no-less than seven non-vanishing Yukawa entries are required, so that the matching between 
 fundamental parameters and observables is one-to-one, but within the SM there are no predictions. Conversely,  
 in the lepton sector equipped with the type-I seesaw for neutrino masses, the same strategy yields  precise 
 numerical predictions for the leptonic Dirac phase and for the absolute scale of neutrino masses~\cite{Bjoorkeroth:2019ndr}.}
Serendipitously, it was  found that two specific    
PQ symmetries able to enforce  maximal parameter reduction were also characterised by  
peculiar cancellations among the PQ charges of the  heavier quarks resulting in $N_2=N_3=0$. 

The second possibility with non-vanishing anomaly coefficients for the heavier generations
was thoroughly explored 
in Ref.~\cite{DiLuzio:2017ogq}.\footnote{In this reference  additional 
possibilities in which the contribution to the anomaly of two generations vanish identically
were also identified.}
With just two Higgs doublets,  in order to have 
all three quark generations with different charges, some of the SM Yukawa operators 
are necessarily forbidden, so that  zero textures are present in the quarks mass matrices~\cite{Bjorkeroth:2018ipq}.
In contrast, by imposing  that the PQ symmetry does not forbid any of the SM 
Yukawa operators, then two quark generations must have the same charges,  so that 
 for example $N_1 =N_2 = - N_3$.  
The PQ charge assignments  that satisfy this condition, and that can thus be compatible with nucleophobia, 
were classified in Ref.~\cite{DiLuzio:2017ogq}.

As regards  the second condition, to enforce   $C_{ap}-C_{an}\approx 0$ in 
 \eqn{eq:cpMcn} we need $c^0_u - c^0_d  = f_{ud} $ 
 where  a value $f_{ud} = \frac{m_d - m_u}{m_d + m_u}\approx \frac{1}{3}$ corresponds to
 $\frac{m_d}{m_u}\approx 2$.
 Let us  denote
by $\tan\beta= {v_2}/{v_1}$ the ratio of the VEVs of  $H_{1,2}$, and
introduce the shorthand notation $s_\beta=\sin\beta$,
$c_\beta=\cos\beta$. The ratio $\mX_1/\mX_2=-\tan^2\beta$ is fixed by the
requirement that the PQ Goldstone boson is orthogonal to the Goldstone eaten
up by the $Z$-boson, as in the standard DFSZ axion model (cf.~\sect{sec:DFSZ}).
Labelling as $H_1$ the Higgs doublet that couples to the up and charm quarks 
and considering the case in which  all the generations have a non-vanishing anomaly but 
there is a cancellation  $N_2+N_3=0$.   
The anomaly coefficient of the light quarks  can be written as 
$2 N_\ell=  \mX_1-\mX_2$, and 
from this we obtain  $c^0_u-c^0_d=
\frac{1}{2N}(\mX_1+\mX_2)=s^2_\beta-c^2_\beta$. 
The second condition for nucleophobia that  requires $c^0_u-c^0_d  = f_{ud} $ is then realised for
 $s^2_\beta \approx  2/3$ that is  $\tan^2\beta =  \frac{m_d}{m_u}\approx 2$   
(a value for which the top Yukawa coupling remains safely perturbative up to the Planck scale). 
In summary, while $C_{ap}+C_{an}\approx 0$ is enforced just by charge assignments and does not require
any tuning of the parameters,  $C_{ap}-C_{an}\approx 0$ requires a specific choice  $\tan\beta \approx \sqrt{2}$.

It is also worthwhile mentioning that  in nucleophobic  models the requirement that a single generation 
of quarks contributes  to the QCD anomaly also allows straightforwardly for  $N_{\rm DW} = 1$. 
There are  in fact two ways in which $H_{1,2}$  can be coupled to the PQ symmetry breaking field: 
$H^\dagger_2 H_1\Phi$ in which case $|\mX_\Phi| = 2N_\ell=2N$, the axion field has the same periodicity than
the $\theta$ term, and the number of domain walls is $N_{\rm DW}=1$. The other possibility 
is $H^\dagger_2 H_1\Phi^2$, in which case $|\mX_\Phi| = N_\ell=N$ and  $N_{\rm DW}=2$.
In contrast, in  conventional DFSZ models one gets respectively  $N_{\rm DW} = 3$ and $6$. 




\subsection{Enhanced axion couplings and astrophysics}
\label{sec:Enhanced_couplings_and_astrophysics}

As seen in \sect{sec:Astro_bounds}, limits from stellar evolution strongly constrain the allowed axion parameter space. 
The strongest astrophysical limit on the axion-photon coupling is derived from observations of the R-parameter in globular clusters.
This is also known as the HB bound and constraints the axion-photon coupling down to $g_{a\gamma}\lesssim 0.65\times 10^{-10}\,{\rm GeV}^{-1}$ (cf. \sect{sec:Astro_bounds_gag}).
The axion coupling to electrons is, instead, strongly constrained by RGB 
stars and by WDs. 
As discussed in \sect{sec:Astro_bounds_gae}, these bounds amount to roughly $g_{ae}\lesssim 2-3\times 10^{-13}$, depending on the particular observable and analysis. 

Interestingly, these astrophysical bounds do not depend on the pseudoscalar mass and are valid up to masses of the order of the stellar interior temperature (several keV).  
Therefore, such bounds can be readily applied to axion models beyond benchmark, like the ones discussed above in this section.

\begin{figure}[t]
\begin{center}
\includegraphics[width=.7\textwidth]{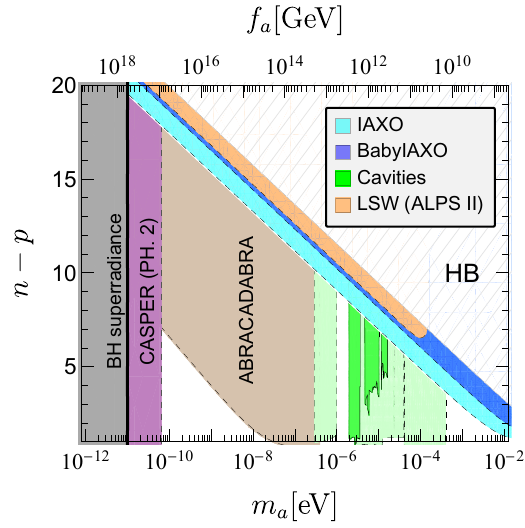}
\caption{\label{fig:KSVZClockWork}
Parameter space (mass and $n-p$) for the clockwork construction (KSVZ type) discussed in  \sect{sec:gag-clockwork}. 
We remind that $3^n$ and $3^p$ are, respectively, the PQ charges of the 
scalar singlets that give masses to the exotic quarks and exotic leptons.
Notice that $n-p$ is an integer $\geq 1$.
Following the convention used in \sect{sec:Experiments}, we are shading with a lighter green the expected sensitivity of next generation axion cavity searches and in darker green the reported results of current haloscopes. 
Moreover, we are using a dashed contour line for sensitivity and a continuous line for experimental results.
In light green, from left to right, we find KLASH, CULTASK (combined sensitivity of CAPP-12TB and CAPP-25T), and MADMAX. 
}
\end{center}
\end{figure}

Let us first notice that the WD/RGB bound dominates over the HB constraint in all models in which $C_{ae}$ and $C_{a\gamma}$, defined in \eqn{eq:gagammagaf}, satisfy the condition $C_{ae}/C_{a\gamma}\gtrsim 10^{-2}$.
This is easily satisfied in the KSVZ-like model discussed in \sect{sec:gag-clockwork} and \ref{sec:gaeKSVZenhanced}, in which the coupling with electrons is suppressed by a loop factor $\sim 10^{-4}$ (cf. \eqn{eq:Cae}).
Hence, for such model the astrophysical bounds on the axion-electron coupling can be ignored, much the same as in the benchmark KSVZ axion. 
The parameter space is shown  in \fig{fig:KSVZClockWork}.
Notice that, thanks to the strong enhancement of the axion-photon coupling, the model is accessible to a large number of the axion experiments expected to take data in the near future, including ALPS II and BabyIAXO (cf. \sect{sec:Experiments}). 
The maximal value of $n$ shown in figure can be easily accommodated and is well below  the threshold required from the LP condition presented in \sect{sec:KSVZ-like}.

\begin{figure}[t]
\begin{center}
\includegraphics[width=.45\textwidth]{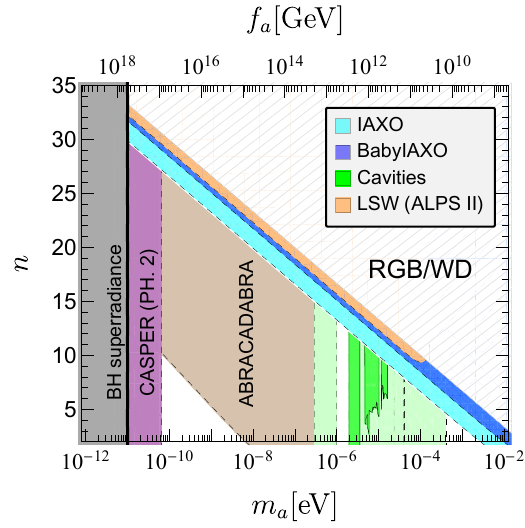}
\hspace{0.3 cm}
\includegraphics[width=.45\textwidth]{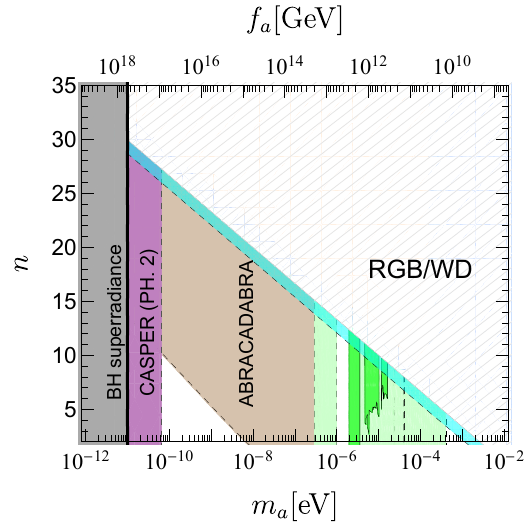}
\caption{\label{fig:DFSZClockWork}
DFSZ model with enhanced axion-photon and axion-electron couplings, discussed in \sect{sec:DFSZ-like} and \sect{sec:gaeDFSZenhanced} for $C_{ae}/C_{a\gamma}$ minimal (left) and maximal (right).
Following the convention used in \sect{sec:Experiments}, we are shading with a lighter green the expected sensitivity of next generation axion haloscope and in darker green the reported results of current haloscopes. 
Moreover, we are using a dashed contour line for sensitivity and a continuous line for experimental results.
In light green, from left to right, we find KLASH, CULTASK (combined sensitivity of CAPP-12TB and CAPP-25T), and MADMAX. 
}
\end{center}
\end{figure}

A similar enhancement of the axion-photon coupling is expected in the DFSZ-type models discussed in \sect{sec:DFSZ-like}.
However, in this case the axion electron coupling is also naturally very large and the condition $C_{ae}/C_{a\gamma}\gtrsim 10^{-2}$ is always  expected, making the HB bound irrelevant. 
To see this, let us notice that from \eqn{eq:ENclockwork} and \eqn{eq:Xeclockwork},
it follows that (for large $n$)
\begin{align}
\label{eq:clockworkCeCgamma}
\frac{C_{ae}}{C_{a\gamma}} \simeq 
 \frac{1}{6 }\[\frac{v_d^2}{v^2} + \sum_{k=1}^{n} \(1-\frac{1}{2^{n+1-k}}\) \frac{v_k^2}{v^2}\]\, ,
\end{align}
where  $v^2 = v^2_e+v^2_d+\sum_{k=1}^n v^2_k$ is the square of  the electroweak VEV and $v_1\equiv v_u$.
Mathematically, the minimum of \eqn{eq:clockworkCeCgamma} corresponds to the (unphysical) condition  $v_e=v$, which implies that all the other VEVs are zero and $C_{ae}/C_{a\gamma}=0$.
However, perturbative unitarity bounds on the top Yukawa coupling (cf. \sect{sec:DFSZ}), de facto require the term in parenthesis in \eqn{eq:clockworkCeCgamma} to be always larger than $\sim$ 0.1, implying a finite lower bound on $C_{ae}/C_{a\gamma}\gtrsim 10^{-2}$.
A lower value of  $C_{ae}/C_{a\gamma}$ for such models is, if mathematically possible, very unnatural. 
The axion parameter space for the case of minimal $C_{ae}/C_{a\gamma}$
 is shown in the left panel of \fig{fig:DFSZClockWork}.
The potential of the axion helioscopes is minimally impacted. 
In fact, the increase of the RGB/WD bound is (partially) compensated by a more efficient production of solar axions, as discussed in \sect{sec:Helioscopes}.
The predicted sensitivity of ALPS II is also sufficient to explore part of the region below the astrophysical bounds, though this section is clearly reduced with respect to the KSVZ-like model. 

The axion parameter space for the higher coupling scenario, when the term in parenthesis in \eqn{eq:clockworkCeCgamma} is of order 1, is shown in the right panel of \fig{fig:DFSZClockWork}.
In this case, only a full scale IAXO (not BabyIAXO), among the  planned future helioscopes, would have the capability to probe the axion parameter space allowed by the RGB/WD bounds. 
ALPS II would equally be unable  to probe such model below the region excluded by astrophysics. 

As evident from the figures, the large axion-photon couplings predicted in such models permit an almost complete exploration of their parameter space by the next generation of axion probes. 
In particular, several haloscope searches are expected to have enough sensitivity to probe the parameter region all the way to the minimal value of the axion-photon coupling (corresponding to $n=2$).  

Nucleophobic axion models are also phenomenologically interesting, in particular in relation to the SN 1987A and various NS axion bounds, discussed in \sect{sec:Astro_bounds_gaN}. 
Although the SN and NS constraints on the axion couplings are still uncertain, they seem to restrict efficiently the axion parameter space at masses higher than a few 10 meV, if the standard mass-coupling relation is assumed (see, e.g., \fig{fig_DFSZ_astro_bounds} for DFSZ axions).
These constraints show some tension with the axion interpretation of the stellar cooling anomalies, which favour slightly higher masses~\cite{Giannotti:2017hny}.
In this respect, nucleophobia would improve the significance of the interpretation of the stellar cooling anomalies in terms of axions.  
Moreover, experimentally, nucleophobia opens up large sections of the parameter space to experiments, primarily helioscopes, which are sensitive to the higher axion mass region.

\begin{figure}[t]
\begin{center}
\includegraphics[width=.7\textwidth]{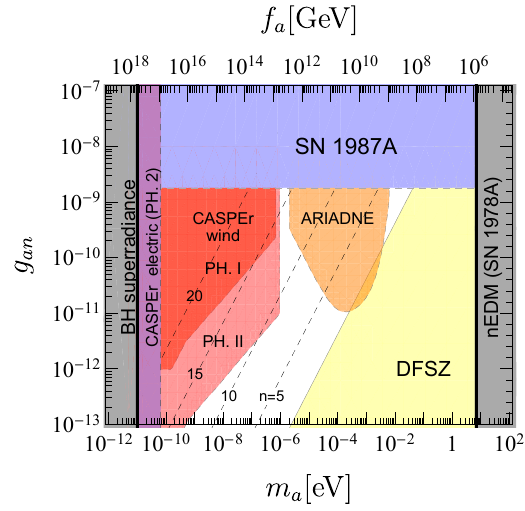}
\caption{\label{fig:nucleophilic}
Current bounds and prospectives on the axion nucleon coupling.
The black dashed lines correspond, from right to left, to $n=5$, 10, 15, and  20 for the model discussed in \sect{sec:gaNenhanced} (see, in particular, \eqn{eq:expPQup}).
}
\end{center}
\end{figure}

Models that enhance the axion nucleon coupling may also find interesting phenomenological applications. 
A recent analysis~\cite{Buschmann:2019pfp}, which ascribes the observed excess of X-rays from a few nearby neutron stars to ALPs coupled to nucleons and photons, could be interpreted in terms of nucleophilic QCD axions of the kind described in \sect{sec:gaNenhanced}.
The interpretation given in Ref.~\cite{Buschmann:2019pfp} asks for a mass below a few $\mu$eV, with $g_{a\gamma}g_{aN}\sim $ a few $10^{-21}\,{\rm GeV^{-1}}$, a value several orders of magnitude larger than what expected, for example, in the DFSZ axion model. 
Such large couplings could, however, be possible in the case of the clockwork models described here, with enhanced coupling to photons and/or nucleons. 
Assuming, for example, a standard axion-photon coupling, the hinted region could be accessible to an axion of the kind presented in 
\sect{sec:gaNenhanced}, with $m_a\sim 1\,\mu$eV and $n\sim 30$ (Cf. \eqn{eq:expPQup}), easily accessible to CASPEr Electric (see \fig{fig:nucleophilic}).

Regardless of the phenomenological motivations, nucleophilic axions are viable axion models that could be probed in the near future, a fact that should encourage the exploration of the axion-nucleon parameter space, still quite poorly probed. 
An enhancement of the axion-neutron coupling by a factor larger than $10^{3}$ with respect to the DFSZ axion, corresponding to $n\sim 10$ in the model in \eqn{eq:expPQup}, would make the model sensitive to CASPEr wind in phase II. 

%
%


%

\subsection{Flavour violating axions}
\label{sec:gaFCNC}

Assuming that axions exist and that the PQ symmetry acts on SM particles,  
the possibility that they couple differently  to fermions of the same 
charge but of different generations is quite plausible.  There is in fact no fundamental reason why  
the global and anomalous PQ symmetry should act universally in flavour space as the gauge interactions do.  
After all, this possibility  is almost  as old as the axion, as it was already 
contemplated in the Bardeen and Tye seminal paper~\cite{Bardeen:1977bd} in 1978.  
This idea may then be daringly expanded  to speculate whether the  PQ symmetry can have 
something to do with flavour, that is with  the  pattern of fermions masses and mixing angles in the SM.  The PQ symmetry could for example 
act as a flavour symmetry, or could emerge from a set of genuine  flavour symmetries, a point of view that had 
been  advocated by Wilczek already in the early 80's~\cite{Wilczek:1982rv}.

In the second half of the 80's the possibility of an axion coupled to the SM fermions in a generation 
dependent way  was fired up  by the observation  in heavy ion collisions at GSI 
of  a sharp peak  at $E(e^+) \sim 300\,$keV  in the positron spectrum~\cite{Schweppe:1983yv,Clemente:1983qh,Cowan:1985cn}. 
The positron data  were later found to be correlated with the simultaneous detection of electrons  
also featuring in their energy spectrum a narrow peak 
at the same energy~\cite{Cowan:1986fj}. The data  were consistent with the production of a particle of mass  $1.6-1.8\,$MeV,  
of probable pseudoscalar character~\cite{Schafer:1985qb}, which then decayed  
into $e^+e^-$. Hence the  characteristics of such a particle were well consistent 
with those of the original WW axion~\cite{Weinberg:1977ma,Wilczek:1977pj}. 
However, this interpretation was challenged by the rather large value of the mass, which implied 
enhanced couplings, e.g.~to either charm or bottom quarks,  in conflict with limits 
from $J/\psi \to \gamma a$ or $\Upsilon \to \gamma a$. These limits, however,  
required the axion to escape from the detectors with no deposit of energy, and could 
have been  circumvented by  short lived axions  $\tau_a \lsim 6\times 10^{-13}\,$s~\cite{Balantekin:1985kk}, 
due to very large values of the axion-electron coupling $\gae$. 
However, for generation independent PQ symmetries  the ratio of the axion-electron and  
axion-muon couplings is determined by the ratio of the lepton masses as $g_{a\mu}/\gae = m_\mu/m_e$, see \eqn{eq:gagammagaf}. 
This predicted a large enhancement of $g_{a\mu}$, in plain conflict  
with existing measurements of $(g-2)_\mu$. 
Generation dependent axion couplings remained the only way out, and a series of interesting papers 
analysing this possibility appeared~\cite{Peccei:1986pn,Krauss:1986wx,Bardeen:1986yb,Krauss:1987ud,Geng:1988nc}.
While eventually new experimental results quickly ruled out the $1.8\,$MeV GSI 
axion~\cite{Mageras:1986nz,Bowcock:1986ig,Brown:1986xs,Hallin:1986gh,Davier:1986qq,Riordan:1987aw,Bjorken:1988as}, 
the road to further develop  generation dependent PQ scenarios, and to explore   
axion-flavour interconnections was paved. 

Recently, several models attempting to relate fermion family symmetries to the PQ  
symmetry have been put forth. 
Well motivated  realisations  identify the U(1)$_{\rm PQ}$ with the
horizontal $U(1)$ symmetry responsible for the Yukawa hierarchies
\cite{Davidson:1981zd,Davidson:1983fy,Davidson:1984ik}.
Models of Froggatt-Nielsen \cite{Froggatt:1978nt} type have recently
regained some attention 
(see e.g.~Refs.~\cite{Ema:2016ops,Calibbi:2016hwq,Bonnefoy:2019lsn,Alanne:2018fns}),
in relation to possible solutions to the strong CP problem. 
They typically predict axion flavour transitions controlled by the CKM matrix,
although subject to built-in $O(1)$ uncertainties, which are  
intrinsic to most  flavour models based on $U(1)$. 
Flavoured PQ symmetries can also arise in the context of 
Minimal Flavour Violation \cite{Albrecht:2010xh,Arias-Aragon:2017eww}  
or in models based on non-abelian horizontal (gauge) symmetries like $SU(3)_F$, 
which can lead to an almost\footnote{For three chiral families 
trilinear PQ symmetry breaking terms need to be forbidden in the 
scalar potential, so although technically natural the PQ symmetry is still imposed by hand. 
On the other hand, in the presence of $n_g$ chiral families the 
gauging of $SU(n_g)$ would have delivered 
a truly accidental axion for $n_g>4$, 
since the first operator breaking the PQ symmetry is of dimension $n_g$.} 
accidental global $U(1)$'s which can play the role of a PQ 
symmetry~\cite{Berezhiani:1985in,Berezhiani:1989fp}. 
In both these types of constructions the resulting axion  corresponds   to a pseudoscalar 
`familon' that can mediate FCNC
transitions much alike the axions of $U(1)$ flavour models. 
 A different motivation for the non-universality of the PQ current,
  that  was advocated in~\cite{DiLuzio:2017ogq,Bjorkeroth:2019jtx}, 
 is that  of  constructing (nucleophobic)  axion models in which the 
 tightest astrophysical bounds can be circumvented,
see the review in~\sect{sec:nucleophobic}. 
A genuinely different approach is the attempt of maximise the predictive power 
of SM flavour data  by  searching for $U(1)$ symmetries that would enforce the maximal 
number of textures zeros in  the fermion mass matrices  (compatibly with non-vanishing 
masses and mixings)~\cite{Bjorkeroth:2018ipq,Bjoorkeroth:2019ndr}.
It was found in Ref.~\cite{Bjorkeroth:2018ipq} that  all $U(1)$ symmetries suitable to realise 
this requirement in the quark sector have a QCD anomaly, and thus correspond to generation 
dependent PQ symmetries. The particular realisations based on this approach are 
interesting because, differently from most  (if not all) other models,    
allow to  fix the structure of the axion couplings to SM fermions 
in terms of the values of quark masses and mixings, including the 
off-diagonal flavour changing ones,  up to the values of 
$f_a$ and $\tan\beta$, as in conventional DFSZ models.
%

Clearly all the constructions listed above do not have `natural flavour conservation', and thus  
predict new sources of FCNC processes. However,  the latter may be still consistent with 
experiments if the scale $f_a$ suppressing all axion couplings were sufficiently large or if 
flavour transitions mainly affected the second and third generation SM fermions.  

\subsubsection{Generation dependent Peccei-Quinn symmetries}
\label{sec:gaFCNCgen}
\smallskip
 
The important modifications in the structure of the axion couplings 
to the SM fermions, in the case when the PQ symmetry is generation 
dependent, were briefly addressed in~\sect{sec:IntroFlavourViolating}. 
There it was shown that the consequences of dropping the assumption of 
 generation independence of the PQ charges were twofold: 
 firstly, flavour violating (FV)  couplings arise, 
and secondly, besides coupling to axial-vector currents, `flavoured'  axions 
couple to vector currents as well. This latter point arises 
because 
the PQ charges of the fermions cannot maintain 
an exact chiral structure ($\mX_L = -\mX_R$), while in general  
$\mX_L +\mX_R \neq 0$ for the PQ symmetry to be anomalous.   

Before discussing the existing bounds on axion FV interactions 
let us introduce some notations.
In the simple scheme discussed in~\sect{sec:IntroFlavourViolating}, with only two Higgs doublets and 
where only two generations were considered, the axion-fermion couplings  could be easily expressed in 
terms of the  PQ Higgs   charges $\mX_{1,2}$. 
Expressing the fermion couplings  in terms of  the charges of the (two or more) Higgs doublets
is  possible also in more complicated scenarios,  however, the connection depends on the specific model.
Hence, for the seek of generality, in this Section we will express the couplings simply in terms of   
the PQ charges of the SM fermions. 
Our starting point is the way the axion couples to the fermion current, as written in the first line of \eqn{eq:Leffaxion4}, 
 except that  $f_{L,R}$ have now to be understood as  vectors of SM fermions of the same electric charge (e.g.~$f=(u_1,u_2,u_3)^T$)  while  $\bX_{\!f}$ in the equation below is the diagonal  matrix 
 of the associated PQ charges.
 Going from the basis in which the PQ charges are well defined  
 to the mass eigenstate basis  $f_{L,R} \to  U^f_{L,R} f_{L,R}$, 
 with unitary matrices $U^f_{L,R}$, yields 
\begin{equation}
\label{eq:LafFC}
\mathcal{L}_{a} =  - \frac{\partial_\mu a}{2 f_a}\frac{1}{N}
\[\bar f_L \gamma^\mu \(U^{f\dagger}_L \bX_{\!f_L} U^f_{L}\)f_L
+\bar f_R \gamma^\mu \(U^{f\dagger}_R \bX_{\!f_R} U^f_{R}\)f_R\]\,.
\end{equation}
 The light quarks mass eigenstates $u,d$  can then be redefined 
  as in \eqn{eq:qtransfa}  in order to  remove the anomalous $G \tilde G$ term
  from the Lagrangian,  and again this results in   adding to the couplings the 
  model independent contribution in~\eqn{eq:cq1}. 
%
  The flavour-diagonal axion-fermion couplings  receive  
corrections from the mixing:  $ c^0_{f_i} \to  c^0_{f_i} + \delta c^0_{f_i}$ where  
\begin{equation}
\label{eq:mixcorrection}
   \delta c^0_{f_i}  = \frac{1}{2N}
\(W_{f_{iL}} -W_{f_{iR}}\) 
    , \qquad 
\text{with}   
\qquad 
W_{f_{iR}}  = \[ U^{f\dagger}_R (\bX_{\!f_R} - \mX_{f_{iR}} \mathbb{I})  U^f_{R}\]_{ii}\,, 
\end{equation}
where $\mathbb{I}$ is the identity matrix in generation space 
and a similar expression folds for $W_{f_{iL}}$ upon swapping $L \leftrightarrow R$. 
Such a correction 
can be invoked  to improve  nucleophobia  by tuning a cancellation 
against  $C_{a,\rm{sea}}$  in \eqn{eq:cpPcn}, or to achieve 
electrophobia  without recurring to a third Higgs  
by cancelling  $c^0_e$ against the correction from mixings, 
as it was done in  Ref.~\cite{DiLuzio:2017ogq}; 
but except for these two cases the $\delta c^0_{f}$ contributions 
do not change much the overall picture. 
 Of leading importance are instead the off-diagonal terms, that 
 give rise to FV axial-vector and vector axion couplings:
\begin{align}
\label{eq:mixFV}
\mathcal{L}_{af_if_j} &= - \frac{\partial_\mu a}{2 f_a}
\[\bar f_i \gamma^\mu \(C^V_{f_if_j} - C^A_{f_if_j} \gamma_5\) f_j\], \\
C^V_{f_if_j}&=  \frac{1}{2N} \(  U^{f\dagger}_L \bX_{\!f_L} U^f_{L} +U^{f\dagger}_R \bX_{\!f_R} U^f_{R} \)_{ij} ,     \\
C^A_{f_if_j}&=  \frac{1}{2N} \(  U^{f\dagger}_L \bX_{\!f_L} U^f_{L} -U^{f\dagger}_R \bX_{\!f_R} U^f_{R}   \)_{ij} .
\end{align}
The crucial point regarding these couplings is that, while the matrices  of charges $\bX_{\!f_{R,L}}$
are presumably fixed in any specific model, little is known about the mixing matrices $U^f_{R,L}$.\footnote{
The models 
discussed in in Ref.~\cite{Bjorkeroth:2018ipq} are a remarkable exception since, thanks to the 
maximal reduction  in the number of free parameters, both the quark mixing matrices remain 
fixed in terms of measured quantities.}
In the quark sector, nothing is known about the RH  matrices $U^f_{R}$, and their structure remains 
completely arbitrary.   The LH mixings are instead constrained to satisfy $V_{\rm CKM} = U^\dag_{u_L} U_{d_L}\approx \mathbb{I}$.
This, however,  only implies  $U_{u_L} \approx U_{d_L}$ but does not provide additional information
on the size of the off-diagonal entries, and in particular it does not forbid large flavour mixings. 
Differently from the quark sector, the lepton sector is characterised by large mixings, but 
one does not know if they originate from the neutrino or from the charged lepton rotation matrices  (or from both), 
so that for example also $U^\ell_L \approx \mathbb{I}$ remains a viable assumption.    


%


\subsubsection{Constraints on flavour violating axion couplings}
\label{sec:gaFCNCexp}

Searches for FV decays involving invisible final states are the main experimental tool  
to probe the off-diagonal axion couplings  $C^{A,V}_{f_if_j}$. 
A general analysis of such flavour-changing processes involving a generic massless 
NGB, 
which holds also for FV axion, can be found  in Ref.~\cite{Feng:1997tn}.  
A recent thorough collection of limits  on the FV couplings 
can be found  in  Ref.~\cite{Bjorkeroth:2018dzu}, and    
brief reviews about the status of the art  are presented in 
Refs.~\cite{Ziegler:2019gjr,DiLuzio:2019mie}.\footnote{Close to the completion of this review, 
Ref.~\cite{MartinCamalich:2020dfe} appeared, where 
the importance of flavour violating transitions for axion searches
was reiterated,  and additional limits  on FV axion couplings 
from   three-body meson decays and  baryonic decays,  including 
the decay  $\Lambda \to na$ that would represent a new cooling 
mechanism  for the  SN1987A,   were derived. 
}
The currently best limits for each type of FV transition  are 
listed in  \Table{tab:LimitsFV}. They coincide with the limits given in Ref.~\cite{Bjorkeroth:2018dzu}
after the correspondence  between their and our  notations 
($V^f_{ij}/{v_{\rm PQ}} \equiv C^V_{f_if_j}/{(2f_a)}$) is accounted for. 

The strongest bounds on FV axion couplings to quarks come from  
meson decays into final states containing invisible particles. 
Note, however, that decays involving   initial and final pseudo-scalar mesons like $P = K,B,D,\pi$  
are only sensitive to the vector part of the FV quark current,
since  $\langle P' |  J_\mu^5  | P \rangle = 0$
by the Wigner-Eckart theorem. 
Searches for $K^+ \to \pi^+ a$ decays provide  the tightest limits.
The current bound from E949/E787~\cite{Adler:2008zza} 
(see   \Table{tab:LimitsFV})  implies   
$m_a < 17 \cdot |C^V_{sd}|^{-1}\mu\text{eV}$ which,  if one assumes  that 
$C^V_{sd}$ 
is not particularly suppressed, is about three orders of magnitude stronger than typical astrophysical bounds.
In the next future, NA62 is expected to improve the limit on ${\rm Br}(K^+ \to \pi^+ a)$ by a factor of
$\sim\,$70~\cite{Anelli:2005ju,Fantechi:2014hqa}, thus
strengthening the axion mass bound by a factor $\sim\,$8.
The  most sensitive processes involving a quark of the third generation are 
$B^\pm \to \pi^\pm a$ and   
$B^\pm \to K^\pm a$.
Present limit from CLEO~\cite{Ammar:2001gi}  imply respectively 
$m_a < 11.4 \cdot 10^{-2} \cdot  |C^V_{bd}|^{-1}\,$eV and
$m_a < 9.5 \cdot 10^{-2} \cdot  |C^V_{bs}|^{-1}\,$eV which, 
 for maximal mixing, are close  to the astrophysics bounds.
Note that the latter  bound could be presumably strengthened by 
a factor $\sim 6$ 
at BELLE II \cite{Abe:2010gxa}.
Processes involving FV transitions between  up-type quarks
 are much less constrained.   Decays of charmed mesons of the
form $D\to \pi a$  are only  subject to the trivial requirement  ${\rm Br}(D\to \pi a)  < 1 $, 
which can be translated into weak bounds,  which are at least 
two orders of magnitude worse than the ones for down-type quarks 
and not competitive with limits from  astrophysics.

As we have said meson decays can only constrain  FV vector couplings. 
In order to set bounds on the axial-vector FV couplings
one has to resort to other flavour changing processes,
as for example neutral meson ($K^0,D^0,B^0_{d},B^0_{s})$ mixing,  since 
meson mass splittings receive from 
axion interactions  an
additional contribution  $(\Delta m)_a$. However, 
in spite of the fact that,  for example,  the measurement  of the mass difference in the neutral kaon system
$\Delta m_K /m_K \simeq 0.7  \times 10^{-14}\,$ gives a number that is four orders of magnitude smaller 
than the limit from kaon decays  Br$(K^+\to \pi^+ a)\lsim 0.7\times 10^{-10}$, the sensitivity to this type of new physics of the former observable 
is not competitive with the sensitivity of the latter.
To understand the reasons for this, let us write the approximate expressions 
for the relevant quantities in the game: 
\begin{align}
\label{eq:DeltaK} 
\frac{(\Delta m_K)_a}{m_K}&\simeq     \frac{f_K^2}{f_a^2} |C^A_{sd}|^2, \\ 
\label{eq:Kpia}
\Gamma(K^+\to \pi^+ a) &\simeq  \frac{m_K^3}{16\pi f_a^2}    |C^V_{sd}|^2, \\  
\label{eq:Kmunu}
\Gamma(K^+\to \mu^+\nu) &\simeq  \frac{m_K}{8\pi} \(G_F f_K m_\mu |V_{us}|\)^2 \,,
\end{align}
where $f_K$  is  the kaon decay constant,  $G_F$ the Fermi constant, and $V_{us}$ the relevant 
element of the CKM matrix.  
The leptonic decay in \eqn{eq:Kmunu} has the largest branching ratio $\simeq 63.4\%$,
so we approximate  $\Gamma_K^{\rm tot} \approx  \Gamma(K^+\to \mu^+\nu)$ and we obtain:
\begin{equation}
\label{eq:BrKpia}
{\rm Br}(K^+\to \pi^+ a) \simeq  \(G_F f_K m_\mu |V_{us}|\)^{-2}  \times   \frac{m^2_K}{2f^2_a}  |C^V_{sd}|^2\,. 
\end{equation}
The prefactor in the RHS of this equation accounts for chirality (and Cabibbo) suppression of the SM decays, and 
it is not present for  $K^+\to \pi^+ a$ decays.  This    factor is huge $\sim 5 \times 10^{14}$, and  enhances the 
effects of $C^V_{sd}$  largely overcompensating for the better precision of $\Delta m_K$. As a result    
the limits on $f_a/ |C^V_{sd}|$  are more than five order of magnitude better than the corresponding 
limits on $f_a/ |C^A_{sd}|$.

Differently from  pseudoscalar  mesons, for two-body charged lepton decays $\ell_i  \to \ell_j\, a$  
 both vector and axial-vector couplings contribute because  the decaying particle has non-zero spin.
 Experimentally it is more convenient to search for decays of anti-leptons. The angular 
differential decay rate for two-body decays is 
\begin{equation}
\label{eq:Gmuea}
\frac{d \Gamma(\ell_i^+ \to \ell_j^+ a)}{d \cos\theta} =
\frac{m_{\ell_i}^3}{128\pi f^2_a}\[|C^V_{\ell_i\ell_j}|^2 +|C^A_{\ell_i \ell_j}|^2 
+ 2 \Re(C^A_{\ell_i \ell_j} C^{V^{\scriptstyle *}}_{\ell_i \ell_j} )P_{\ell_i}\cos\theta\] \, , 
\end{equation}
where $P_{\ell_i}$ is the polarisation vector of the decaying particle.
Strong bounds have been obtained  from searches for  $\mu^+\to e^+ a $ decays, the best 
of which was obtained more than thirty years ago at TRIUMF, giving
Br$(\mu^+ \to e^+ a)< 2.6\cdot 10^{-6}$ \cite{Jodidio:1986mz}.
This bound, however, was obtained by exploring  a kinematical region forbidden for 
$\mu^+ \to e^+ \nu \bar\nu$ SM decays.   Namely, the bound  holds only 
 if  the decay is purely vector or purely axial-vector, 
implying in this case $m_a < 2.5\cdot \big|C^{V(A)}_{\mu e}\big|^{-1}\,$meV.  
However, the bound  would  evaporate  for axion interactions with a  SM-like $V-A$ structure. 
Slightly weaker bounds, which however  do not depend on the chirality properties of the 
coupling, were obtained  by the Crystal Box experiment by searching for the radiative decay  
$\mu^+\to e^+ \gamma a $~\cite{Goldman:1987hy,Bolton:1988af}, see \Table{tab:LimitsFV}. 

Assuming the decays are isotropic, or in case anisotropic decays are 
explicitly searched for,  it is convenient to express the limits in terms of  
an effective coupling 
\begin{equation}
\label{eq:Cetot}
C_{\ell_i\ell_j} = \( |C^V_{\ell_i\ell_j}|^2 + |C^A_{\ell_i\ell_j}|^2\)^{1/2}. 
\end{equation}
Recent searches that explicitly evaluate  limits for anisotropic two body muon decays
have been carried out  by the TWIST collaboration~\cite{Bayes:2014lxz}, and yield 
$f_a < (1.0-1.4) \cdot C_{\mu e}$ depending on the anisotropy of the decay. 
These bounds are  slightly less stringent than the old  limits from  Crystal Box.
In the next future, the MEG \cite{Renga:2018fpd} and Mu3e \cite{Blondel:2013ia} experiments at PSI
are expected to improve the bounds on $\mu$-$e$ transitions by about one order of magnitude. 
Bounds on $\tau^+\to \mu^+ a$ and $\tau^+ \to e^+ a$  FV decays 
have been obtained by the ARGUS collaboration~\cite{Albrecht:1995ht}, and are also reported in ~\Table{tab:LimitsFV}. 
However, they imply limits on the axion mass which remain well below the astrophysical limits.
\begin{table}[t!] 
\renewcommand{\arraystretch}{1.2}
\centering
\begin{tabular}{@{} |l|l|r|l| @{}}
\hline
Decay  &  Branching ratio  & Experiment/Reference  & \qquad $f_a$ (GeV)  \\ 
\hline
\hline
$K^+ \to \pi^+ a$  &  $< 0.73 \times 10^{-10} $ &  E949+E787~\cite{Adler:2008zza}  &  $> 3.4\times 10^{11}\, |C^V_{sd}| $\\ 
$B^\pm \to \pi^\pm a$  &  $< 4.9 \times 10^{-5} $ &  CLEO~\cite{Ammar:2001gi}  &  $> 5.0\times 10^{7}\, |C^V_{bd}| $\\ 
$B^\pm \to K^\pm a$  &  $< 4.9 \times 10^{-5} $ &  CLEO~\cite{Ammar:2001gi}  &  $> 6.0\times 10^{7}\, |C^V_{bs}| $\\ 
\hline
$D^\pm \to \pi^\pm a$  &  $< 1 $ &  &  $> 1.6\times 10^{5}\, |C^V_{cu}| $\\ 
\hline
$\mu^+ \to e^+  a$  &  $< 2.6 \times 10^{-6} $ & TRIUMF~\cite{Jodidio:1986mz}  &  $> 4.5\times 10^{9}\, |C^{V(A)}_{\mu e}| $\\ 
$\mu^+ \to e^+\gamma  a$  &  $< 1.1 \times 10^{-9} $ & Crystal Box~\cite{Bolton:1988af}  &  $> 1.6\times 10^{9}\, C_{\mu e} $\\ 
$\tau^+ \to e^+ a$  &  $< 1.5 \times 10^{-2} $ & ARGUS~\cite{Albrecht:1995ht}  &  $> 0.9\times 10^{6}\, C_{\tau e} $\\ 
$\tau^+ \to \mu^+  a$  &  $< 2.6 \times 10^{-2} $ & ARGUS~\cite{Albrecht:1995ht}  &  $> 0.8\times 10^{6}\, C_{\tau \mu} $\\ 
\hline
  \end{tabular}
  \caption{
  \label{tab:LimitsFV} 
Limits on FV axion couplings to the SM fermions. 
The TRIUMF limit on $\mu^+\to e^+ a$ holds only for purely vector or purely axial-vector interactions. 
For the other leptonic transitions the total coupling $C_{\ell_i\ell_j}$ is defined in \eqn{eq:Cetot}.
} 
\end{table}

\subsection{Extending the mass region region for  dark matter axions}
\label{sec:extendingma}

As we have discussed in \sect{sec:misalignment}, the contribution to
DM from the axion misalignment mechanism is evaluated by
solving a second order differential equation for the misalignment
angle $\theta(t)$ with time dependent coefficients $H(t)$ and
$m_a(t)$. 
In pre-inflationary scenarios the axion field is homogenised over
distances much larger than the horizon, and the spatial derivative term in 
\eqn{eq:equation_motion1} can be dropped. 
In post-inflationary scenarios field modes with wavelength $\lambda(t) \ll t$ 
are quickly redshifted away, and restricting to the relevant  
super-horizon modes we can 
again drop spatial derivatives, so that~\eqn{eq:equation_motion1} becomes
\beq
	\label{eq:equation_motion_t}
	\ddot \theta(t) + 3 H(t) \dot \theta(t) +  m^2_a(T(t)) \theta(t) = 0\,.
\eeq
 %
If the coefficients $H(t)$ and $m_a(t)$ have
a power-law dependence on time, then \eqn{eq:equation_motion_t} admits
an exact solution. For example, let us assume that $R(t) \propto t^p$,
where $p > 0$ is a new constant describing the cosmological model
($p=1/2$ for radiation-domination). We also assume that
$m_a(T) \propto T^{-\gamma}$, see \eqn{eq:QCDaxion_mass}, and $\gamma$
is related to the exponent governing the temperature dependence of the
topological susceptibility. Assuming that the entropy   (see~\eqn{eq:rho_and_s}) 
is conserved within a comoving volume,
$d(s R^3)/dt = 0$, and neglecting the change in the number of degrees of freedom
$g_S(T)$, we have $T \propto 1/R \propto t^{-p}$ and
$m_a(t) \propto t^{\gamma p}$. Under these conditions,  
the general solution of \eqn{eq:equation_motion_t} is 
\beq
	\label{eq:sol_equation_motion_t}
	\theta(t) = \( \frac{m_a \,t^\alpha} {2\alpha}   \)^\beta \[
	C_1   \Gamma(1+\beta)   J_\beta \( \frac{m_a \, t^\alpha}{\alpha}\) +
	C_2  \Gamma(1-\beta)    J_{-\beta}\( \frac{m_a \, t^\alpha}{\alpha}\)
	\]\,,
\eeq
where $\alpha=(2+\gamma p)/2$, $\beta = (1-3p)/(2+\gamma p)$,
$\Gamma(x)$ is the Euler gamma function of argument $x$,
$J_\kappa (x)$ is the Bessel function of the first kind of order
$\kappa$, and $C_{1,2}$ are integration constants. The case studied in
\sect{sec:section2} corresponds to a cosmological evolution during
radiation domination, that is $p=1/2$, with the topological susceptibility
obtained from lattice simulation, which roughly corresponds to
$\gamma \approx 4$ for $T>T_C$ and $\gamma \approx 0$ for $T\ll T_C$, see
the discussion below \eqn{eq:QCDaxion_mass}. We have also set the
initial conditions $\dot\theta =0$ and $\theta = \theta_i$ at the time
at which the PQ symmetry breaking occurs, see
\sect{sec:misalignment}. Several studies have been carried out under
these assumptions, which suggest that the DM energy density is 
saturated for an axion mass lying within a window
$m_a \sim \(10-100\)\,\mu$eV.
This mass window can be significantly altered by various non-standard 
conditions:
%
%

\begin{enumerate} 

\item Assuming $\theta_i \ll O(1)$ in pre-inflationary PQ breaking
  scenarios. This possibility has been already discussed in
  \sect{sec:misalignment};

\item Assuming a non-standard cosmological evolution, that is by
  modifying the evolution of the Hubble parameter $H(t)$. This
  possibility will be explored in \sect{sec:modifiedcosmo};

\item Altering the functional dependence of $m_a(t)$ by appealing
  to beyond-the-SM particle physics.  Some examples that exploit this
  possibility will be reviewed in \sect{sec:mirror}; 

\item Assuming a sufficiently large initial value $\dot\theta_i \neq 0$, see
  \eqn{eq:dottheta-limit}.

  \item Axion production  via  parametric   resonance in the decay of the PQ field radial mode. 
\end{enumerate} 
  The last two possibilities will be reviewed in  \sect{sec:thetadot}.

%


\subsubsection{Non-standard cosmological evolution}
\label{sec:modifiedcosmo}
%
%
The evolution of the Hubble parameter in the early Universe can be modified in different ways, 
and this can alter the mass region in which the relic density of axions  saturates 
the DM density with respect to the conventional scenario reviewed in~\sect{sec:axionmassbounds}.
We will now review some of these possibilities. \\

\noindent
{\it Entropy generation:}
The standard computation of the present axion energy density relies on
the conservation of the entropy in a comoving volume, see 
\eqn{eq:numberdensity0}. Suppose now that a new species $X$, like a
massive scalar field with $g_X$ degrees of freedom, is present in the
early Universe and decays into thermalised products after the axion
field starts to oscillate at $T_{\rm osc}$, but prior to 
BBN. 
The present amount of entropy is increased by a
factor $\Delta \equiv 1 + g_X/g_S$, where $g_S$ is the number of
entropy degrees of freedom in the SM at $T_{\rm osc}$. In this
scenario, the present axion energy density in \eqn{eq:VRMaxions} would
be lowered by the entropy injection~\cite{
Steinhardt:1983ia, Lazarides:1987zf,  Lazarides:1990xp,Kawasaki:1995vt}  
by the same factor $\Delta$.
If we demand that the axion energy density matches that of  
CDM  in spite of entropy dilution, then  from~\eqn{eq:standarddensity} we obtain  that 
the mass of the axion has to be lower  with respect to the conventional value 
 derived in~\sect{sec:misalignment}  as 
%
\beq
	m_a \to  m_a \,\Delta^{-\frac{2+\gamma}{3+\gamma}}\,.
	\label{eq:axionmass_diluted}
\eeq
In supersymmetric theories in which the axion forms a supermultiplet
this scenario can be particularly motivated.  
Late  decays of the scalar superpartner of the axion can 
release a large entropy at a late epoch of the Universe’s 
evolution~\cite{Choi:1996vz,Hashimoto:1998ua},  diluting  the axion energy density and  raising 
the upper bound of the axion decay constant up to $f_{a}\sim 10^{15}\,$GeV
without conflicting with the observed amount of DM. Further studies of the supersymmetric realisation 
of  the entropy dilution mechanism and specific models can be 
found in Refs.~\cite{Kawasaki:2011ym,Baer:2011eca,Bae:2014rfa,Co:2016vsi}.\\



\noindent
{\it Unconventional cosmologies:}
If the evolution of the Universe at temperatures around or below
$\sim 1\,$GeV is characterised by a period of non-canonical expansion,
the onset of axion oscillations would be altered and that would lead to a
different axion relic density today. Thus,  non-standard cosmologies
can enlarge the mass range for axion DM.

Such a scenario might occur for example if the evolution of the
Universe is described by a scalar-tensor gravity
theory~\cite{jordan1955schwerkraft, Fierz:1956zz, Brans:1961sx} rather
than by general relativity. Scalar-tensor theories benefit from an
attraction mechanism which at late times makes them flow towards
standard general relativity, so that discrepancies with direct
cosmological observations can be avoided. Some consequences of a
modified expansion rate due to a scalar-tensor theory have already
been discussed in the literature in relation to a possible large
enhancements of the WIMP DM relic
density~\cite{Catena:2004ba, Catena:2009tm, Meehan:2015cna,
  Dutta:2016htz, Dutta:2017fcn}, or to lower the scale of leptogenesis
down to the TeV range~\cite{Dutta:2018zkg}.

Another possibility consists in enlarging the particle content of the
SM, by including a new particle which at early times
dominates the expansion rate of the Universe. This possibility also
results into a modification of the cosmic evolution that departs from
the $\Lambda$CDM predictions. Scenarios of this type have
been extensively discussed in the literature, in relation to the
expected abundance of WIMPs~\cite{Gelmini:2006pq, Gelmini:2006mr,
  Gelmini:2008sh, Erickcek:2011us, Redmond:2017tja, Visinelli:2017qga}
and their free-stream velocity~\cite{Visinelli:2015eka}. A similar possibility 
has been explored in relation to axion physics both
for hot~\cite{Grin:2007yg} and cold relics~\cite{Visinelli:2009kt,
  Visinelli:2017imh, Draper:2018tmh, Ramberg:2019dgi, Blinov:2019rhb}. 
In brief, we assume that a new field $\phi$ coexists along with the SM particles and
comes to dominate the energy density in the early Universe, down to a
decay temperature $T_{\rm dec}$ that marks the transition to the
standard cosmological scenario. If we assume that the new particle has
an equation of state $w_\phi$ and energy density $\rho_\phi$, and
decays generating an extra radiation density $\rho_\r^{\rm ex}$ at a
rate $\Gamma_\phi$, the transition can be modelled
as~\cite{Ramberg:2019dgi} 
\begin{align}
\dot\rho_\phi + 3(1 + w_\phi)H\rho_\phi &= -\Gamma_\phi\rho_\phi\,, 
\label{eq:Kinetic_psi}\\
\dot\rho_\r^{\rm ex} + 4H\rho_\r^{\rm ex} &=
\Gamma_\phi\rho_\phi\,, \label{eq:Kinetic_Rad} 
\end{align} 
where a dot indicates a derivation with respect to cosmic time $t$. If
it is assumed that the decay products are light SM particles, then the
energy density of the surrounding plasma is increased.

\begin{figure}[t!]
\begin{center}
\begin{multicols}{2}
	\includegraphics[width=\linewidth]{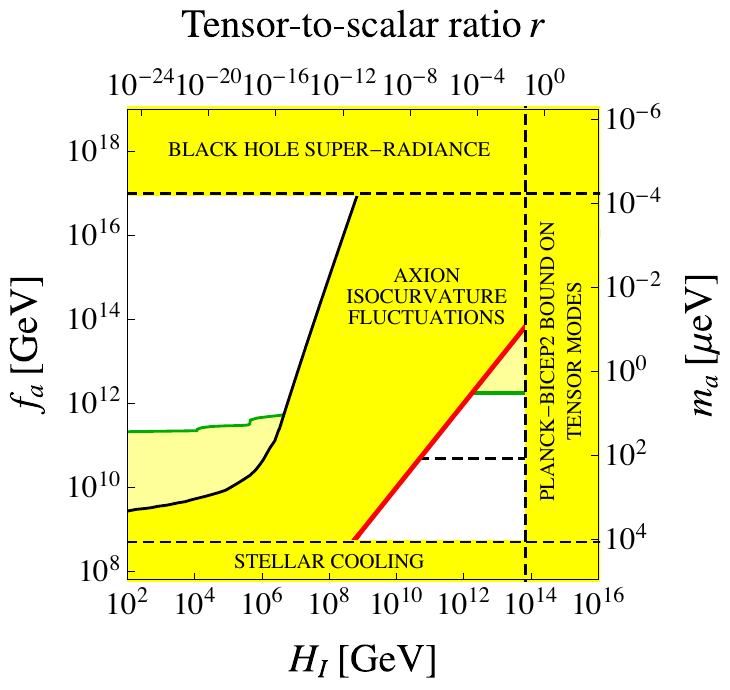}
	\includegraphics[width=\linewidth]{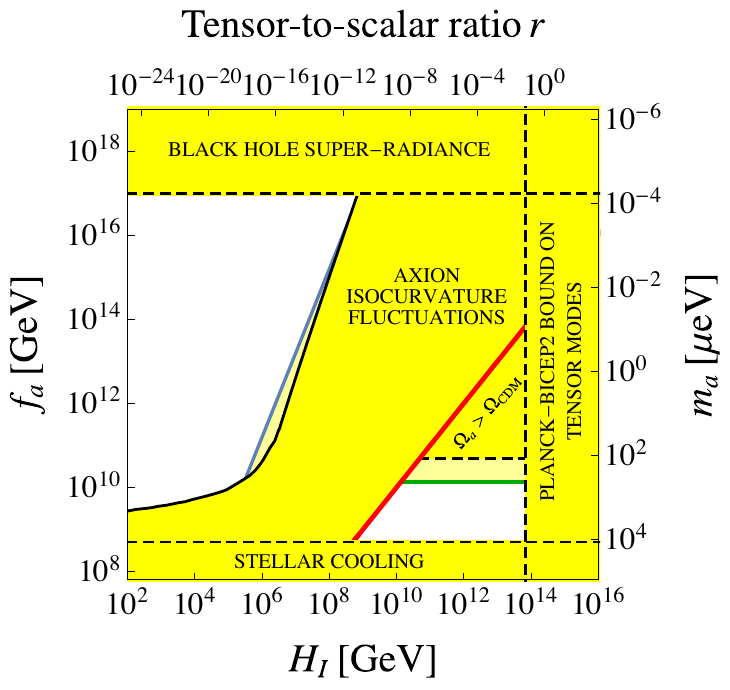}
\end{multicols}
\caption{Region of axion parameter space where the axion constitutes
  the totality of the DM observed for an early
  matter-dominated period (left) and for an early kination period
  (right), see text for additional details. For each figure, axes and
  bounds have been described in Fig.~\ref{fig:bound_standard}.}
	\label{fig:bound_NSC}
\end{center}
\end{figure}

A non-standard cosmology evolution alters the moment at which the
coherent oscillations of the axion field commence and possibly dilutes
their energy density thereafter, modifying the present axion energy
density for a given axion mass and initial misalignment angle. Models
that have been considered in the literature include an early
matter-dominated period~\cite{Dine:1982ah, Turner:1983he,
  Steinhardt:1983ia, Scherrer:1984fd},
for which $w_\phi = 0$, and a
period of dominance by a ``fast-rolling'' kination
field~\cite{Barrow:1982ei, Ford:1986sy}, with $w_\phi = 1$. We have
summarised these modifications in Fig.~\ref{fig:bound_NSC}, where we
show the effects of an early matter-dominated period (left) or a
kination domination (right) on the axion parameter space, 
assuming a decay temperature $T_{\rm dec} =
100\,$MeV.
These results have to be compared with what has been obtained for the
standard cosmological model in Fig.~\ref{fig:bound_standard}. In the
pre-inflationary scenario described in \sect{sec:misalignment}, the
bounds from the non-detection of axion isocurvature fluctuations give
more stringent constraints on the allowed parameter space with respect
to the result in the standard cosmology, bound in each of the figure
by the contour in green and with additional yellow shading where
excluded. For the matter-dominated scenario with $\theta_i \simeq 1$
and for a decay temperature $T_{\rm dec} \ll 1\,$GeV, the axion can be
the DM for values of the axion decay constant that are
generally larger than what obtained in the standard scenario, that is
$f_a \gg 10^{11}\,$GeV for a decay temperature
$T_{\rm dec} \ll 1\,$GeV. Comparing to the standard scenario, the
value of the DM axion mass is generally larger when an early
kination period occurs, and smaller when in the presence of an early
matter-dominated period. These changes are ultimately due to the
effects of entropy dilution and to the production of a different
number of axions from the altered $T_{\rm osc}$ in
Eq.~\eqref{eq:condition_oscillations}, as explained in detail in
Ref.~\cite{Visinelli:2009kt}. 
Unconventional cosmologies, such as early matter domination, 
can also modify the small scale structure of axion DM and lead to a linear growth of 
cosmological perturbations at early times, offering further 
opportunities for astrophysical tests \cite{Blinov:2019jqc}. 

%
%


\subsubsection{Modifying  the  $m_a$-$f_a$ relation or the axion mass  function $m_a(T)$ }
\label{sec:mirror}
\smallskip


%

 The axion abundance would also change if
the axion mass/decay constant or the mass dependence on the temperature were modified.
We will now explore some scenarios that realise this possibility.\\

\noindent
{\it Axion dark matter  with a non-conventional $m_a$-$f_a$ relation:} 
The estimate of the axion energy density due to  the misalignment mechanism 
carried out in \sect{sec:misalignment}   led to 
\eqn{eq:standarddensity}, which shows that besides a dependence on the 
value of the initial misalignment angle, the axion contribution to the CDM crucially 
depends on the value of the axion mass.  
In the conventional axion model of  \sect{sec:strongCPprob},
the product of the zero temperature axion mass $m_a$ and the PQ
constant $f_a$ is fixed by the value of the QCD topological susceptibility 
at zero temperature, $(m_a f_a)^2 = \chi(0)$ 
for which the standard value is $\chi(0) = \(75.5{\rm \,MeV}\)^4$.
However, it is possible to conceive models in which the relation between 
the axion mass and its decay constant is modified, a condition that 
can be  conveniently  expressed as $(m_a f_a)^2 = \alpha_\chi\, \chi(0)$,
with $\alpha_\chi \neq 1$.  
This would change the fraction of
axion DM for a given mass and initial misalignment angle or,
equivalently, the value of the mass required for the axion to account 
for all of  the DM. 
Assuming that the misalignment mechanism dominates the cosmological axion 
production, to keep the energy density in axions constant it is easy to see from 
\eqns{eq:Tosc1}{eq:rhoa_analytic} that the mass should 
be rescaled by the factor
$\alpha_\chi^{(2+\gamma)/(3+\gamma)}\simeq \alpha_\chi^{6/7}$, where  the
numerical exponent in the last relation corresponds to the canonical  value 
$\gamma=4$. Hence,  if $\alpha_\chi>1$ a larger axion mass is needed to account for the
totality of CDM in axions, while the opposite is true if $\alpha_\chi<1$.     

An example of a model in which  $m_a$ is decreased from its standard value
was proposed in Ref.~\cite{Hook:2018jle}. This model relies on a $Z_N$ symmetry under
which $a \to a + \frac{2 \pi f_a}{N}$, and furthermore the axion
interacts with $N$ copies of QCD whose fermions transform under $Z_N$
as $\psi_k \to \psi_{k+1}$.  Surprisingly, adding up the contributions
of all the sectors one finds that cancellations occur in the axion
potential with a high degree of accuracy. As a result, the axion mass
gets exponentially suppressed with respect to the standard case,
and in the large $N$ limit one obtains  
\begin{align}
	\label{eq:ZN}
	m_a^{(N)} \approx \frac {m_a}{2^{(N-4)/2}} \,,
\end{align}
which corresponds to  $\alpha_\chi \approx 2^{-N} < 1$ and, at constant relic density, 
 to a rescaled axion mass $m_a \to 2^{-\frac{6N}{7}} m_a$.
 This construction can be particularly relevant for axion searches in the very 
low mass region. \\
Another way to decouple  $m_a$ from  $f_a$ relies on 
an extra dimensional scenario with a large compactification
radius~\cite{Dienes:1999gw}. The axion mass is bounded 
as $m_{a}\lsim {\rm min}\[\frac{1}{2R_{c}}, \frac{m_{\pi}f_{\pi}}{f_{a}}\]$
where $R_{c}$ is the compactification radius, and the second term inside square brackets represents  
the usual PQ axion mass.
Present limits on deviations from gravitational Newtonian low at short distances  
constrain $R_{c}$ to values well below $\sim 0.1\,$mm~\cite{Geraci:2008hb,Sushkov:2011zz},
so that this mechanism can yield axions lighter than expected only 
for relatively small values of the axion decay constant $f_{a}\lsim 3\cdot 10^{9}\,$GeV.\\


%
%

%

\noindent
{\it Axion dark matter  with a modified mass function $m_a(T)$:} 
%
%
%
Another interesting possibility are scenarios in which the zero
temperature axion mass is unchanged but its temperature dependence
is non-standard.  In this case, the temperature at which  the axion field oscillation  
commence differs from 
\eqn{eq:Tosc1}.  
Since the axion energy density scales 
as\footnote{This can be seen as follows: $\rho_a=m_a\, n_a$
  with $ n_a\propto m_a(T_{\rm osc})\,T_{\rm osc}^{-3}\propto H(T_{\rm osc})T_{\rm osc}^{-3}\propto T_{\rm osc}^{-1} $.
  }
\begin{align}
\label{eq:axion_abundance_ma}
\rho_a\propto m_a  T_{\rm osc}^{-1}\,,
\end{align}
shifting the temperature at which the oscillations start would change
the energy density yield in axions even if the axion mass is unchanged.
%
%
%
%
%
This effect is expected in the cosmological scenario predicted by the
theory of the \textit{mirror world}~\cite{Berezhiani:2000gw}, extended
to include the axion~\cite{Giannotti:2005eb}.

The mirror world idea is very
old~\cite{Lee:1956qn,Holdom:1985ag,Glashow:1985ud,Khlopov:1989fj} and
based on the assumption that the gauge group is the product of two
identical groups, $ G\times G^{\prime} $.  In the simplest possible
model, $ G $ is the SM gauge group and $ G^{\prime} $ an
identical copy of it.  Standard particles are singlets of
$ G^{\prime} $ and mirror particles are singlets of $ G $.  This
implies the existence of mirror particles, identical to ours and
interacting with our sector only through gravity 
(and possibly via other renormalisable portal couplings, 
which are assumed to be small). Since the
gravitational interaction is very weak, mirror particles are not
expected to come into thermal equilibrium with ordinary particles.
Hence, there is no reason to expect that the standard and mirror
Universe have the same temperature.\footnote{In the exact $Z_2$ symmetric case this requires an inflationary dynamics yielding different reheating temperatures in the standard and mirror sectors 
\cite{Berezhiani:2000gw}.}
In fact, cosmological
observations require a colder mirror Universe in order to reduce the
radiation energy density at the time of the
BBN~\cite{Berezhiani:2000gw,Roux:2020wkp}.  Denoting with
$ x =T^{\prime}/T$ the ratio of the mirror and standard temperatures,
one finds that $ x\lesssim 0.4 $ is needed to accommodate the most
recent combined analysis of the Cosmic Microwave Background from the
Planck collaboration and observations of the Baryon Acoustic
Oscillations~\cite{Aghanim:2018eyx}.

The possibility to implement the PQ mechanism in the mirror world
scenario was proposed in
Refs.~\cite{Rubakov:1997vp,Berezhiani:2000gh,Gianfagna:2004je}.  The
general feature is that the total Lagrangian must be of the form
$\mathcal{L}+\mathcal{L}^\prime+\lambda \, \mathcal{L}_{\rm int}$,
where $\mathcal{L}$ represents the ordinary Lagrangian,
$\mathcal{L}^\prime$ is the Lagrangian describing the mirror world
content, and $\mathcal{L}_{\rm int}$ is an interaction term with a
coupling $\lambda$ which is taken to be small enough to 
ensure that the two sectors do not come into thermal equilibrium.
The simplest realisation of the mechanism restricts the interaction to
the Higgs sector. Ordinary and mirror world have each two Higgses,
which interact with each other. The axion emerges as a combination of
their phases in a generalisation of the Weinberg-Wilczek mechanism.
For $\lambda=0$, the total Lagrangian contains two identical
$U(1)_{\rm axial}$ symmetries, while the $\mathcal{L}_{\rm int}$ term breaks
them into the usual $U(1)_{\rm PQ}$, so that only one axion field
results.

As long as the mirror-parity is an exact symmetry, the particle
physics is exactly the same in the two worlds, and so the strong CP
problem is simultaneously solved in both sectors.  In particular, the
axion couples to both sectors in the same way and their
non-perturbative QCD dynamics produces the same contribution to the
axion effective potential.  Hence, the total zero temperature axion
mass, which includes the mirror world contributions, is
$m_{\rm tot}(0) = \sqrt{2}\, m_a$, only slightly larger than the standard
zero temperature axion mass $m_a$, given in Eq.~\eqref{eq:axionmass}.
However, at temperature $T \sim 1\,$GeV the axion mass could be
considerably larger than its standard value. Assuming the same
confinement temperature $T_C$ in the two 
sectors, and neglecting a possible dependence of
the exponent $\gamma$ 
on the temperature so that $\gamma(T)\approx \gamma(xT) = \gamma \approx 4$, the
expression in \eqn{eq:QCDaxion_mass} for $T \gtrsim T_C$ gives
\begin{align}
\label{Eq:maMirror}
m_{tot}^2(T)=
m_a^{2}\left[ \left( \dfrac{T_C}{T}\right)^{2\gamma} 
+ \left( \dfrac{T_C}{x\,T}\right)^{2\gamma} \right] = 
m_a^{2}(T)\( 1 + \frac{1}{x^{2\gamma}} \)\,,
\end{align}
where $m_a(T)$ is the standard temperature dependent axion mass.  Of
course, this different temperature dependence implies a different
value for $T_{\rm osc}$.
To estimate the contribution of the axion  energy density, let us neglect the first term in the parenthesis in
\eqn{Eq:maMirror}.  This approximation is justified for $x\ll 1$.  In
this case, it is easy to see that the oscillation temperature in the
mirror world scenario, $T^{\prime}_{\rm osc}$, is related to the
standard oscillation temperature as
\begin{align}
\label{eq:Tosc_mirror}
T^{\prime}_{\rm osc}\simeq x^{-\gamma/(\gamma+2)}T_{\rm osc}\,.
\end{align}
The effect of this  on the expected axion abundance can be inferred 
from~\eqn{eq:axion_abundance_ma}:
\begin{align}
\label{eq:rho_mirror}
\rho_a^{\rm mirror}\simeq \sqrt{2}\, x^{\gamma/(\gamma+2)}\rho_a\,,
\end{align}
%
%
where the $\sqrt{2}$  accounts for the  modification
of the zero temperature axion mass. Although in this case  the 
modification to the zero temperature axion mass is only a minor
effect, corresponding to  a factor $\alpha_\chi=2$,  the above result shows
that in the mirror world scenario the present  energy density in axions 
can be considerably smaller than expected, and this allows to saturate $\Omega_{\rm DM}$   
for large values of the PQ constant, a result that can be of interest for experiments  searching  
for DM axions in mass regions well below the conventional window. \\ [-3pt]

A different  way to  deplete the energy density of the QCD axion, thus allowing for 
 larger $f_a$ and smaller axion DM masses, was put forth in
 Ref.~\cite{Agrawal:2017eqm}. 
It is assumed that the axion couples to a massless $U(1)'$ dark photon
via an $ a F' \tilde F'$ term similar to the axion-photon coupling. 
The dark photon, however,  is decoupled from the SM and does not interact 
with the thermal bath, a condition that has to be enforced to 
maintain it massless also at finite temperature.  
When the axion  starts oscillating  certain modes of the dark photon become 
tachyonic and start growing exponentially, and in this regime 
energy is efficiently transferred from the axion into the dark photons,
leading to an exponential suppression of the axion density and 
drastically reducing its contribution to the DM, and opening a 
window for low mass values as small as $m_a\sim 10^{-10}\,$eV.  
However, the previous conclusion was based on a linear analysis and did not take into account 
the backreaction of the produced hidden photons on the axion dynamics, 
which becomes significant in the non-linear regime. Including the latter in a detailed 
lattice calculation Ref.~\cite{Kitajima:2017peg} found that the axion 
abundance can be suppressed at most by a factor of $\mathcal{O}(10^2)$. \\ [-3pt]


Another mechanism that brings in a hidden Abelian gauge field
 exploits the  Witten effect~\cite{Witten:1979ey} of hidden monopoles on the QCD axion dynamics.
Long time ago Witten has shown that in the presence of a CP violating $\theta$-term, 
monopoles acquire a non-zero electric charge and become dyons.
 When a dynamical axion field replaces
 $\theta$, its potential receives additional contributions from
 interactions with the monopoles~\cite{Fischler:1983sc} and because of this the field
 oscillations begin much before the epoch of the QCD phase
 transition.  This scenario does not work with QED monopoles,  
because of the extremely  tight observational constraints on their abundance.
 For this reason  Refs.~\cite{Kawasaki:2015lpf,
 Kawasaki:2017xwt} attempted to implement the same mechanism  
exploiting monopoles of a  hidden $U(1)'$ symmetry.  
 The axion abundance turns out to be inversely proportional to the abundance of hidden
 monopoles, and  when the density of monopoles is sufficiently 
 large to  make up a significant  fraction of the DM,
the abundance of  axions with decay constant smaller 
than about $10^{12}\,$GeV gets suppressed.  
 While this mechanism does not seem to be able to extend by   
much  the axion window towards  low  mass values, 
and moreover predicts that DM is accounted for by hidden monopoles, rather than by axions, 
it has some interesting features, as for example that of  suppressing axion 
isocurvature perturbations, and  of  disposing of the domain wall problem.  \\ [-3pt]

Another way to allow for larger values of the axion decay constant for axion DM was discussed 
in Ref.~\cite{Baratella:2018pxi}.
It is based on the possibility of having a long period of supercooling in the early Universe proceeding  
down to temperatures $\lsim T_\o$,  
so that after start  oscillating the axion quickly relaxes to the minimum of its potential.
At the end of supercooling, the Universe reheats to $T \sim \mathcal{O}$(TeV). 
A standard evolution begins, but now  with 
an initial value of  $\theta_i \ll 1$. 
Hence, this mechanism provides a dynamical way to arrange 
for small values of $\theta_i$ as  initial condition  also for post-inflationary scenarios.


\subsubsection{Alternative mechanisms for axion dark matter production}
\label{sec:thetadot}
\smallskip

In this section we review some mechanism for axion DM production that are 
alternative  to the misalignment mechanism. Also in these cases the axion mass for which 
the DM density is saturated can be pushed to values sensibly larger 
than the conventional ones. \\

\noindent
{\it Axion dark matter from  initial velocity.}\ 
As we have seen in \sect{sec:misalignment}, oscillations of the axion
field can start when the age of the Universe $t_U\sim H^{-1}$ is of
the order of the oscillation period $\sim m_a^{-1}$.  However, one
additional condition must also be satisfied: at $t_{\rm osc}$, as
defined by \eqn{eq:condition_oscillations}, the axion kinetic energy
$\dot a^2/2$ should not exceed the potential barrier $2 m_a^2 f_a^2$,
otherwise the axion field keeps rotating and oscillations are
prevented.  This implies that if  $\dot\theta_i \gsim 2 m_a(t_{\rm osc})$, 
the conventional misalignment scenario is not realised. 
%
In this case the axion DM scenario  would be genuinely different from the
conventional misalignment scenario, hence it is worthwhile to investigate in
some detail the consequences of this possibility, a task that was
carried out in Refs.~\cite{Co:2019jts, Chang:2019tvx}.
%
%
The axion velocity $\dot\theta$ can be related to the density of the
PQ charge associated with the symmetry $\Phi \to e^{i\alpha} \Phi$ of
the field $\Phi$ introduced in Eq.~\eqref{eq:PhidecKSVZ}:
\begin{equation}
\label{eq:PQdensity}
n_\theta = i\(\Phi \dot\Phi^\dagger - \Phi^\dagger\dot\Phi\)  =  \dot\theta f_a^2\,, 
\end{equation}
where the last expression holds after replacing $\Phi \to f_a/\sqrt{2}$.
The charge density  in a comoving volume is conserved, hence the scaling 
$\dot\theta \sim R^{-3}$ in \eqn{eq:initial_condition_dottheta}, 
and  it is then convenient to  introduce the (constant) comoving density  $Y_\theta = n_\theta/T^3  $. 
Saturating the condition $\dot\theta \sim  m_a$ at $T_{\rm osc}$   
the critical value for departing from the conventional misalignment scenario is obtained:
\begin{equation}
\label{eq:Ycritical}
Y_\theta^c =\frac{n_\theta}{T^3} \simeq 
\frac{m_a(T_{\rm osc}) f^2_a}{T_{\rm osc}^3} \simeq \frac{f^2_a}{\mP T_{\rm osc} }  \,, 
\end{equation}
where $m_a(T_{\rm osc} )\sim H(T_{\rm osc})\sim T^2_{\rm osc}/\mP$
has been used.  The
energy density in axions can be written as
$\rho_a = m_a Y_\theta^c T^3 $ with $m_a$ the zero temperature axion
mass, and it saturates the DM density if at the temperature $T_\e$ of
matter-radiation equality $\rho_a \sim T_\e^4$, which yields
$m_a f_a^2 \sim T_\e T_{\rm osc} \mP$.  For $Y_\theta > Y_\theta^c$
oscillations are delayed until some lower temperature
$T_*<T_{\rm osc}$ when the kinetic energy is insufficient to overcome
the potential barrier. In this case relic axions are overproduced and
one has to lower the scale $f_a$ to match $\rho_a \approx \rho_{DM}$.
The numerical studies in Refs.~\cite{Co:2019jts,Chang:2019tvx} indeed
confirm that at fixed values of $f_a$ the axion kinetic mechanism can
produce more DM than the conventional misalignment scenario, and this
opens up an interesting mass window for axion DM in the range
$m_a \in [10^2,10^5]\,\mu$eV.
 
 In post-inflationary scenarios, in terms of quantities defined at the PQ scale,   
 the requirement  that the the kinetic energy  overcomes the potential barrier  
 ($\dot\theta_i \gsim 2 m_a(t_{\rm osc})$)  translates   into the condition
$\big|\dot\theta_{\rm PQ}\big| /H_{\rm PQ} \gsim 10^{11}\(f_{a}/10^{11}\,{\rm GeV}\)^{7/6}$~\cite{Chang:2019tvx}
(see also \eqn{eq:dottheta-limit}), that is  rather large values of  $\dot\theta_{\rm PQ}$ are needed.
In the models of Ref.~\cite{Co:2019jts}  the PQ symmetry is broken during inflation, the scaling 
$\dot\theta \sim R^{-3}$ is delayed 
until the radial mode settles to the minimum due to new dynamics of the rotation,
which can occur at temperatures well below $f_{a}$,  and the condition  on 
$\big|\dot\theta\big| /H$ is accordingly relaxed. 
The basic mechanism  to generate  $\dot\theta_{\rm PQ} \neq 0$ 
 is to introduce at the high scale an explicit breaking of the PQ symmetry via higher
dimensional operators similar to the ones discussed
in~\sect{sec:PQquality}. In this way the potential gets tilted, and
the axion starts moving towards the potential minimum acquiring a
velocity.  Of course an explicit breaking can shift the axion from the
CP conserving minimum, so that while it must be effective in the early
Universe, it must become negligible at lower temperatures.  This can
be arranged by assuming very flat potentials, so that the expectation
value of the radial mode is initially large
$\langle \varrho_a\rangle \gg f_a$ enhancing the effects of the operators
that at later times, when eventually $\varrho_a$ relaxes to $f_a$, become
sufficiently suppressed~\cite{Co:2019jts}.\\

\noindent
{\it Axion dark matter  from parametric resonance.}\ 
This mechanism  represents another  alternative way for producing axion DM,
and allows  saturating the DM relic density  for $f_{a} \ll 10^{12}\,$GeV, that is 
for relatively large axion mass values $m_{a}\in [10^{-3}$-$10^{-1}]\,$eV~\cite{Co:2017mop}.
As in the previous mechanism, it is assumed that the axion radial mode $\langle\varrho_{a}\rangle$  
has a large initial value $\langle\varrho_{a}\rangle \gg f_{a}$. Axion production  becomes efficient when 
the field $\varrho_{a}$ starts  oscillating near the minimum of its effective potential, and rapidly  decays into 
 axions  due to a broad  parametric resonance,  in a way similar to  particle production 
by the oscillating inflaton field  in the preheating stage after inflation~\cite{Kofman:1994rk,Shtanov:1994ce,Kofman:1997yn}. 
Axions are initially produced with momenta of order of the mass of the radial mode and thus, 
in contrast to the misalignment mechanism, the axions produced in this way are initially relativistic. 
They can redshift sufficiently to be CDM, although for  large values of the initial 
momentum they can remain sufficiently warm to leave a signature in structure  formation. 
Even warmer axions are produced in the model of Ref.~\cite{Harigaya:2019qnl}
where the PQ phase transition and the parametric production of axions is delayed 
to temperature much below the PQ scale $f_{a}$. 
Axion production via parametric resonance  in a supersymmetric model, due to  oscillation of the 
axion superpartner, was  studied in Ref.~\cite{Ema:2017krp}. Although some cosmological consequences like 
the production of an axion dark radiation component  or  the possibility of detecting gravitational
waves from explosive axion production were considered,  possible consequences for 
axion CDM were not addressed.

\subsection{Super-heavy axions}
\label{sec:heavy}

In \sects{sec:gag}{sec:gaN}, we showed how axions can be made more or
less strongly coupled to SM fields through the dynamics of additional
fields.  In such models, the modification of the interaction strength
was realised by altering the $C_{af}$ parameters, defined in
\Eqn{eq:Laint1}, while preserving the relation between the axion mass
and decay constant $f_a$ given in \eqn{eq:axionmassfa}.
In this Section we consider instead the possibility that the
$m_a$--$f_a$ relation is modified (still keeping the solution of the
strong CP problem), thus allowing for $m_a \gtrsim 100$ keV axions
(the 100 keV threshold is actually needed in order to evade (most) of
the astrophysical constraints).  We denote the latter
\emph{super-heavy} axions, in contrast to the canonical heavy axion
regime up to $m_a \lesssim 0.1$ eV.  It should be first noted that
such super-heavy axions are cosmologically unstable. For instance, the
decay channel $a\to \gamma+\gamma$ yields \beq
\label{eq:atogammagamma}
\Gamma (a \to \gamma+\gamma) = \frac{g^2_{a\gamma} m_a^3}{64 \pi} = 
\frac{E/N - 1.92}{0.318 \ \text{s}} \( \frac{m_a}{100 \ \text{keV}} \)^5 \, ,
\eeq
where in the last step we used the numerical values 
of the axion-photon coupling and the standard QCD $m_a$--$f_a$ relation\footnote{For 
fixed $f_a$, this provides a lower bound on the decay rate, compared to the case where 
$m_a$ is enhanced.} 
(cf.~\eqn{eq:gagammagaf}, (\ref{eq:axionmassfa}) and (\ref{eq:Cangamma})). 
Hence, barring an unrealistic cancellation in the axion-photon coupling, 
an axion with $m_a \gtrsim 100$ keV cannot be DM.  

Relaxing the relation between $m_a$ and $f_a$ 
requires some modifications of the gauge structure of the SM,
particularly of the strong interaction sector.  Yet, models which
envisioned a large axion mass for some fixed couplings appeared early
on
(see, e.g.,
Refs.~\cite{Yang:1978gq,Dimopoulos:1979pp,Tye:1981zy,Holdom:1982ex,Holdom:1985vx,Dine:1986bg,Flynn:1987rs}).
The reason can be perhaps traced, at least in part, in the desire of
avoiding the requirements of extremely small (experimentally
prohibitive) axion couplings, which seemed an inevitable consequence
of the experimental and astrophysical constraints.
Early experimental tests of the original WW axion model (cf.~footnote
(\ref{foot:WWruledout}) in \sect{sec:UVcomp}) ruled out the range
$f_a\lesssim 10^{4}\,{\rm GeV}$.  According to the standard QCD
relation in \eqn{eq:axionmassfa}, this constraint implies an axion
mass below 0.1 keV.  However, light axions with masses below a few
keV, can be easily produced in stars and impact their evolution
possibly beyond what is observationally allowed
(cf.~\sect{sec:Astro_bounds}).  As it turns out, astrophysical
constraints are considerably more restrictive than the original
bounds, pushing the range of excluded couplings all the way up to
$f_a \sim 10^{8}\ {\rm GeV}$, implying extremely weakly interacting
axions.  Hence, somewhat surprisingly, the relatively weak bounds that
ruled out the WW model brought to a rather dramatic result.  The only
two viable options left were either to make the axion extremely weakly
coupled (invisible axion) or to make it heavier.\footnote{Another
  non-trivial option would have been to relax the hypothesis of the
  universality of the PQ current (see for instance the discussion at
  the beginning of \sect{sec:gaFCNC}).  More recently, it was shown in
  Ref.~\cite{Alves:2017avw} that an $\mathcal{O}(10\ \text{MeV})$
  axion with the standard $m_a$--$f_a$ QCD relation could still be
  viable under the following conditions: $i)$ the axion couples only
  to first generation fermions, with dominant decaying channel into
  electrons $ii)$ the axion-pion coupling is suppressed (pion-phobia)
  and $iii)$ large hadronic uncertainties in rare $K$ decays are
  invoked.  The UV completion of such an axion is however non-trivial,
  and it requires several extra fields beyond those of benchmark axion
  models.}
Either way, the axion could avoid the astrophysical bounds.  Both
roads were pursued.  Models with a larger mass for fixed coupling have
the additional appeal to be more easily tested in the laboratory while
invisible axions were, at the time, truly thought to be beyond the
foreseeable experimental potential (this, of course, before the
seminal paper of Sikivie \cite{Sikivie:1983ip}).

Additional motivations for super-heavy axions can be found in
connection with the PQ quality problem
\cite{Georgi:1981pu,Dine:1986bg,Barr:1992qq,Kamionkowski:1992mf,Holman:1992us,Ghigna:1992iv}
(cf.~\sect{sec:PQquality}), since explicit PQ-breaking terms would
produce a shift, in some cases very large, to the axion mass (for an
application in the context of Gamma Ray Bursts see
\cite{Berezhiani:1999qh}).
This, however, would generically also shift the minimum of the axion potential 
away from its CP conserving point, thus spoiling 
the solution of the strong CP problem.\footnote{Turning up the argument, 
one can observe 
that generic PQ-breaking effective operators would spoil the axion
solution for $f_a$ above a few GeV or, if we forbid $d=5$ operators,
for $f_a$ a few TeV.  Hence, such scales are interesting for a
theoretical point of view. However, unless we make the axions
super-heavy, $f_a\sim 1$ TeV implies $m_a\sim 1$ keV which is excluded
by stellar argument.  This is a strong motivation to consider
super-heavy axions \cite{CidVidal:2018blh}.}
Indeed, some of the original models did not satisfy this condition. 

An even earlier strategy to raise the axion mass was put forth by
Holdom and Peskin~\cite{Holdom:1982ex} (see also \cite{Holdom:1985vx}), which pointed out that the
contribution of the small colour instantons to the axion potential
could be made sizeable by some nontrivial dynamics above the
electroweak scale which reversed the sign of the colour $\beta$
function at some mass above 1 TeV or so.  This idea was later
reconsidered in Refs.~\cite{Dine:1986bg,Flynn:1987rs} in the context
of GUTs, where it was argued that in the presence of new sources of
chiral symmetry breaking at high energies there is no generic reason
for the small instanton contribution to the axion potential to be
aligned to the long-distance QCD contribution.  Hence, also in this
case raising the axion mass could have spoilt the solution of the
strong CP problem.
Interestingly, the idea that small instantons could contribute to the axion mass was resuscitated very recently in Ref.~\cite{Gherghetta:2020keg}, 
in a model where QCD is embedded in a theory with one compact extra dimension. 
In this scenario, it was shown that the contribution of the small instantons to the 
axion mass can be larger than the usual large instantons contribution, though it is required that the theory is close to the non-perturbative limit.
At any rate, the $\theta$ angle is not shifted in this construction and hence the
axion solution to the strong CP problem naturally preserved.  




Another, perhaps less economic, strategy is to assume that additional
mass contribution emerges from the $U(1)_{\rm PQ}$ current anomaly,
related to some hidden gauge sector with a confinement scale larger
than $\Lambda_{\rm QCD}$.
%
%
%
One of the earliest attempts in this direction is the mirror world
axion scenario, already introduced in \sect{sec:mirror}.  In this
case, a whole new SM sector is introduced, with mirror and ordinary
particles interacting with each other only gravitationally and,
possibly, through some other very weak couplings, insufficient to
bring the two sectors in thermal equilibrium during the cosmological
evolution (cf.~\sect{sec:mirror}).  The first example of such
models~\cite{Rubakov:1997vp} was based on a mirror extended GUT
theory, with gauge group $ SU(5)\times SU(5)^{\prime} $, each one with
its own PQ symmetry.  A $Z_2$ symmetry (\emph{mirror parity}) for
exchange of ordinary and mirror particles guaranties the equality of
the $\theta$ parameters in the two sectors.  The $U(1)\times U(1)'$
symmetry is reduced to just one PQ symmetry through the introduction
of an $SU(5)\times SU(5)^{\prime}$ singlet complex scalar field, of PQ
charge different from zero, that interacts with both ordinary and
mirror Higgs.  As explained in \sect{sec:mirror}, in such a symmetric
model the contribution to the axion mass from the hidden mirror sector
would be a factor of $\sqrt 2$ larger than in the standard case.  To
have a larger mass gain, one has to assume a breaking of the mirror
symmetry which does not, however, spoil the solution of the strong CP
in both sectors.  The way to do that is to have the mirror parity
broken softly in the Higgs sector, without affecting the structure of
the Yukawa couplings which has to remain the same in the two sectors
(so that $\arg\det Y_u Y_d = \arg\det Y'_u Y'_d$).  In
Ref.~\cite{Rubakov:1997vp}, it was assumed that the mirror parity is
broken by soft terms and that the breaking of $SU(5)^\prime$ to the
mirror SM happens at a much lower scale
than in the ordinary sector.  Since the coupling constant of $SU(5)$
runs faster than that of $SU(3)$, it results that
$\Lambda_{\rm QCD}^{\prime}\gg \Lambda_{\rm QCD}$.  Hence, the axion
mass takes most of its contribution from the mirror sector.

Somewhat simpler models, which did not require a GUT but just a
SM$\times $SM$^{\prime}$ gauge group, were considered in
Ref.~\cite{Berezhiani:2000gh} and in its supersymmetric
extension~\cite{Gianfagna:2004je}.  The mirror parity is spontaneously
broken and induces a larger electroweak scale in the mirror sector,
without affecting the Yukawa sectors.  The higher mirror Higgs VEV
generates heavier mirror fermions and, consequently, a faster
renormalisation group evolution and a larger confinement scale
$\Lambda_{\rm QCD}^{\prime}\gg \Lambda_{\rm QCD}$.  These
constructions were also motivated by the need to avoid Planck induced
corrections which afflicted the model in Ref.~\cite{Rubakov:1997vp}
and could hence spoil the solution of the strong CP.

More recently, Ref.~\cite{Fukuda:2015ana} considered a mirror KSVZ model.
The construction  assumes a SM$\times $SM$^{\prime}$ gauge group,
with a softly broken $Z_2$ symmetry to ensure the alignment  of the effective $\theta$ angles in the two sectors.
Just like in the models in Ref.~\cite{Berezhiani:2000gh,Gianfagna:2004je}, 
described above, soft breaking terms induce a larger electroweak symmetry breaking scale (and hence heavier quarks) in the mirror 
sector.
In order to get a considerable enhancement  of the axion mass, however, the model in Ref.~\cite{Fukuda:2015ana} requires the addition of scalar quarks in both sectors. 
The choice of scalar, rather than fermionic, quarks is motivated by the fact that they do not contribute to the effective $\theta$ angles. 
The ratio of the mirror and standard confinement scales turns out to be roughly equal to the ratio of the mirror and standard scalar masses. 
The somewhat more complicated construction permits a larger mass for a fixed PQ scale than in the case of the models in Ref.~\cite{Berezhiani:2000gh,Gianfagna:2004je}, with masses of $ 100\,{\rm MeV}$ or so achievable for $f_a\sim 10^{3-5}\, {\rm GeV}$.

In recent times, several more models (see
e.g.~\cite{Hook:2014cda,Dimopoulos:2016lvn,Hook:2019qoh}) reconsidered the $Z_2$
symmetry to ensure CP conservation in super-heavy axion models with
a scale $f_a$ all the way down to $\sim$ TeV, hence accessible to
detector searches such as ATLAS and CMS~\cite{Dimopoulos:2016lvn}. For
instance, in the latter construction
there is only one massless exotic quark transforming under $SU(\mN)$
and an axion whose mass is induced by $SU(\mN)$ instantons, so that in
the limit of vanishing QCD coupling it remains massive. The problem
with the two vacuum angles $\theta_{\rm QCD}$ and $\theta_\mN$ is again 
solved  by imposing a $Z_2$ symmetry that implies vacuum angles
alignment.  

A critical analysis of the $Z_2$ symmetry mechanism to guarantee the solution 
of the strong CP problem in models with a hidden QCD sector is presented 
in Ref. \cite{Albaid:2015axa}.
Indeed, the requirement of an additional parity symmetry can be circumvented and
mechanisms that avoid its use have been proposed in recent years. 
In Ref.~\cite{Gherghetta:2016fhp}, it is assumed that the $SU(3)_c$ group is a subgroup of 
an enlarged QCD colour group, $SU(3+\mN)$, which breaks into $SU(\mN) \times SU(3)_c$.
After the breaking, the unique $\theta$ angle of the $SU(3+\mN)$ group becomes 
a common factor of the $G\tilde G$ terms of $SU(\mN)$ and of $SU(3)_c$.
Thus, the axion dynamics can take care of the CP violation in both sectors,
while the axion mass gets most of its contribution from the $SU(\mN)$ sector, 
which can have a substantially larger confinement scale than the QCD.

Another proposal that avoids the introduction of the
discrete $Z_2$ symmetry was put forth in~\cite{Agrawal:2017ksf} (see
also \cite{Agrawal:2017evu,Fuentes-Martin:2019dxt,Csaki:2019vte}).  At high energies the
gauge group is a product of factors $SU(3)^N$ and $SU(3)_c$ is the
diagonal subgroup that survives at low energies.  Non-perturbative
effects in each individual $SU(3)$ factor generate a potential for the
corresponding axion. The vacuum is naturally aligned to ensure
$\theta = 0$, while the masses of these axions can be much larger than
for the standard QCD axion, reaching values well above the
GeV. Gaillard et al.~\cite{Gaillard:2018xgk} also consider a solution
with massless new fermions. The solution is based on an enlarged but
{\it unified} colour sector, where unification solves the issue of the
different $\theta$ parameters that arise in the presence of two or
more individual confining groups.  The unified colour group breaks
spontaneously to QCD and to another confining group.  Instantons of
the unified group with a large characteristic scale contribute to
breaking of a PQ symmetry and provide a source of large masses for the
axions, so that no light pseudoscalar particles remain at low scales.
This construction can yield both, dynamical and fundamental axions,
with masses that can lie in the several TeV range.

\section{Axions and... }
\label{sec:beyond QCD}

This Section is devoted to the collection of various topics in which axion physics can be related to other open issues of the SM, such as massive neutrinos (\sect{sec:connection_neutrinos}), the baryon asymmetry (\sect{sec:axionBAU}) and inflation (\sect{sec:axion_inflation}).  In \sect{sec:axion_GW} we touch on  possible observational signals of the PQ phase transition 
from the detection of   gravitational waves (GW). 
 We then discuss possible solutions of the DW issue (\sect{sec:AxionsDW}) and of the PQ quality problem (\sect{sec:protectingPQ}),
as well as  the embedding of axions in UV motivated frameworks, such as composite dynamics (\sect{sec:CompositeAxions}), grand unified theories (GUTs) (\sect{sec:axionGUTs}) and String Theory (\sect{sec:AxionStringT}). Every topic is briefly sketched, with the scope of mainly redirecting to the relevant literature.

\subsection{Axions and neutrino masses}
\label{sec:connection_neutrinos}

Axions and neutrinos share various properties: 
they are both extremely lighter than charged leptons 
and possess a feeble coupling to SM fermions. 
In fact, the idea of connecting massive neutrinos with a spontaneously broken $U(1)_{\rm PQ}$ comes 
a long way. 
Early studies (such as \cite{Kim:1981jw,Mohapatra:1982tc,Shafi:1984ek,Berezhiani:1985in}) were actually 
motivated by the natural emergence of intermediate mass scales in grand-unified theories. 
In particular, the axion-neutrino connection has been largely explored
in the context of the type-I seesaw \cite{Shafi:1984ek, Langacker:1986rj, Shin:1987xc, He:1988dm,
Dias:2005dn, Dias:2014osa, Salvio:2015cja, Clarke:2015bea, Ballesteros:2016euj, Ballesteros:2016xej, Ballesteros:2019tvf}, in which the heavy RH  neutrinos $N_R$ obtain their mass $M_R$ 
from a coupling $N_R N_R \Phi$ to the PQ symmetry breaking scalar singlet~$\Phi$. 
This connection is soundly motivated by the fact that the RH neutrino and the PQ symmetry breaking scales naturally fall in the same intermediate range $M_R \sim f_a\sim 10^9 \div 10^{12}$ GeV,  
and further supported by the possibility of naturally producing a cosmological baryon asymmetry of the correct size via leptogenesis~\cite{Langacker:1986rj}. 
Considering instead only scalar extensions of the SM a simple setup based on the Zee model for radiative neutrino masses was discussed in Refs.~\cite{Bertolini:1990vz,Arason:1990sg}, while extensions based on the type-II (III) seesaw and other radiative neutrino mass models were explored later on~\cite{Bertolini:2014aia,Ahn:2015pia,Bertolini:2015boa}. Scenarios implementing Dirac neutrinos have also been discussed~\cite{Chen:2012baa,Gu:2016hxh, Carvajal:2018ohk, Peinado:2019mrn}. 
The latter, however, miss the main motivation behind the axion-neutrino connection, 
that is the identification of the PQ scale with the scale suppressing neutrino masses. 

The constructions above often aim at providing a minimal SM extension addressing most of the shortcomings of the SM. It is fair to say, however, that they often lack in 
predictivity being the collection of somewhat orthogonal ingredients. The most genuine signature of the axion-neutrino connection would in fact 
be an axion coupling to neutrinos, of the type
$
\mathcal L \supset 
(m_\nu/f_a) \, a \, \bar \nu i \gamma_5 \nu$,
which is clearly beyond any experimental accessibility due to the 
huge $m_\nu / f_a$ suppression. 
Hence, any chance for predictivity beyond the single ingredients in isolation (e.g.~Type-I seesaw and PQ mechanism) can only arise indirectly as a self-consistency of the whole setup. 

A non-trivial step in this direction was achieved recently, in the context of the 
SMASH
model of Refs.~\cite{Ballesteros:2016euj, Ballesteros:2016xej, Ballesteros:2019tvf}, in which the modulus of the PQ scalar is also the key ingredient for successful inflation. A robust prediction of this setup is that the PQ symmetry is broken after inflation and never restored after it, thus making the range for axion DM in principle calculable. 


A different predictive approach, involving also flavour, was instead pursued recently 
in Ref.~\cite{Bjoorkeroth:2019ndr}, which classified all the generation dependent $U(1)$ symmetries which, in the presence of two leptonic Higgs doublets, can reduce the number of independent high-energy parameters of type-I seesaw to the minimum number compatible with non-vanishing neutrino mixings and CP violation in the leptonic sector. This setup leads to definite predictions for the charged leptons and neutrino mass matrices and, if extended to the quark sector, necessarily leads to a QCD anomalous $U(1)_{\rm PQ}$~\cite{Bjorkeroth:2018ipq}, thus predicting the existence of a QCD axion, with couplings to SM fermions fixed in terms of SM fermion masses and mixings.


\subsection{Axions and the cosmological baryon asymmetry}
\label{sec:axionBAU}

The strong CP-violating parameter must be extremely small today, 
as required by the non-observation of the nEDM. 
However, if the smallness of the theta angle were due to the cosmological evolution of an axion field, 
it is plausible that $\theta$ was $\mathcal{O}(1)$ in the early Universe 
and, conceivably, such source of CP violation could have played a role 
for baryogenesis. This idea was first put forth by 
Kuzmin, Tkachev and Shaposhnikov in Ref.~\cite{Kuzmin:1992up}, 
which considered the possible effects on electroweak baryogenesis
 of strong CP-violation  related to an axion field with a large background value. 
The conclusions of this work were, however, negative: 
SM baryon number violating processes are only effective at temperatures above $T_{\rm EW}\sim 100$ GeV, 
however, at these temperatures  strong CP violating effects are suppressed by an exceedingly small   
exponential factor $\exp(-8\pi^2/g^2_s)$ where $g_s$ is the strong gauge coupling. 
They concluded that the only possibility was to diminish the temperature 
of the electroweak phase transition down to $\Lambda_{\rm QCD}$, and to require that
no entropy was injected in the plasma after the phase transition to avoid  diluting the baryon asymmetry.  
This, however, also implied that the Universe got over-dominated by axion DM, contrary to observations.

More recently, this problem was reconsidered in Ref.~\cite{Servant:2014bla} 
in the context of cold electroweak baryogenesis (see 
e.g.~Refs.~\cite{GarciaBellido:2003wd,Tranberg:2003gi}),
a scenario that can lead to a very efficient production of baryon number 
if  the electroweak symmetry breaking is triggered through a fast tachyonic instability
(`Higgs quenching') and if it occurs in a range of  temperatures 10 MeV$\lsim T_{\rm EW}\lsim$1 GeV.   
In this scenario baryon production occurs strongly out of equilibrium so that there are no washout effects 
and the efficiency can be very large.  The  corresponding  rates can be described in terms of an effective 
equilibrium  temperature $T_{\rm eff}$ much larger 
than $T_{\rm EW}$ and one or two orders of magnitude larger than the reheating temperature 
of the plasma after the the phase transition $T_{\rm RH}$. The baryon-to-photons ratio  scales as   
$(T_{\rm eff}/T_{\rm RH})^3$ so that it can easily reach the observed value.  
Axion oscillations start well after reheating, and are not delayed with respect  to the conventional 
scenario,  so that the cold DM energy density from axion misalignment remains as in the standard case.  
%
%

A different  approach  to overcome the no-go of Ref.~\cite{Kuzmin:1992up} was  
put forth more recently in Refs.~\cite{Ipek:2018lhm,Croon:2019ugf}. Instead of delaying the electroweak phase transition 
down to 1\,GeV or below, the idea is to increase the QCD confinement scale  $\Lambda_{\rm QCD} \gsim T_{\rm EW}$.  
This is achieved by promoting the  strong coupling to a dynamical quantity, which evolves through the vacuum 
expectation value of a singlet scalar field that mixes with the Higgs field. 
QCD confinement and electroweak symmetry breaking  occur simultaneously 
close to the TeV scale, providing  large CP violation from an axion field value $\theta (T)$ 
of $\mathcal{O}(1)$ 
together with baryon number violation and the out-of-equilibrium condition from the phase transitions, 
which are expected to be  first order. 

Finally, Ref.~\cite{Co:2019wyp} proposed a mechanism wherein the cosmological matter-antimatter asymmetry
stems from  initial conditions in which the axion field is  fastly rotating, much alike 
in the alternative scenario for axion DM production reviewed in \sect{sec:thetadot}. 
A $\dot\theta \neq 0$ corresponds to an asymmetry-density  of the PQ charge, 
which is converted into the baryon asymmetry via QCD and electroweak sphalerons. 
This rotation can be induced at the PQ  scale by effective operators that break explicitly the symmetry,  
tilting  the bottom of the mexican-hat potential. However, these operators must  fade away  rapidly enough at lower temperatures 
not to spoil the solution of the strong CP problem, 
that is the zero temperature minimum of the axion potential  
must remain determined by the QCD non-perturbative  effects.
The mechanism encounters difficulties because to preserve the baryon asymmetry from being washed out, 
$\dot\theta$ must remain large down to the temperature when sphaleron transitions get out of equilibrium.
This, however, implies that at the QCD phase transition the axion kinetic energy would dominate the potential energy, 
thus delaying the onset of oscillations which results in an unacceptable overproduction of DM. The proposed way out 
is to engineer a way to increase the temperature of the electroweak phase transition in order to suppress 
sphalerons transitions at much earlier times.  


\subsection{Axions and inflation}
\label{sec:axion_inflation}

A period of accelerated expansion in the very early Universe called {\it inflation}~\cite{Starobinsky:1980te, Guth:1980zm} is usually invoked to address various problems of the standard cosmology, namely the absence of monopoles~\cite{Guth:1979bh, Guth:1980zm, Einhorn:1980ik} and domain walls~\cite{Sato:1981ds}, the fact that the Universe appears to be homogeneous and isotropic (the {\it horizon problem})~\cite{Kazanas:1980tx}, and the fact that the Universe appears to possess fine-tuned initial conditions that lead to its exceptionally flatness already at recombination (the {\it flatness problem}). One class of inflationary models relies on the dynamics of a field that is responsible for the inflationary period, the {\it inflaton}, which evolves under the influence of a nearly flat potential. 
In the simplest model of {\it single field} inflation
the equation of motion for the inflaton field (denoted by $\phi$) evolving under the potential $V(\phi)$ is similar to \eqn{eq:equation_motion}, 
\beq
	\ddot\phi + 3H\dot\phi - \frac{1}{R^2}\nabla^2\phi + \frac{dV(\phi)}{d\phi} = 0\,.
	\label{eq:equation_inflaton}
\eeq
For super-horizon modes, the spatial derivative can be neglected. A nearly flat potential grants an inflationary period in which the inflaton evolves under a {\it slow-roll} dynamics if $i)$ $\ddot\phi \ll H\dot\phi$, and $ii)$ $\dot\phi^2 \ll V(\phi)$. Condition $i)$ leads to $3H\dot\phi \simeq -dV(\phi)/d\phi$, while condition $ii)$ gives the equation of state
\beq
	w_\phi = \frac{\frac{1}{2}\dot\phi^2 - V(\phi)}{\frac{1}{2}\dot\phi^2 + V(\phi)} \simeq -1\,.
\eeq
If both conditions $i)$ and $ii)$ are satisfied, the energy density of the $\phi$ field is constant during inflation, which yields a quasi-exponential growth of the scale factor according to \eqn{eq:friedmann1}.

Different models have been considered so far in order to embed inflation 
into an axion framework. 
One possibility consists in considering the dynamics of the PQ 
complex field during inflation~\cite{Linde:1991km} 
and identify the inflaton field with the radial mode $\varrho_a$ of the 
PQ complex field 
(see \eqn{eq:PhidecKSVZ}). 
At sufficiently high temperatures, 
the potential of $\varrho_a$ in Eq.~\eqref{eq:VPhiKSVZ} can be approximated by a quartic potential. However, such a form of inflaton potential has been excluded to a high level of confidence by the measurements of the CMB spectra by the \textit{Planck} mission~\cite{Martin:2013nzq, Gerbino:2016sgw, Kinney:2018nny}. For this reason, Ref.~\cite{Fairbairn:2014zta} considered 
a non-minimal coupling to gravity 
so that the potential at large values of $\varrho_a$ is flattened out 
(see e.g. Ref.~\cite{Linde:2011nh}) and the model reconciles with observations.
This model also circumvents the problem that, for relatively high values of the Hubble rate during inflation $H_I$, axion isocurvature fluctuations during inflation are too large with respect to what is allowed by measurements~\cite{Akrami:2018odb}, see \sect{sec:isocurvature}. 
The reason being that 
the radial field has not yet relaxed to its minimum value during inflation 
and it evolves in the regime $\varrho_a \gg f_a$, thus suppressing isocurvature fluctuations.\footnote{Another way to circumvent isocurvature bounds 
in low-scale models of hybrid inflation 
has been proposed in Ref.~\cite{Schmitz:2018nhb}.} 

A less minimal kind of embedding has been carried out in the SMASH model~\cite{Ballesteros:2016euj, Ballesteros:2016xej, Ballesteros:2019tvf}, 
in which the inflaton field is a linear combination of the radial modes 
of the PQ field $\Phi$ and the Higgs doublet $H$, 
both coupled non-minimally to gravity. 
The model shares some similarities with 
the Higgs inflation model of~\cite{Bezrukov:2007ep}, 
but thanks to the mixed embedding of the inflaton field 
it provides a solution of the unitarity problem of standard Higgs inflation \cite{Burgess:2009ea,Barbon:2009ya}, thus making inflationary predictions 
more reliable. 

Within the single field slow-roll inflation model, it is possible to obtain predictions for the scalar and tensor power spectra on super-horizon scales. As discussed in \sect{sec:isocurvature}, tensor modes for single-field inflation are defined in terms of the tensor-to-scalar ratio $r$ which, if measured in the future at the level of 
$r \sim 10^{-3}$, would shut off the pre-inflationary scenario completely~\cite{Fox:2004kb, Hertzberg:2008wr, Visinelli:2009zm, Visinelli:2014twa, Marsh:2014qoa, Fairbairn:2014zta}, see also Fig.~\ref{fig:bound_standard}. It has recently been shown~\cite{Tenkanen:2019xzn} that a model of inflation with the SM Higgs field, in which the Higgs field is coupled non-minimally to {\it Palatini} gravity~\cite{Bauer:2008zj} leads to an inflation energy scale $H_I \sim 10^8\,$GeV and a low tensor-to-scalar ratio $r\sim 10^{-13}$. In this model, axion isocurvature fluctuations that would evade the current bounds can be accommodated, and it is then possible to realise axion DM in the pre-inflationary scenario with a value of the axion energy scale of the order of $\sim 10^{14}\,$GeV.

The QCD axion itself could have driven inflation, evolving via a series of tunnelling events in the so-called chain inflation model~\cite{Freese:2004vs, Freese:2005kt}. 
Assuming an axion model in which the continuous shift symmetry is broken 
down to a discrete $\mathbb{Z}_{N_{\rm DW}}$ symmetry, with a large  
number $\mathcal{O}(100)$ of local minima $N_{\rm DW}$, 
and a tilting term parametrised via 
a soft breaking contribution with height $\eta$,  
the potential reads
\beq
	V(a) = V_0\,\[1-\cos\(\frac{N_{\rm DW} \, a}{v_a}\)\] + \eta \cos\(\frac{a}{v_a}\)\,.
	\label{eq:potential_chain}
\eeq
The axion evolves in one of the false vacua described by the potential in Eq.~\eqref{eq:potential_chain}, providing about one $e$-fold of inflation before tunnelling to the next false vacuum and approaching the true vacuum of the theory. Chain inflation is a viable model for solving the horizon, entropy and flatness problem of standard cosmology and for generating the right amount of adiabatic cosmological perturbations~\cite{Chialva:2008zw, Chialva:2008xh}, although a considerable effort in model building is necessary since a large number of tunnelling events, $\gtrsim 10^8$ per Hubble volume per $e$-fold, are required~\cite{Cline:2011fi}.
%
Finally, two  classes of inflation models which come from superstring axion 
scenarios (see \sect{sec:AxionStringT}) are axion inflation~\cite{Svrcek:2006hf, Svrcek:2006yi, Grimm:2007hs,
  Long:2014dta} where the inflationary potential is given by the
standard cosine potential, and axion
monodromy~\cite{Silverstein:2008sg, McAllister:2008hb,
  Marchesano:2014mla}, in which the potential includes extra terms like a linear term.

\bigskip


%
%



\subsection{Gravitational waves from the Peccei-Quinn phase transition}
\label{sec:axion_GW}

The  detection of GW by LIGO~\cite{Abbott:2016blz} 
has rendered clear that a new powerful tool is now available 
for  the exploration of the Universe. In particular,  first-order phase transitions 
in the early Universe  can produce  stochastic GW  signals which could be potentially observed.
The LIGO/VIRGO frequency band corresponds to first-order phase transitions
which could have happened  at  temperatures around $10^{7}$-$10^{8}\,$GeV. 
This is intriguingly close to the  lowest possible energy scale where the PQ 
symmetry can be broken~\cite{Dev:2016feu}. 
Although  the astrophysical bounds reviewed  in \sect{sec:Astro_bounds}  require $f_{a}$ to lie  above 
the scales at which LIGO/VIRGO are most sensitive, in the presence of a certain amount 
of supercooling  the nucleation temperature of the PQ phase transition can be actually 
smaller than $f_{a}$  by an order of magnitude or more~\cite{DelleRose:2019pgi,vonHarling:2019gme}, 
thus motivating  studies  of  the potential of  GW   experiments to detect the imprint  of 
the phase transition that gave birth to the axion. 
A GW signature could be detected only if the PQ phase transition 
occurred after the end of inflation and if it is of strong first  order. 
In weakly coupled models the transition, however,  is typically second order, 
except in the region of parameters where the PQ symmetry is broken
through the Coleman-Weinberg mechanism, while  in strongly coupled realisations 
of axion models the transition is often first order. This restricts the type of models 
that can produce the sought signals. 
 Recent studies of the energy density stored in  stochastic GW from a first order PQ transition 
and of the corresponding peak frequency 
have been presented in~\cite{DelleRose:2019pgi,vonHarling:2019gme}.
Ref.~\cite{Croon:2019iuh}  studies the GW signal in a dynamical 
super-heavy axion models 
where the phase transition is associated with a new   colour-like gauge group
confining around the TeV scale. In this case, however, the signal frequency remains well below 
the LIGO/VIRGO sensitivity range. 
A study of the axion (and ALP) parameter space which may be probed by future GW detectors
can be found in Ref.~\cite{Machado:2019xuc}.


\subsection{Solutions to the domain walls problem}
\label{sec:AxionsDW}

The  DW problem arises from the fact that  the axion field $a$, being an angular variable, takes 
values in the interval $[0, 2\pi v_a)$.  The axion potential is periodic
in $a$ with period $\Delta a = 2 \pi v_a / N_{\rm DW}$, 
and  thus it enjoys 
an exact  $\mathbb{Z}_{N_{\rm DW}}$ discrete symmetry.  
Once at $T\sim \Lambda_{\rm QCD}$ the  non-perturbative QCD effects  lift the axion potential, 
$N_{\rm DW}$ 
degenerate vacua appear.  
In general  the initial value of   $a$ at the bottom of the originally flat potential  is 
randomly selected and it  differs  in different patches of the Universe,  
so that in each  patch the axion  will eventually flow towards a different minimum,
breaking spontaneously  $\mathbb{Z}_{\NDW}$.   DWs will then form at the boundaries between 
regions of different vacua.   The cosmological DW problem~\cite{Sikivie:1982qv}  consists in the fact 
that the energy density  of the DWs would largely overshoot the critical density of the Universe.
There are, however,  two  scenarios in which there is no DW problem: 
(i) The first   corresponds to  the pre-inflationary scenario in which 
an initial patch characterised by some value of  $\theta_i$ gets exponentially
inflated to super-horizon scales, so that  after inflation  the whole observable Universe 
is characterised by  a  unique minimum of the axion potential, and is thus free from topological defects.  
(ii) The second scenario encloses the models in which   $N_{\rm DW}=1$, so that  there is a single 
value of $\theta$  where the potential has a minimum.  Although the vacuum is unique, a particular type of DWs still form,   
but they are harmless. The reason can be pictured as follows: 
when  $U(1)_{\rm PQ}$ gets broken, strings form, and in circulating around a string 
$\theta$  changes  by $2\pi$. 
When the temperature  approaches the QCD scale and the axion potential gets tilted,  
the  unique minimum $\theta=0$ is selected.  Still, in a two dimensional region attached to the string 
 the phase must jump from $0$ to $2\pi$ and  back to $0$, and this region corresponds to 
 a wall with one edge attached to the string.  However, a  configuration of  walls  bounded by strings  tends to 
 rapidly collapse~\cite{Vilenkin:1984ib}, and  eventually the whole system of walls and strings disappears 
 without ever coming to dominate   the energy density of the Universe.
Hence,   axion models with $N_{\rm DW} =1$ are safe with  respect  to the DW problem 
also in  post-inflationary scenarios.

A first example of a construction with $N_{\rm DW} =1$ is the original KSVZ 
model \cite{Kim:1979if,Shifman:1979if} that we have reviewed in \sect{sec:KSVZ},   
where a single pair of electroweak singlet exotic quarks in the fundamental of $SU(3)_c$ yields 
a colour anomaly $2N = N_{\rm DW} = 1$. 
 Georgi and Wise~\cite{Georgi:1982ph} considered instead 
the possibility of cancelling part of the 
QCD anomaly of DFSZ-type of models by introducing suitable representations 
of exotic quarks  of  KSVZ-type with   PQ charge of the opposite sign,  so 
that a total  anomaly $2N=1$ eventually results. 

A different kind of construction features an apparent value $\NDW>1$, while 
the physical number of DW is in fact $\NDW=1$.
These constructions  rely on the introduction of extra  symmetries, in such a way that 
the degenerate vacua of the axion potential are  connected 
by symmetry transformations. 
The first realisation of this mechanism is due to Lazarides and Shafi (LS)~\cite{Lazarides:1982tw}
which observed that the DW problem does not exist if the discrete subgroup 
$\mathbb{Z}_{\NDW}$  in $U(1)_{\rm PQ}$  can be embedded in the centre of  
a continuous gauge group, and they provided   a neat  example  based on the GUT symmetry 
$SO(10)\times U(1)_{\rm PQ}$.
A different possibility was proposed in Ref.~\cite{DiLuzio:2017tjx}. 
The axion arises from an accidental $U(1)$  enforced 
on the   potential of a scalar multiplet  $Y_{\rm LR}$ 
by a   gauge symmetry  $SU(\mathcal{N})_L \times SU(\mathcal{N})_R$. 
 $Y_{\rm LR}$  transforms  under the group 
as $(\mathcal{N},{\mathcal{\overline{N}}}) $
and is responsible for the spontaneous breaking of the gauge group down to 
$SU(\mathcal{N})_{L+R}$.  Although  the construction gives a QCD anomaly 
with coefficient $N = \mathcal{N}/2$, 
all the minima can be connected by gauge transformations corresponding to the centre   
$\mathbb{Z}_\mathcal{N}$ of the unbroken group, and hence they are gauge equivalent.
While in the original LS model $\mathbb{Z}_{\NDW}$ was 
embedded into a local group,  embedding in global groups  can  also yield the same result.  
For example,  models  with global family groups were considered in Ref.~\cite{Barr:1982bb}.

A different type of constructions  in which  $\NDW=1$  can be engineered, 
rely on the presence of  more than one global $U(1)$  symmetry~\cite{Barr:1982uj}.
The anomalous PQ will in general correspond to a combination
of the various Abelian groups, and it can be arranged so that QCD 
effects  break this specific combination to the trivial subgroup $\mathbb{Z}_{1}$.

A horizontal realisation of the PQ symmetry that, together with $B-L$ global invariance, 
can solve the DW problem for any arbitrary number of fermion generations was discussed in
Ref.~\cite{Davidson:1983tp}. 
Other models enforce  $\NDW=1$ making use in different ways  of  generation dependent PQ symmetries. 
For example  a partial cancellation of the anomaly contributions between 
different generations can be arranged~\cite{DiLuzio:2017ogq,Bjorkeroth:2019jtx}, 
or the PQ charges are chosen in such a way that two generations give vanishing contributions
to the anomaly~\cite{Bjorkeroth:2018ipq}, or more in general it can be assumed that some SM quark flavours 
have a special status with respect to the PQ symmetry~\cite{Geng:1988nc}, for example it can happen  that 
only one or two  out of all  the SM quarks  contribute to the anomaly~\cite{Hindmarsh:1997ac}.  
Of course, as was discussed in \sect{sec:gaFCNC}, all  these models feature  flavour violating 
axion couplings,  and can be  thoroughly  tested  by  searching  for flavour changing processes.

Models  for which none of the  above two conditions (i) and (ii) are satisfied, that is 
the  PQ symmetry is broken after inflation, and  $\NDW>1$ gives rise to  
the same number of physically inequivalent  degenerate vacua,   
 can also remain viable,  but additional assumptions are needed.  The DW problem can
be disposed of in a simple way by introducing an explicit  breaking of the PQ symmetry so that  
the degeneracy between the  different vacua is removed and there is a unique 
minimum of the potential. This can be done either by introducing 
an explicit breaking of the PQ symmetry via Planck-suppressed effective operators~\cite{Sikivie:1982qv} 
or by non-perturbative potential terms induced 
by a new confining gauge group \cite{Barr:2014vva,Reig:2019vqh,Caputo:2019wsd}. 
Breaking explicitly the PQ symmetry is, however, a
delicate issue: sufficiently large breaking effects are needed to guarantee that 
regions trapped in false vacua  will cross over to the  
 true vacuum before DWs start dominating the Universe energy
density.\footnote{A stronger bound actually originates from requiring that 
the axions produced from the collapse of DWs do not have an abundance larger than the DM one \cite{Hiramatsu:2010yn,Hiramatsu:2012sc,Kawasaki:2014sqa}. 
An early matter domination era around the MeV scale can help in 
relaxing this bound \cite{Harigaya:2018ooc}.} 
However,  at the same time they should not be too large, 
otherwise they would spoil the  PQ solution. The present limit $\theta\lsim 10^{-10}$ still leaves 
a viable region in parameter space where these two
conditions can be simultaneously matched~\cite{Sikivie:2006ni,Marsh:2015xka,Caputo:2019wsd}.



\subsection{Solutions to the Peccei-Quinn quality problem}
\label{sec:protectingPQ}


As we have discussed  in~\sect{sec:PQquality}  in order to ensure   the effectiveness of the PQ  
mechanism in solving the strong CP problem a strong requirement needs to be satisfied: 
the PQ symmetry  must  remain a good symmetry well beyond the level of renormalizable  operators, 
and arguably up to effective operators of dimensions $d > 10$.   
However, the PQ symmetry is a global symmetry, and in QFT global symmetries are not 
deemed  to be  fundamental or exact. Moreover, the PQ symmetry is anomalous, 
and as such  at the quantum level it  is not even a real symmetry. Analogies with 
baryon and lepton $U(1)$ symmetries in the SM, which are also global and anomalous, 
but whose origin as accidental symmetries is well understood, naturally lead to speculate whether  
the PQ symmetry might also arise accidentally, in the sense that other fundamental symmetries 
(gauge and Lorentz)  might forbid, up to the required dimension,  operators that do not conserve 
the PQ charge.
Several construction have been put forth to realise this idea. 
Probably the simplest possibility  is that of a discrete gauge
symmetry. For instance a $\mathbb{Z}_n$ acting on  
$\Phi$ as $\Phi \to e^{i 2\pi / n} \Phi$ would forbid 
all effective operators $(\Phi^\dag \Phi)^m\Phi^{k}$ with $k < n$. 
A possible way to generate  discrete gauge symmetries 
in 4-dimensional QFT  was suggested in Ref.~\cite{Krauss:1988zc}.
Consider a $U(1)$ gauge symmetry under which the axion multiplet  
$\Phi$ carries charge 1, while a second field $\xi $ carries 
charge $n$: $\Phi \to e^{i \alpha} \Phi$,  $\xi \to e^{i n \alpha} \xi$.
Suppose that $\xi$ undergoes condensation at some high-energy scale
$\vev{\xi} \gg f_a$.  
The invariance of the VEV $\vev{\xi}\to  e^{i n \alpha} \vev{\xi} $  
with    $\alpha =2\pi/n$ corresponds to a  $\mathbb{Z}_n$ discrete gauge 
symmetry, unbroken above $f_a$,  under
which   $\Phi$ transforms as $\Phi \to e^{i\alpha} \Phi = e^{i 2\pi / n}
\Phi$. This ensures that the first allowed operator   breaking
PQ explicitly is $\Phi^n$.
Mechanisms with large local discrete symmetries  have been for example invoked in 
Ref.~\cite{Dias:2002gg,Carpenter:2009zs,Harigaya:2013vja,Dias:2014osa,Harigaya:2015soa}.

Other models rely on the introduction of a new local Abelian  symmetry $U(1)'$ 
that, as in Ref.~\cite{Barr:1992qq},  can be directly added to enlarge the SM gauge group. 
In this reference two singlet scalars  $\Phi$ and  $S$ are also introduced,  with  $U(1)'$ charges 
respectively $p$ and $q$, and these charges are chosen in such a way that 
the lowest order effective operator that breaks the PQ symmetry, namely $(S^\dagger)^p \Phi^q$,  is of 
the required high dimension $d=p+q$. 
Ref.~\cite{Holman:1992us} considers instead a supersymmetric GUT extension 
 $E_6\times U(1)'$ , where  the usual trilinear superpotential Yukawa coupling $\mathbf{27}^3$ is  
 forbidden by the new Abelian symmetry, and is  replaced by  a coupling 
involving a new unconventional representation  $\mathbf{27}_1 \mathbf{27}_{-1} \mathbf{\overline{351}}_0$, 
where subscripts refer to the $U(1)'$ gauge charges.
A global PQ symmetry under which $\mX(\mathbf{27})=1$  and 
$\mX(\mathbf{\overline{351}})=-2$  arises accidentally, and 
 the lowest order  gauge invariant PQ violating operators  $\mathbf{27}^6 $ or 
$\mathbf{\overline{351}}^6$ are of the required large dimension. 
Soft supersymmetry breaking effects might, however, endanger this solution \cite{Dobrescu:1996jp}. 
Ref.~\cite{Fukuda:2017ylt} considers instead the possibility of embedding the PQ 
symmetry into a gauged $U(1)'$, which is rendered non anomalous by 
adding exotic fermions coupled to the SM only via gauge interactions, so that 
the subset of gauge rotations acting only on the SM quarks can be interpreted as an
accidental (global and anomalous)  $U(1)\subset U(1)'$. 
Ref.~\cite{Duerr:2017amf} promotes baryon number to  a local gauge symmetry by 
canceling the anomaly  via the introduction of exotic fermions which play the role of  KSVZ  exotic quarks.  
The authors  show that the PQ symmetry remains sufficiently protected by this gauged baryon number. 
Finally, Ref.~\cite{Bonnefoy:2018ibr} 
addresses the axion quality problem in a 4-dimensional clockwork model, 
with the PQ global symmetry arising accidentally due to a gauged $U(1)^N$ 
symmetry.

Other  approaches  exploit instead new non-Abelian gauge groups. 
Georgi, Hall and Wise \cite{Georgi:1981pu} provided the first example\footnote{Incidentally, 
this is also the first paper where the issue of the PQ quality problem was clearly laid down.} 
of an accidental PQ symmetry arising from a grand-unified gauge 
group based on $SU(9)$, which incorporates the usual $SU(5)$ GUT 
as a subgroup. 
The construction is non-trivial, since it involves several 
$SU(9)$ representations (left-handed fermions in a $\mathbf{36} \oplus 5 \times \mathbf{\overline{9}}$ 
and scalar fields in a $\mathbf{80} \oplus 5 \times \mathbf{9} \oplus \mathbf{126}$). 
Writing the most general renormalizable Lagrangian allowed by the $SU(9)$ gauge symmetry 
enforces a QCD anomalous U(1) symmetry that is spontaneously broken at the unification scale. 
In this model the PQ symmetry is explicitly broken at the $d=5$ level.
The complexity of this construction shows that it is highly 
nontrivial to get an accidental PQ symmetry in GUTs (without resorting to extra gauged $U(1)$'s) 
and we are not aware of other successful attempts, apart for the original one in Ref.~\cite{Georgi:1981pu}. 
Instead the construction discussed in Ref.~\cite{DiLuzio:2017tjx} takes  inspiration from  
a type of  flavour models in which the SM Yukawa couplings are promoted to dynamical 
scalar fields $Y(x)$ transforming in the bi-fundamental of a gauge group $SU(3)_L\times SU(3)_R$. 
As it  was remarked in Ref.~\cite{Nardi:2011st},   
gauge invariant operators  in the scalar potential for $Y(x)$ are Hermitian and thus respect   
an accidental  $U(1)$ rephasing invariance, with the sole exception of $\det Y(x)$ which 
is non-Hermitian and of dimension $d=3$.  Then, generalising the construction to an
 $SU(\mN)_L\times SU(\mN)_R$ invariant potential with  $\mN>4$,  the dimension of the symmetry breaking operator 
is promoted to  $d=\mN$, so that the quality of the accidental $U(1)$ remains 
determined by the dimension of the gauge group which can be arbitrarily chosen. 
An even simpler possibility based on the same approach is to replace the 
 SM flavour symmetry   $SU(3)_L\times SU(3)_R$  with a different one 
$ SU(M)_L \times SU(N)_R$ ($M\neq N$) 
 and require  that the Yukawa field  transforms as 
 $(M,\overline{N})$
 (for $M$ and/or $N >3$ this can be done  by introducing new exotic vectorlike 
 quarks)~\cite{LucLucaEN:inprep2}. In this case the dimension of the first flavour-singlet 
non-Hermitian scalar operator  is at least as large as 
the least common multiple of $M$ and $N$, so that large operator dimensions for the PQ breaking operators 
can be obtained with group factors of relatively small degree.
Finally, scalar-gauge theories in which $Y(x)$ transforms as the 
symmetric (anti-symmetric) of $SU(\mN)$ broken into $SO(\mN)$ ($Sp(\mN)$) 
also lead to an accidental Goldstone for $\mN > 4$ \cite{Buttazzo:2019mvl}, 
and they might be used to construct models for the PQ protection along the lines of \cite{DiLuzio:2017tjx}.  

Ref.~\cite{Lee:2018yak}  introduces an exotic   sector equipped with an
 $SU(\mN)$  gauge symmetry where a  hidden baryon number  $U(1)_{B_\mN} $ 
 appears accidentally,  similarly to what happens in the SM with ordinary $U(1)_B$.
However, differently from the SM where $B$ violating effective operators are allowed 
already at $d=6$,  the  $B_\mN$ violating operators of lowest dimension are the gauge (and, assuming $\mN$ even, 
also Lorentz)    invariant  $SU(\mN)$ singlets   
$\epsilon_{\alpha_1\alpha_2\dots\alpha_\mN}\mathcal{Q}^{\alpha_1}\mathcal{Q}^{\alpha_2} \dots \mathcal{Q}^{\alpha_{\mN}}$  
with dimension $d=3\mN/2\gg3$.  In the presence of fermion species chiral under $SU(\mN)\times SU(2)_C$ 
a $U(1)_{B_\mN}$-$SU(3)^2_C$ anomaly can arise, so that the  new baryon symmetry 
can act in the SM as  a PQ symmetry.
In Ref.~\cite{Cheung:2010hk} instead the SM quark flavour symmetry has been used 
directly, and the required level of suppression for the PQ breaking operators is obtained by 
assuming that quark masses are generated radiatively, so that the Yukawa couplings 
do not correspond to fundamental symmetry breaking scalars,  but 
result from a more complicated set of spurions that allow to keep the PQ symmetry exact up to $d=12$. 

Composite axion models (reviewed in \sect{sec:CompositeAxions}) are also well suited to 
arrange for  approximate PQ symmetries preserved up to operators of high dimension.
One of the first constructions exploiting compositeness with this aim was the Randall model~\cite{Randall:1992ut}. 
The relevant gauge symmetry is $SU(\mN)\times SU(m)\times SU(3)_C$.
Exotic fermions  transform under the group factors with suitable chiral 
assignments in such a way that an accidental global $U(1)$  arises, which is non-anomalous 
with respect to the first two factors but has an $SU(3)_C$ anomaly. 
$SU(\mN)$  becomes strong at a large scale breaking spontaneously $SU(m)$
(much alike $ SU(3)_C$ condensates  in the SM  break spontaneously $SU(2)_L$) 
and breaking also the global $U(1)$,   but  not $SU(3)_C$.
In this model the lowest dimensional operator consistent with the gauge and Lorentz 
symmetries,  but not respecting the $U(1)$, can be built 
with  $2m$  uncoloured fermions. Hence the quality of the PQ symmetry is controlled 
by the dimension of the $SU(m)$ group.  
Other composite models for PQ symmetry protection exploiting similar ideas can be found in 
Refs.~\cite{Dobrescu:1996jp,Redi:2016esr,Lillard:2018fdt,Gavela:2018paw}. 

Extra-dimensional setups are also well suited to protect the PQ symmetry. 
For instance, in the proposal of Ref.~\cite{Choi:2003wr} the axion is identified 
with the 5-th component of a gauge field in a 5D orbifold field theory compactified on 
$S^1/Z_2$, for which all PQ breaking terms other than the QCD anomaly 
are naturally suppressed by the higher-dimensional gauge symmetry 
and the 5D locality. A recent geometrical solution of PQ quality 
problem was considered in Ref.~\cite{Cox:2019rro}, by 
modelling the axion with a bulk complex scalar field in a slice of AdS$_5$, 
where the $U(1)_{\rm PQ}$ symmetry is spontaneously broken in the bulk but explicitly broken on the UV brane. By localising the axion field towards the IR brane, gravitational violations of the PQ symmetry on the UV brane are sufficiently suppressed by the warp factor.


\subsection{Axions and composite dynamics}
\label{sec:CompositeAxions}


Composite axion models were originally motivated by the possibility 
of dynamically explaining the hierarchy $f_a \ll \mP$, as a consequence 
of a confining non-abelian gauge theory, 
largely inspired by pions and chiral symmetry breaking in QCD 
as well as techni-colour models for the electroweak scale.

For instance, the original model of Kim \cite{Kim:1984pt} 
(see also \cite{Choi:1985cb}) was based on a new gauge group $SU(\mathcal{N})$ 
that confines at a scale $\Lambda_{SU(\mathcal{N})} \sim f_a$, and comprises 
the following vector-like fermion content 
with transformation properties under $SU(\mathcal{N}) \times SU(3)_c$:
 $\psi_{L,R} \sim (\mathcal{N}, 3)$ and $\xi_{L,R} \sim (\mathcal{N}, 1)$.  
In the limit of zero fermion masses and for 
$g_s \to 0$ the model has an $SU(4)_L \times SU(4)_R \times U(1)_V$ 
global symmetry, which is spontaneously broken down to 
$SU(4)_{L+R} \times U(1)_V$ by the fermion condensates 
$\vev{\bar \psi_{L} \psi_{R}} = \vev{\bar \xi_{L} \xi_{R}} \sim \Lambda^3_{SU(\mathcal{N})}$. The resulting 15 
NGB transform as $1+3 + \bar 3+8$ under $SU(3)_c$. The colour singlet, 
with a $(\bar \psi_L \psi_R - 3 \, \bar \xi_L \xi_R)$ content, 
is identified with the composite axion  
and $U(1)_{\rm PQ}$ corresponds to the $[(T^{15})_L \times \mathbb{I}_R 
- \mathbb{I}_L \times (T^{15})_R]/\sqrt{2}$ 
broken generator of $SU(4)_L \times SU(4)_R$, with $T^{15} = 
\tfrac{1}{2\sqrt{6}} \text{diag}(1,1,1,-3)$
belonging to the $SU(4)$ Cartan sub-algebra, 
so that the $U(1)_{\rm PQ}$ is anomalous under QCD but not under $SU(\mathcal{N})$. 
When $g_s$ is turned on, 
the $SU(4)_{L+R}$ global symmetry is explicitly broken down to the gauged $SU(3)_c$. 
The axion gets a tiny mass from QCD instantons, since the associated current 
is QCD anomalous
\beq 
\label{eq:compaxcurr}
J^\mu_{\rm PQ} = - \frac{1}{2 \sqrt{6}} \[  \bar \psi \gamma^\mu \gamma_5 \psi
- 3 \bar \xi \gamma^\mu \gamma_5 \xi \]  
\qquad \Longrightarrow \qquad
\partial_\mu J^\mu_{\rm PQ} = \frac{g_s}{16 \pi^2} \frac{\mathcal{N}}{2 \sqrt{6}} G \tilde G
\, ,
\eeq
while the other pseudo NGB in non-trivial colour representations 
get masses of order
$g_s \Lambda_{SU(\mathcal{N})} / (4\pi)$
via perturbative gluon loops 
(similarly to the QED contribution to the mass of the charged pion of order 
$e \Lambda_{\rm QCD} / (4\pi)$). 
The axion couplings to SM fields in Kim's composite axion model follow from 
the expression 
of the PQ current in \eqn{eq:compaxcurr} and from 
the definition of the `axi-pion' constant \cite{Kaplan:1985dv} 
$\langle 0 | J^\mu_{\rm PQ}  | a \rangle = i F_a p^\mu$ 
(with $F_a$ of order $\Lambda_{SU(\mathcal{N})}$, 
analogously to $f_\pi \sim \Lambda_{\rm QCD}$). 
Matching with the axion effective Lagrangian in \eqn{eq:Leffaxion} 
one gets $f_a = \sqrt{6} F_a / \mathcal{N} \sim \sqrt{6} \Lambda_{SU(\mathcal{N})} / \mathcal{N}$. 
Note that similarly to KSVZ-like models where SM fields ($f$) are uncharged under the PQ, 
the model-dependent axion coupling to SM fermions vanish, 
$c^0_f = 0$, 
and also $g^0_{a\gamma} = 0$. The latter actually depends on the specific choice 
of the exotic fermion representations in Kim's model. 
By assigning to them non-trivial electroweak quantum numbers one can also get 
direct EM anomaly contributions to $g^0_{a\gamma} \neq 0$ \cite{Kaplan:1985dv},  
as discussed in \sect{sec:KSVZ-like}. 

It should be noted that in Kim's original composite axion model, 
similarly to standard KSVZ constructions, 
vector-like exotic fermion masses need to be 
forbidden 
in order not to spoil the whole 
framework. 
In fact, bare mass terms, $m_{\psi,\,\xi}$, 
even if $\ll \Lambda_{SU(\mathcal{N})}$, would still misalign the minimum 
of the QCD axion potential unless extremely suppressed,  
since they contribute to the composite NGB axion mass as 
$m_a \sim \sqrt{\Lambda_{SU(\mathcal{N})} m_{\psi,\,\xi}}$.  
The model is also prone to the PQ quality problem (see \sect{sec:PQquality}), 
since Planck-suppressed $U(1)_{\rm PQ}$ breaking 
operators, which are in principle allowed by gauge symmetries, 
could spoil the solution of the strong CP problem. 
This motivated the construction of composite axion models in which an extra, 
chiral gauge symmetry avoids the presence of exotic fermion mass terms and 
protects as well from higher-order PQ symmetry breaking operators. 
Models of this type were briefly reviewed towards the end of \sect{sec:protectingPQ}. 



\subsection{Axions and GUTs}
\label{sec:axionGUTs}

Unified gauged theories provide a rationale for the large value of the axion decay 
constant $f_a$, which for phenomenological reasons must lie between $10^{8}$ 
GeV and the Planck scale. 
In fact, if the axion field is embedded into 
a scalar representation responsible for the breaking of the GUT symmetry (e.g.~in the global phase of a 
complexified adjoint representation) 
one has $f_a  = v_a/N_{DW} \approx M_G / N_{\rm DW}$, where $M_G \gtrsim 10^{15}$ GeV 
denotes the scale of GUT breaking, which is bounded from below by the non-observation 
of proton decay.  
Models of this type, in which the global $U(1)_{\rm PQ}$ commutes 
with the GUT group, were first proposed in the context of 
$SU(5)$ \cite{Wise:1981ry} 
and $SO(10)$ \cite{Lazarides:1981kz} 
(for modern variants see also 
\cite{Ernst:2018bib,DiLuzio:2018gqe,Ernst:2018rod,FileviezPerez:2019fku,FileviezPerez:2019ssf}). 
A remarkable feature of these models is the connection between proton decay 
and standard axion searches, which will start to probe the GUT-scale axion in the coming decade 
(cf.~the reach of Casper-Electric and ABRACADABRA in \sect{sec:Haloscopes}). 
For instance, the minimal $SU(5) \times U(1)_{\rm PQ}$ model of Ref.~\cite{DiLuzio:2018gqe}, 
based on the original Georgi-Glashow \cite{Georgi:1974sy} field content plus a 
$24_F$ representation that fixes both neutrino masses and gauge coupling unification \cite{Bajc:2006ia,Bajc:2007zf,DiLuzio:2013dda}, 
provides a rather sharp prediction (arising from gauge coupling unification and proton decay constraints) 
for the value of the axion mass  $m_a\sim  5\,$neV, setting a well-definite target for axion DM experiments. 
Regarding the connection with axion DM 
let us recall that, in order not to over-produce axion DM, GUT-scale values of $f_a$ require the PQ symmetry 
to be broken before inflation (pre-inflationary PQ breaking  scenario) together with a certain tuning 
of the initial misalignment angle to small values  
 (for $f_a \sim 10^{15}$ GeV, $\theta_i \sim 1\%$).\footnote{This condition can, however, be circumvented 
 by appealing for example to the mechanism of axion dilution by late entropy production 
 reviewed in \sect{sec:modifiedcosmo},
as is done e.g. in the supersymmetric GUT model of Ref.~\cite{Co:2016vsi}.}
 Note that values of $f_a\ll M_G$ can still be obtained for $v_a \sim M_G$, but these 
 require values of  $N_{\rm DW}$ is unrealistically large. 
Another class of models which allows  for $f_a \ll M_G$ and that at the same time can be compatible 
with post-inflationary PQ breaking scenarios is based on $SO(10)$ GUTs in which the axion is embedded into a representation responsible 
for an intermediate symmetry breaking stage, possibly related to the scale of 
right-handed neutrinos \cite{Mohapatra:1982tc,Holman:1982tb,Bajc:2005zf,Bertolini:2013vta,Altarelli:2013aqa,Ernst:2018bib}. In such a case, however, the scale $f_a$ turns out to be ``sliding'', since it is only weakly constrained by gauge coupling unification \cite{Ernst:2018bib}. 

Regarding the structure of axion couplings in GUTs, these are fixed up to some 
minor model-dependent factors, thus providing a well-defined and motivated 
benchmark for axion searches. 
Most notably, the value of the group theory factor  that determines the  axion coupling to photons is 
fixed to be $E/N = 8/3$. This can be seen from the explicit expression   
\beq 
\label{eq:EoNGUTs} 
\frac{E}{N} = \frac{\Tr Q^2}{\Tr T^2_C} 
= \frac{\Tr (T^3_L)^2 + \frac{5}{3} \Tr T_Y^2}{\Tr T^2_C} 
= \frac{8}{3} \, ,
\eeq
where $Q = T^3_L + \sqrt{5/3} \,T_Y$, 
in terms of the GUT-normalised hypercharge generator $T_Y$. This result follows from the fact that  
all the generators of the GUT group, and in particular $T_C, T^3_L$ and $T_Y$, 
have the same normalisation.\footnote{Different values of $E/N$ are instead possible 
in the `unificaxion' framework of Ref.~\cite{Giudice:2012zp}, 
which assumes that the anomaly factors are due to 
intermediate-scale KSVZ-like fermions which form 
incomplete GUT multiplets and assist gauge coupling unification. 
It should be noted, however, that such scenarios are not easily motivated from a 
UV point of view. Indeed, as long as the $U(1)_{PQ}$ symmetry commutes with the GUT group, 
the full GUT representation contributes to the anomaly coefficients, yielding the result 
in \Eqn{eq:EoNGUTs}. Moreover, all the fragments of the original GUT multiplet 
obtain mass of order $f_a$, and thus do not improve  gauge coupling unification.\label{foot:unificaxion}} 

The Yukawa sector of an axion GUT model is similar (as far as concerns global PQ charges)
to that of the DFSZ-I model with Yukawa Lagrangian in \eqn{eq:yukDFSZ}. 
E.g.~in minimal $SU(5)$ one has (neglecting for simplicity neutrino masses 
and corrections to the charged fermion mass relations)
\beq 
\label{eq:LYSU5}
\mathcal{L}^Y_{\rm SU(5)} \supset 
- Y_{10} 10_F 10_F 5_{H_u}
- Y_5 \bar 5_F 10_F \bar 5_{H_d} \, ,   
\eeq
which after projecting onto the SM components matches \eqn{eq:yukDFSZ}, 
with the following GUT-scale boundary values for the Yukawa matrices: $Y_{10} = Y_U$ and $Y_{5} = Y_E = Y_D^T$. 
Hence, we can take over the derivation of the axion couplings to SM fermions in the 
DFSZ-I model (cf.~\Eqns{eq:axuu}{eq:axee}), which yields 
%
\beq 
\label{eq:c0fGUTdef}
c^0_{u_i} = \frac{1}{N} \cos^2\beta \, , \qquad 
c^0_{d_i} = \frac{1}{N} \sin^2\beta \, , \qquad 
c^0_{e_i} = \frac{1}{N} \sin^2\beta \, ,  
\eeq
where we used $v_a = f_a (2N)$
and with the important difference (compared to DFSZ-I) 
that the colour anomaly $N$ is a model-dependent factor 
that can be computed only after specifying the full GUT model. 

While we have exemplified this derivation in the context of $SU(5)$, 
similar results apply to the $SO(10)$ case, 
where however care must be taken in order to properly orthogonalise the 
physical axion field with respect to the Goldstone fields 
of all the broken gauge generators 
(see Ref.~\cite{Ernst:2018bib} for a detailed account). 
Another peculiarity of GUTs like is $SO(10)$ is also the fact that some vacua 
of the axion potential can be connected by gauge transformations, 
and hence the naive identification $N_{\rm DW} = 2 N$ does not hold \cite{Lazarides:1982tw,Ernst:2018bib}.


\subsection{Axions from superstrings}
\label{sec:AxionStringT}

In this Review we have focussed on QFT axion solutions to the strong
CP problem for which a spontaneously broken PQ symmetry, of which the
axion is the resulting pNGB, is a necessary ingredient.  As already
noted, the PQ symmetry is global and anomalous, hence with good reason
it can be considered unsatisfactory to impose PQ as a fundamental
symmetry of a Lagrangian. Indeed, a more sound possibility is that it
arises automatically as a consequence of other fundamental principles.
 
Another avenue that has been ventured is  exploring whether axions can
arise naturally from a more fundamental theory  as for
example superstring theory, supergravity or M-theory.  At first glance, candidates featuring
the qualitative properties of the QCD axion are superabundant in
string theory. They arise from ten-dimensional antisymmetric gauge
field tensors that, upon compactifying six internal coordinates
$M_{10}\to M_{4}\times V_{6}$, with $V_{6}$ some compact manifold,
behave like pseudoscalars in the 4D effective theory.  Each string
theory realisation has antisymmetric $p$-forms that can host the
axion, and that go under different names: the Ramond-Ramond (RR)
$C$-fields for Type II strings, the $C_{3}$-form for supergravity, the
NS-NS $B_{2}$-field for the heterotic string (NS stands for
Neveu-Schwartz), the RR $C_{2}$-field for the Type I string.  Their
relations with possible QCD axions in four dimensions have been studied
and reviewed in several papers, see for example
Refs.~\cite{Witten:1984, Svrcek:2006hf, Svrcek:2006yi, Conlon:2006tq}.
Let us consider for definiteness the gravity supermultiplet, that is
present in all string theories.  Besides the 10D graviton $g_{MN}$, it
contains an antisymmetric tensor $B_{MN}$ and the dilaton.  After
compactification, the fields in $B_{MN}$ can be categorised into a set
of tangential fields $B_{\mu\nu}$ with indices in Minkowski space $M_{4}$,
and a set of fields $B_{ij}$ with indices in the internal space
$V_{6}$.  Let us consider $B_{\mu\nu}$ first: it has only one
transverse degree of freedom so that, denoting by
$H_{\mu\nu\sigma}$
its field strength tensor, one can define a scalar
dualisation $H_{\mu\nu\sigma}\propto f_{a}
\epsilon_{\mu\nu\sigma\rho}\partial^{\rho} a$.\footnote{One 
defines $A^\mu= \frac{1}{6} \epsilon^{\mu\nu\alpha\beta} H_{\nu\alpha\beta}$, then   
from the equation of motion 
$\partial^\mu H_{\mu\alpha\beta}=0$  for the 2-form field strength tensor 
$H^{\mu\alpha\beta} = \frac{1}{2}  \partial_{ [\mu }B_{ \alpha\beta ]}$,
one has  $\partial_\mu A_\nu - \partial_\nu A_\mu =0$, that is  
$A_\mu = f_a \partial_\mu a$ for some scalar $a$ and for some scale $f_a$ 
that normalises canonically the scalar field $a$ in 4D.}
In the effective
theory coming from the heterotic string the pseudoscalar $a$ has the
qualitative features of an axion: since it originates from a gauge field it
has no potential at the renormalisable level, and hence enjoys a shift
symmetry. However, since the shift symmetry is anomalous, the field $a$ 
also
couples to the QCD $G\tilde G$ term and it acquires
non-perturbatively a small mass from QCD instantons (and in general
also a much larger mass from other string-related non-perturbative
effects, in which case, however, it would not serve as a QCD
axion, see below).  Since $B_{\mu\nu}$ can be discussed independently
of the particular string compactification, it is called the Model
Independent (MI) axion. A MI axion exists also in type-I strings, it
arises from a RR two form $C_{2}$ in a way similar to the MI heterotic
string axion, to which is in fact related by  $S$-duality.  
Discrete symmetries arising from string compactification 
might also help in obtaining an accidental, intermediate-scale axion 
(see e.g. Refs.~\cite{Kim:2017age,Kim:2017tdk}).

If the MI pseudoscalar fails to incarnate a useful QCD axion, one can
resort to the $B_{ij}$ components. 
These can be decomposed as $B_{ij} = \sum_n a_n (x_\mu)\, b^n_{ij}(y_k)$ 
with  $x_\mu \in M_4$, $y_k \in V_6$,
$b^n_{ij}$ are harmonic two-form on $V_6$, 
 and  $a_n$ correspond to scalars living in 4D.  
Massless pseudoscalars  arise  from the $b^n_{ij}$ zero modes, 
which correspond   to   $\int_C d\Sigma^{ij} b_{ij}$   
where the integral over two-manifolds $C$ in $V_6$     
does not vanish if $C$ has  non-trivial topology. 
The properties of these states will clearly depend on the compactification 
scheme, and for this reason the corresponding axions are called Model 
Dependent (MD).  The number of zero modes corresponds to the number of 
non-trivial two-cycles in $V_6$, which is the  Hodge number of the compact 
manifold~\cite{Witten:1984}. When considering manifolds
sufficiently complicated to have some chance of resulting in an
effective theory containing the SM, the Hodge number could be as large
as $\mathcal{O}\(100\)$~\cite{Taylor:2012dr}.

Plenty of others axion candidates are found in Type II A and B
superstrings as well as in 11D supergravity.\footnote{This 
abundance of pseudoscalars, a large number of which could remain 
ultralight, has led to concoct  the so called 
 {\it  string axiverse}~\cite{Arvanitaki:2009fg,Acharya:2010zx,  Cicoli:2012sz},  
wherein hundreds of  light axion-like particles populate  a logarithmically 
distributed mass spectrum  possibly down to the  Hubble scale $\sim 10^{-33}\,$eV. 
Constraints on this scenario from cosmological considerations  are discussed 
 e.g.~in Refs.~\cite{  Stott:2017hvl, Visinelli:2018utg}.}
Nevertheless, obtaining
a phenomenologically acceptable QCD axion from these theories is not
an easy task. Superstring axion constructions encounter a certain
number of obstacles, among which
the most serious   are the following:\\

\noindent {1.\ \it String-related contributions to the axion  potential.}\ 
Although at the renormalisable level string axions are
massless because they originate from 10D gauge fields, the shift
symmetry of the effective Lagrangian can be violated by all sort of
string-related effects: worldsheet
instantons~\cite{Dine:1986zy,Wen:1985jz}, brane
instantons~\cite{Becker:1995kb} gravitational instantons, gauge
instantons from other factors of the gauge group. These effects can
easily overwhelm the breaking induced by low energy QCD instantons,
and this would exclude the possibility that these pseudoscalars could
serve as QCD axions.  This is basically the same issue that we have
discussed in \sect{sec:PQquality}: suppose that the axion couples to a
string instanton of action $S$ with a natural mass scale $M\sim \mP$.
Such instantons with their anti-instantons will then generate a
contribution to the axion potential
\begin{equation}
\label{eq:string_instantons}
V_S(\theta) = -2 M^{4} e^{-S} \cos (\theta + \xi)\,,
\end{equation}
where the phase $\xi$ is unrelated with QCD or with complex quark mass
phases.  Upon minimising $V_S$ together with the QCD potential
$V(\theta)\sim - m^{2}_{\pi}f^{2}_{\pi}\cos\theta$ 
it is  found that in order not to spoil the axion solution to the strong CP
problem one needs to require
\begin{equation}
\label{eq:StringCP}
M^{4} e^{-S }  \lesssim 10^{-10} f^{2}_{\pi} m^{2}_{\pi}\,. 
\end{equation}
This  implies $S \gtrsim 200$ which is a  serious constraint on string axion 
models~\cite{Banks:1996ea}.
A few solutions to this problem have been proposed. 
If supersymmetry survives down to 
a scale $\mu$  lower than $M$ then the contribution in the LH side of~\Eqn{eq:string_instantons}
could be reduced as $M^{4} \to M^{2}\mu^{2}$~\cite{Witten:1984,Svrcek:2006hf, Svrcek:2006yi}. 
For  MD axions of  type IIB superstring, 
 Ref.~\cite{Cicoli:2012sz} (see also Ref.~\cite{Ringwald:2012cu}) 
 put forth the possibility that in a large volume scenario, characterised by an exponentially
large volume of the extra dimensions,  a strong suppression 
 $M\sim \mP/ \sqrt{V_6}$, with $V_6\gg 1 $ in units of the string length, 
 can be engineered.
\\

{2.\ \it Cosmological constraints on the axion scale $f_{a}$.}\ As we
have seen in \sect{sec:benchmark}, there is a phenomenologically
preferred window for the axion decay constant $10^{9}\lsim f_{a}/{\rm
  GeV} \lsim 10^{12}$.  The upper bound on $f_{a}$ leads to some
tension with string theory which naturally predicts much larger
values.  For example, for the MI axion of the heterotic string the
value $f_{a} \simeq 1.1 \times 10^{16}\,$GeV was first obtained in
Ref.~\cite{Choi:1985je}, and later confirmed by a more refined
computation in Ref.~\cite{Fox:2004kb}.  Analogous predictions for the
MD heterotic string axion as well as for other types of string models
were derived in Ref.~\cite{Svrcek:2006yi}, which concluded that it is
generally difficult to push $f_{a }$ drastically below $ \sim
10^{16}\,$GeV,  while more natural values are  close to the reduced Planck
mass $(8\pi G_{N})^{-1/2}\sim 2.4\times 10^{18}\,$GeV. This  would
imply a very large overproduction of axion DM. 
Some  possible ways out have been proposed, as for example the early attempt of
Barr within the $E_{8}\times E_{8}$ heterotic string~\cite{Barr:1985hk} and more 
recently Conlon's construction  in Ref.~\cite{Conlon:2006tq},  which are both able 
to arrange for values  of $f_a$ within the canonical window.\\

Despite the promising initial conditions for a string origin of the axion, 
it appears surprisingly  hard to construct explicit string theoretic examples 
with a successful QCD axion candidate.  It seems then fair to say  that, given the   
present status of the art, there are no sufficient reasons to theoretically 
disfavour the QFT axion with respect to their superstring  homologues.

\bigskip

\section{Concluding remarks and \em{desiderata}} 
\label{sec:desiderata} 

Axion physics is  witnessing an exponential growth 
of interest  from the particle physics community. 
According to the inSPIRE database, during the first lustrum of this millennium   
less than 250 papers were published containing the word {\it axion} in the title. 
In the following years,  the  number of publications has steadily grown, 
reaching the stunning number of more than 1,250 papers  published 
in the last quinquennium, 
and there is no hint that this growth is going to diminish in the forthcoming years. 
It is an experimental fact that particle physicists interests and wishes 
do not render a theory correct (as LHC results have recently reconfirmed).
However,   scientific attention gets naturally focussed 
by theories that, besides being highly consistent and remarkably elegant, 
are able to explain some longstanding  theoretical conundrums and are also particularly promising 
as far as regards the possibility of  experimental verification. 
Undoubtedly axion physics  belongs to this class of theories. For this reason we are convinced that 
any effort to develop further crucial theoretical issues in axion physics 
is soundly justified while, at the same time, experimental axion searches 
should be  pursued along all viable pathways. 

A major aim of this Review was that of  motivating  experimental colleagues 
to explore  all the accessible regions in the axion parameter space. We 
have shown how axions can hide in regions that lie well beyond 
the boundaries of canonical windows. Axions can, for example,  embody the whole 
DM also for  mass values much larger or much smaller than what it is generally  assumed. 
Axion searches which exploit their couplings to nucleons and 
electrons are complementary to traditional experimental searches which exploit their  
coupling to photons. As we have stressed,  each type of  axion couplings to SM particles 
could be suppressed well below their benchmark values,
while leaving the other couplings  substantially unaffected.
In particular, it is possible to  decouple the axion  from the photon, 
in which case experimental searches that rely on the couplings to nucleons 
and electrons would play a crucial role. It is also conceivable  that  a strong suppression 
would instead occur for the axion-nucleon and axion-electron interactions. 
This would relax the tightest astrophysical bounds rendering viable 
regions in the $m_a$-$\gag$ parameter space that are generally regarded as excluded. 
Conversely, large enhancements of a single type of coupling are also 
possible. For this reason,  new leading-edge experiments based on novel search techniques, 
  which in many cases are characterised by  limited sensitivities, can still  contribute 
  to circumscribe the landscape of phenomenologically viable axion models.


 While the current blossoming of the  field of axion physics 
is certainly driven by experiments, there exist important calls 
also for the theory community.
We have collected in a  (personal)  list of   {\it desiderata} those theoretical advancements that  
we consider crucial  for further developments  of  axion physics, and that is conceivable that could 
be  accomplished  in the forthcoming  years.\\

{\it Origin of the Peccei-Quinn symmetry.}\ 
Undoubtedly the PQ symmetry can be considered the `standard model' 
for generating  in a QFT  the effective  axion-gluon interaction  
needed  to solve the strong CP problem. 
However,   a `standard model'   for explaining  the origin of the PQ symmetry is still missing. 
The remarkably large number of model realisations  that have been reviewed in this work 
provides the best evidence of this statement.
Candidate models should in first place provide a cogent explanation for  
how  a global PQ symmetry arises, and  should also naturally embed some mechanism 
to preserve it at an exceptionally good level. However, 
they would acquire real credibility if, with no additional or ad hoc theoretical inputs,  
they could automatically shed light on some other unsettled 
issue of the SM, like for example the flavour problem, the origin of neutrino masses, etc. 
We believe that any progress in this direction would represent an important milestone
to  strengthen the plausibility of the axion hypothesis. \\

{\it Topological defects.}\ 
Assessing the axion contribution to the DM arising from axion-related topological defects 
(in post-inflationary PQ breaking scenarios, and within $N_{\rm DW} = 1$ models)  
is, at the time of writing,  a major unsettled question.
Tackling this problem requires extensive numerical simulations 
of the string network evolution and an extrapolation through several orders 
of magnitude between  the PQ scale down to the Hubble scale at the time the axion 
acquires its mass. 
Although this appears as a remarkably difficult task, the importance of converging towards a 
reliable estimate cannot be overstated, since it would drive  the scanning strategy of 
axion DM experiments. \\

{\it Temperature dependence of the axion mass.}\  
Lattice studies of the temperature dependence of the axion mass have made remarkable 
progresses in the last few years. However, it seems that universal consensus 
on the behaviour of the mass function $m_a(T)$ has not yet been reached. 
A final convergence on this issue is highly  desirable in order to refine 
estimates of the misalignment contribution  to  axion DM. We believe that it is realistic to expect that 
 this goal  will be achieved in the not so far future.\\

{\it Axion emissivity from supernova cores.} \ 
Our  understanding of  nuclear matter in the extreme conditions of 
proton-neutron stars in the cooling phase, shortly after a SN explosion,  
is still limited. Also this issue has recently  witnessed important advancements. However,  a detailed and reliable treatment of axion emissivity 
is, at least in part, still lacking.  
Any improvement  in this direction  would be  of utmost importance. 
In fact, although the bound   has been reassessed more than one time since first established, 
constraints from the observed neutrino signal from SN1987A  still provide one of  the strongest  
limits for  axion models, and this in spite of the fact that  the collected data were very sparse.  
A new galactic SN explosion would produce an immensely  richer  data set,  
which  could be optimally interpreted only  if a better understanding of this issue will be available. \\

We do not know if the axion exists. What we do know, is that 
this hypothetical particle has  been able to focus an enormous amount of theoretical 
and experimental efforts in the attempt of understanding its properties and arrive at its  discovery.  
 It might be deemed surprising that so much commitment 
 is devoted to prove what remains, admittedly, essentially a theoretical speculation. 
 We believe that the explanation lies in the beauty and in the elegance 
 of the theoretical construction.  Whether these two paradigms can really provide 
 insight into the way nature works, only future experiments will tell. 
For the moment axion physics, in all its aspects, is healthy and frisky. May it remain so 
until the axion is discovered.

%
\setcounter{secnumdepth}{0}

\section{Acknowledgments}

%
We thank 
Claudio Bonati, 
Andrea Caputo,
Luc Darm\'e, 
Giovanni Grilli di Cortona, 
Axel Lindner, 
David J.~E.~Marsh, 
Pablo Qu\'ilez 
and
Sunny Vagnozzi,
for reviewing parts of the manuscript.  
We acknowledge 
Stefano Bertolini, 
Dmitry Budker, 
Marco Gorghetto, 
Igor G.~Irastorza, 
Derek Jackson Kimball,
Giacomo Landini, 
Federico Mescia, 
Alessandro Mirizzi,
Fabrizio Nesti,
Alessio Notari,  
Michele Redi,
Javier Redondo,
Andreas Ringwald, 
Giuseppe Ruoso, 
Oscar Straniero,
Daniele Teresi, 
Andrea Tesi
and 
Giovanni Villadoro
for discussions and inputs,
and  
Michael Wiescher for permission to reproduce 
the picture in the left panel of \fig{fig_HR_diagram}. 
We thank 
Hai-Yang Cheng,
Raymond Co, 
Jordy de Vries,
Patrick Draper,
Keisuke Harigaya,  
Gioacchino Piazza, 
Mario Reig 
and 
Armen Sedrakian
for comments and feedbacks on the first arXiv version 
of this review. 
%
%
L.D.L., M.G.~and L.V.~acknowledge the INFN Laboratori Nazionali di
Frascati, where this project was first laid down, and  where relevant portions of 
their work were performed,  for hospitality and
partial financial support. 
A large part of the work of L.D.L., M.G.~and
E.N.~was performed at the Aspen Center for Physics, which is
supported by National Science Foundation grant PHY-1607611.  
The participation of L.D.L.~and E.N.~at the Aspen Center for Physics 
was supported in part by a grant from the Simons Foundation. 
L.D.L.~acknowledges hospitality from the 
Theoretical Physics Group at the University of Padova 
during the final stages of this work. 
E.N.~acknowledges hospitality and support from the Munich 
Institute for Astro- and Particle Physics  (MIAPP),  
which is funded by the Deutsche Forschungsgemeinschaft 
(DFG, German Research Foundation) under Germany' s Excellence 
Strategy – EXC-2094 – 390783311, where this project 
was brought to completion. 
L.V.~thanks the kind hospitality of the Leinweber Center for Theoretical Physics, 
the University of Michigan, and Barry University, where part of 
this work was carried out.
%
L.D.L.~is supported by the Marie Sk\l{}odowska-Curie Individual
Fellowship grant AXIONRUSH (GA 840791). 
E.N.~is supported by the
Italian Istituto Nazionale di Fisica Nucleare (INFN) through the
`Theoretical Astroparticle Physics' project TAsP.  
L.V.~acknowledges
support by the Vetenskapsr{\aa}det (Swedish Research Council) through
contract No.~638-2013-8993 and the Oskar Klein Centre for
Cosmoparticle Physics. The work of L.V.~is part of the research
program `The Hidden Universe of Weakly Interacting Particles' with
project number 680.92.18.03 (NWO Vrije Programma), which is (partly)
financed by the Dutch Research Council (NWO). L.V.~acknowledges
support by the Department of Physics and Astronomy, Uppsala
University, by Nordita, KTH Royal Institute of Technology and
Stockholm University, by GRAPPA University of Amsterdam. 

\newpage


\setcounter{secnumdepth}{0}
\section{Appendix: Tables of notations and of acronyms}
\label{sec:notations}

\begin{table}[!ht]
\begin{center}
\begin{tabular}{||c|c|c||}
\hline &&\\[-10pt]
\hline \\ [-7pt]
Symbol &  Meaning & Equation \\ [3pt]
\hline 
$\phantom{^\big|}$$\theta$  & CP violating QCD angle & \ref{eq:QCDLag} \\
$\epsilon^{\mu\nu\rho\sigma}$ & Levi-Civita symbol ($\epsilon^{0123} = -1$) & \ref{foot:epsilon} \\
$N$ & $U(1)_{\rm PQ}$-$SU(3)_{\rm QCD}$ anomaly coefficient &        \ref{eq:dJPQAnomal} \\
$E$ & $U(1)_{\rm PQ}$-$U(1)_{\rm QED}$ anomaly coefficient &       \ref{eq:dJPQAnomal} \\
$\Phi$ &  Axion multiplet $\Phi(x) = \frac{1}{\sqrt{2}} (\va + \varrho_a(x))\, e^{i  \frac{a(x)}{\va}}$ & \ref{eq:PhidecKSVZ} \\
$\varrho_a$ & Axion multiplet radial mode   & \ref{eq:PhidecKSVZ} \\
$a$ & Axion field: axion multiplet orbital mode   & \ref{eq:PhidecKSVZ} \\
$\va$  & Peccei-Quinn symmetry breaking VEV $\va =\sqrt{2}\langle \Phi \rangle $ & \ref{eq:PhidecKSVZ} \\
$\fa$ & Axion decay constant $\fa = \va/{(2N)}$   & \ref{eq:favsvadef} \\ 
$\NDW$ & Domain wall number $\NDW = 2N$   & \ref{eq:DWdef}\\ 
$m_a$ & Axion mass & \ref{eq:axionmassfa} \\
$C_{a\gamma} $& Axion-photon coupling & \ref{eq:Cagamma} \\
$C_{ap} $& Axion-proton coupling & \ref{eq:Cap} \\
$C_{an} $& Axion-neutron coupling & \ref{eq:Can} \\
$ C_{ae}$& Axion-electron coupling & \ref{eq:Cae} \\
$ C_{a\pi}$& Axion-pion coupling & \ref{eq:Capi} \\
$ C_{an\gamma}$& Axion coupling to the neutron EDM & \ref{eq:Cangamma} \\
$g_{a\gamma}$ & Dimensional  axion-photon coupling (GeV$^{-1}$)& \ref{eq:gagammagaf} \\
$ g_{af}$ & Rescaled axion-fermion coupling  & \ref{eq:gagammagaf} \\
$ g_{d}$ & Rescaled axion coupling to the neutron EDM & \ref{eq:gagammagaf} \\
$ g^S_{aN}$ & CP-violating scalar axion-nucleon coupling  & \ref{eq:fromthetaefftogaN} \\
$\chi$ & Topological susceptibility & \ref{eq:defK} \\
$R$ & Cosmological scale factor   & \ref{eq:line_element} \\ 
$H$ & Hubble parameter $H=\dot R/R$   & \ref{eq:friedmann1} \\ 
$\mP$ & Planck mass & \ref{eq:friedmann1}\\
$\rho,\,s $ & Energy and entropy density  ($(\rho,s)\simeq (\rho,s)_{\rm rad}$ in rad.~domination) & \ref{eq:rho_and_s} \\
$g_*\,,g_S $ & Effective energy and entropy degrees of freedom  & \ref{eq:geff_gSeff} \\
$m_a(T)$ & Axion mass at  temperature $T$ & \ref{eq:chiT} \\
$t_{\rm osc}$ & Time of the onset of axion oscillations & \ref{eq:condition_oscillations}  \\
$\theta_{\rm PQ},\,\dot\theta_{\rm PQ}$ & Misalignment angle, axion velocity at the PQ scale & \ref{eq:initial_condition_theta} \\
$\theta_i,\,\dot\theta_i$ & Misalignment angle, axion  velocity at $t_{\rm osc}$ & \ref{eq:theta-i} \\
$\Omega_a^{\rm mis} $ & Fractional axion energy density from misalignment & \ref{eq:standarddensity} \\
$\varepsilon$ & Energy-loss rate per unit mass & \ref{eq:Primakoff_approx} \\
$\mathcal{R}$ &  Ratio of number of stars in Horizontal and Red Giant branches & \ref{eq:R-parameter}\\
$g_{e 12}$, $g_{\gamma 10}$  & Respectively $g_{ae}/10^{-12}$ and $g_{a\gamma}/10^{-10}\,{\rm GeV^{-1}}$  & \ref{eq:ge12gg10}\\
[-5pt]
 &&\\[-2pt]
\hline
\hline
\end{tabular} 
\label{tab:notations}
\caption{Notations introduced in the text, their meaning, and
equations where they are defined. 
}
\end{center}
\end{table}

\begin{table}[!ht]
\begin{center}
\begin{tabular}{||c|c||}
\hline &\\[-10pt]
\hline \\ [-7pt]
Acronym &  Meaning 
\\ [3pt]
\hline 
ABC & Atomic recombination and de-excitation, Bremsstrahlung, Compton \\ 
ALP & Axion-like Particle \\
BBN & Big-Bang Nucleosynthesis \\
(C)DM & (Cold) Dark Matter \\
CL & Confidence Level \\
CMB & Cosmic Microwave Background \\ 
CMD & Color Magnitude Diagram \\ 
CP & Charge Parity \\
$\chi$PT & Chiral Perturbation Theory \\
DFSZ & Dine-Fischler-Srednicki-Zhitnitsky \\
DIGA & Dilute Instanton Gas Approximation \\
DW & Domain Wall \\ 
EFT & Effective Field Theory \\
FCNC & Flavour Changing Neutral Currents \\ 
FLRW & Friedmann-Lema\^itre-Robertson-Walker \\ 
FV & Flavour Violating \\
GUT & Grand Unified Theory \\ 
GW & Gravitational Waves  \\ 
HB & Horizontal Branch \\ 
HDM & Hot Dark Matter \\
HR & Hertzsprung-Russell \\
IR & Infrared \\ 
KSVZ & Kim-Shifman-Vainshtein-Zakharov   \\
LP & Landau Pole \\ 
LSW & Light Shining Through a Wall \\ 
MI(D) & Model Independent (Dependent) \\ 
(n)EDM & (neutron) Electric Dipole Moment \\
(p)NGB & (pseudo) Nambu-Goldstone Boson \\
NMR & Nuclear Magnetic Resonance \\ 
(N)LO & (Next-to) Leading Order \\
NS & Neutron Star \\ 
OPE & One Pion Exchange \\ 
PQ & Peccei Quinn \\ 
QCD & Quantum Chromo Dynamics  \\
QED & Quantum Electro Dynamics \\ 
QFT & Quantum Field Theory \\
RGB & Red Giant Branch \\
SM & Standard Model \\
SMASH & Standard Model--Axion--Seesaw--Higgs inflation portal \\ 
SN & Supernova \\ 
UV & Ultraviolet \\ 
VEV & Vacuum Expectation Value \\ 
WD(LF)& White Dwarf (Luminosity Function) \\ 
WIMP & Weakly Interacting Massive Particle \\
WW & Weinberg-Wilczek \\
[-5pt]
& \\[-2pt]
\hline
\hline
\end{tabular} 
\label{tab:acronyms}
\caption{Acronyms used in the text and their meaning. 
}
\end{center}
\end{table}

\begin{table}[!ht]
\begin{center}
\renewcommand{\arraystretch}{0.95}
\begin{tabular}{||c|c||}
\hline &\\[-10pt]
\hline \\ [-7pt]
Experiment & Meaning 
\\ [3pt]
\hline 
\phantom{$^{|}$}ABRACADABRA \cite{Kahn:2016aff} & A Broadband/Resonant Approach to Cosmic Axion Detection \\ 
ADMX \cite{Asztalos:2003px} & Axion Dark Matter Experiment \\ 
ALPS (II) \cite{Bahre:2013ywa} & Any Light Particle Search (II) \\ 
ARGUS \cite{Albrecht:1995ht} & A Russian-German-United States-Swedish Collaboration \\
ARIADNE \cite{Arvanitaki:2014dfa} & Axion Resonant InterAction Detection Experiment \\ 
ATLAS \cite{Aad:2008zzm} & A Toroidal LHC ApparatuS \\
AXIOMA \cite{Santamaria_2015} & AXIOn dark MAtter detection \\
BEAST \cite{McAllister:2018ndu} & Broadband Electric Axion Sensing Technique \\
BELLE II \cite{Abe:2010gxa} &  \\
CAPP-8TB \cite{Lee:2019mfy} & Center for Axion and Precision Physics \\
CASPEr \cite{Budker:2013hfa} & Cosmic Axion Spin Precession Experiment \\
CAST \cite{Anastassopoulos:2017ftl} & CERN Axion Solar Telescope \\ 
CLEO \cite{Ammar:2001gi} &  \\ 
CMS \cite{Chatrchyan:2008aa} & Compact Muon Solenoid \\
Crystal Box \cite{Bolton:1988af} &  \\ 
CULTASK \cite{Chung:2016ysi} & CAPP’s Ultra Low Temperature Axion Search in Korea \\ 
DARWIN \cite{Aalbers:2016jon} & Dark Matter WIMP Search With Liquid Xenon \\
E949+E787 \cite{Adler:2008zza} &  \\ 
Fermi LAT \cite{Meyer:2016wrm} & Fermi Large Area Telescope \\
GAIA \cite{Gaia} & Global Astrometric Interferometer for Astrophysics \\
GNOME \cite{Pustelny:2013rza} & Global Network of Optical Magnetometers for Exotic Physics \\
HAYSTACK \cite{Brubaker:2016ktl} & Haloscope at Yale Sensitive to Axion CDM \\
HST \cite{Freedman:2000cf}  &   Hubble Space Telescope   \\
IAXO \cite{Armengaud:2014gea} & International Axion Observatory \\ 
KLASH \cite{Alesini:2019nzq} & KLoe magnet for Axion SearcH \\ 
LHC \cite{Evans:2008zzb} & Large Hadron Collider \\ 
LIGO \cite{Barish:1999vh} &  Laser Interferometer Gravitational-Wave Observatory \\ 
LSST \cite{0912.0201} & Large Synoptic Survey Telescope \\ 
LUX \citep{Akerib:2017uem} & Large Underground Xenon \\ 
LZ \cite{Akerib:2017uem} & LUX-ZEPLIN \\
MADMAX \cite{TheMADMAXWorkingGroup:2016hpc} & MAgnetized Disc and Mirror Axion eXperiment \\ 
ORGAN \cite{McAllister:2017lkb} & Oscillating Resonant Group AxioN \\ 
ORPHEUS \cite{Rybka:2014cya} &  \\ 
OSQAR \cite{Ballou:2015cka} & 
Optical Search for QED vacuum birefringence, Axions, and photon Regeneration \\ 
PandaX \citep{Fu:2017lfc} & Particle aND Astrophysical Xenon experiment \\
PVLAS \cite{DellaValle:2015xxa} & Polarisation of Vacuum with LASer \\ 
QUAX \cite{Barbieri:2016vwg} & QUaerere AXion \\
RADES \cite{Melcon:2018dba} & Relic Axion Detector Exploratory Setup \\ 
SCSS \cite{Rowell:2011wp} & SuperCOSMOS Sky Survey \\ 
SDSS \cite{DeGennaro:2007yw} & Sloan Digital Sky Survey \\ 
SKA \cite{Bacon:2018dui}  &  Square Kilometer Array \\
STAX \cite{Capparelli:2015mxa}
& Sub-THz-AXion \\ 
TASTE \cite{Anastassopoulos:2017kag} & Troitsk Axion Solar Telescope Experiment \\ 
TOORAD \cite{Marsh:2018dlj} & Topological Resonant Axion Detection \\ 
TRIUMF \cite{Jodidio:1986mz} & TRI University Meson Facility \\
VIRGO \cite{Accadia:2012zzb} &  \\
VMB@CERN \cite{LOI2018} & Vacuum Magnetic Birefringence experiment at CERN \\
WMAP \cite{Bennett:2003ba}  &  Wilkinson Microwave Anisotropy Probe \\
XENON100 \citep{Aprile:2014eoa} & 100 kg liquid Xenon target \\  [-8pt]
& \\[-2pt]
\hline
\hline
\end{tabular} 
\label{tab:experiments}
\caption{Experiment acronyms and their meaning. 
Blank entries correspond to experiment names that are not acronyms.}
\end{center}
\end{table}

\clearpage

\newpage


 \section{References}

\bibliography{bib0_Introduction,%
bib1_StrongCP,%
bib2_PQSolution,%
bib3_AxionSearches,%
bib4_AxionLandscape,%
bib5_HeavyAxions,%
bib6_BeyondQCD,%
bib7_Conclusions,%
bib9_Tables}{}


 \bibliographystyle{elsarticle-num}


\end{document}